\documentclass[manuscript]{aastex}

\usepackage{natbib}
\usepackage{lscape}
\usepackage{color}

\shorttitle{Organic species in Infrared Dark Clouds}
\shortauthors{Vasyunina et al.}

\begin{document}


\title{Organic species in Infrared Dark Clouds\thanks{Based on observations carried out with the IRAM 30m Telescope. IRAM is supported by INSU/CNRS (France), MPG (Germany) and IGN (Spain). 
This publication is based on data acquired with the Atacama Pathfinder Experiment (APEX). 
APEX is a collaboration between the Max-Planck-Institut f\"ur Radioastronomie, 
the European Southern Observatory, and the Onsala Space Observatory.
The 22-m Mopra antenna is part of the Australia Telescope,
which is funded by the Commonwealth of Australia for operations as a National
Facility managed by CSIRO. The University of New South Wales Digital Filter
Bank used for the observations with the Mopra Telescope was provided with
support from the Australian Research Council.}}


\author{T. Vasyunina\altaffilmark{*}, A. I. Vasyunin, Eric Herbst\altaffilmark{**}}
\affil{Department of Chemistry, University of Virginia,
              Charlottesville, VA 22904 USA}


\author{Hendrik Linz}
\affil{Max Planck Institute for Astronomy (MPIA), K\"onigstuhl 17, D-69117 Heidelberg, Germany}
\email{linz@mpia.de}

\author{Maxim Voronkov}
\affil{Australia Telescope National Facility, CSIRO Astronomy and Space Science, 
PO Box 76, Epping, NSW 1710, Australia}
\email{maxim.voronkov@csiro.au}

\author{Tui Britton}
\affil{Department of Physics and Astronomy, Macquarie University, NSW 2109, Australia}
\email{Tui.Britton@csiro.au}

\author{Igor Zinchenko}
\affil{Institute of Applied Physics of the Russian Academy of Sciences, Ulyanova 46, 603950 Nizhny Novgorod, Russia}
\email{zin@appl.sci-nnov.ru}

\author{Frederic Schuller}
\affil{European Southern Observatory, Alonso de Cordova 3107, Casilla 19001, Santiago 19, Chile}
\email{fschull@eso.org}

\altaffiltext{*}{Current address: Max Planck Institute for Radioastronomy (MPIfR),  Auf dem H\"ugel 69, D-53121 Bonn, Germany}
\altaffiltext{**}{Also: Departments of Astronomy and Physics,
University of Virginia, Charlottesville, VA 22904 USA}




\begin{abstract}

It is currently assumed that infrared dark clouds (IRDCs)  represent
the earliest evolutionary stages of high-mass stars ($>$ 8 M$_{\odot}$).
Submillimeter and millimeter-wave studies performed over the past 15 years show
that IRDCs possess a broad variety of properties, and hence a wide range of
problems and questions that can be tackled. In this paper, we report an investigation of
the molecular composition and chemical processes in two groups of IRDCs.  
Using the Mopra, APEX, and IRAM radio telescopes over the last four years, we have
collected molecular line data for CO,   H$_2$CO, HNCO, CH$_3$CCH,
CH$_3$OH, CH$_3$CHO, CH$_3$OCHO, and CH$_3$OCH$_3$.   For
all of these species 
we estimated molecular abundances.  We then undertook
chemical modeling studies,
concentrating on the source IRDC028.34+0.06, and compared observed and modeled abundances. This
comparison showed that to reproduce observed abundances of 
complex organic molecules (COMs), a 0-D gas-grain model with constant physical conditions is not sufficient. 
We achieved greater success with the use of a warm-up model, in which warm-up from 10 K to 30 K occurs following a cold phase.


\end{abstract}


\keywords{ISM: clouds, ISM: molecules, Radio lines: ISM, Stars: Formation}



\section{Introduction}


Infrared dark clouds (IRDCs) were  
first identified by the 
\emph{Infrared Space Observatory \/} \citep[\emph{ISO\/};][]{1996A&A...315L.165P} and the
\emph{Midcourse Space Experiment\/} \citep[\emph{MSX\/};][]{1998ApJ...494L.199E}.
Follow-up studies at millimeter, submillimeter, and infrared wavelengths 
showed that these objects are cold, dense, and have masses from  hundreds to 
thousands of solar masses
\citep[e.g.,][]{2000ApJ...543L.157C, 2006ApJ...641..389R,2009A&A...505..405P, 2009A&A...499..149V, 
2009ApJ...696..484B,2011ApJ...736..163R, 2012A&A...547A..49R}.
Thus, some IRDCs  have the potential to harbor not only clusters of low- and 
intermediate mass stars but also high-mass stars with M$>$8M$_{\odot}$ \citep{2010ApJ...723L...7K}.

Many aspects of IRDCs have now been discussed in the literature.
Matters of special attention include the chemical composition
and chemical processes in these clouds.
The first  molecular line study of IRDCs had raised a question concerning
the difference between the chemical composition of low-mass pre-stellar cores and IRDCs  \citep{1998ApJ...508..721C}.
In that survey H$_2$CO was detected in 10 clouds, thus confirming the presence of dense gas. 
Using LVG modeling, \citet{1998ApJ...508..721C} estimated the fractional H$_2$CO abundance to be about 10$^{-10}$. This is a factor 
of 50 lower in comparison with low-mass prestellar cores \citep{1992IAUS..150..171O}. 
The results of this study  tentatively indicate a different chemistry in high- and low-mass star forming regions.
However,  to distinguish differences and similarities of the chemical composition, we require observational data for 
more than one molecular species. 

Multi-line surveys towards IRDCs have started to appear recently  
\citep[e.g.][]{2008ApJ...678.1049S, 2010ApJ...714.1658S, 2011A&A...527A..88V, 2012ApJ...756...60S, 2013arXiv1309.3570H}.
\citet{2008ApJ...678.1049S} observed the N$_2$H$^+$(1-0), HC$_3$N(5-4),
CCS(4$_3$-3$_2$), NH$_3$(1,1), (2,2), (3,3), and CH$_3$OH(7-6) lines toward massive clumps
associated with IRDCs in order to determine their evolutionary status and to study the chemical
conditions within them. Analyzing the CCS/N$_2$H$^+$ abundance ratio, 
 which is expected to decrease with time,
they concluded
that infrared dark clouds are chemically more evolved than low-mass pre-stellar cores.
In a comprehensive molecular line study,  \citet{2011A&A...527A..88V}  
performed observations toward 15 IRDCs in the frequency 
range between 86 and 93 GHz using the Australian 22-m Mopra radio telescope. In total, the 13 molecular species 
N$_2$H$^+$, $^{13}$CS, CH$_3$CN, HC$_3$N, HNC, HCO$^+$, HCN, HNCO, C$_2$H, SiO, H$^{13}$CO$^+$,
H$^{13}$CN, and CH$_3$CCH were searched for in all targets. 
Where possible, \citet{2011A&A...527A..88V} determined molecular
abundances and made a comparison with previously obtained abundances for low-mass pre-stellar cores and high-mass protostellar objects
(HMPOs). 

HMPOs, in comparison with IRDCs, represent a more evolved stage of high-mass star formation \citep{2007prpl.conf..165B}
where the central protostar has been already formed, but continues accreting material.
HMPOs have higher temperatures and appear in emission at mid- and far-IR wavelengths
\citep{2002ApJ...566..931S,2007ApJ...668..348B}. 
In their molecular line study,  \citet{2011A&A...527A..88V} discovered a tendency for IRDCs to have molecular abundances 
similar to low-mass pre-stellar cores rather than to HMPO
abundances on the linear spatial scales probed by the single-dish observations.
The study also showed that a direct comparison between observed abundances depends on many factors and uncertainties.
Even if data are obtained with the same instrument, and molecular abundances obtained using similar
assumptions, the results would not guarantee the emergence of a clear trend among different source types. 
As a consequence, to seriously compare the chemistry in HMPOs, low-mass pre-stellar cores, and IRDCs, it is necessary to employ chemical modeling.

Molecular line information collected  during the last decade aided the 
construction of the first, simple, IRDC chemical models
\citep{2009ApJ...705..123G, 2012ApJ...747..140S}.
Still, the results of these models could
only be compared with the abundances of a few observed molecules.
\citet{2012ApJ...751..105V}  presented the first chemical model for IRDCs
that uses at least 7 species to constrain the modeling abundances. 
For that study,  a simple 0-D model with a gas-grain reaction network and constant homogeneous physical conditions was used.
The modeling was performed for two IRDCs: IRDC013.90-1, which has a 
lower 
ammonia kinetic temperature of 13~K,
and IRDC321.73-1, 
which possesses a slightly elevated ammonia kinetic temperature of 25 K. 
This study showed that observed molecular abundances based on single-dish observations
can be reproduced even with a model that contains only a simple physical structure. 
Moreover, we were able to distinguish a special temperature range between 20 and 30 K,
where grain surface reactions have a significant impact on the gas-phase chemistry. 
Clouds that have these slightly elevated temperatures can indeed feature a different chemistry 
in comparison not only with HMPOs, but also with low-mass pre-stellar cores.
The success of the 0-D model in reproducing single-dish observations of 7 species
confirms that it is sufficient to use simple physical models for explaining these types of data.
It can be insufficient, however, if we add more complex species.

In the current paper, we concentrate on the analysis of molecular line observations towards IRDCs, and in particular on 
the behavior of organic species at early stages of high-mass star formation.  Organic species with 6 or more atoms are commonly referred to as COMs (complex organic molecules), and we use this acronym. 
Surveys by \citet{2008A&A...482..179B,2009A&A...499..215B, 2001ApJ...560..792I} and \citet{2002ApJ...576..264R,2003ApJ...590..314R}
reveal the presence of COMs in hot cores such as SgrB2, Orion KL, and NGC6334f,  which are high-mass star forming regions 
that are much more evolved than IRDCs. 
Also, COMs  were detected
by \citet{2008ApJ...689.1141R, 2011ApJ...741..120R} in  hot cores associated with IRDCs,
and  in the surveys by \citet{2012ApJS..202....1H} in the vicinity of so-called
Green Fuzzies, or Extended Green Objects (EGOs),
which are special regions in IRDCs characterized by an excess of extended 4.5 $\mu$m emission  in {\it Spitzer} IRAC imaging,  
due to H$_2$ shock-excited emission associated with molecular outflows  \citep[e.g.][]{2008AJ....136.2391C, 2009ApJS..181..360C}. 

To explore the possibility of detecting COMs and smaller carbon-containing molecules at earlier stages of high-mass star formation,
we collected molecular line data for two groups of IRDCs  - northern and southern - 
in the frequency ranges 71-115 GHz, 152-155 GHz and 213-245 GHz
with the Mopra, IRAM, and APEX radio telescopes.
In this survey, five  COMs were detected along with  three smaller species. 
In southern sources, we detected CO (carbon monoxide), HNCO (isocyanic acid), H$_2$CO (formaldehyde), 
CH$_3$CCH (methyl acetylene), and CH$_3$OH (methanol), while
 in addition to these five species, 
the high sensitivity of IRAM receivers allowed us to detect 
 CH$_3$CHO (acetaldehyde),  CH$_3$OCH$_3$ (dimethyl ether), and in one case CH$_3$OCHO (methyl formate) in northern sources.
Molecular abundances for 
CO, H$_2$CO, HNCO, CH$_3$CCH and CH$_3$OH
have already been reported for IRDCs in, e.g.,  \citet{2006ApJS..166..567R, 2008ApJ...678.1049S} and \citet{2012A&A...540A.104M}.
Estimated abundances of CH$_3$CHO, CH$_3$OCHO and CH$_3$OCH$_3$  in IRDCs
are reported   for the first time.



The paper is structured as follows.
The source selection and  observations with the Mopra, APEX and IRAM telescopes
are described in Section~\ref{observations}.
The observed transitions  
and estimation of the molecular abundances are presented in 
Section~\ref{results}.
In Section~\ref{modeling}, we compare the results of the 
previous 0-D pseudo time dependent chemical model 
and a more realistic physical model that includes a warm-up phase with observational values.
We discuss the obtained results in Section~\ref{discussion}
and present a summary and our conclusions in Section~\ref{summary}.

\section{Observations}
\label{observations}

\subsection{Source description}
\label{source_description}

For the present study, two sets of IRDCs were used, as listed in 
Tables~\ref{table:main_Mopra} and \ref{table:main_IRAM}.
All sources were classified according to the criteria described in  \citet{2011A&A...527A..88V}.
Sources that show no emission at 8 and 24 $\mu$m were classified as ``quiescent",
those that appear in emission at both 8 and 24 $\mu$m were identified as ``active".
The rest of the clouds, where only 24 $\mu$m emission appears, were attributed to a ``middle" stage.
In addition to this classification, 
as an indicator of star formation activity
 we mark the presence of SiO emission
 based on \citet{2011A&A...527A..88V} and Linz et al. (in prep.), 
 for every source.

A set of southern sources, shown in Table~\ref{table:main_Mopra},  was selected from \citet{2011A&A...527A..88V}.
There we chose 25 points with $> 3 \sigma$ line detections of the four main (1-0)
90-GHz lines:
N$_2$H$^+$, HNC, HCO$^+$, and HCN. For these southern sources we have estimates
of H$_2$ column density from 1.2 mm Simba/SEST data, temperature estimates 
from ammonia Parkes observations and kinematic distances from N$_2$H$^+$ (1-0) 
Mopra data \citep{2011A&A...527A..88V}.
Here we present new observations of these sources performed with  the Mopra and APEX telescopes.

A set of northern sources, shown in Table~\ref{table:main_IRAM},  
was taken from previous studies
and includes
classical clouds from the first studies about molecular line observations in IRDCs 
\citep[e.g.][]{1998ApJ...508..721C,2006A&A...450..569P,2006ApJ...639..227S} 
as well as  the recent work by \citet{2012A&A...547A..49R}. 
For these objects,  H$_2$ column densities were estimated based on ATLASGAL 870 $\mu$m
maps \citep{2009A&A...504..415S};  
 the details of the calculation are provided in Section~\ref{abundance}.
Temperature values were taken from the literature.
For those objects  where two temperatures are available (for instance, the gas temperature as estimated from ammonia 
observations, and the dust temperature), 
we report both values.
Ammonia kinetic temperatures were taken from 
\citet{2006A&A...450..569P, 2012A&A...544A.146W}, and \citet{2008ApJ...678.1049S}, while
dust temperatures were taken from \citet{2012ApJ...756...60S}.
In the current study for northern sources, we present results from the IRAM and APEX observations.

\begin{figure}
\includegraphics[width=5.3cm]{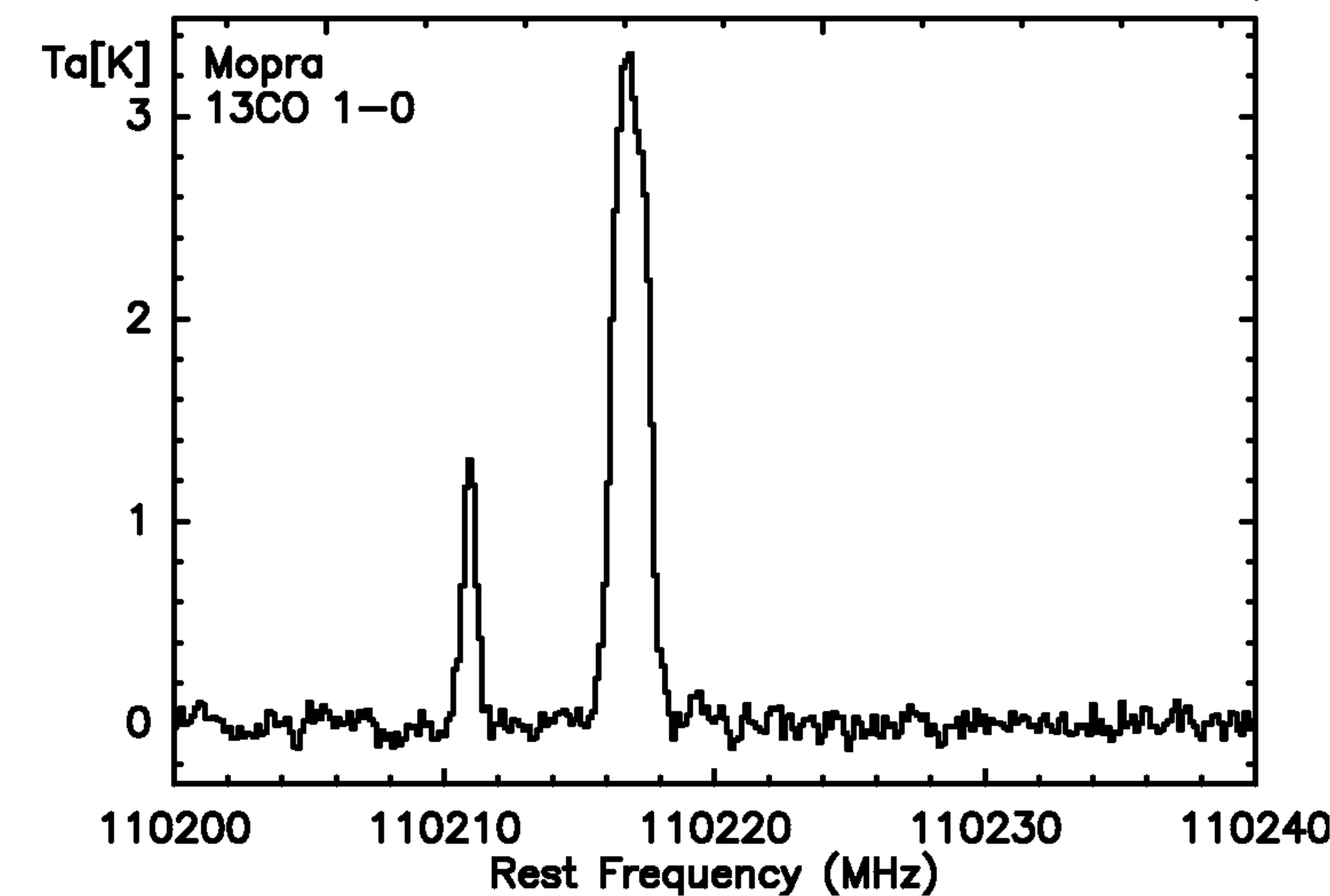}
\includegraphics[width=5.3cm]{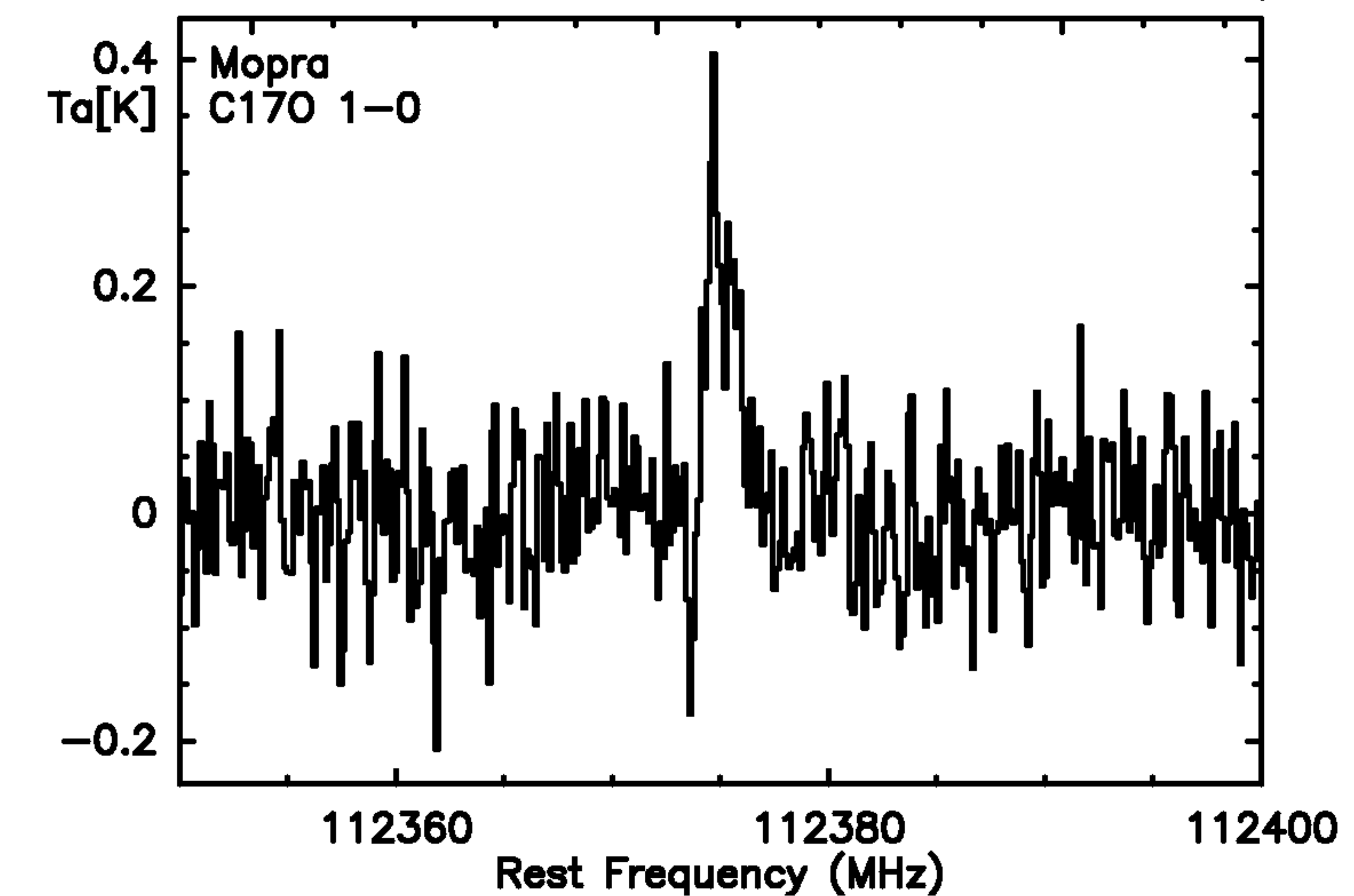}
\includegraphics[width=5.3cm]{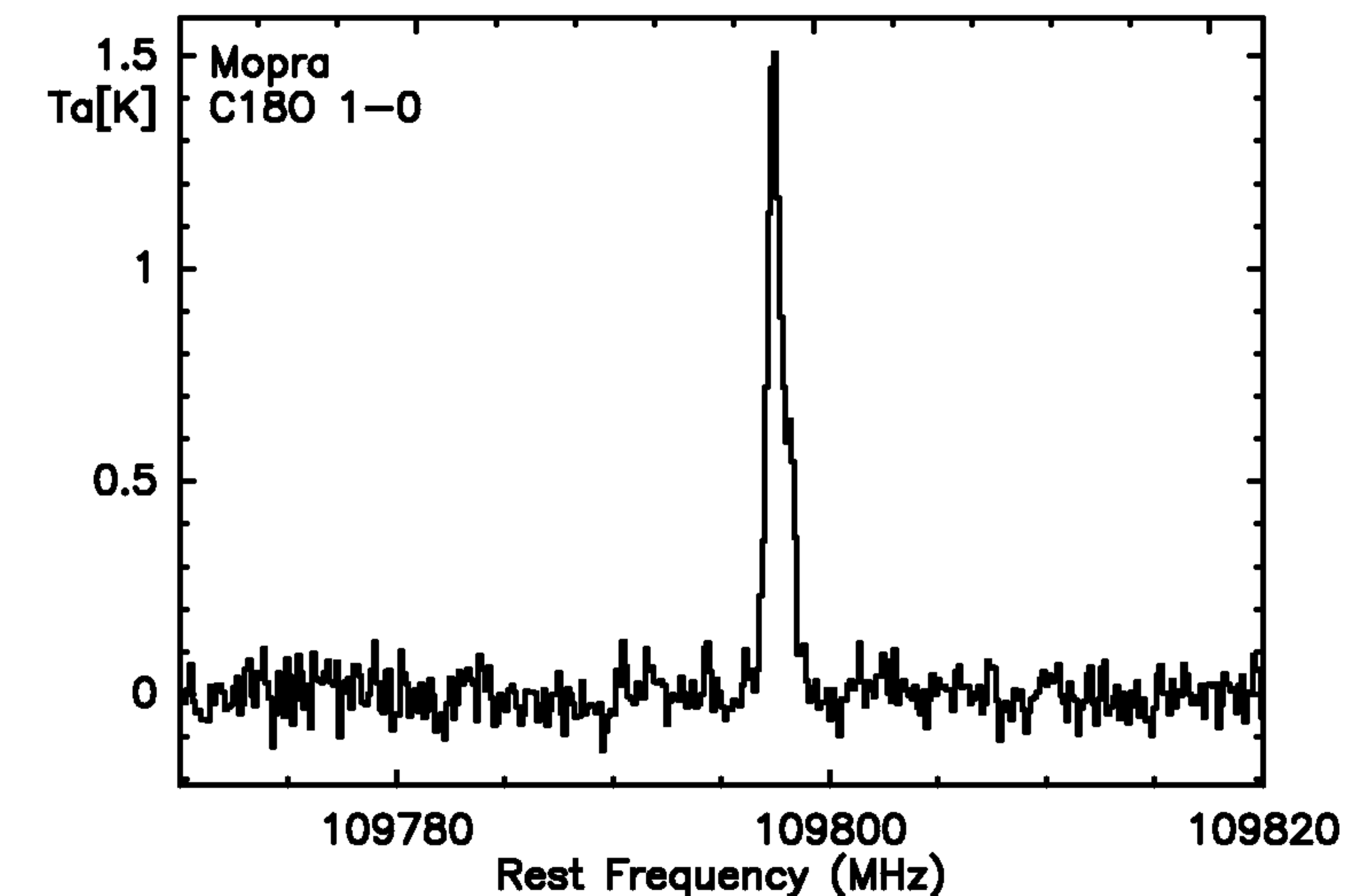}

\includegraphics[width=5.3cm]{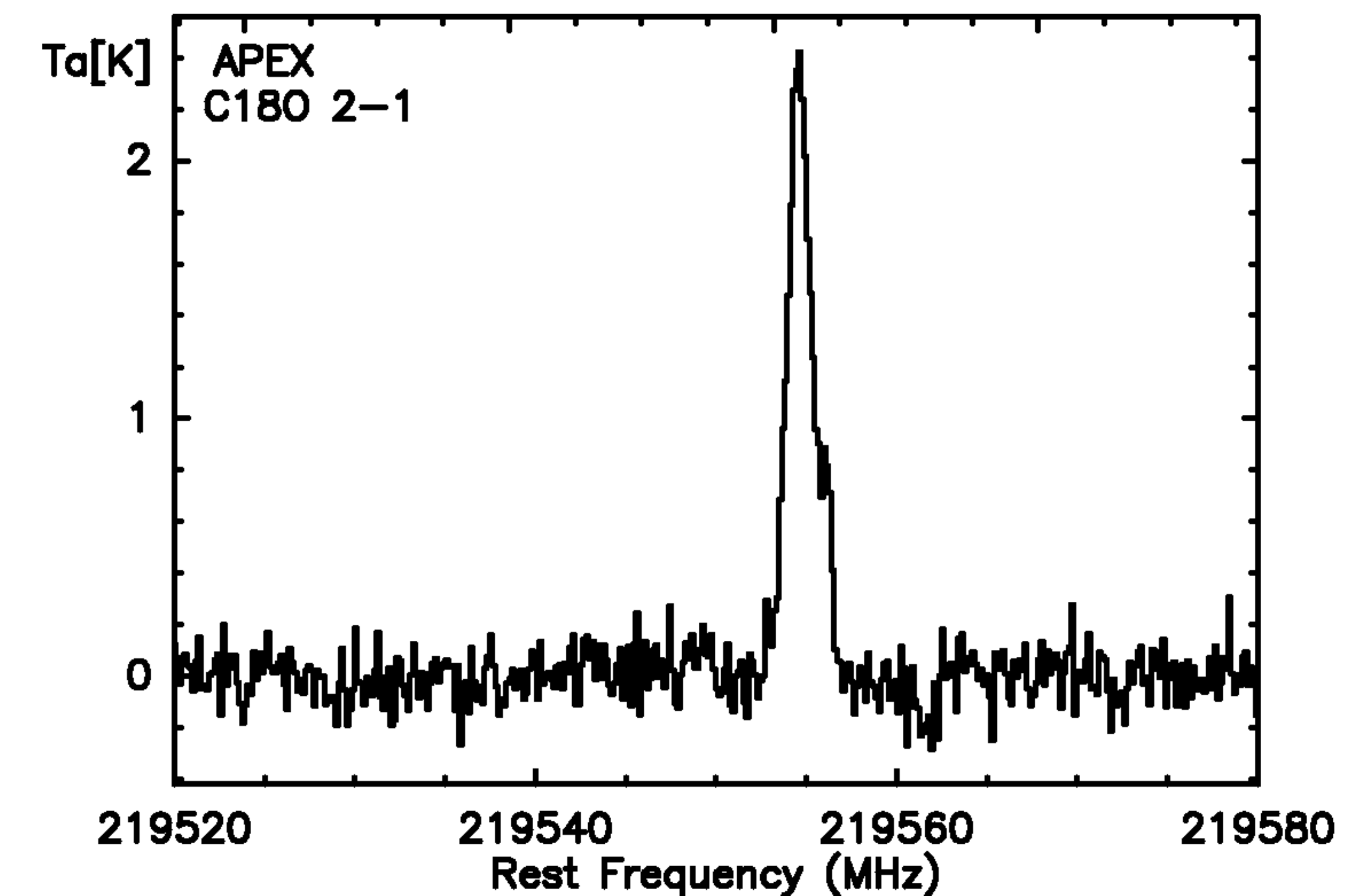}
\includegraphics[width=5.3cm]{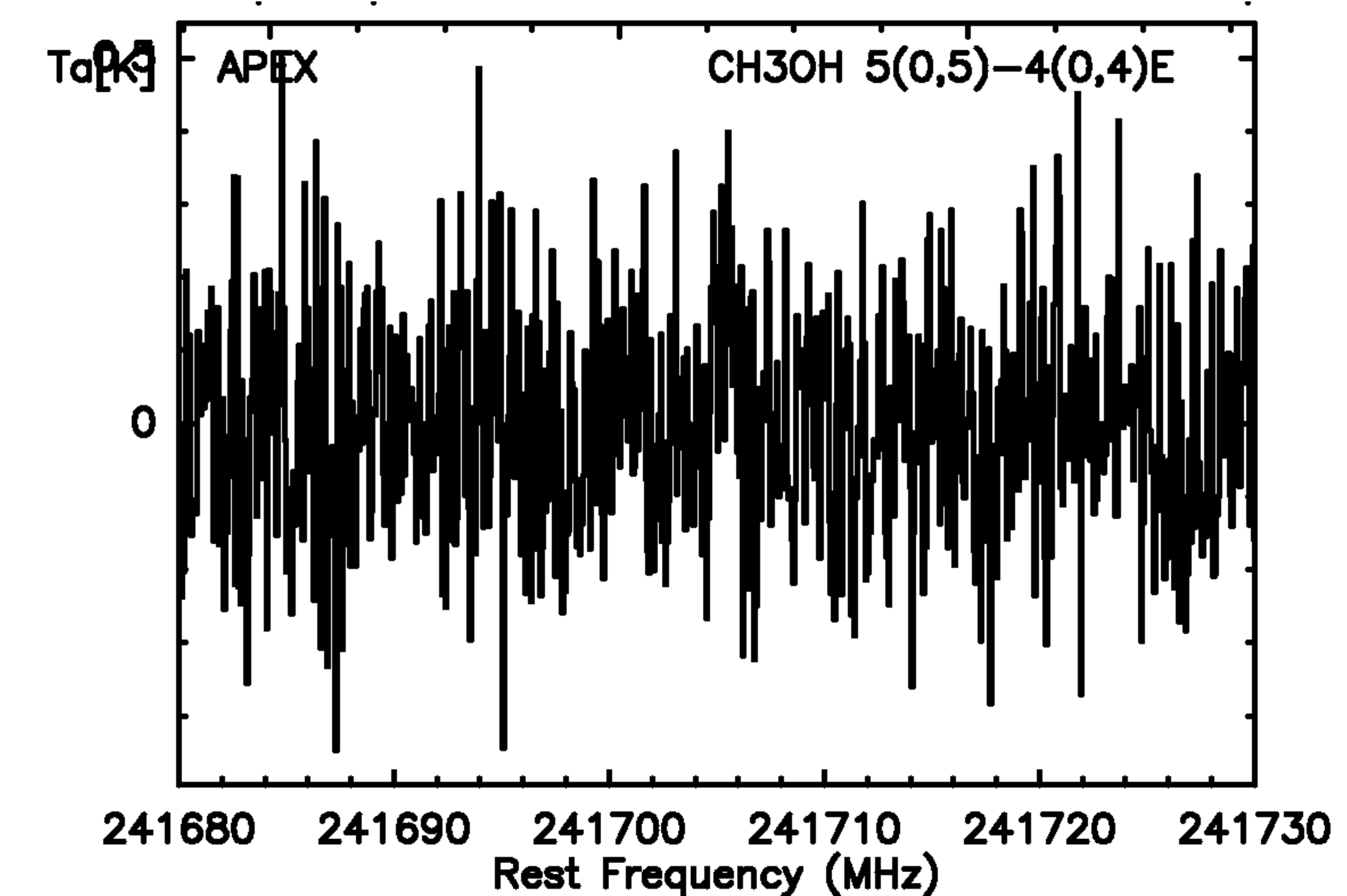}
\includegraphics[width=5.3cm]{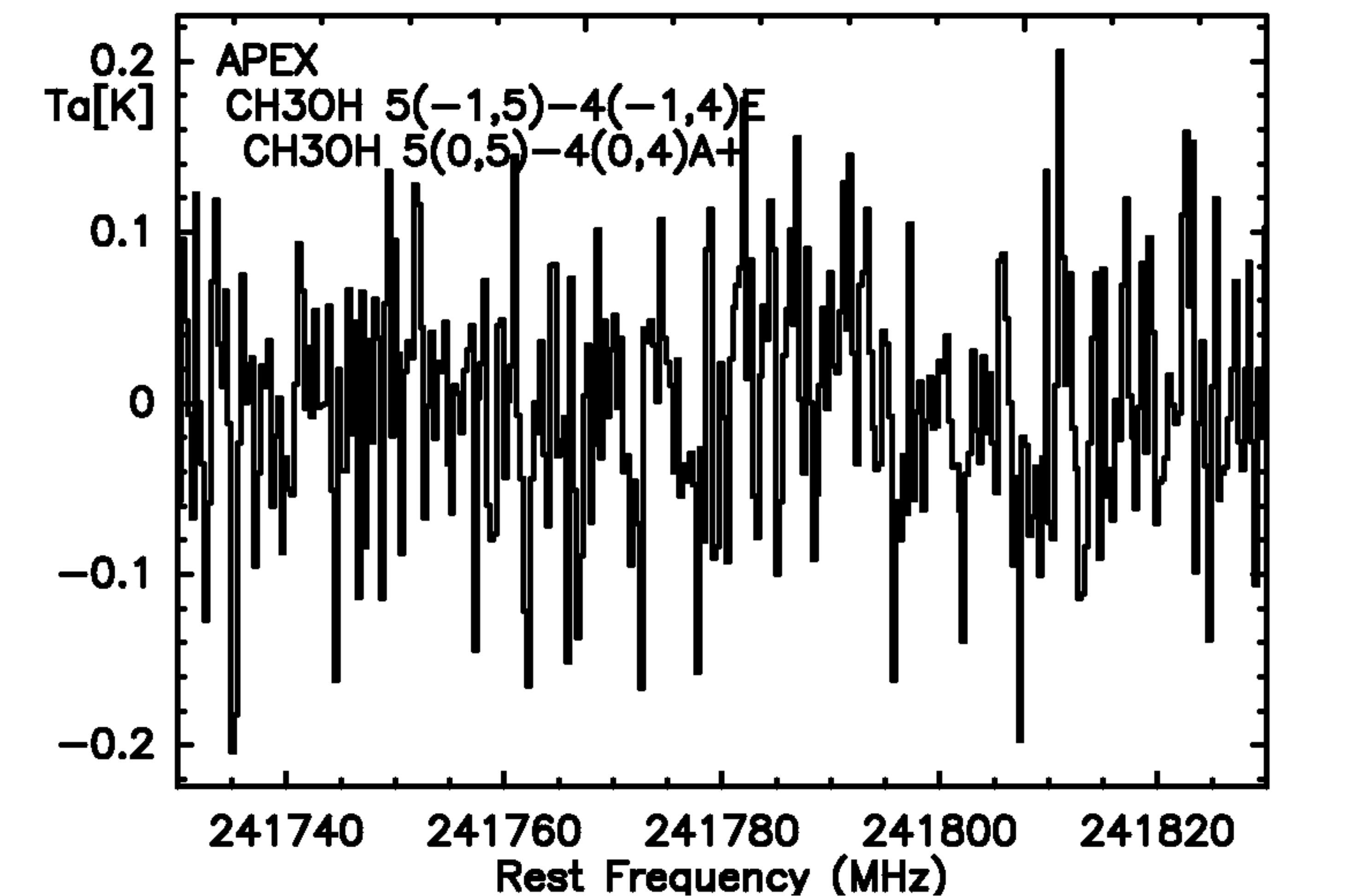}

\includegraphics[width=5.3cm]{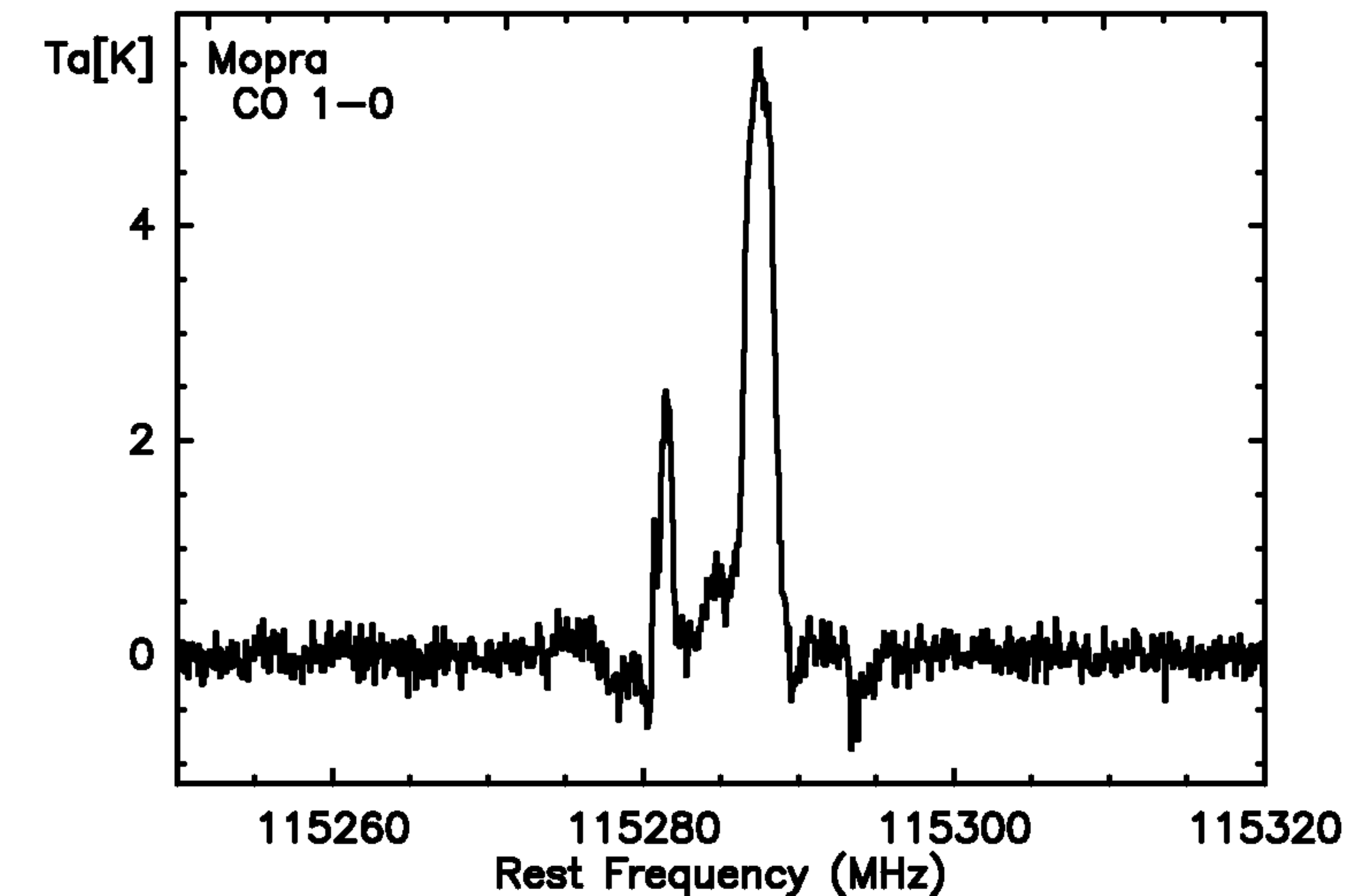}
\includegraphics[width=5.3cm]{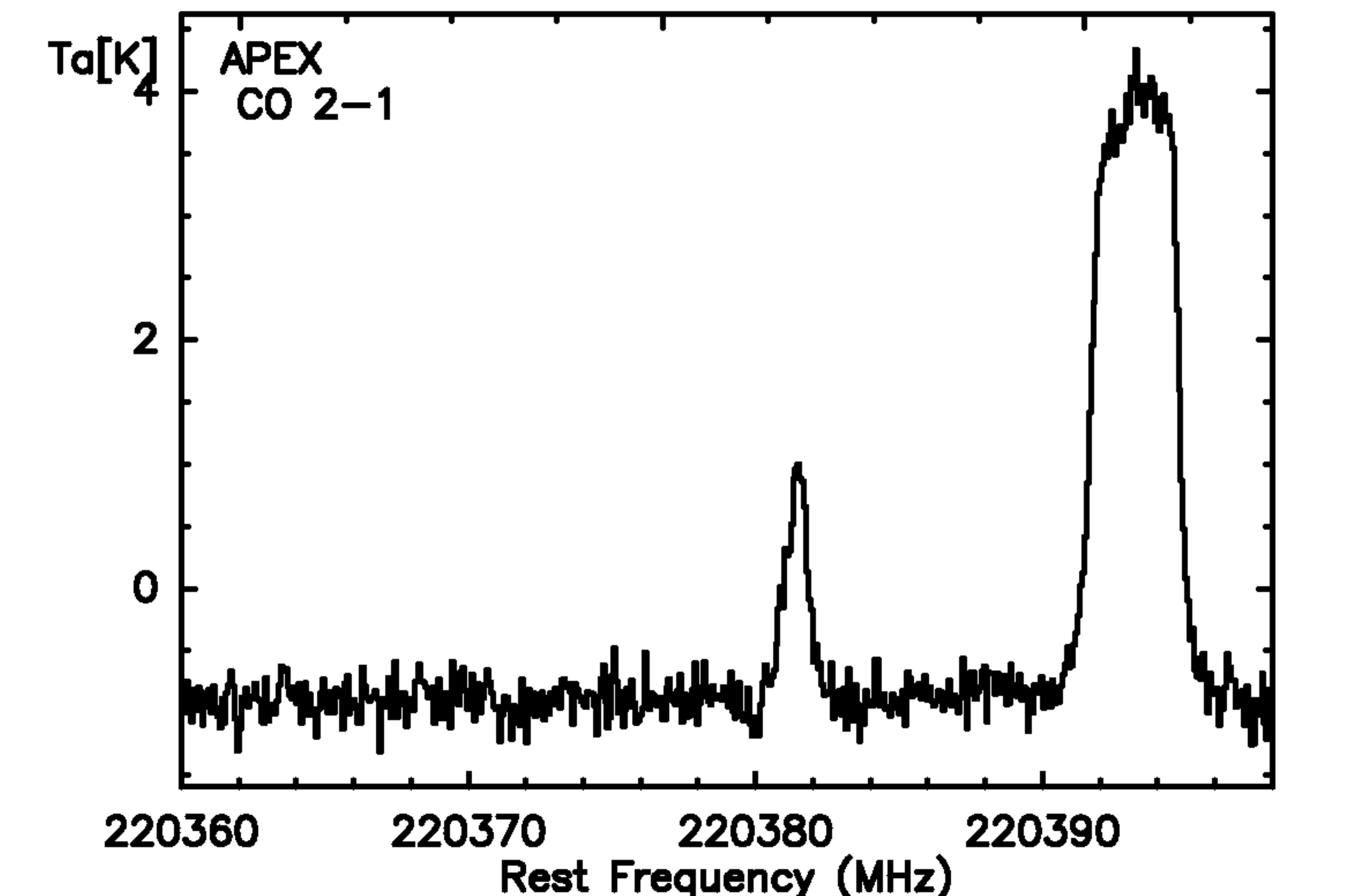}
\includegraphics[width=5.3cm]{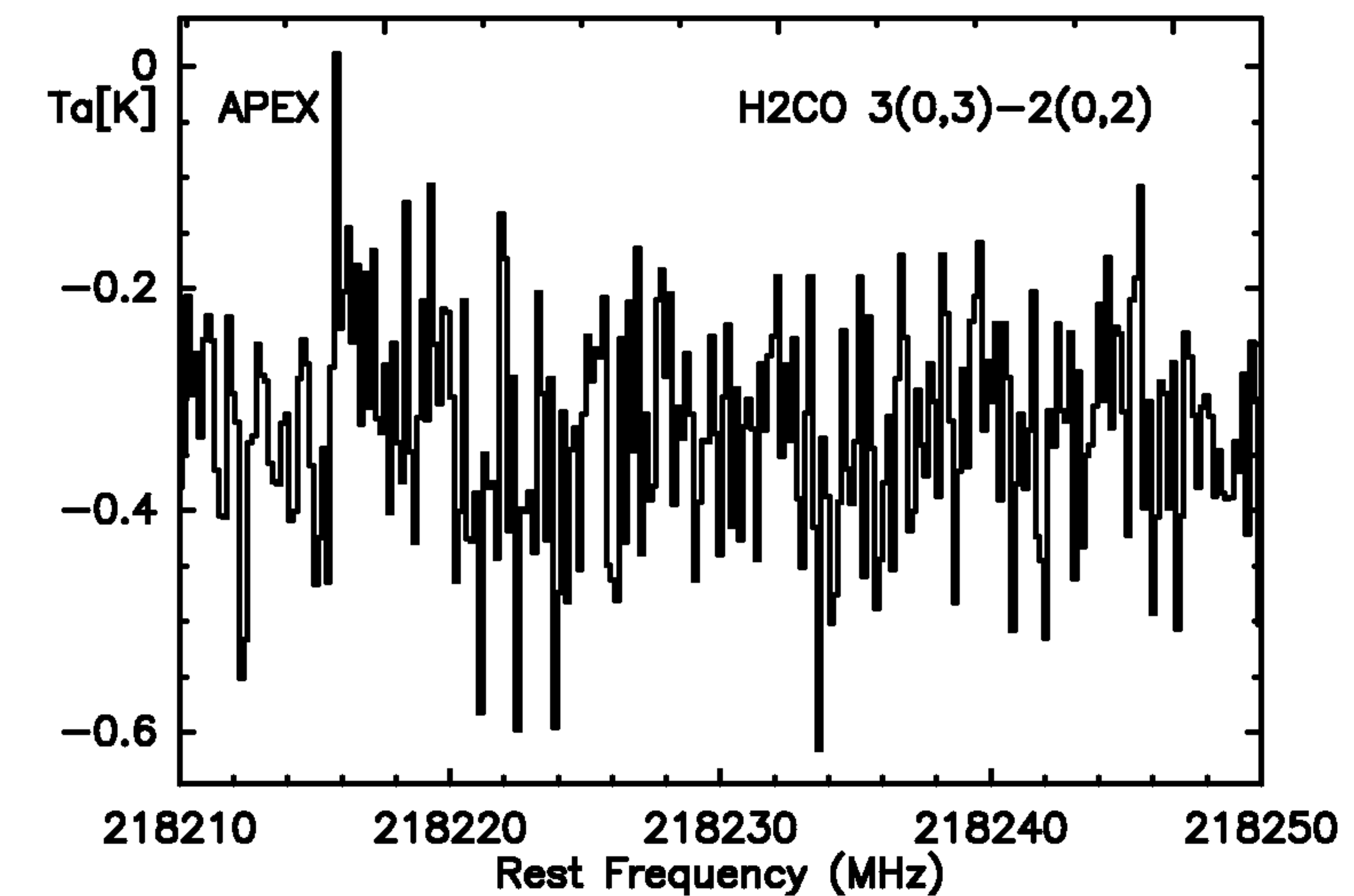}

\includegraphics[width=5.3cm]{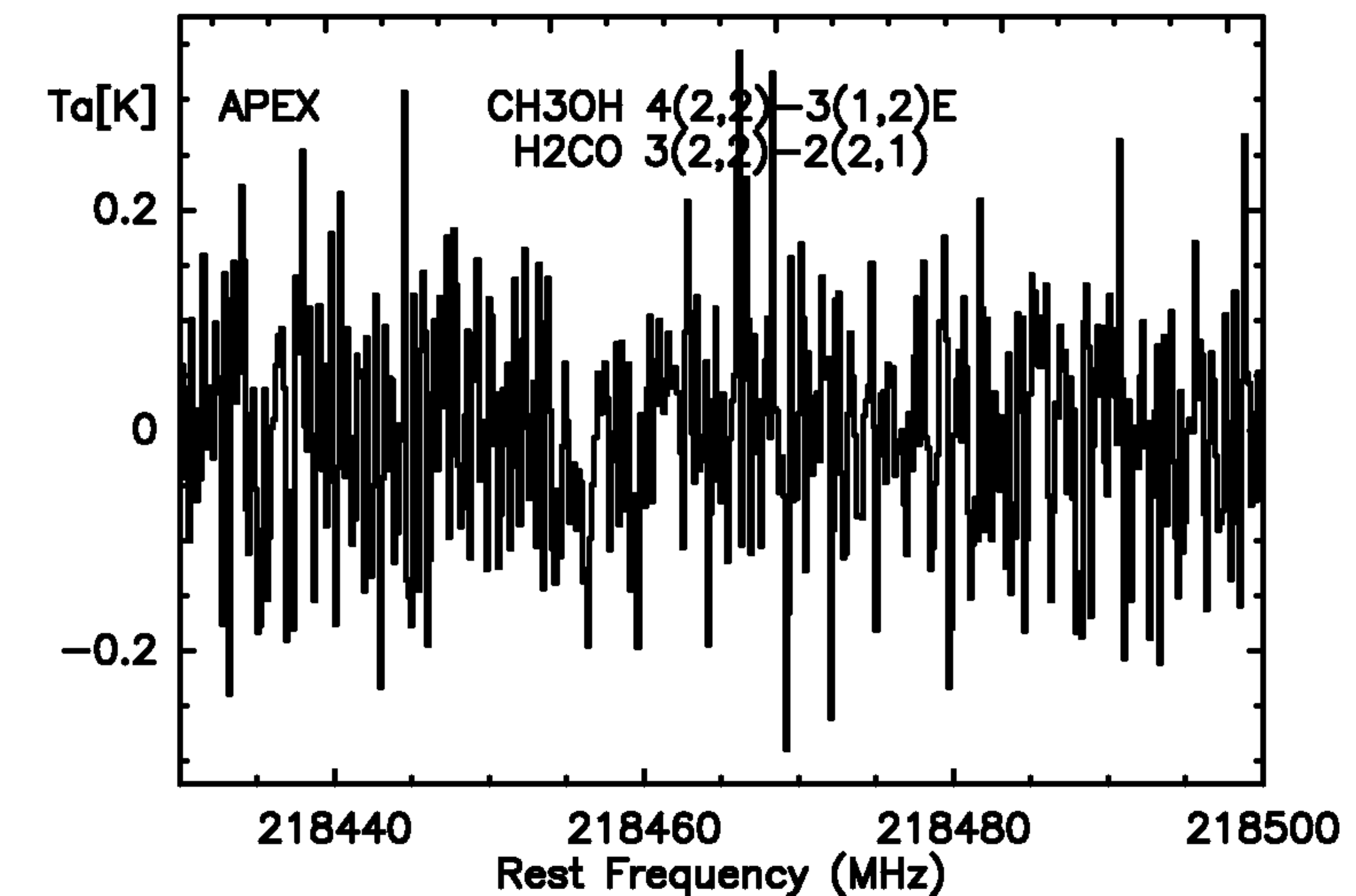}
\includegraphics[width=5.3cm]{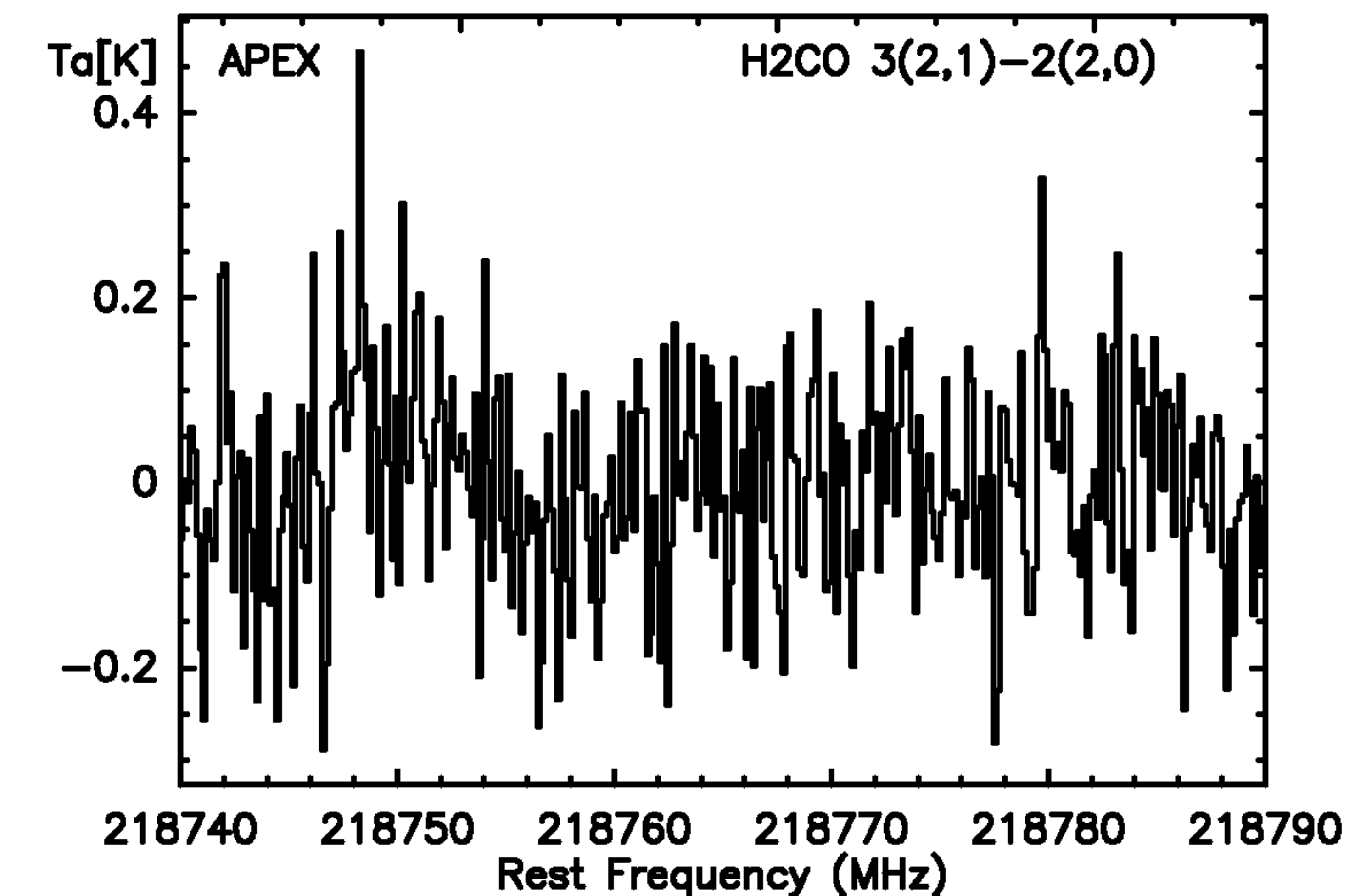}
\caption{Line spectra for IRDC309.37-3.}
\label{spectra1}
\end{figure}

\begin{figure}
\includegraphics[width=5.3cm]{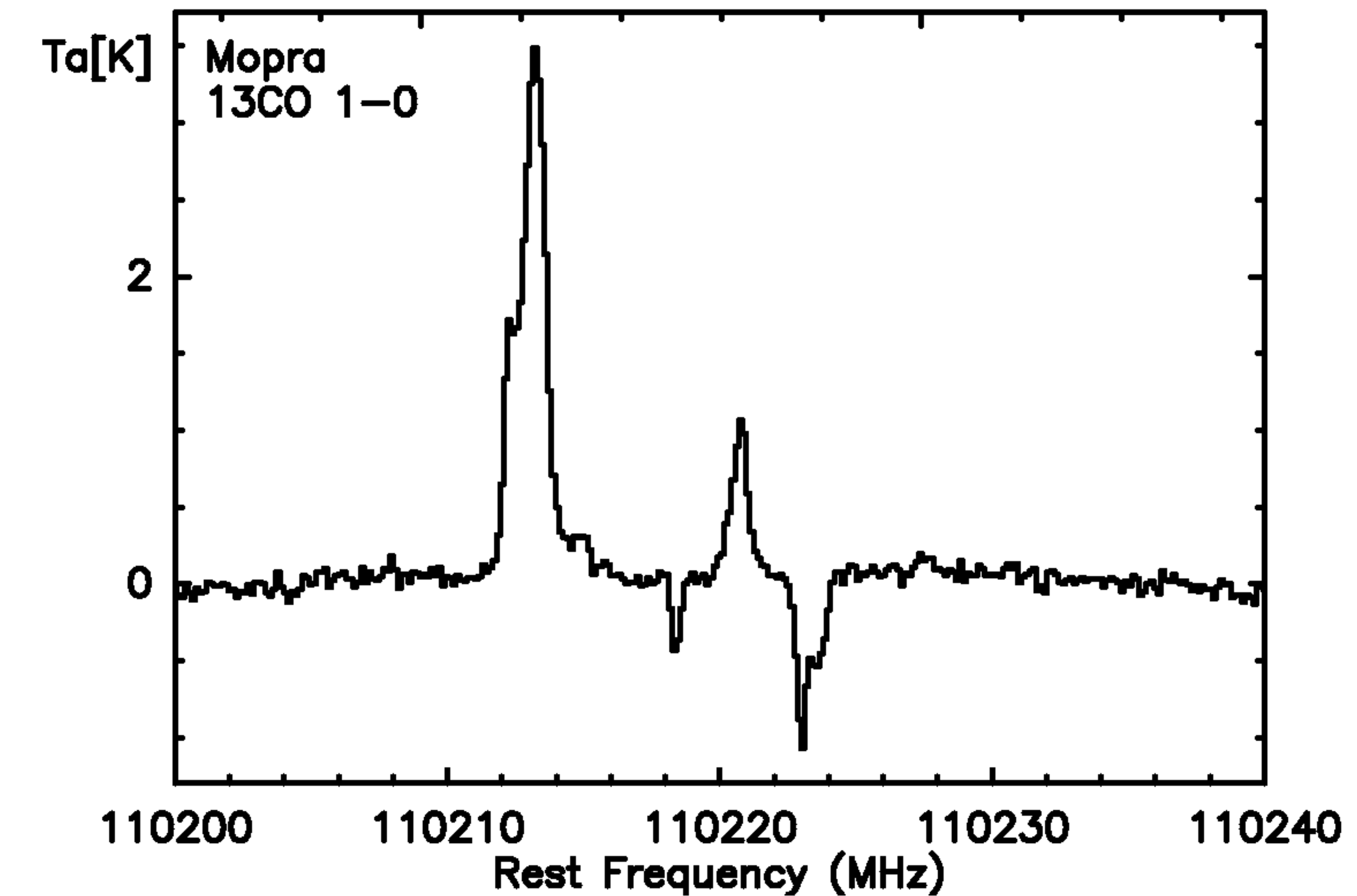}
\includegraphics[width=5.3cm]{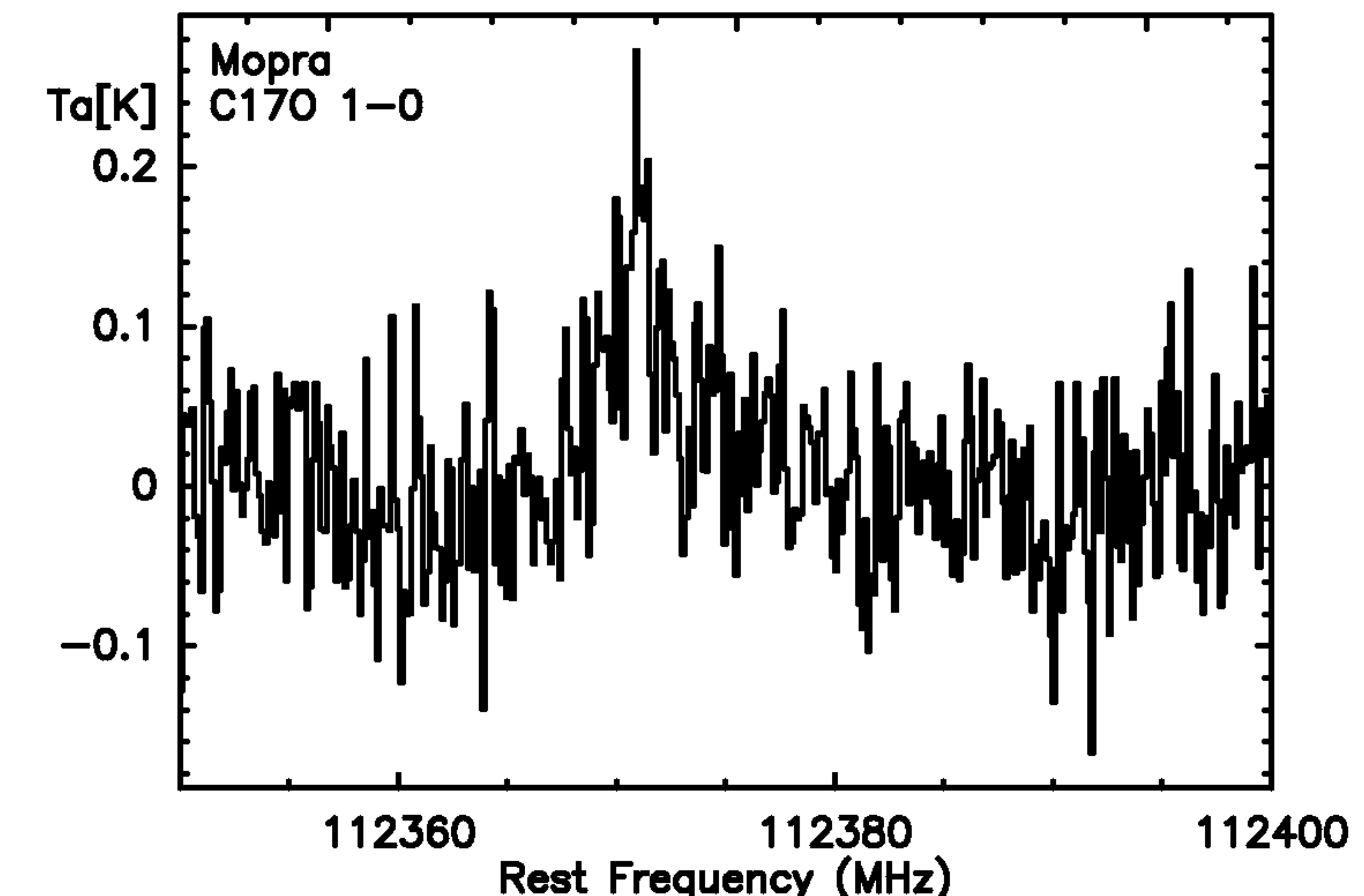}
\includegraphics[width=5.3cm]{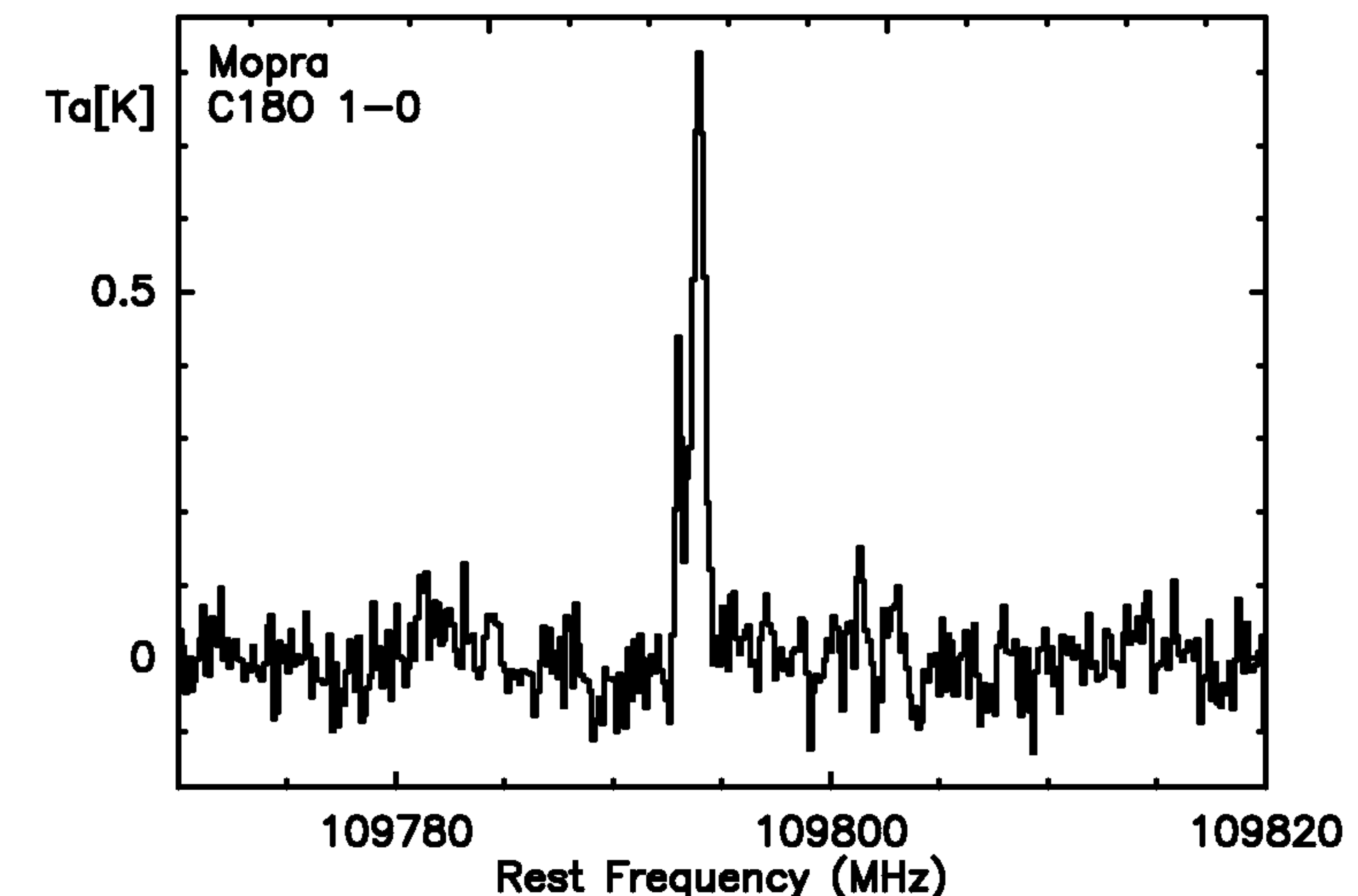}

\includegraphics[width=5.3cm]{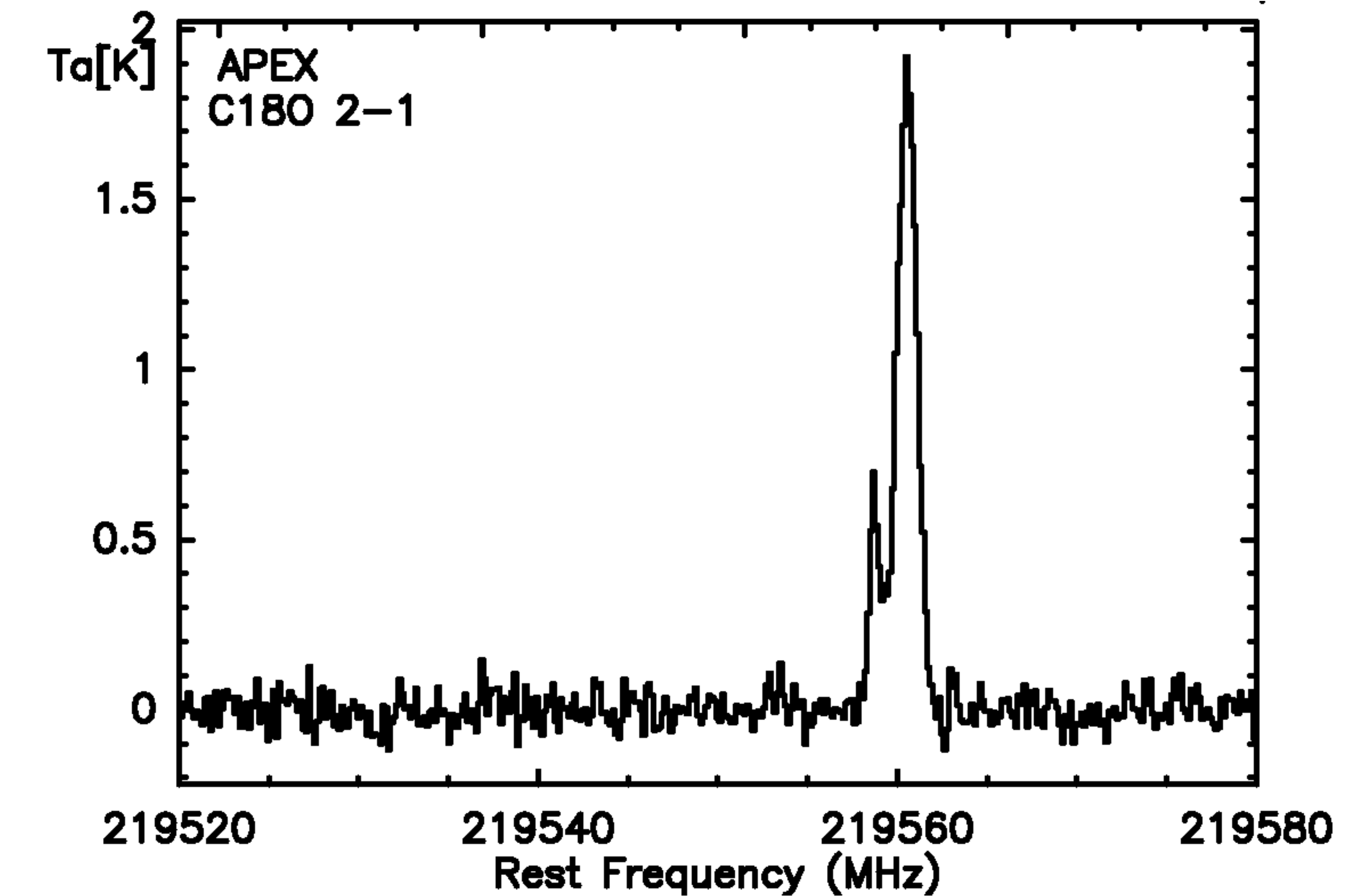}
\includegraphics[width=5.3cm]{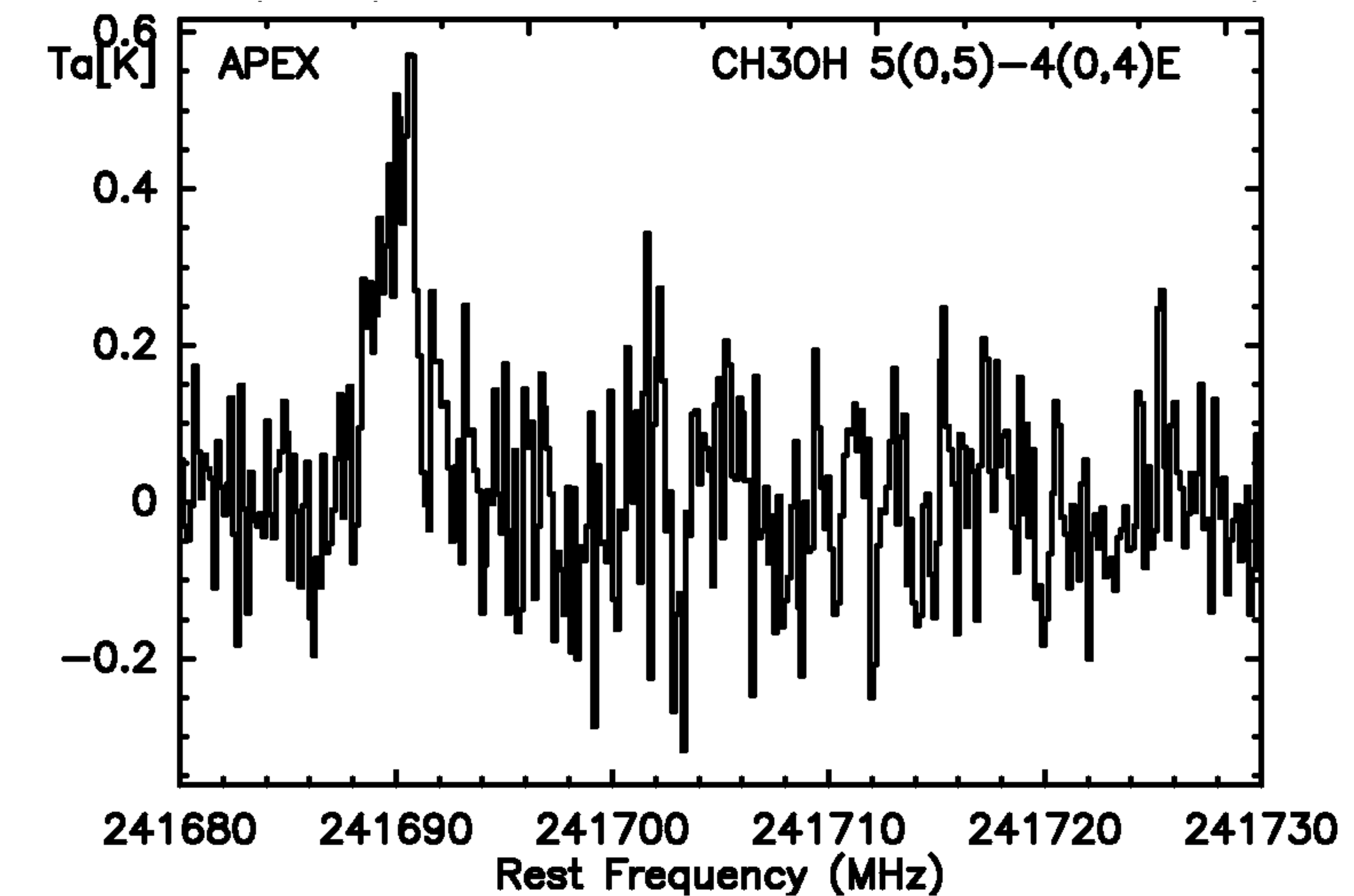}
\includegraphics[width=5.3cm]{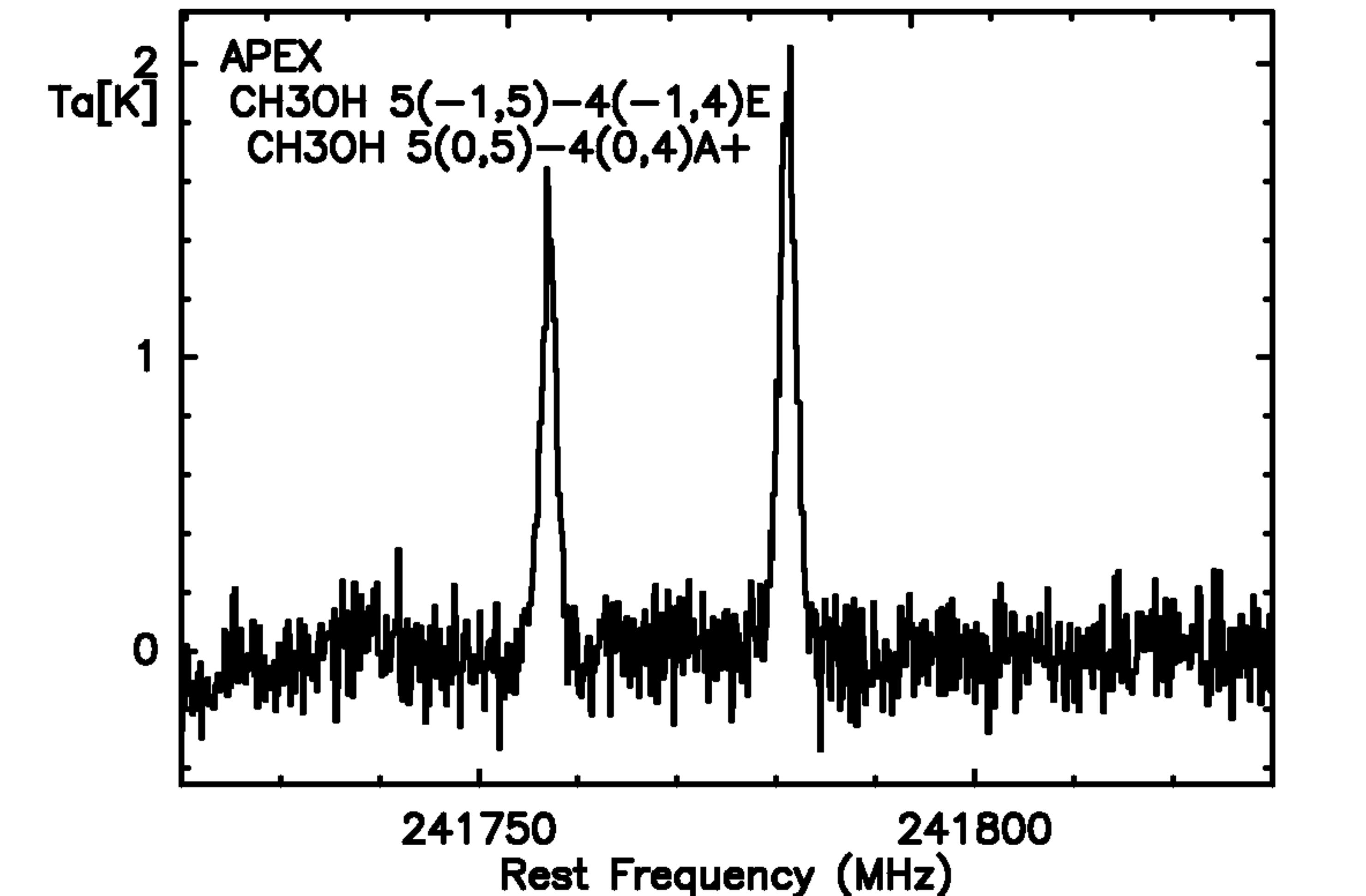}

\includegraphics[width=5.3cm]{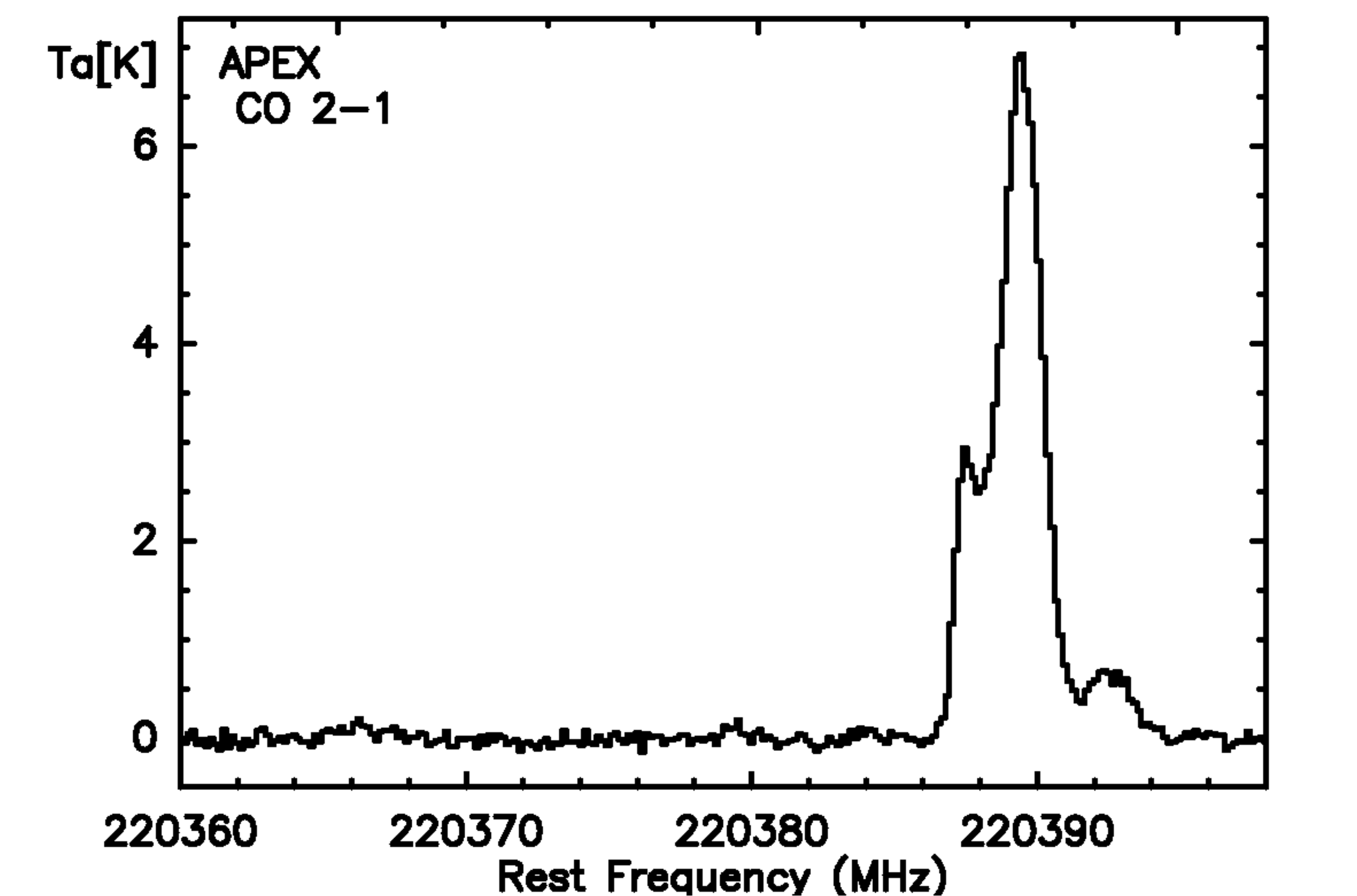}
\includegraphics[width=5.3cm]{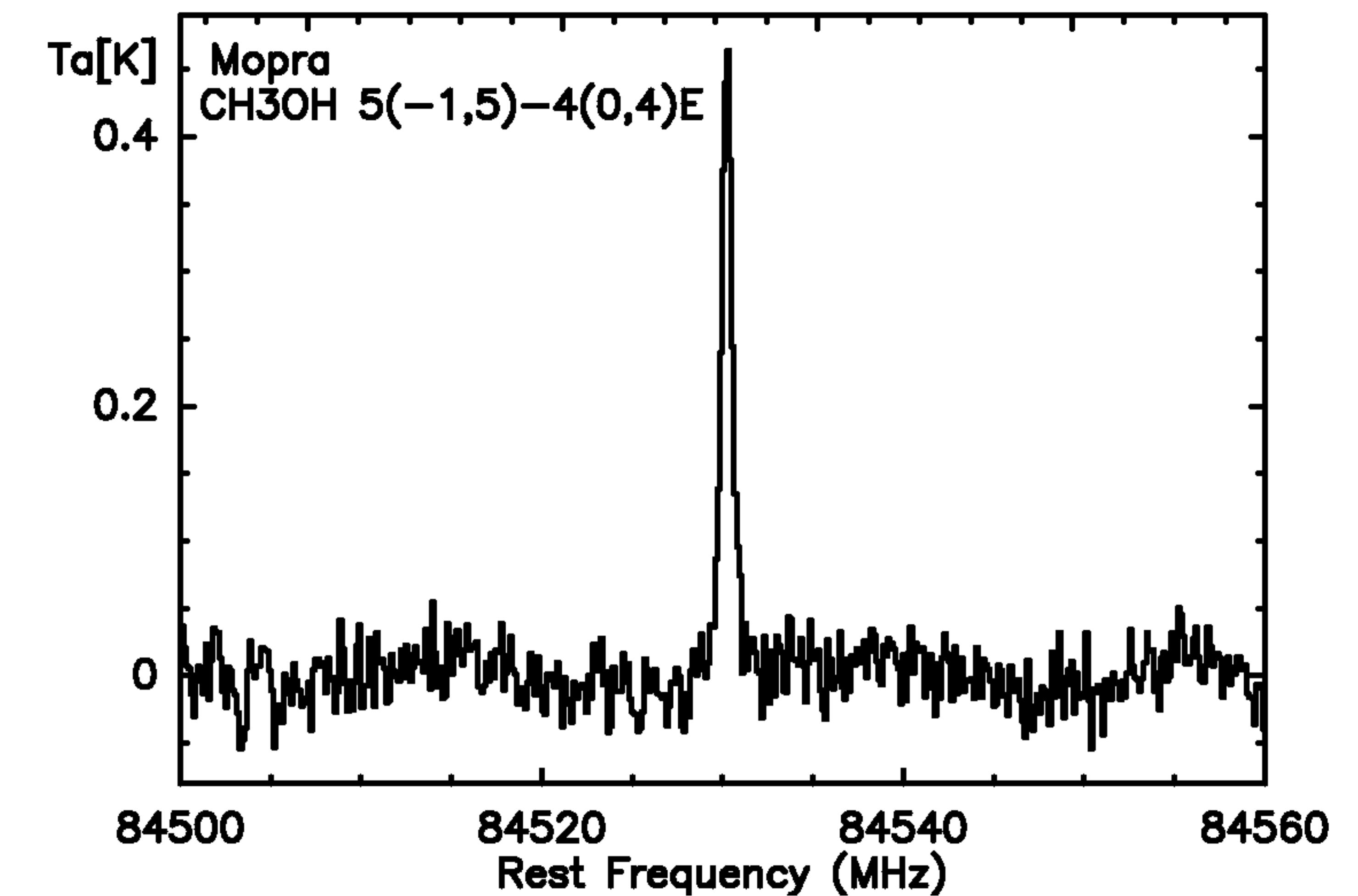}
\includegraphics[width=5.3cm]{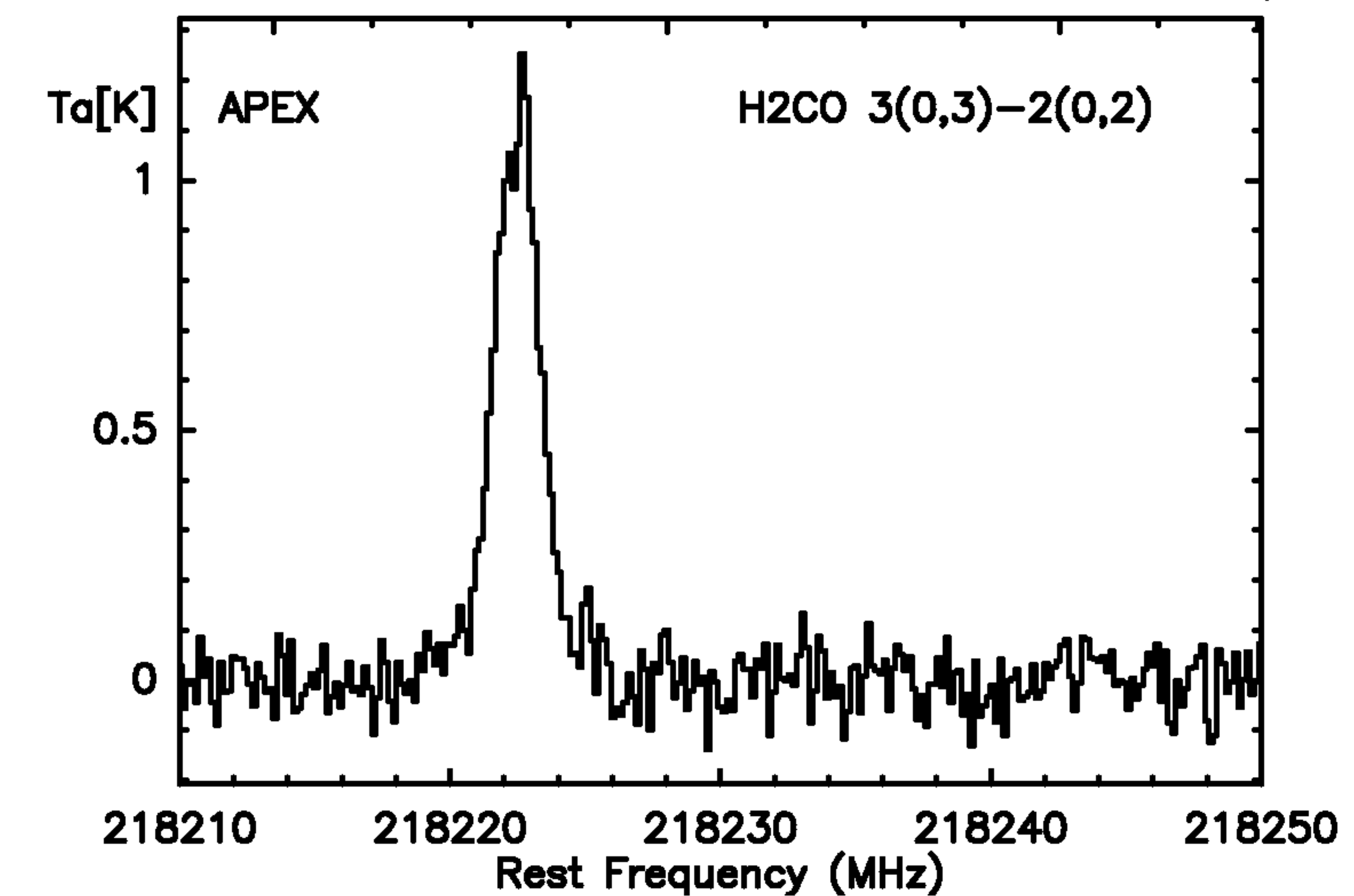}

\includegraphics[width=5.3cm]{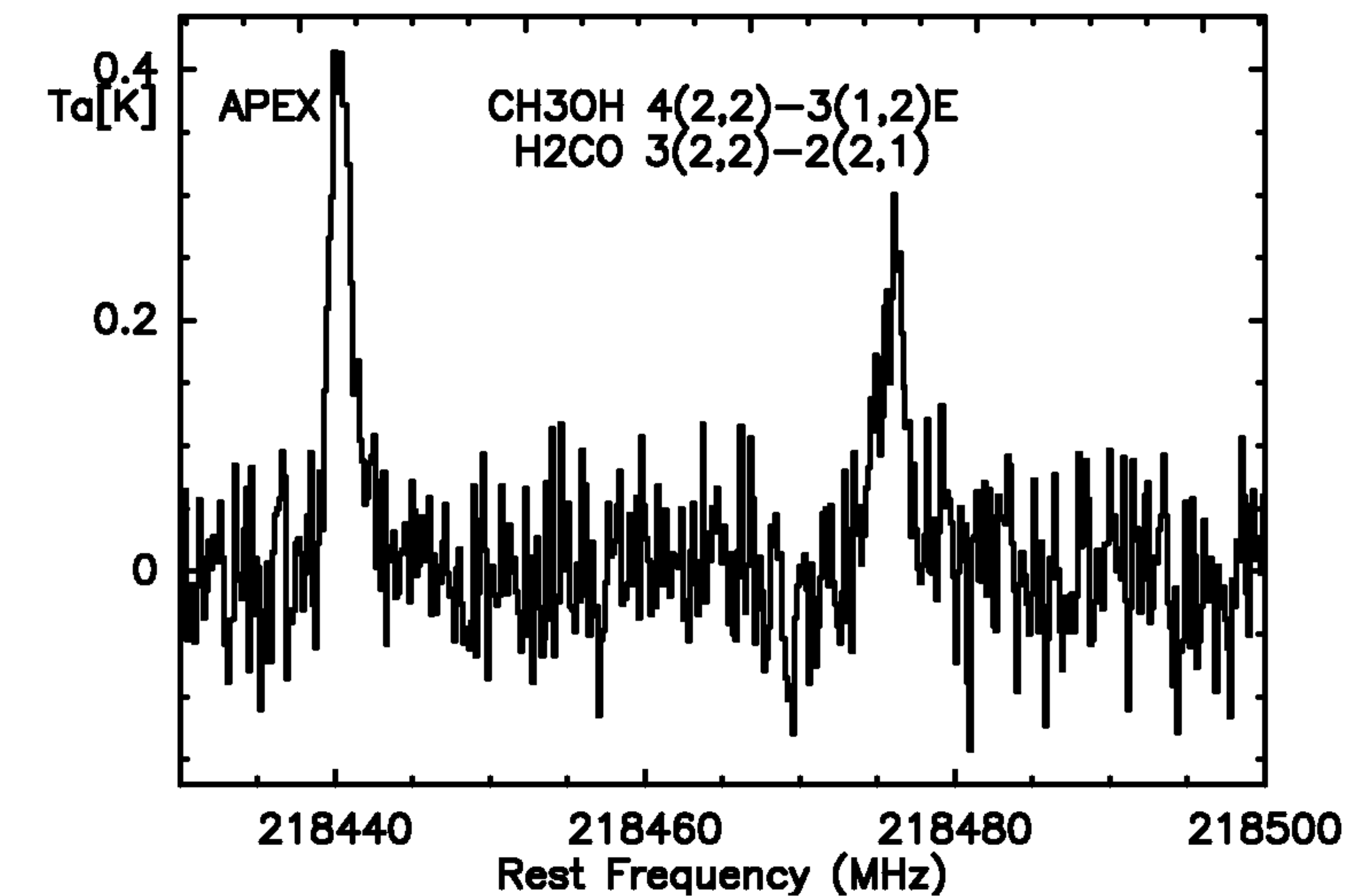}
\includegraphics[width=5.3cm]{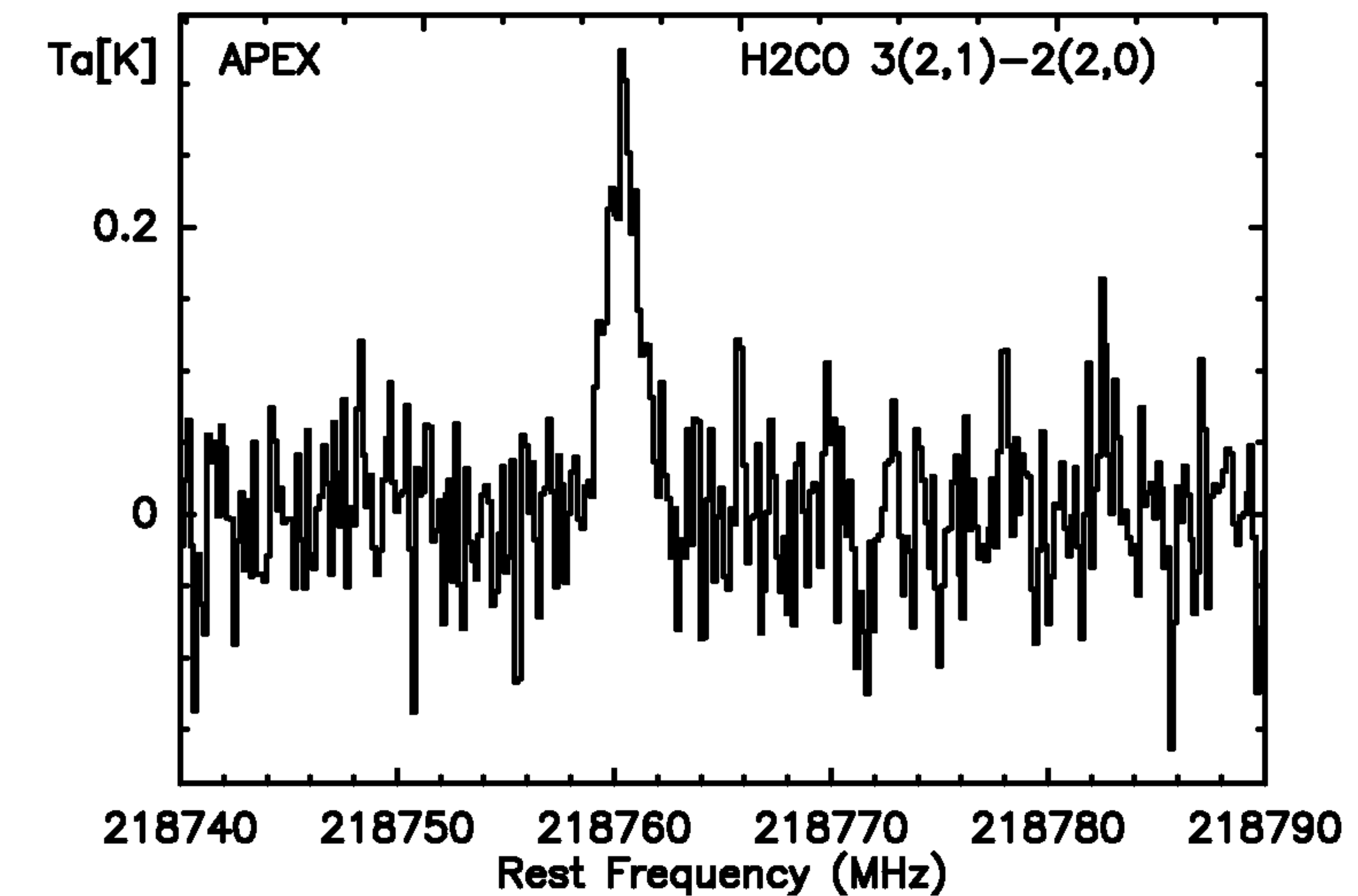}
\includegraphics[width=5.3cm]{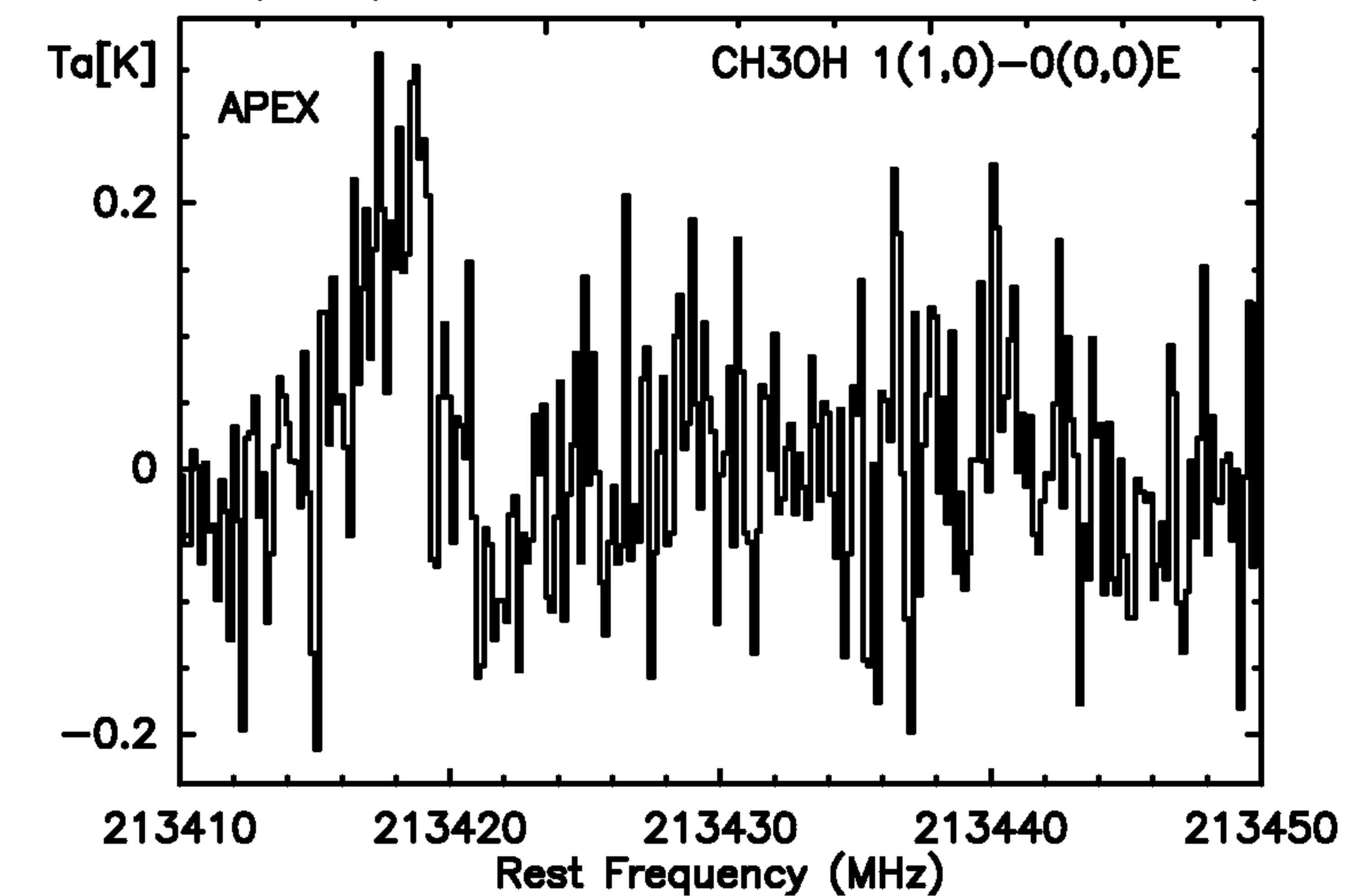}
\caption{Line spectra for IRDC321.73-1.}
\label{spectra2}
\end{figure}

\begin{figure}
\includegraphics[width=5.3cm]{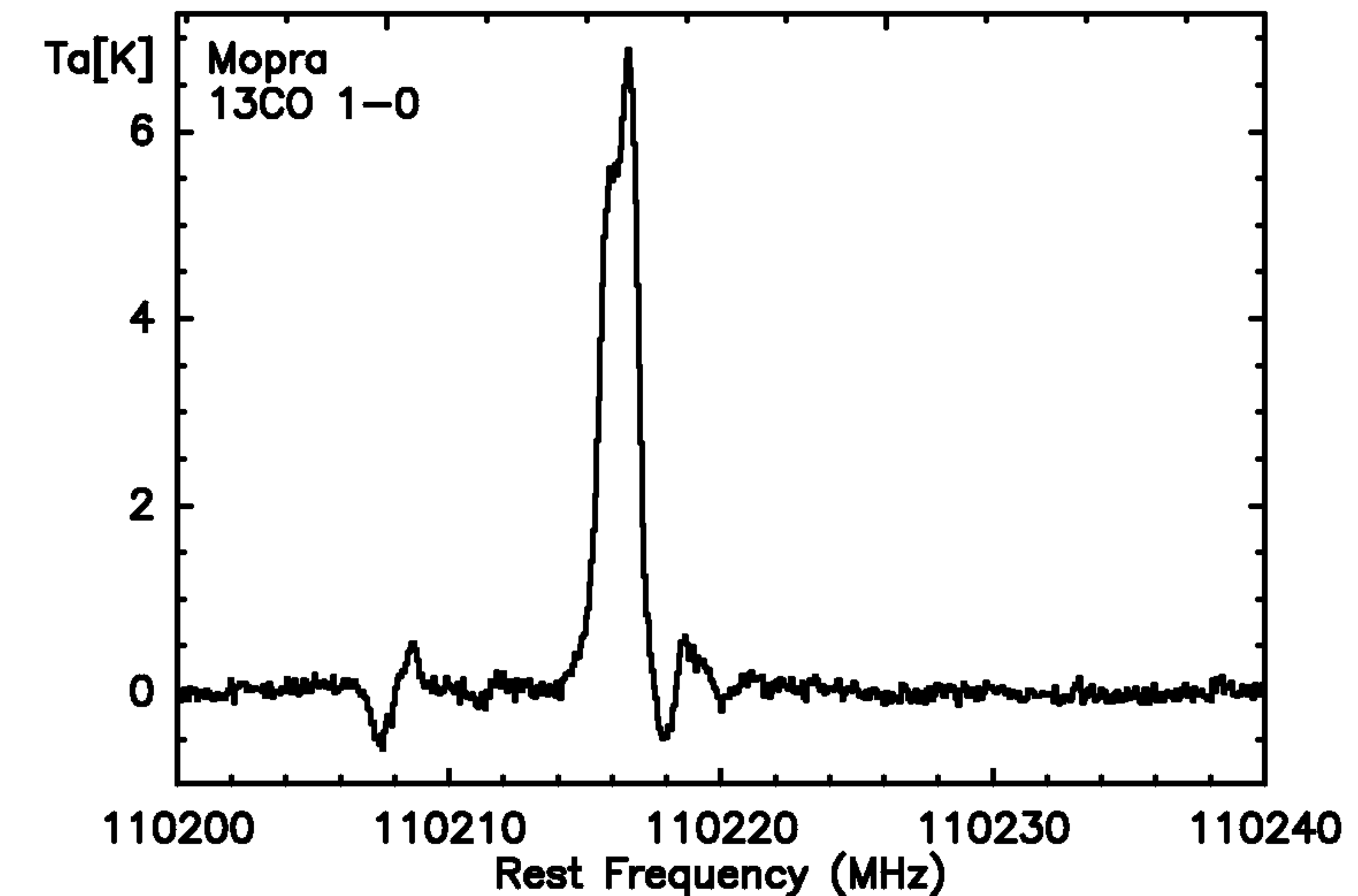}
\includegraphics[width=5.3cm]{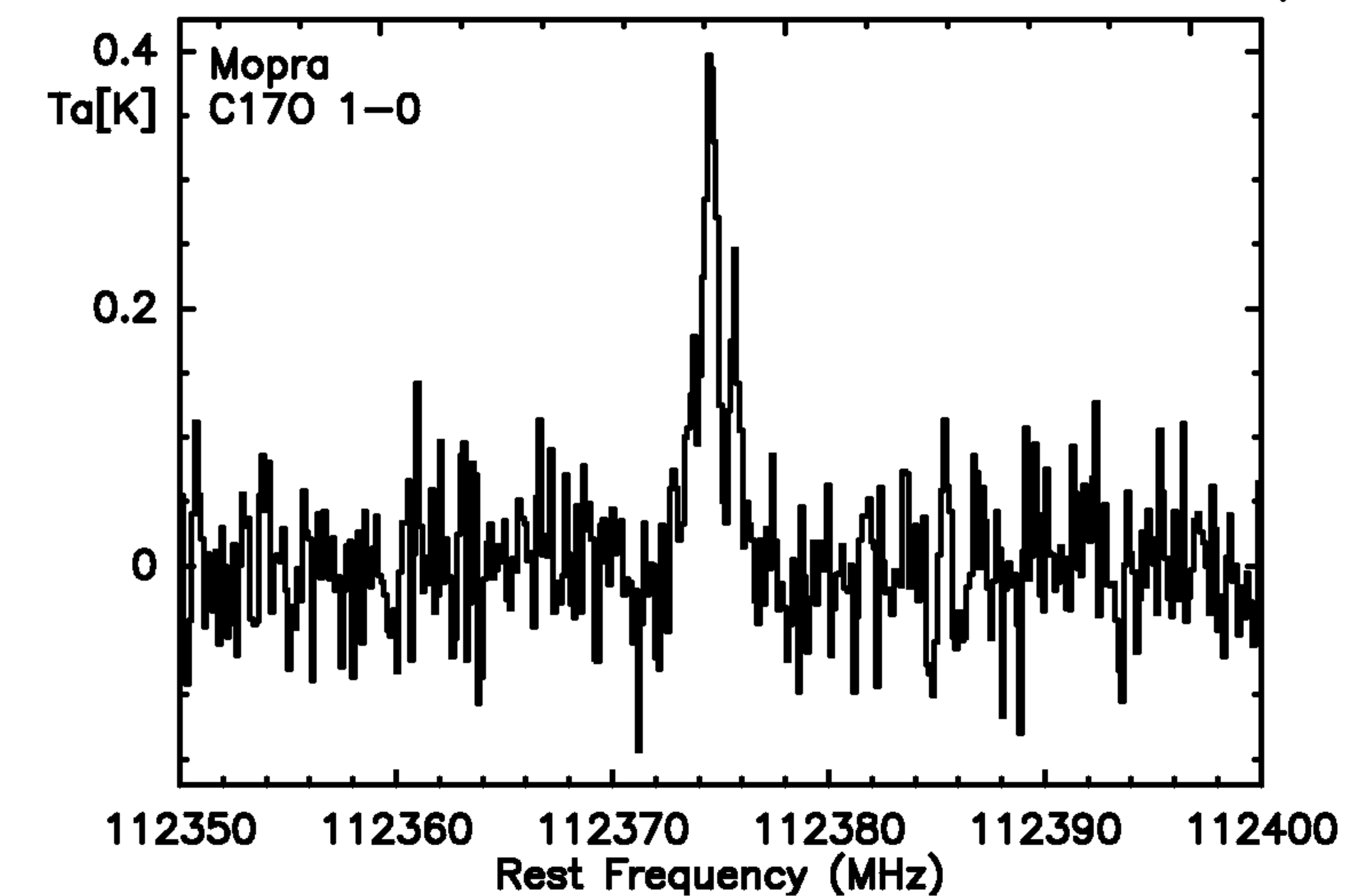}
\includegraphics[width=5.3cm]{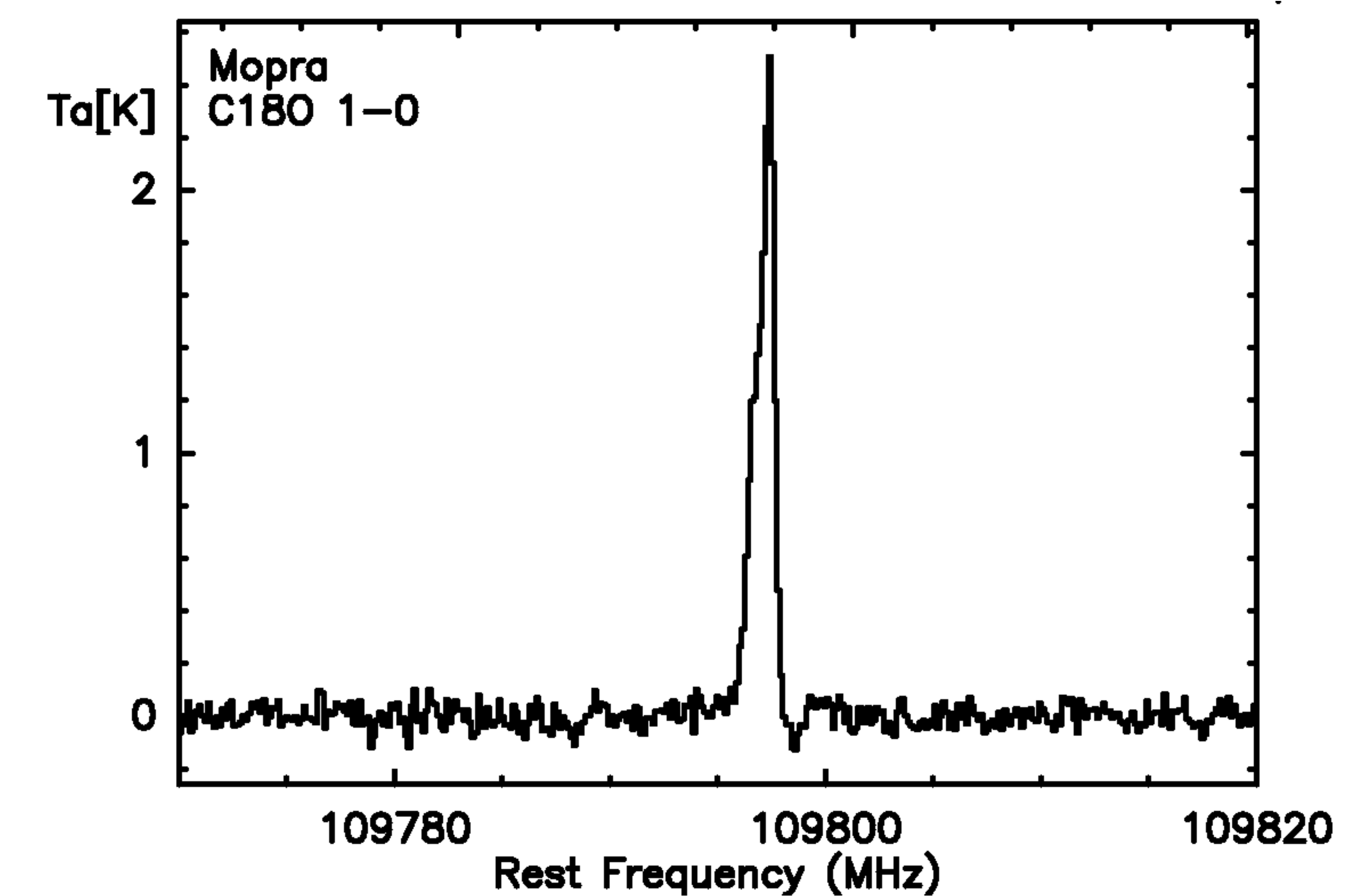}

\includegraphics[width=5.3cm]{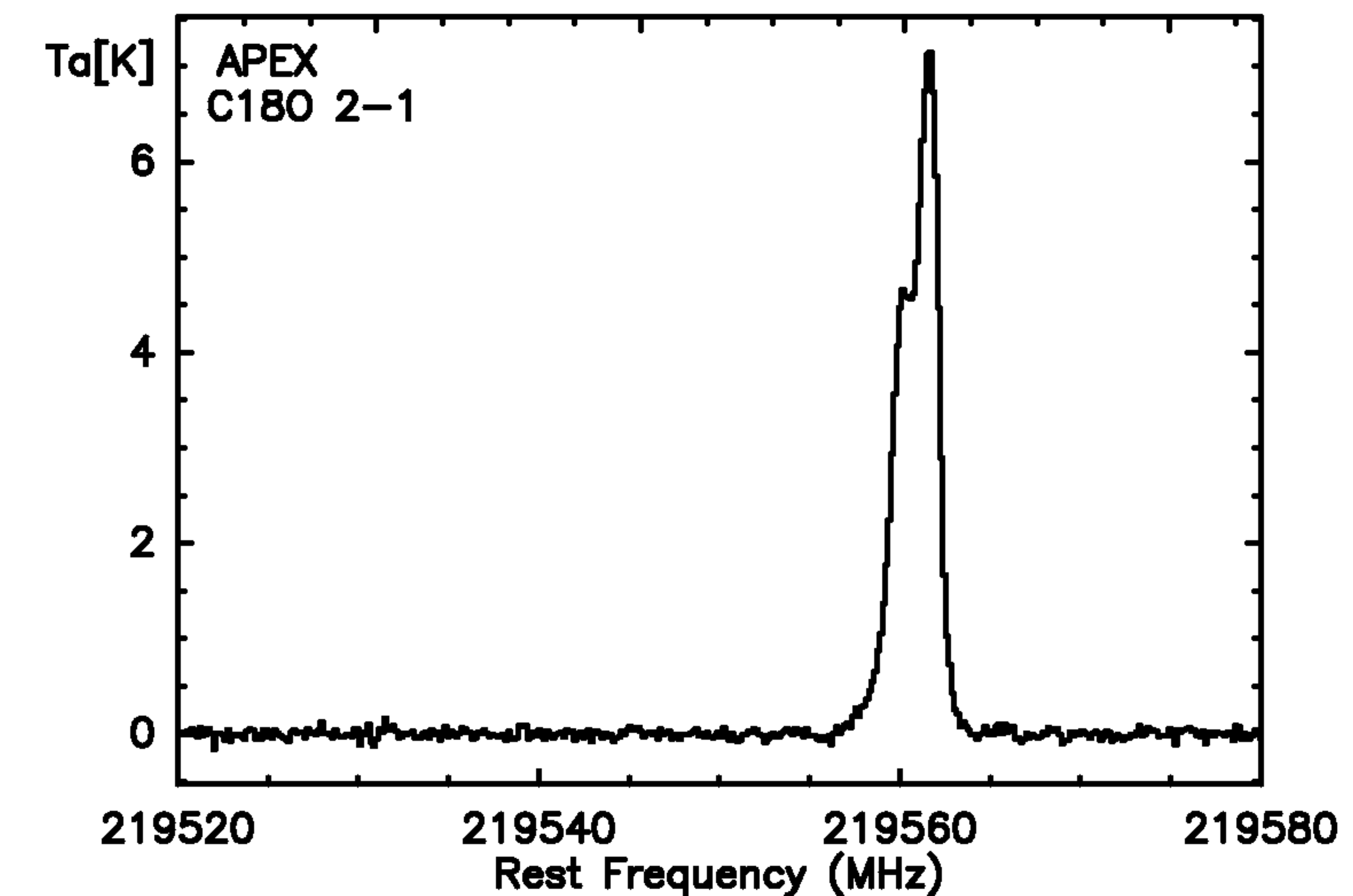}
\includegraphics[width=5.3cm]{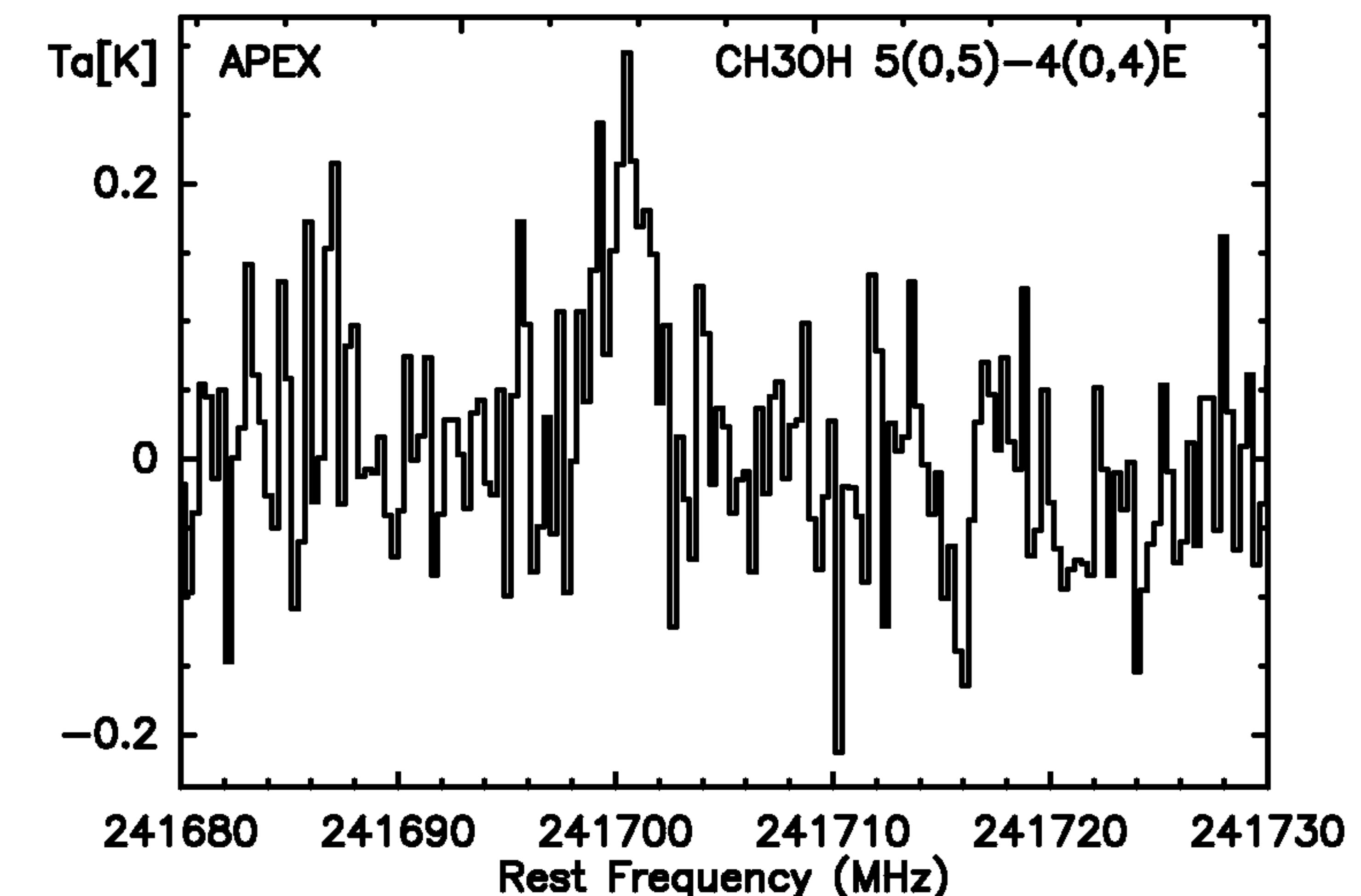}
\includegraphics[width=5.3cm]{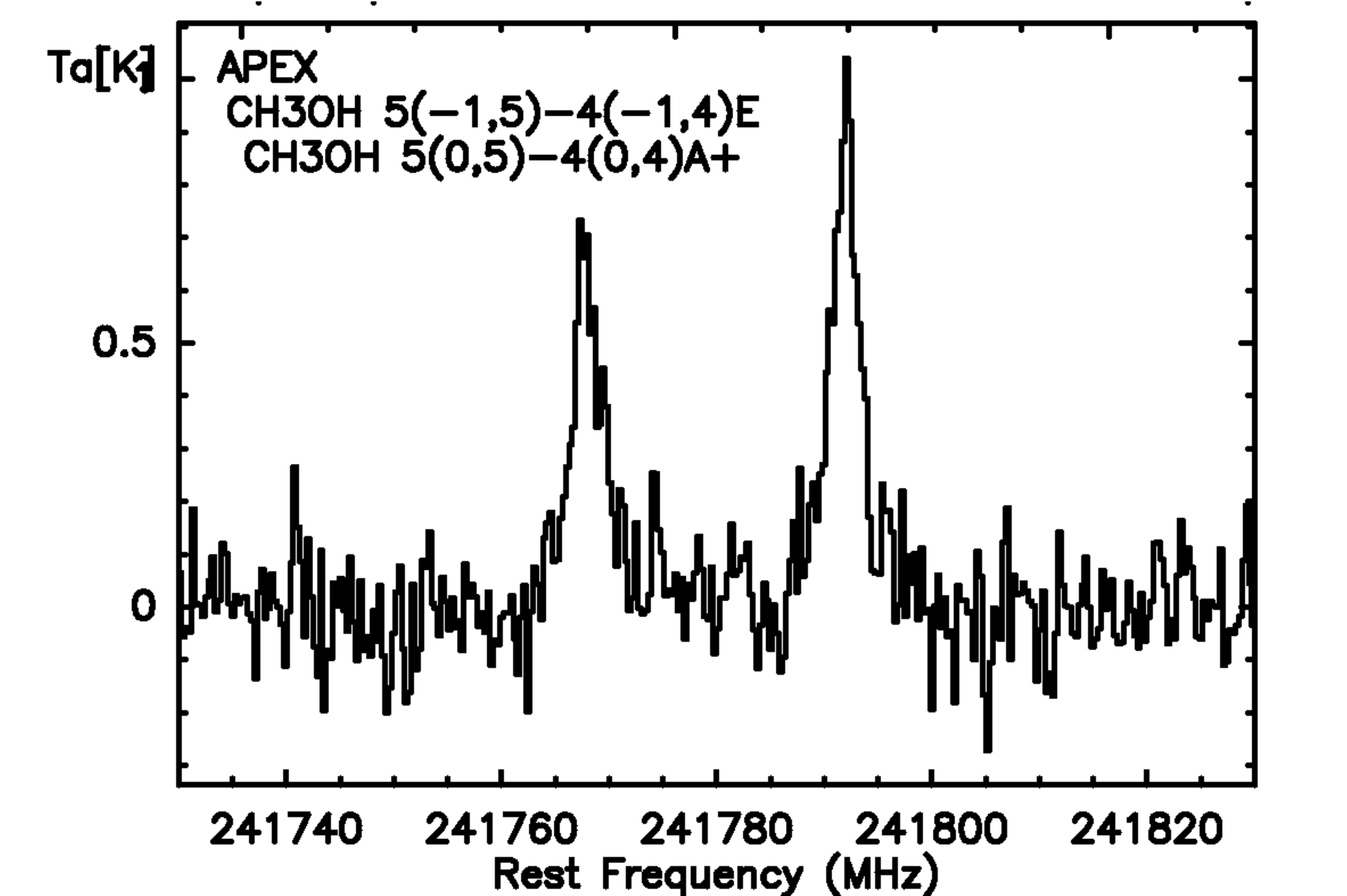}

\includegraphics[width=5.3cm]{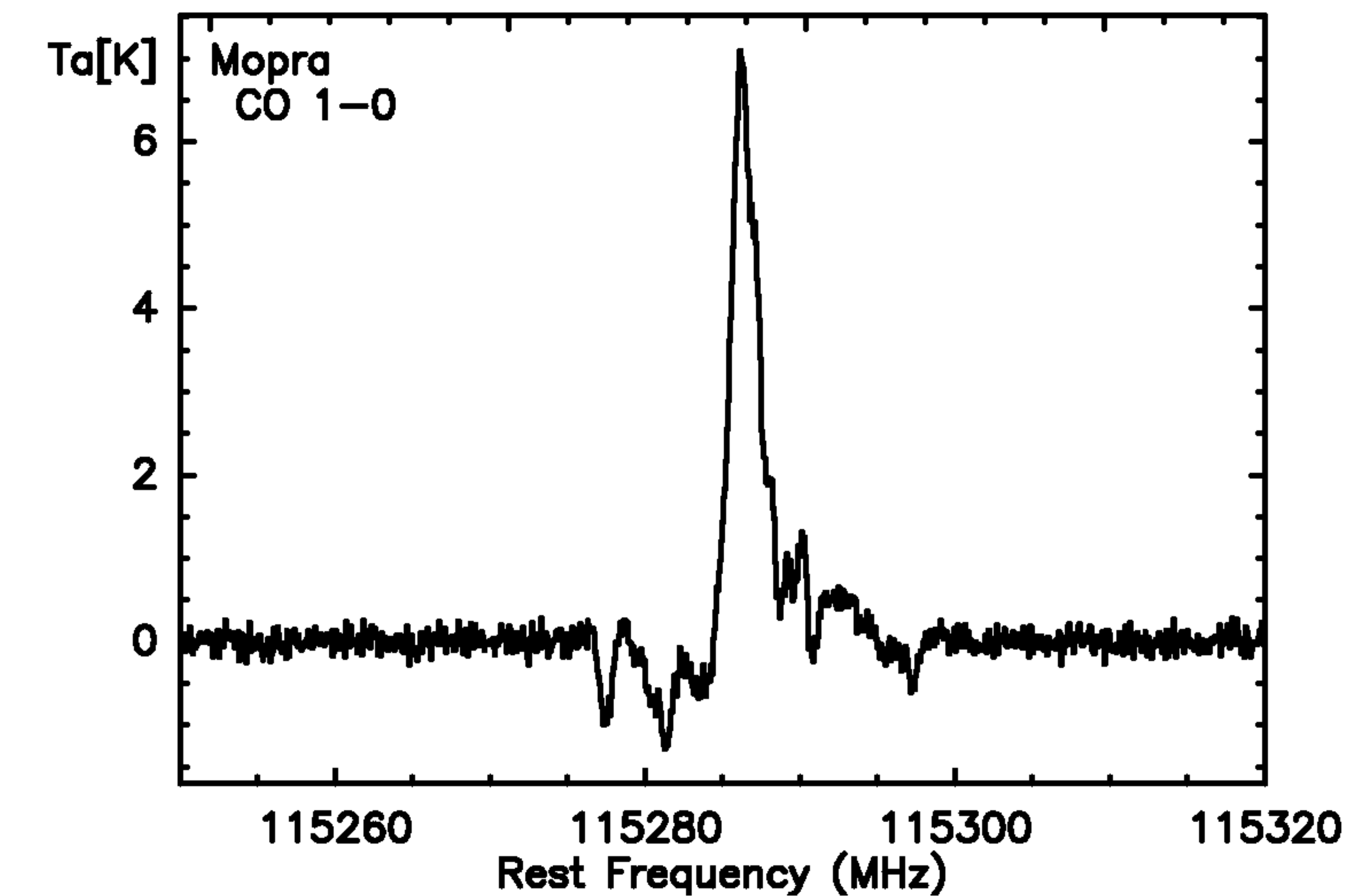}
\includegraphics[width=5.3cm]{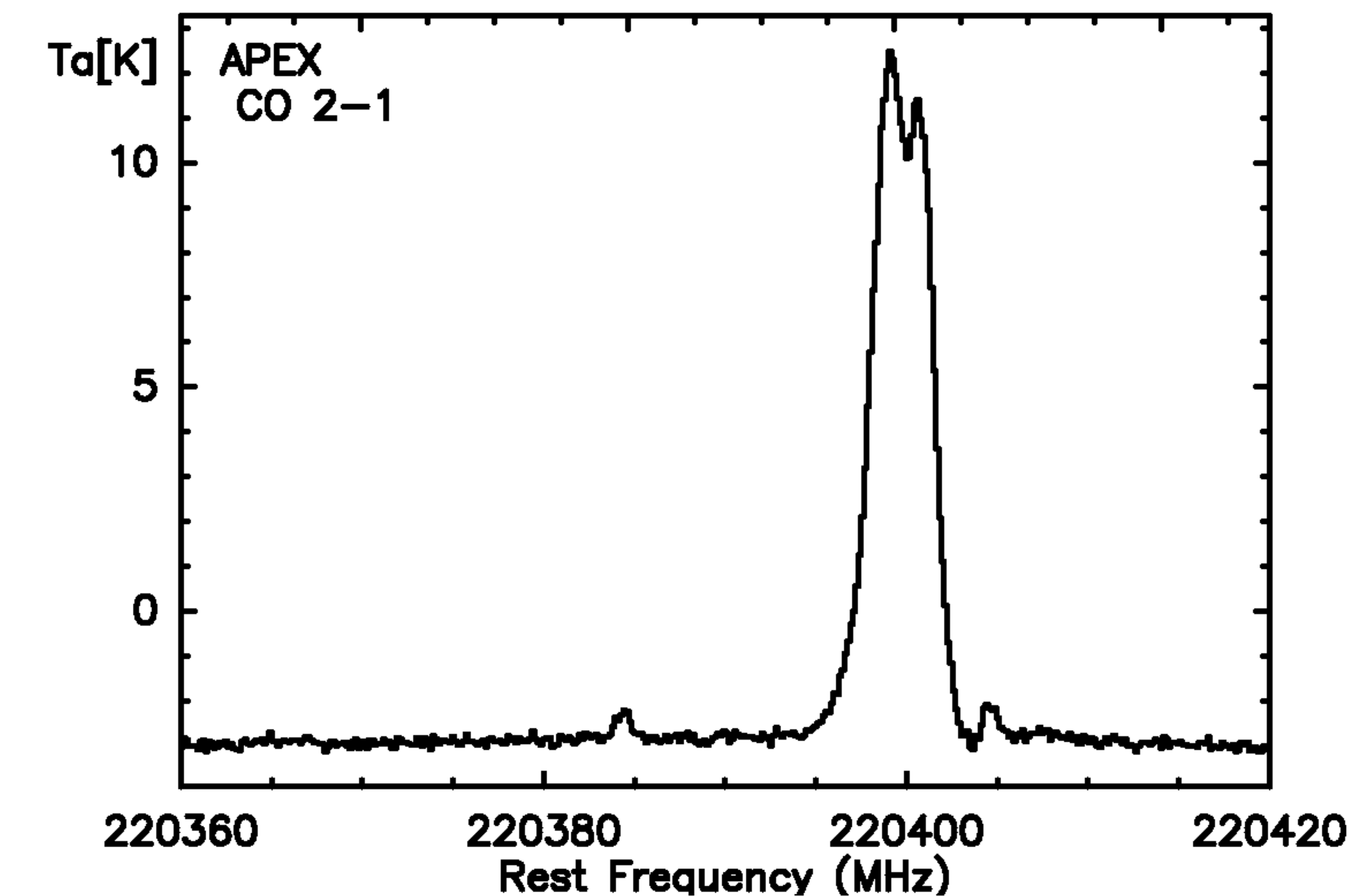}
\includegraphics[width=5.3cm]{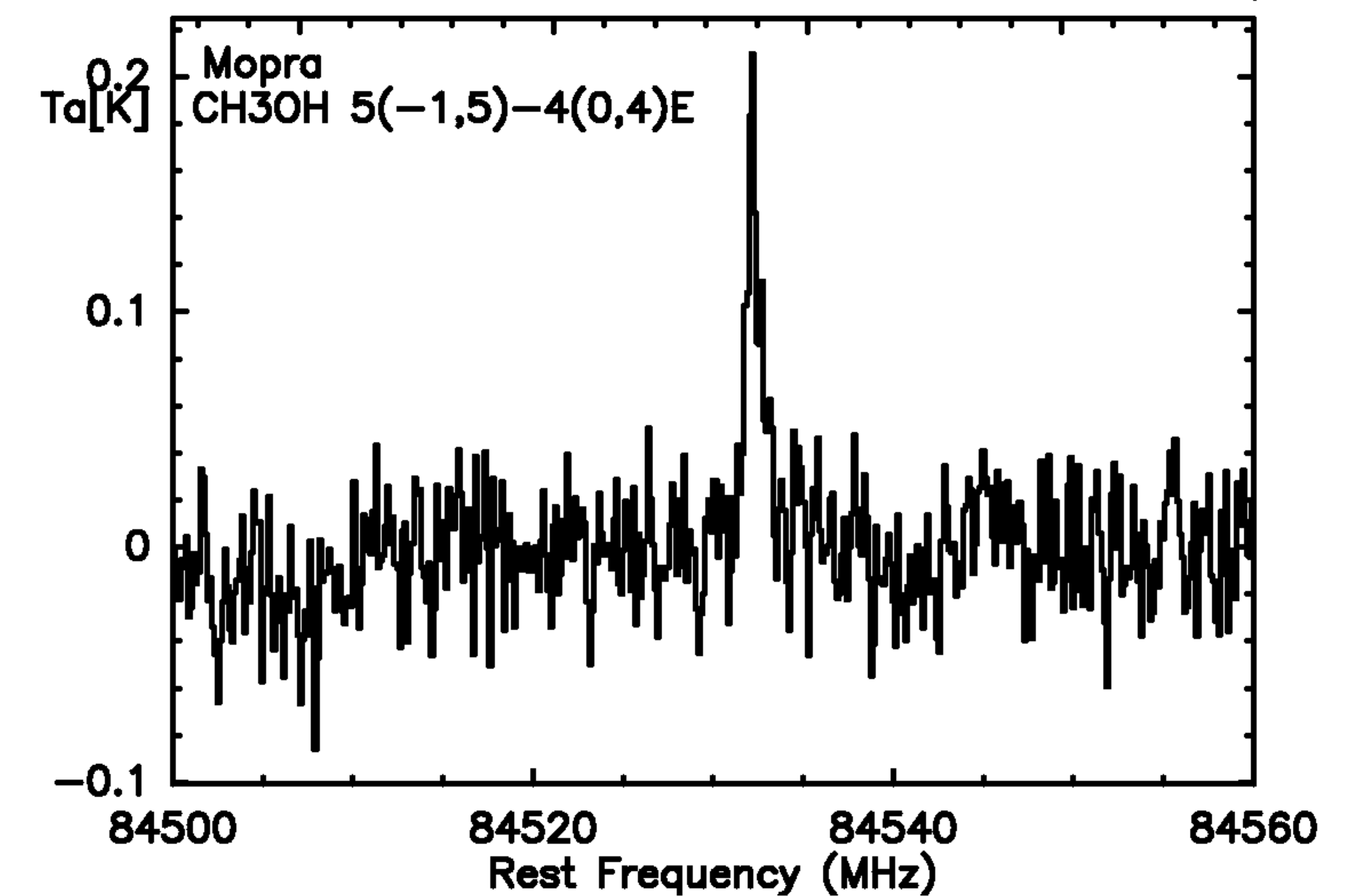}

\includegraphics[width=5.3cm]{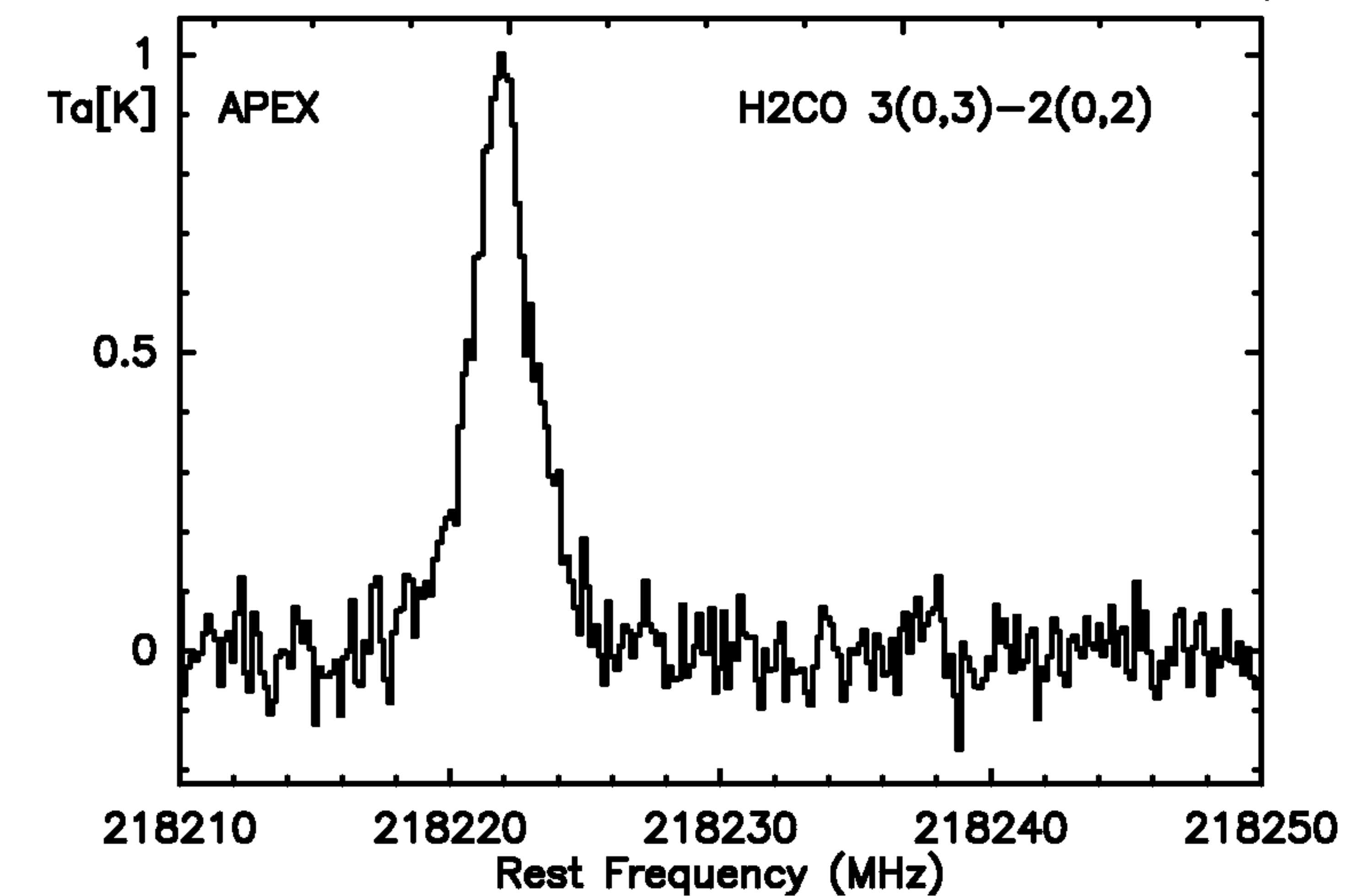}
\includegraphics[width=5.3cm]{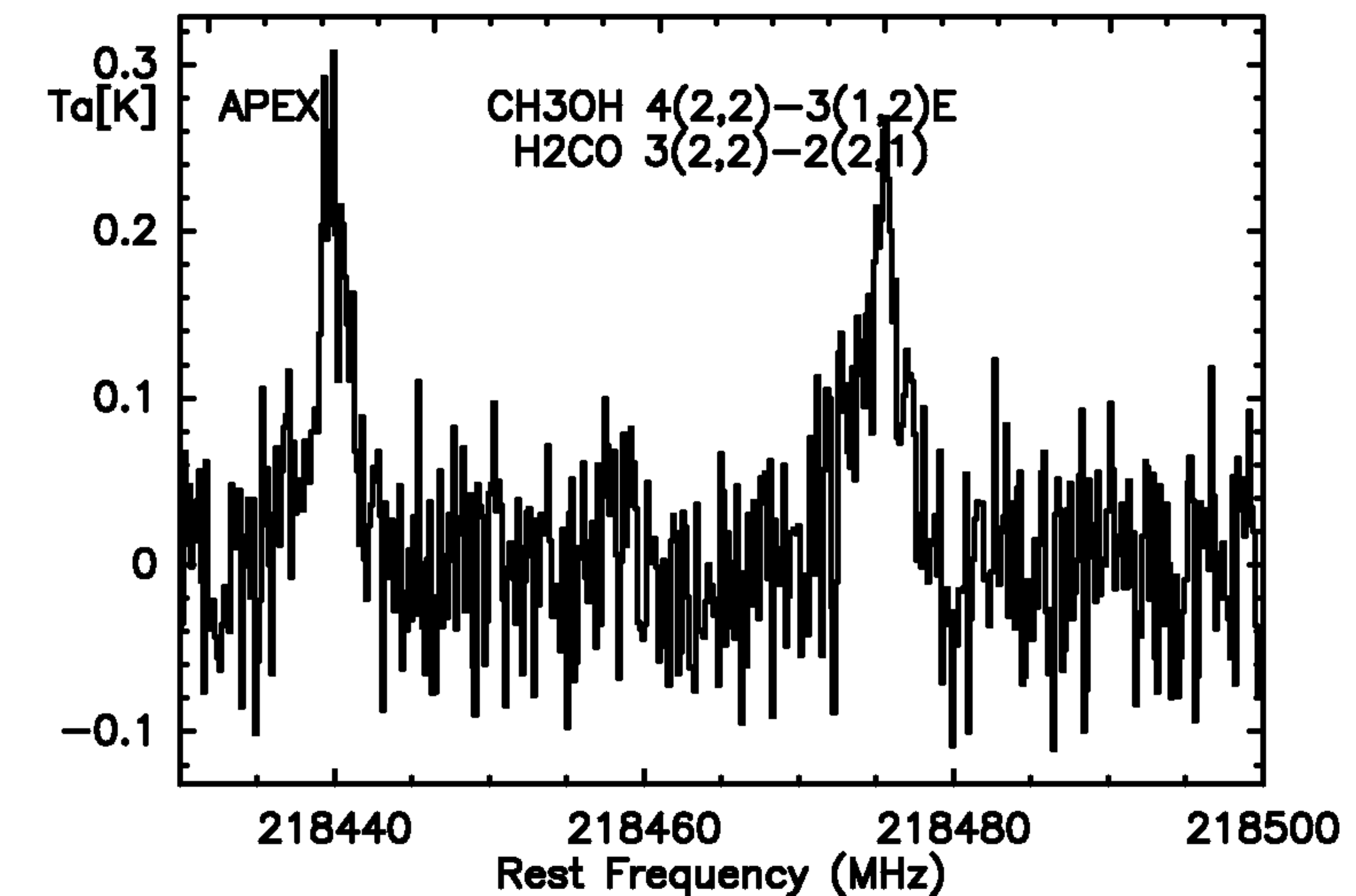}
\includegraphics[width=5.3cm]{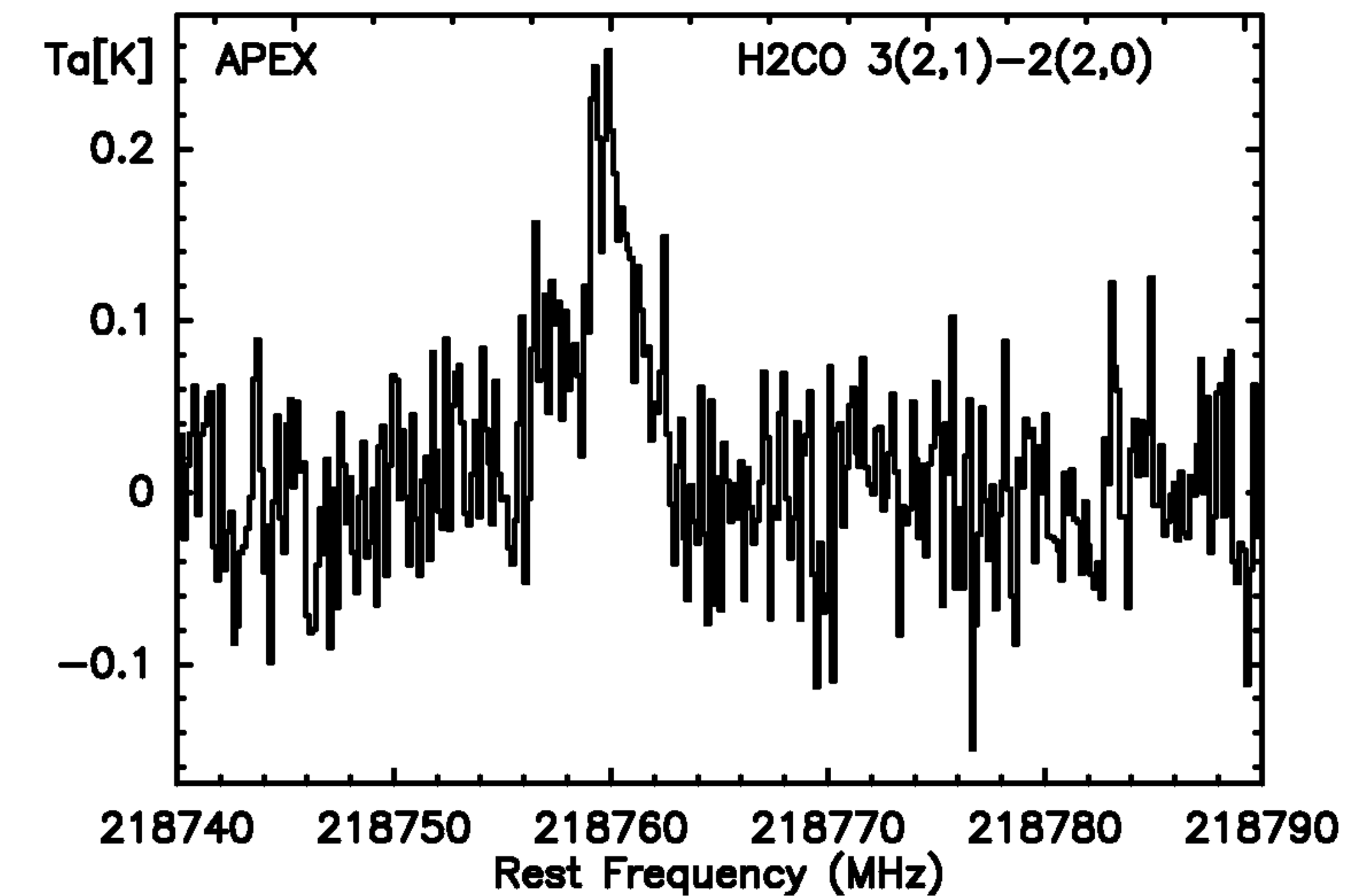}

\includegraphics[width=5.3cm]{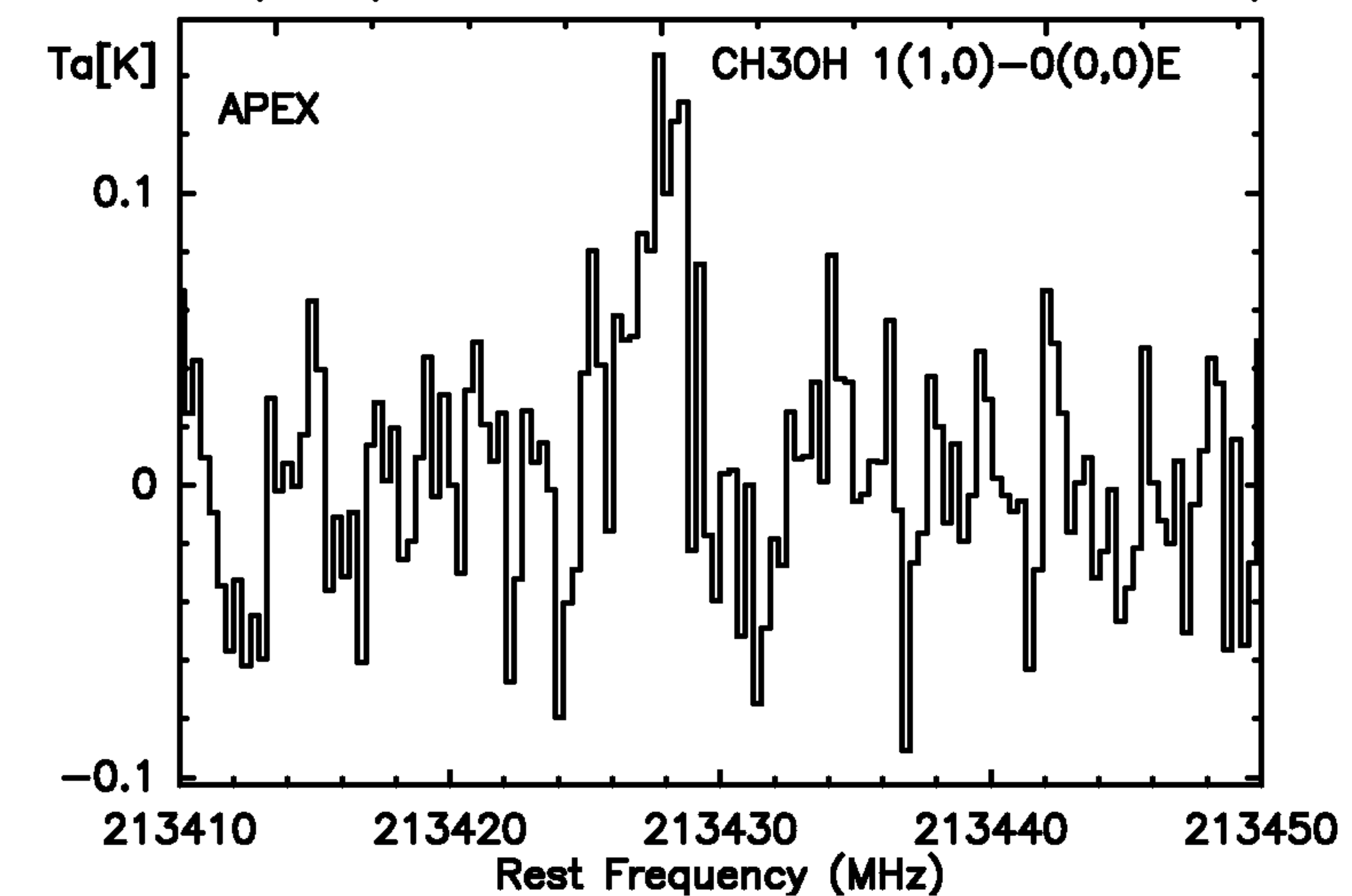}
\caption{Line spectra for IRDC316.76-1.}
\label{spectra3}
\end{figure}

\begin{figure}
\includegraphics[width=5.3cm]{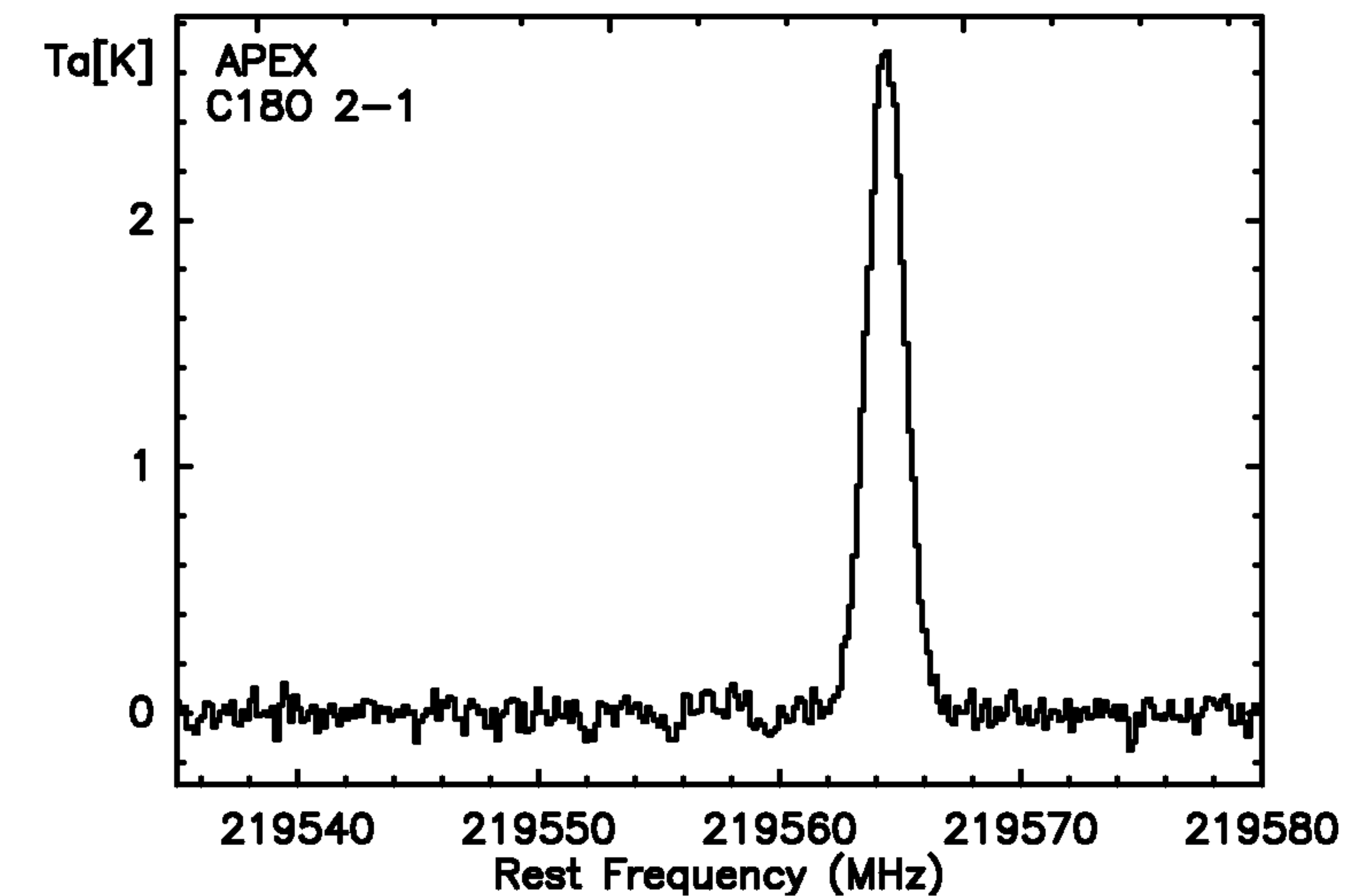}
\includegraphics[width=5.3cm]{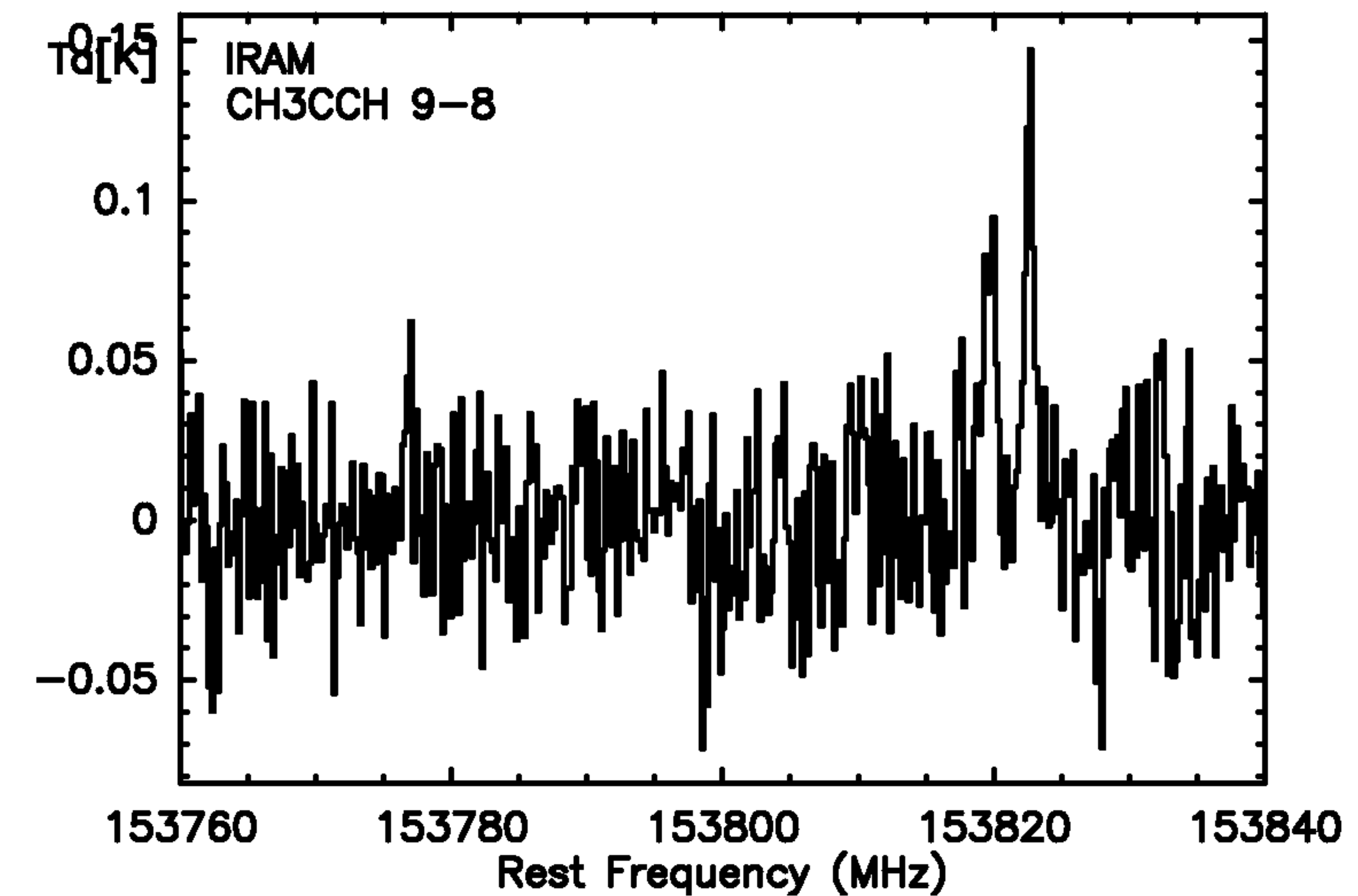}
\includegraphics[width=5.3cm]{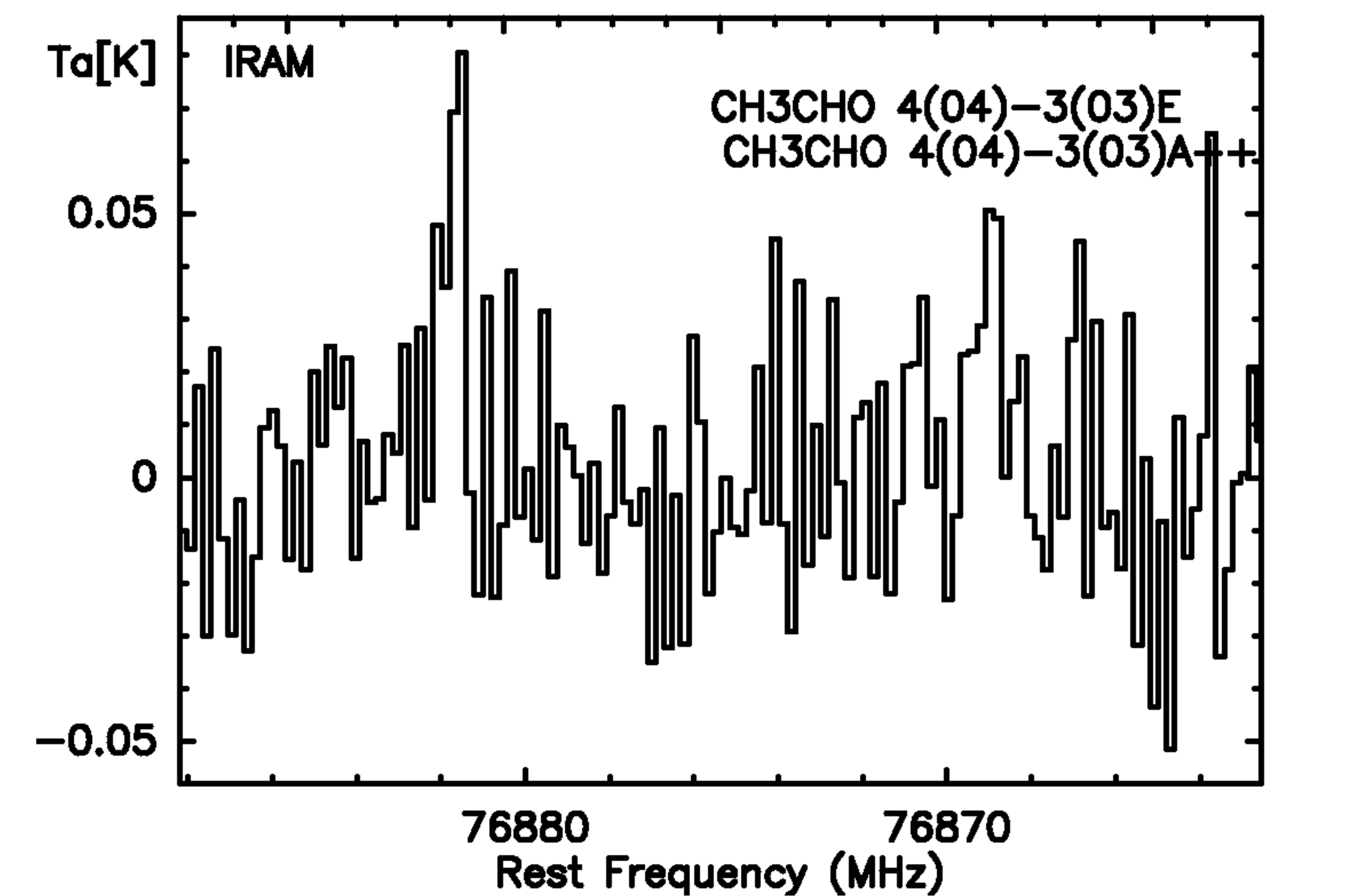}

\includegraphics[width=5.3cm]{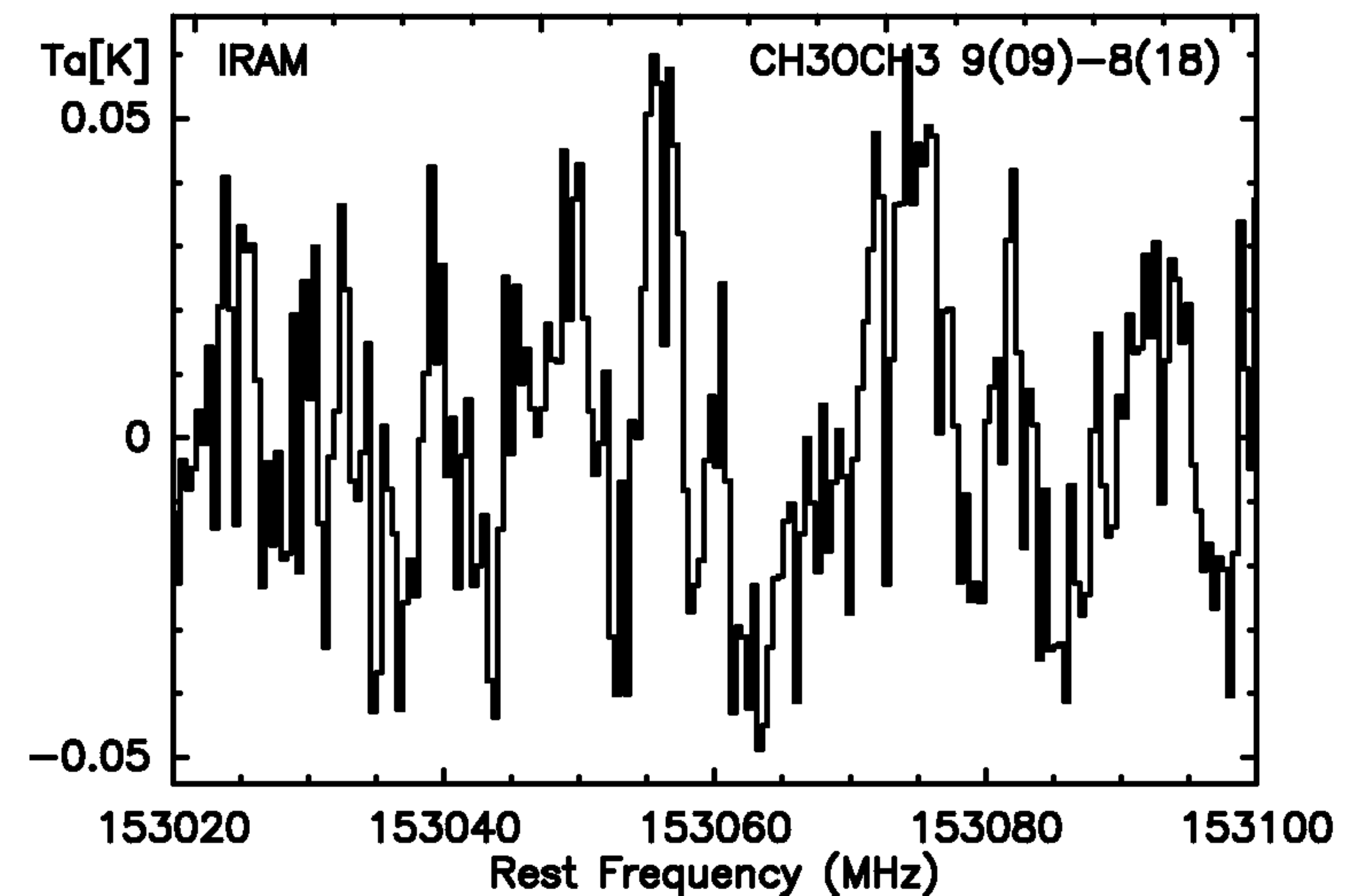}
\includegraphics[width=5.3cm]{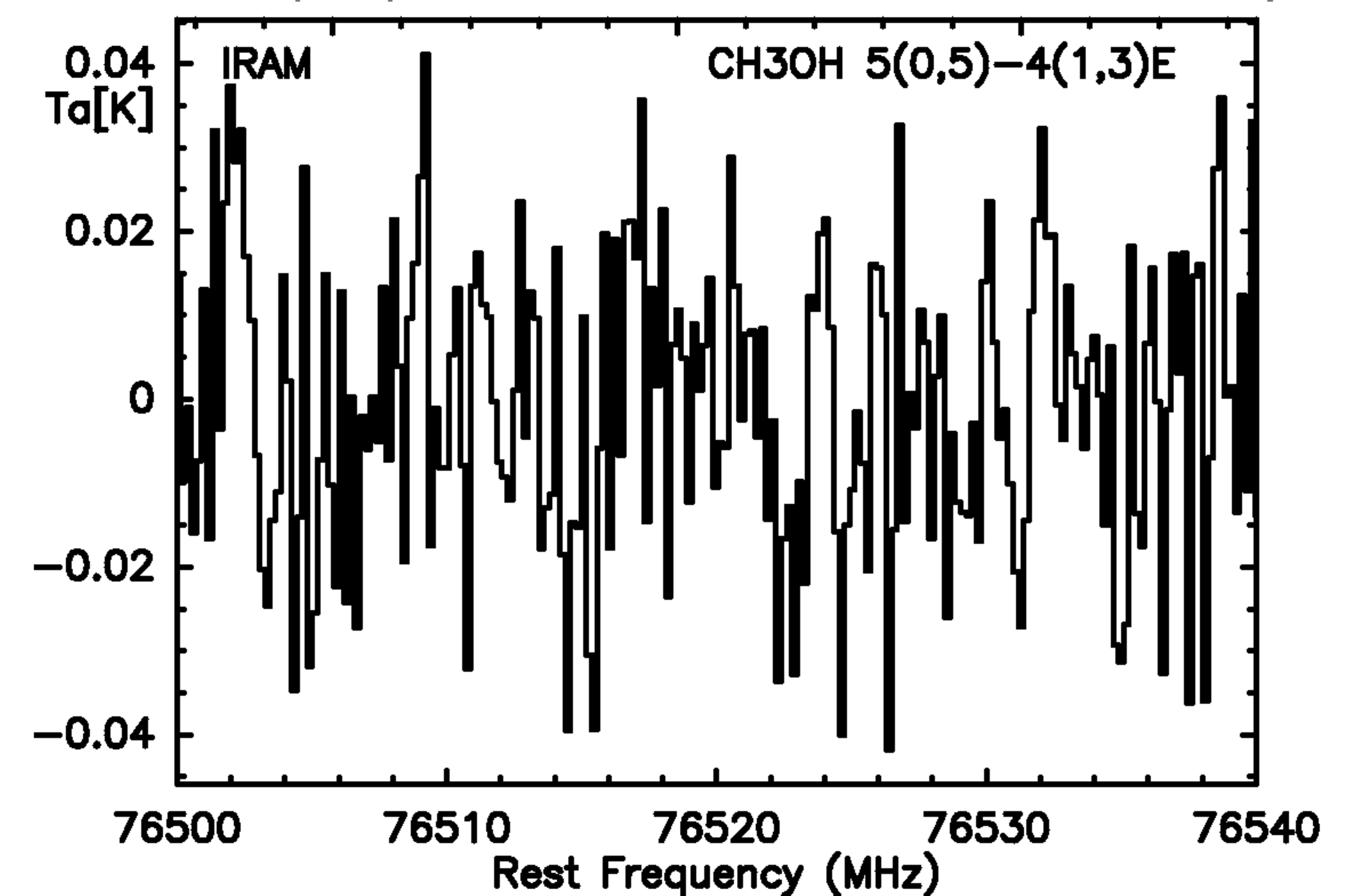}
\includegraphics[width=5.3cm]{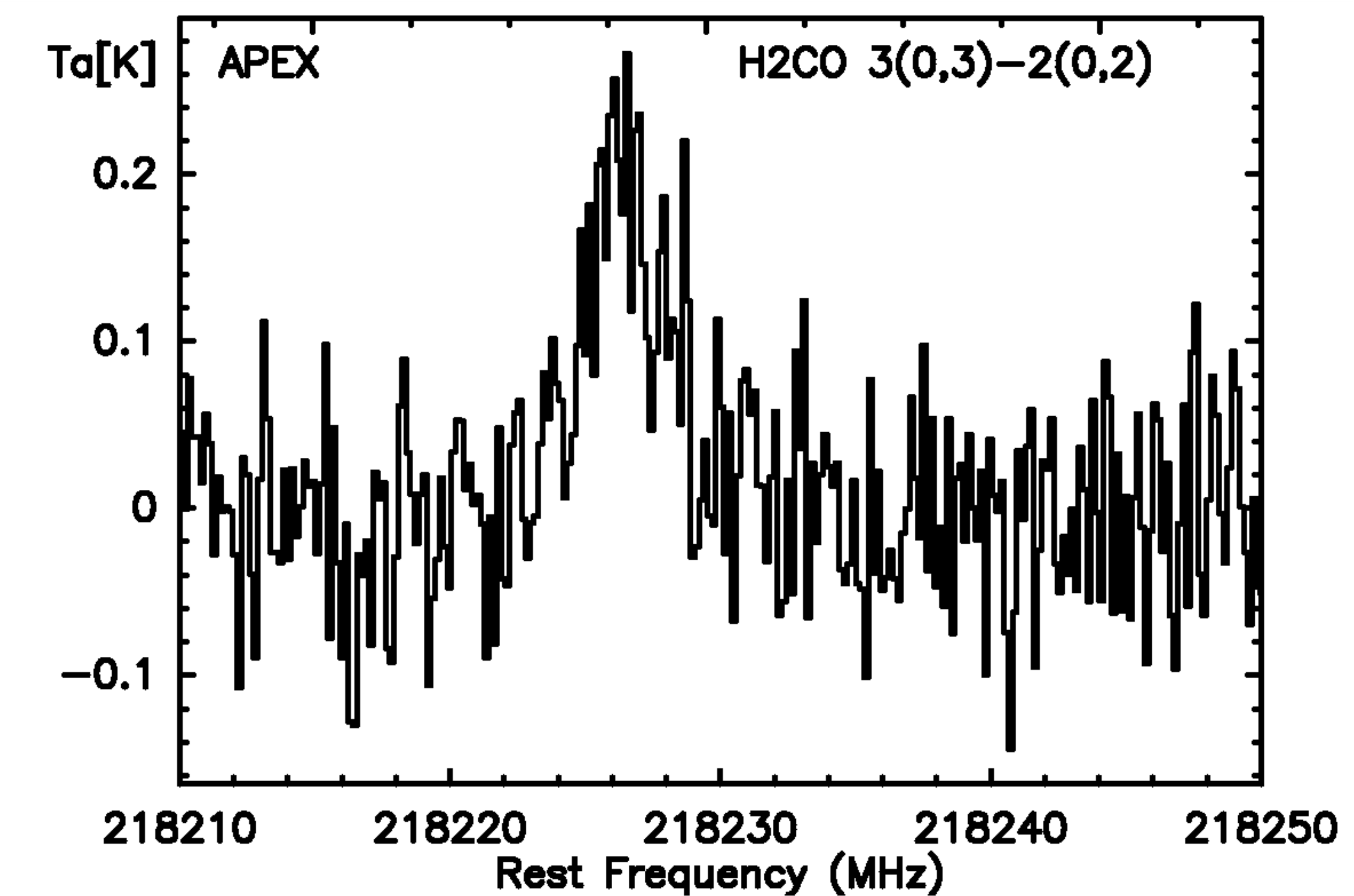}

\includegraphics[width=5.3cm]{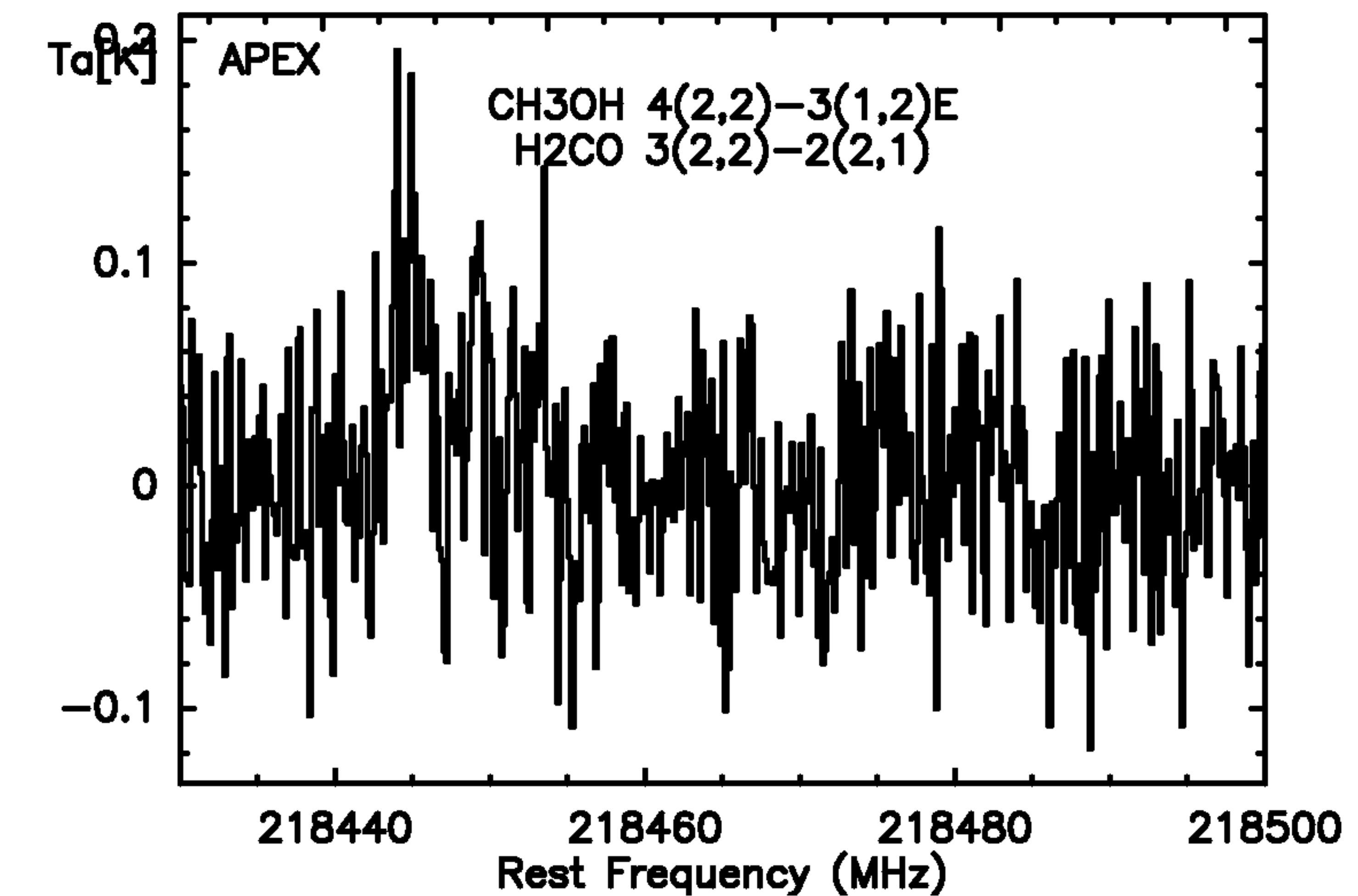}
\includegraphics[width=5.3cm]{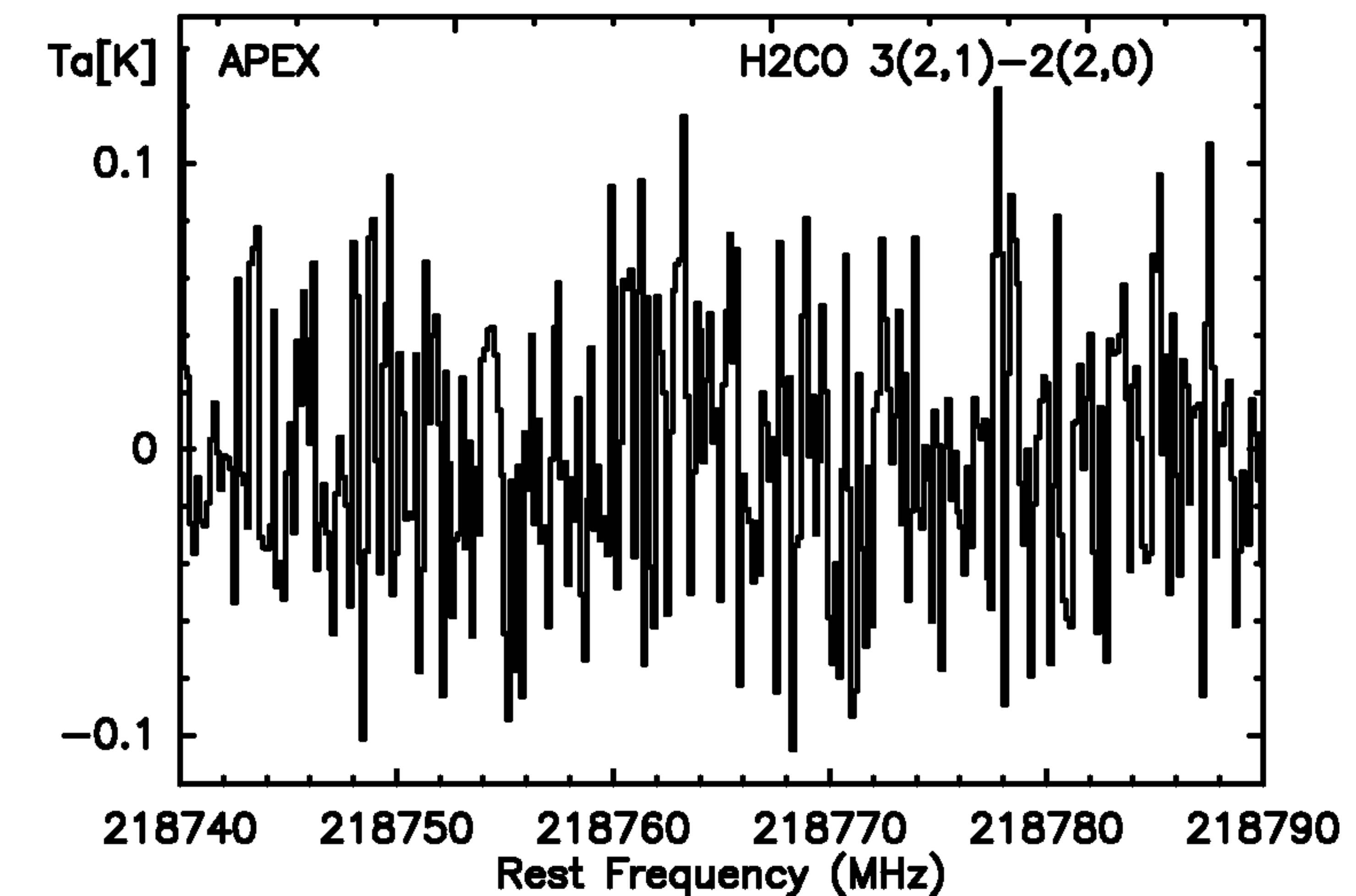}
\includegraphics[width=5.3cm]{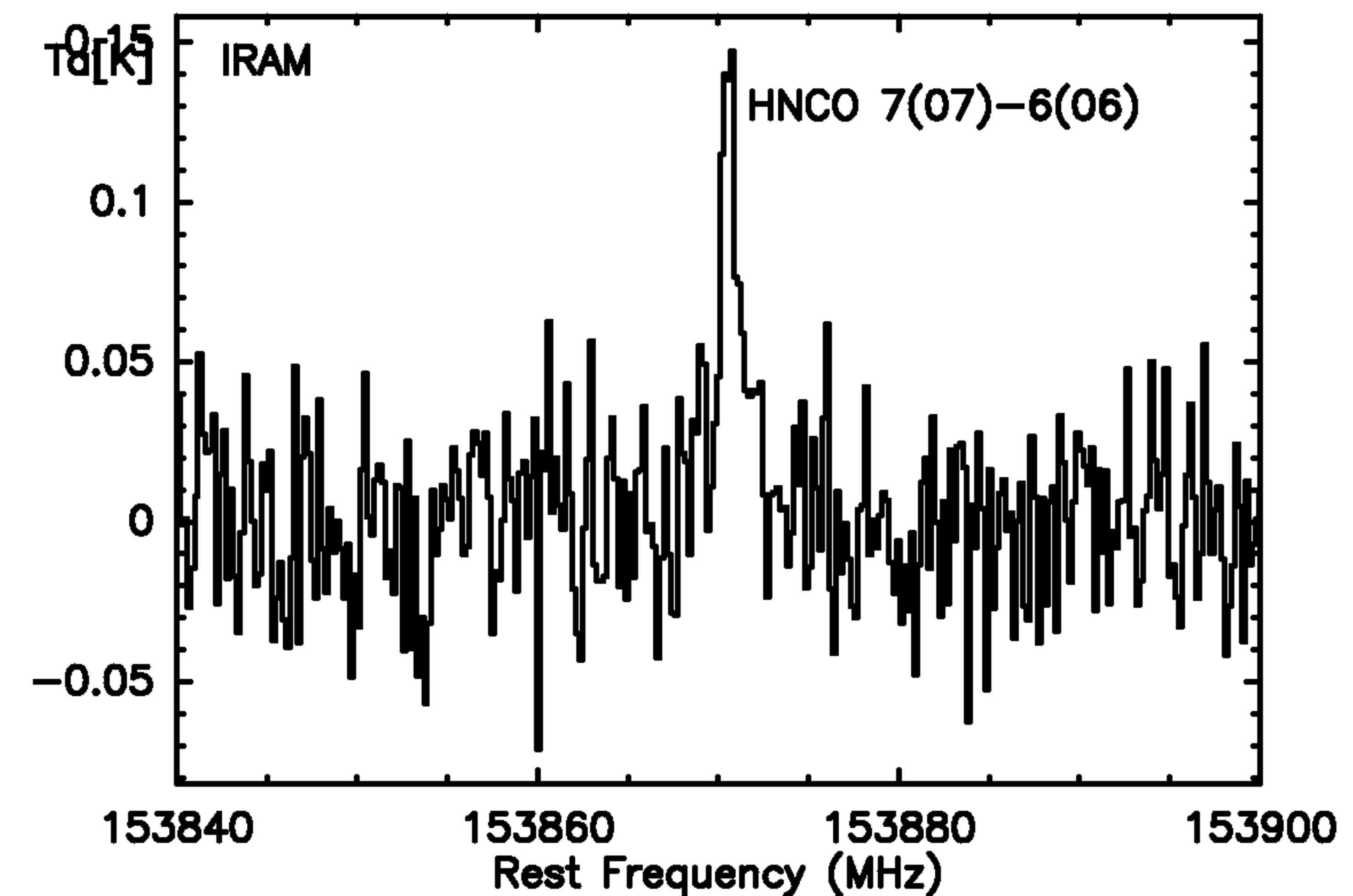}
\caption{Line spectra for IRDC015.05-3.}
\label{spectra4}
\end{figure}

\begin{figure}
\includegraphics[width=5.3cm]{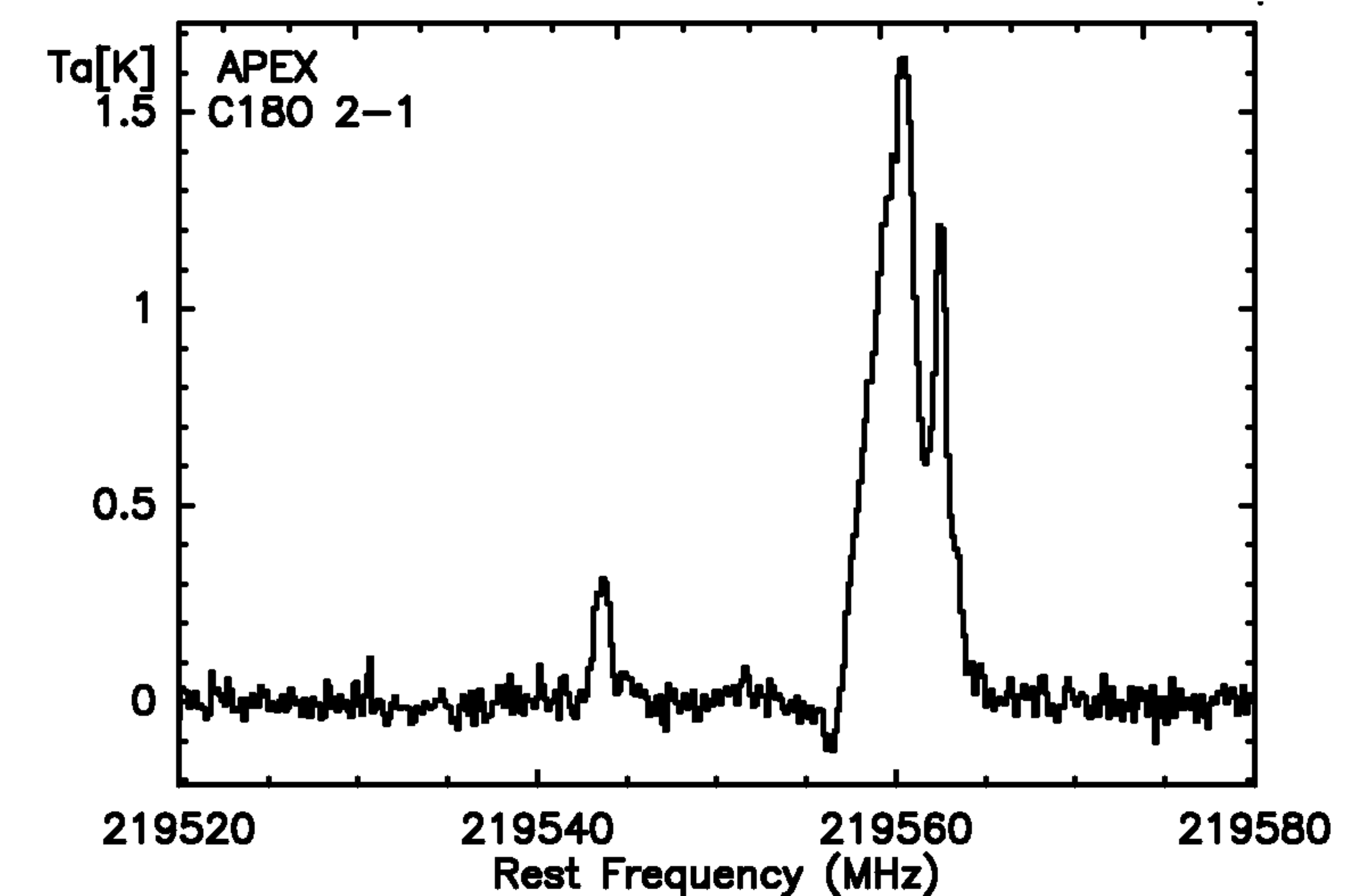}
\includegraphics[width=5.3cm]{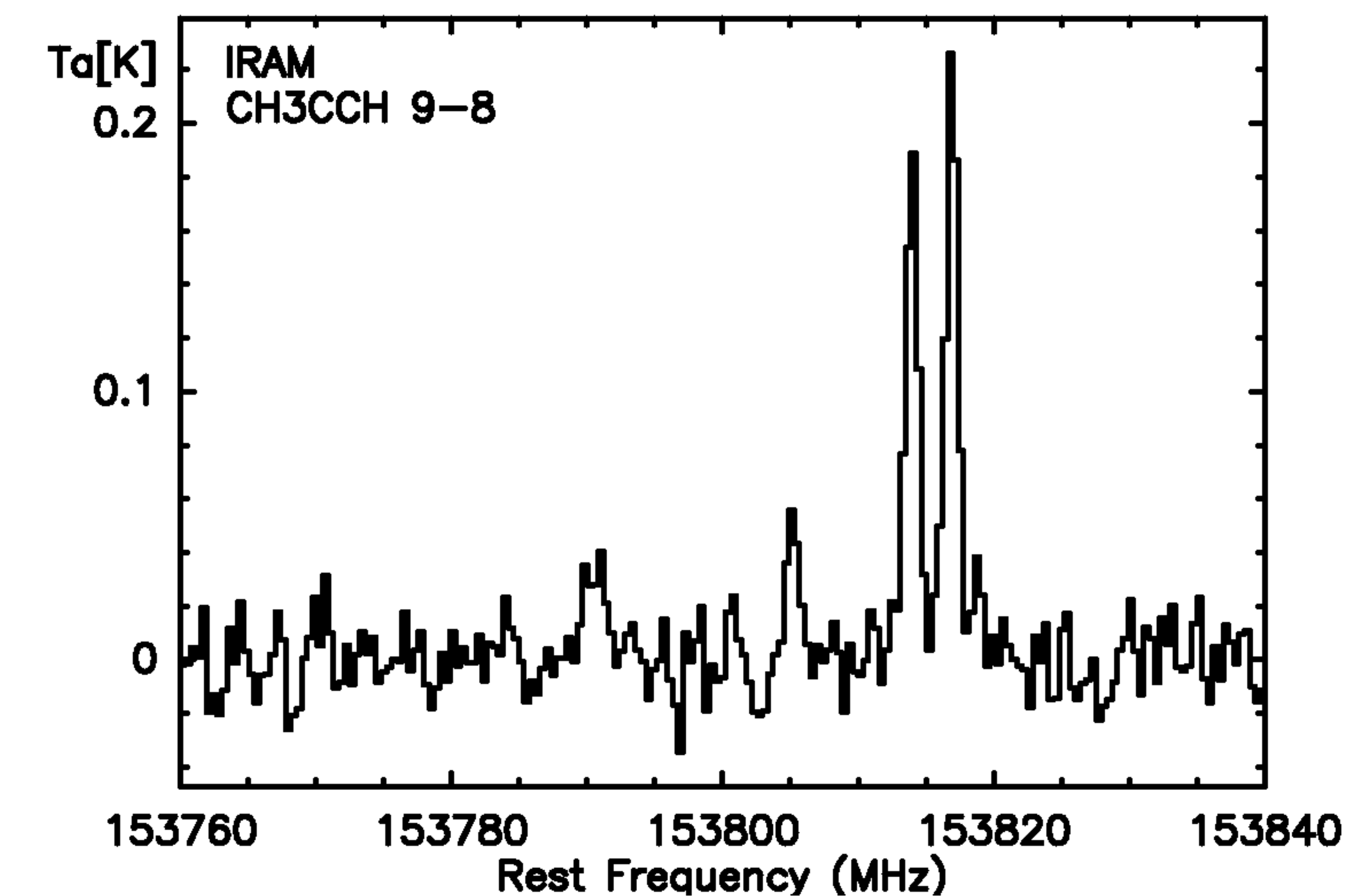}
\includegraphics[width=5.3cm]{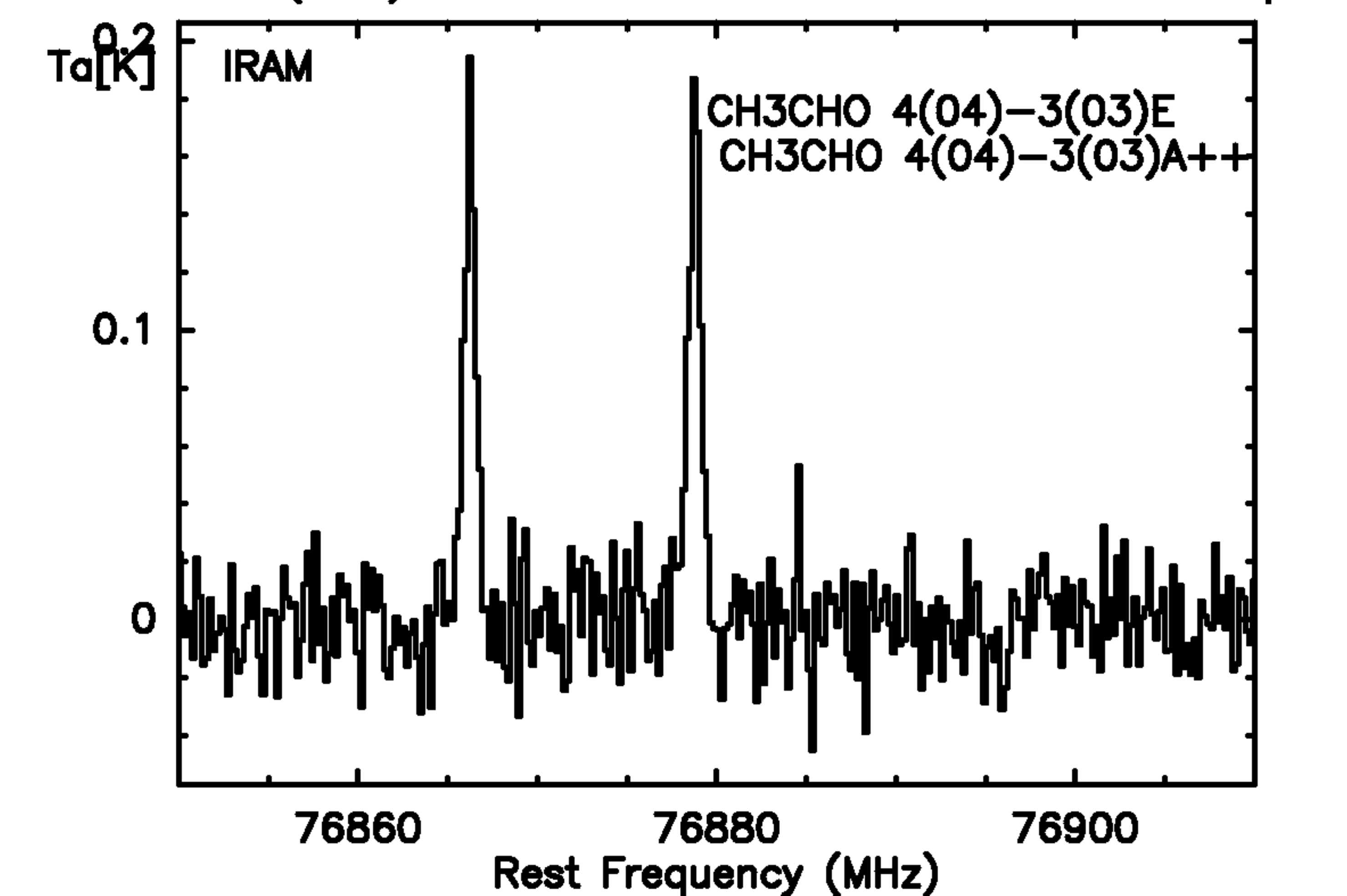}

\includegraphics[width=5.3cm]{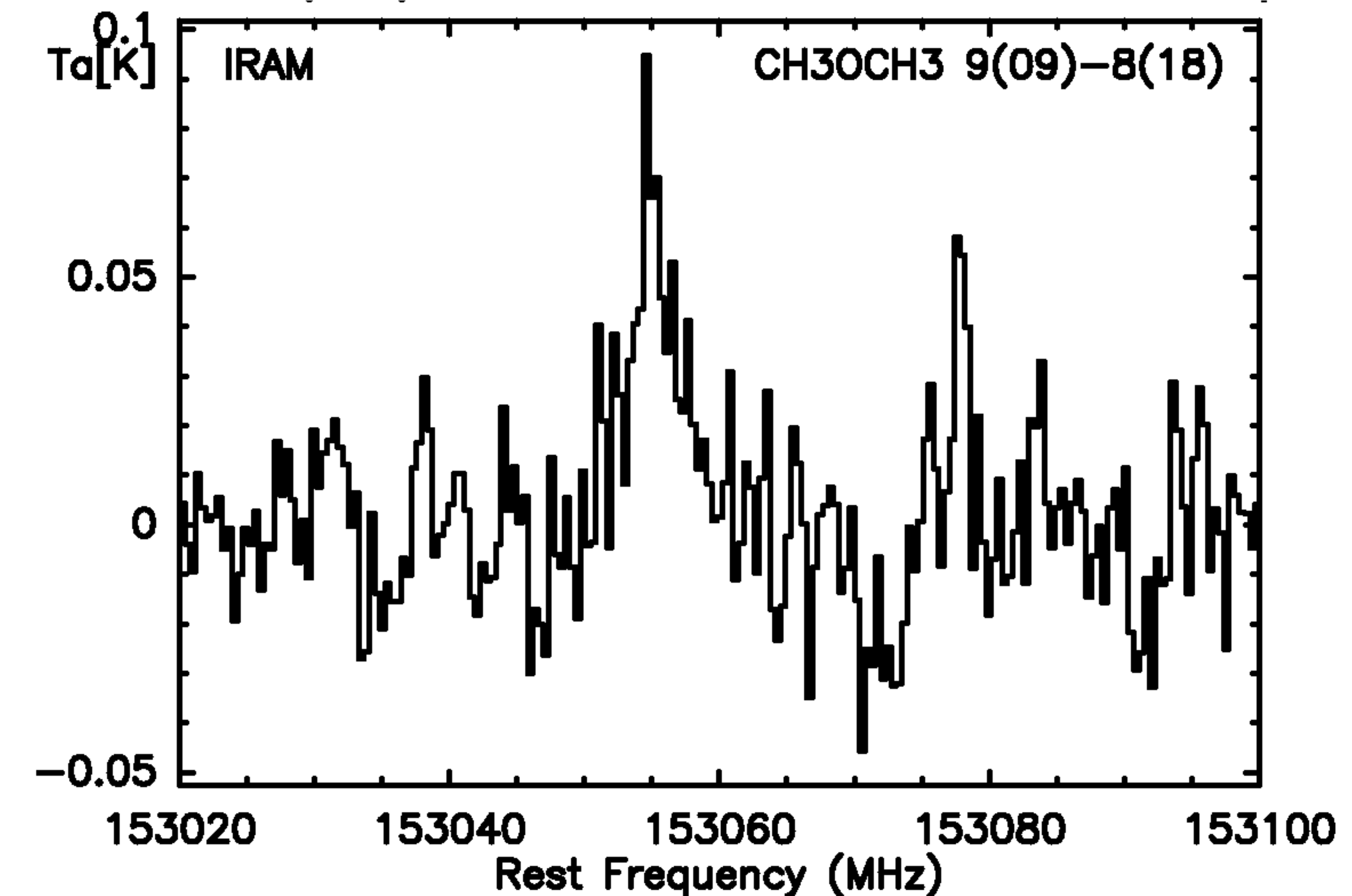}
\includegraphics[width=5.3cm]{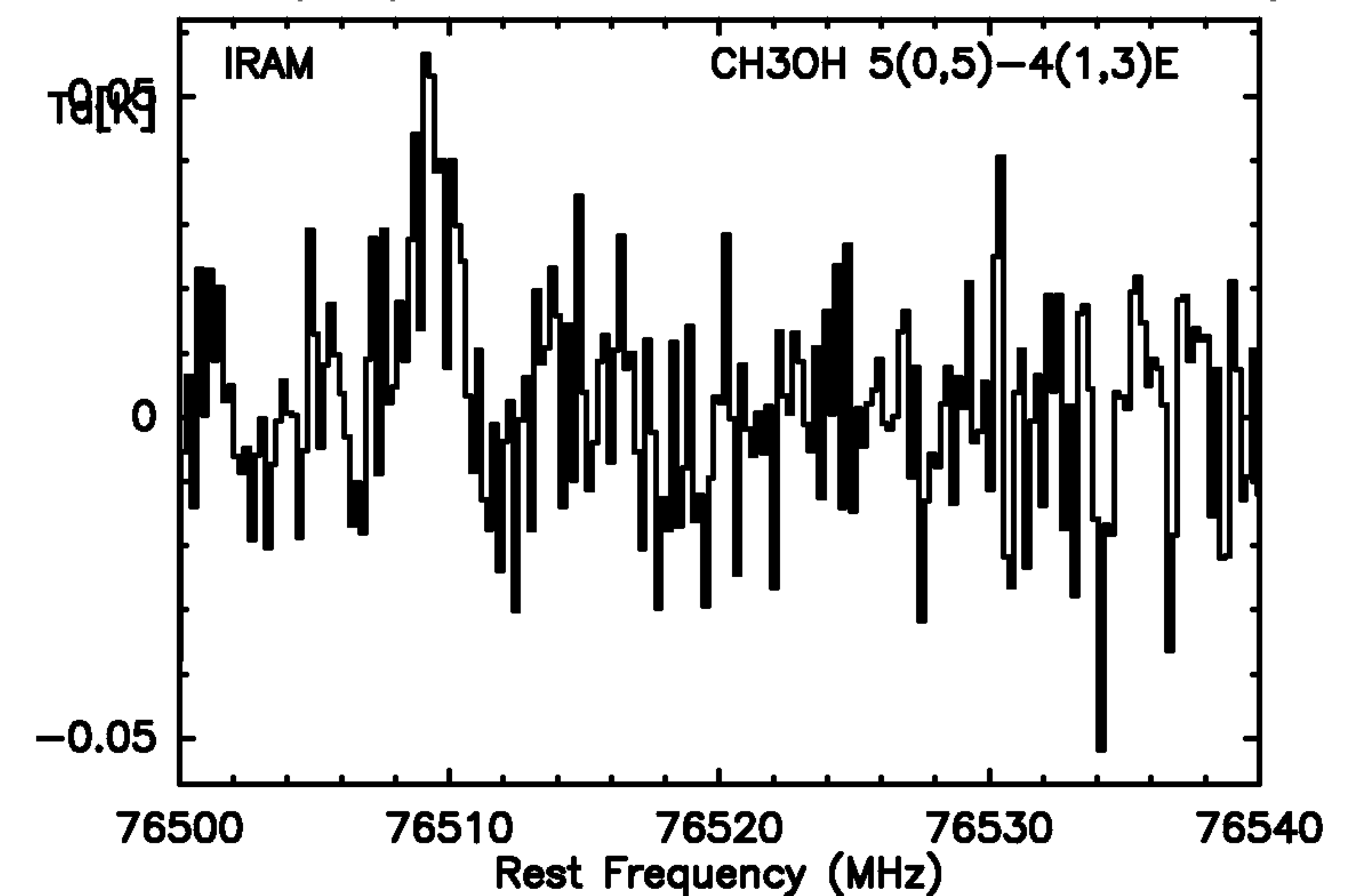}
\includegraphics[width=5.3cm]{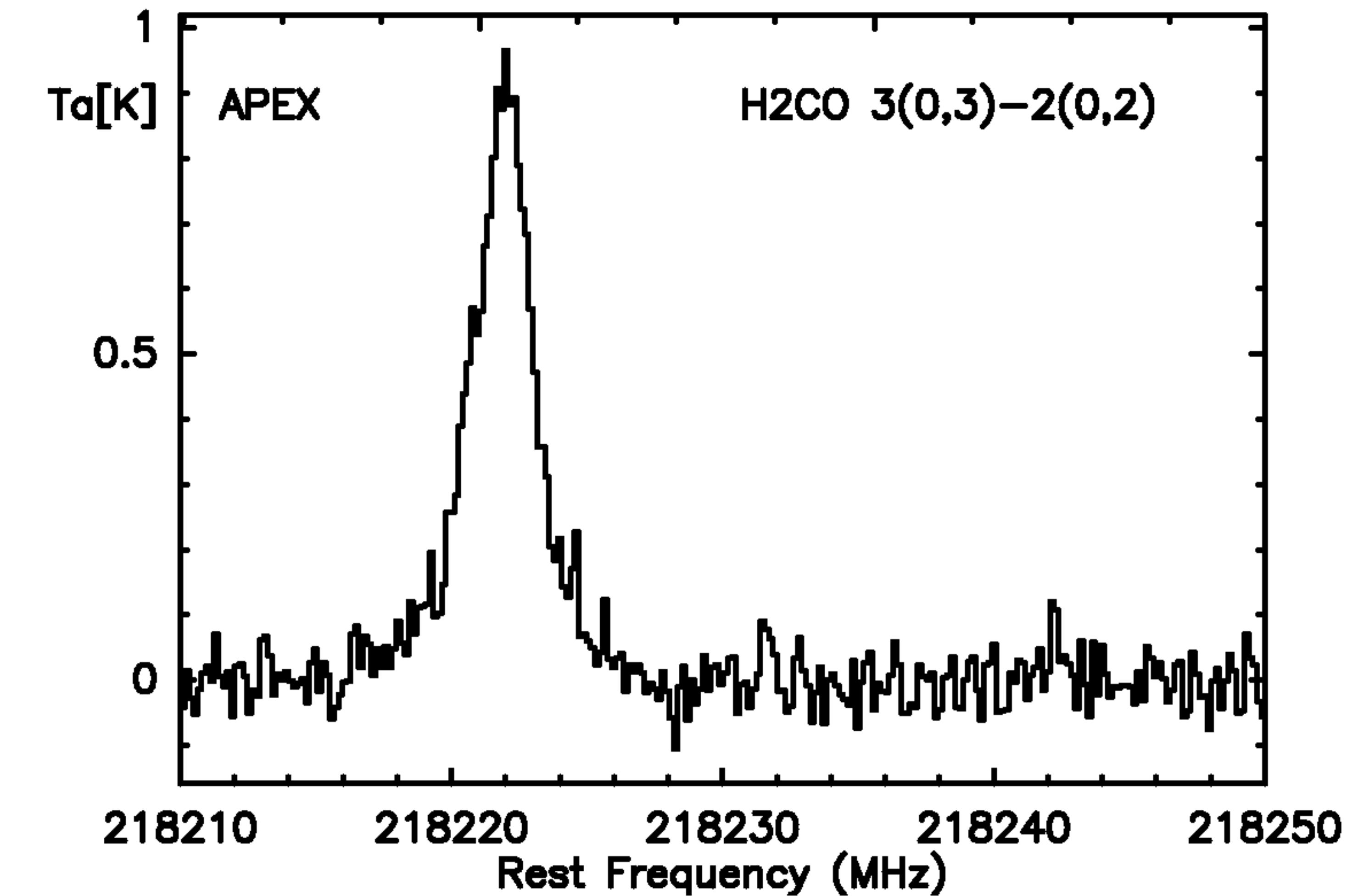}

\includegraphics[width=5.3cm]{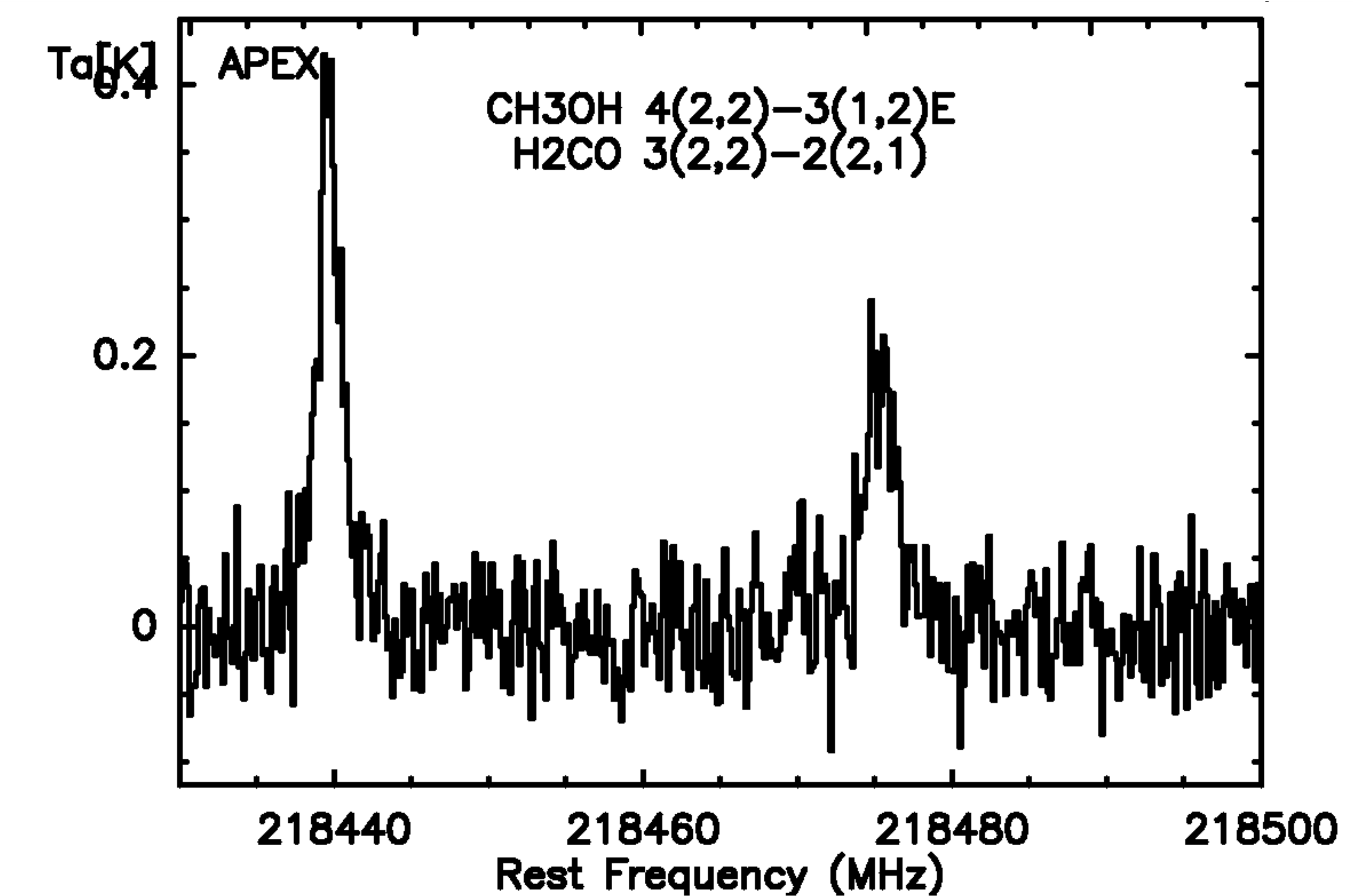}
\includegraphics[width=5.3cm]{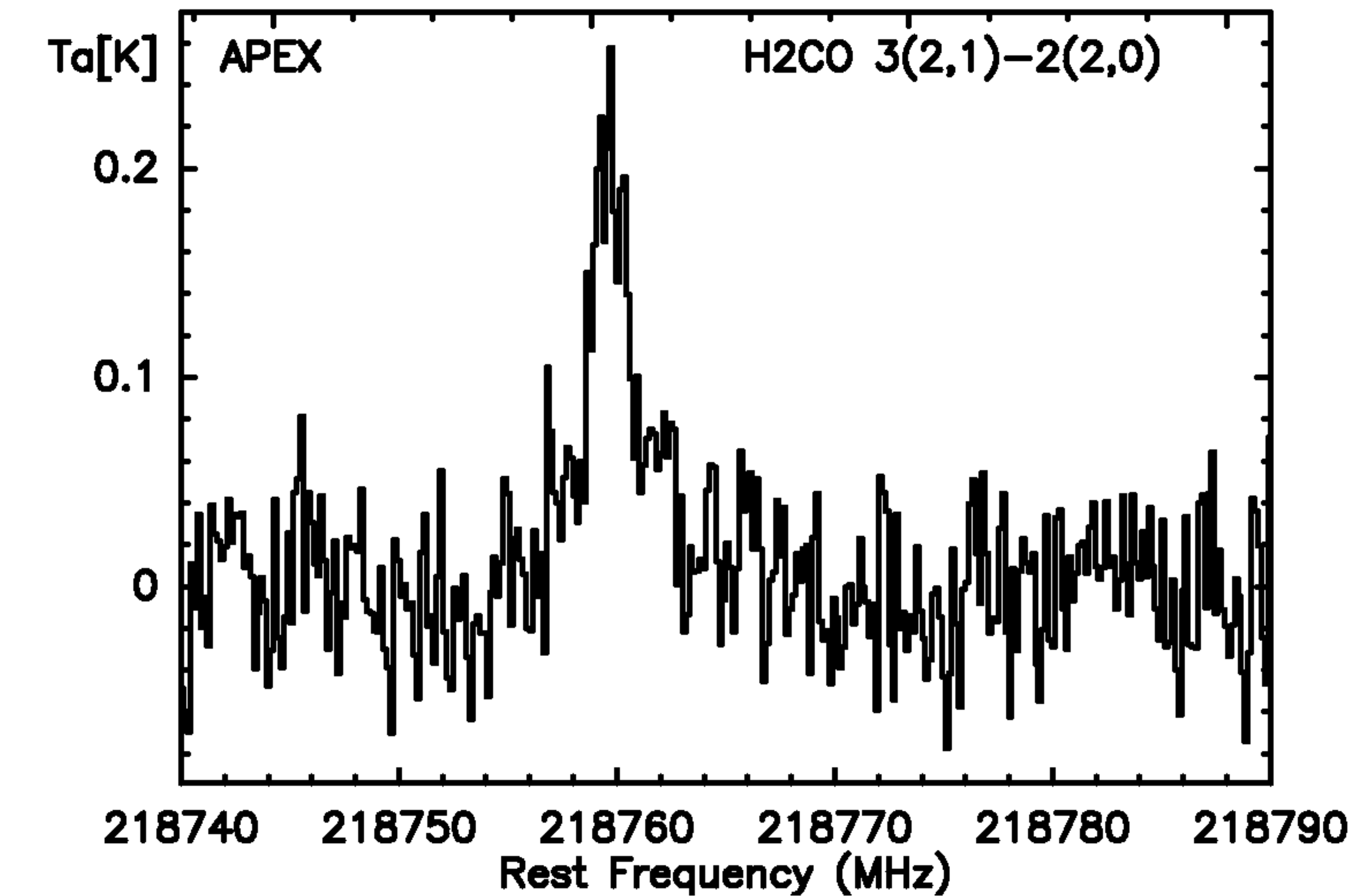}
\includegraphics[width=5.3cm]{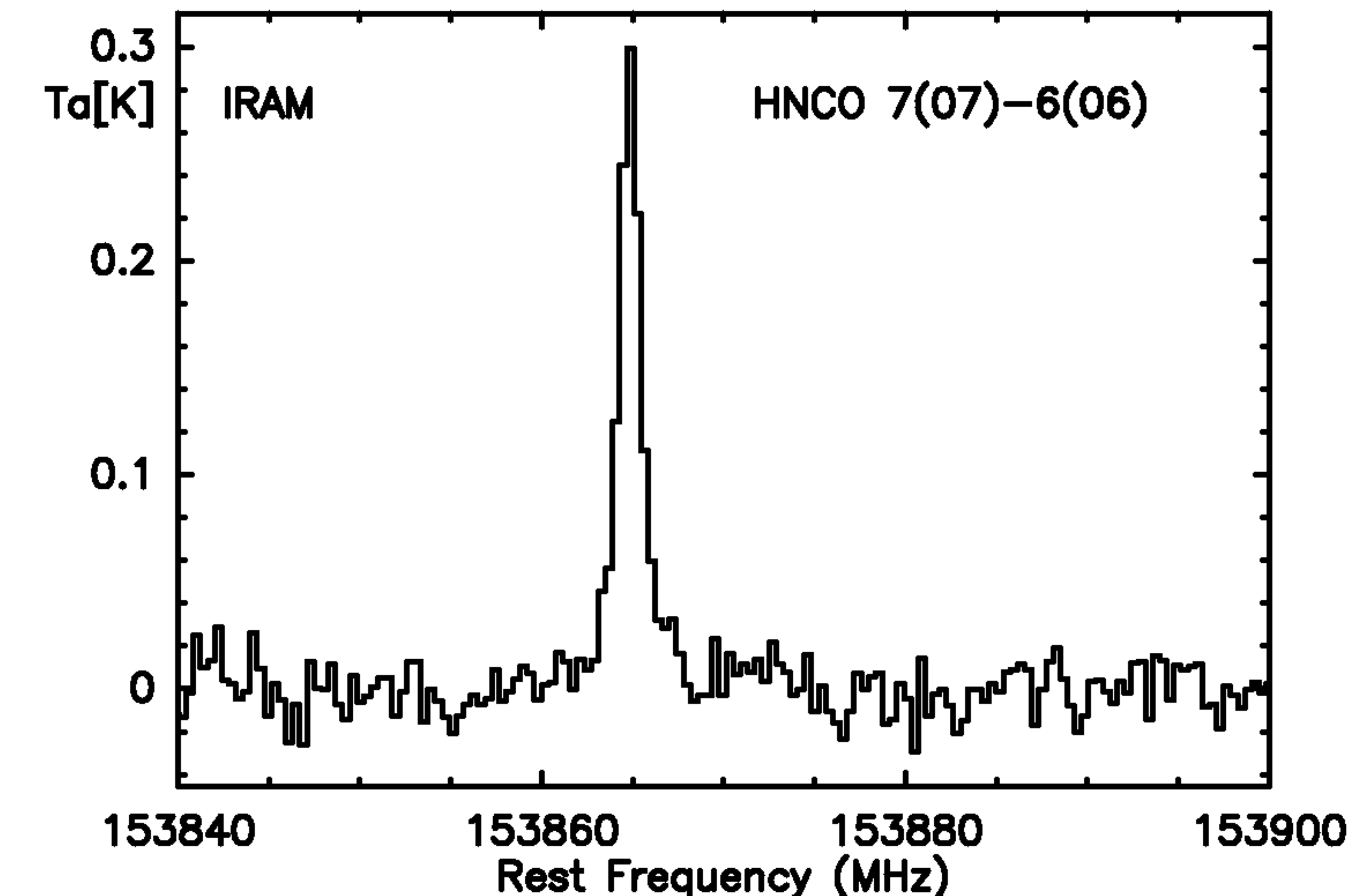}

\includegraphics[width=5.3cm]{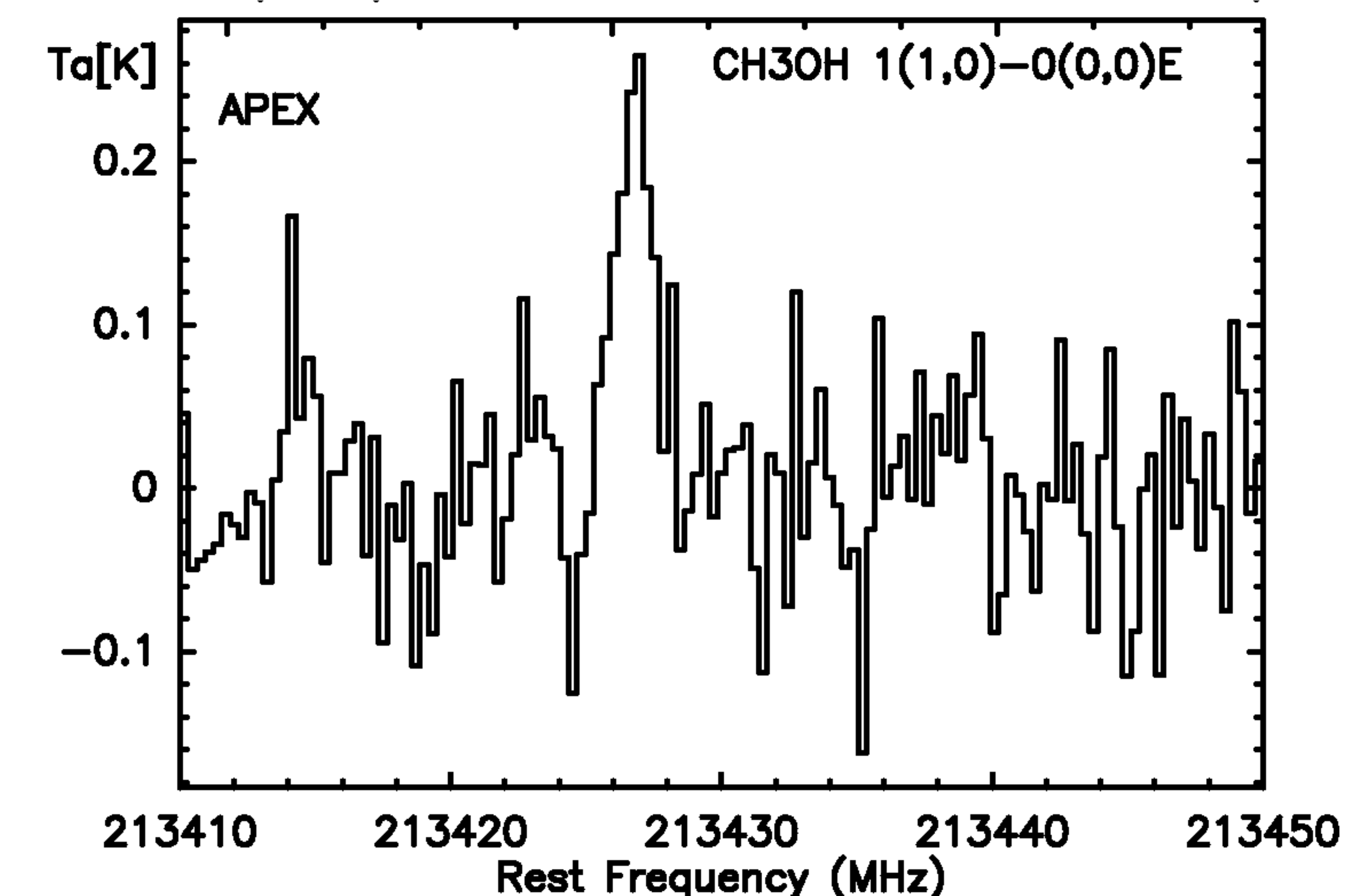}
\caption{Line spectra for IRDC028.34-3.}
\label{spectra5}
\end{figure}

\begin{figure}
\includegraphics[width=5.3cm]{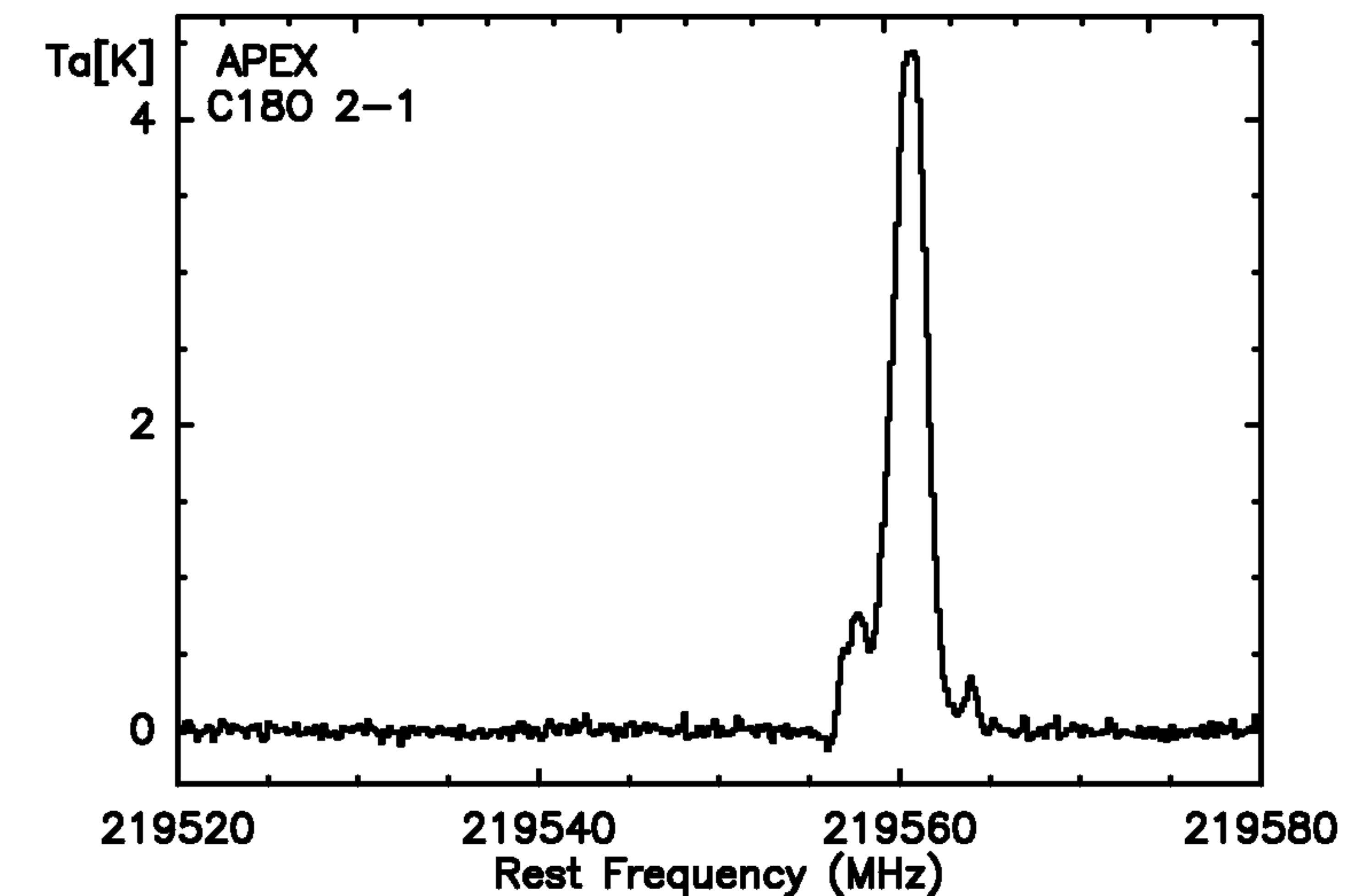}
\includegraphics[width=5.3cm]{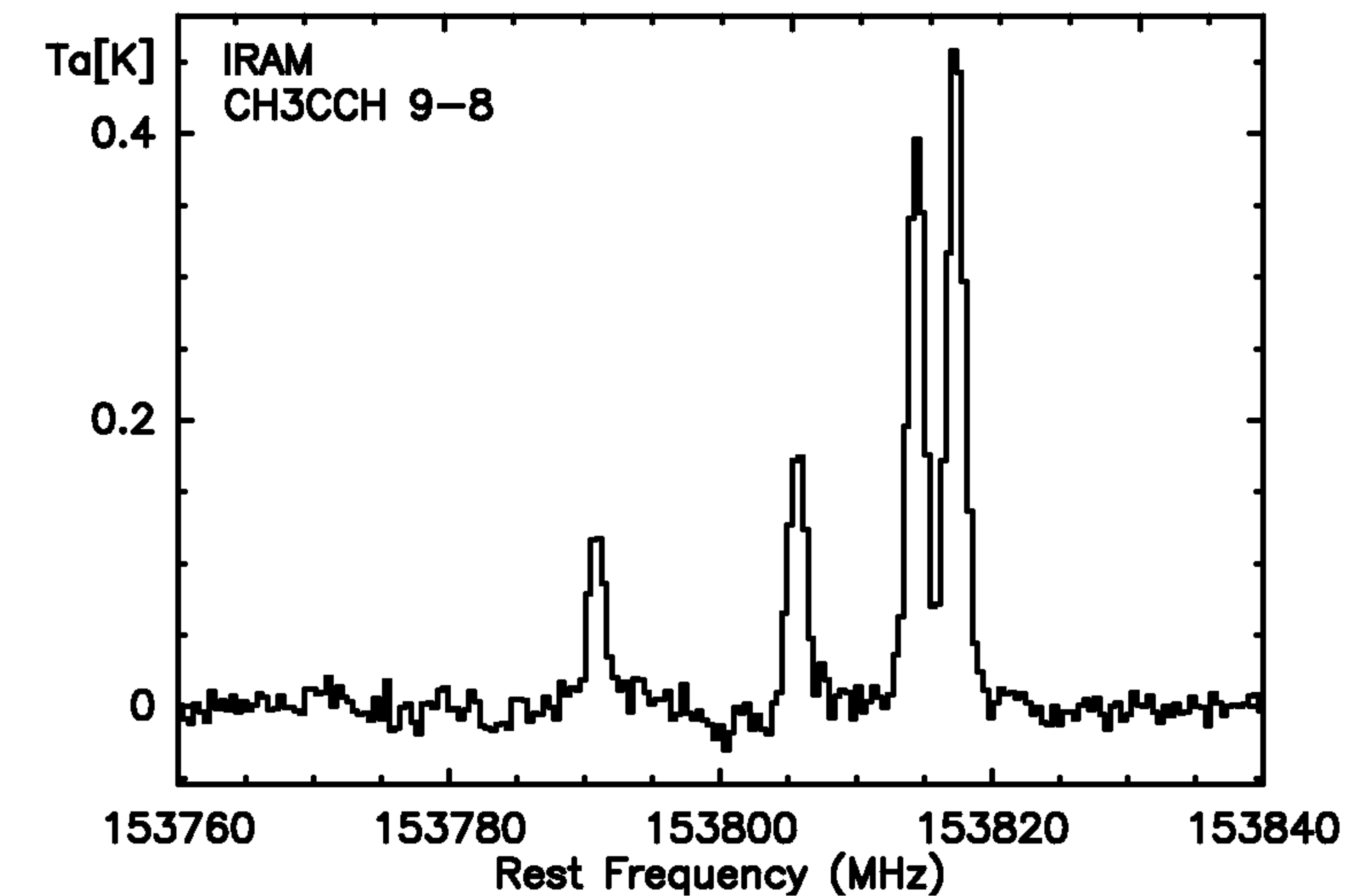}
\includegraphics[width=5.3cm]{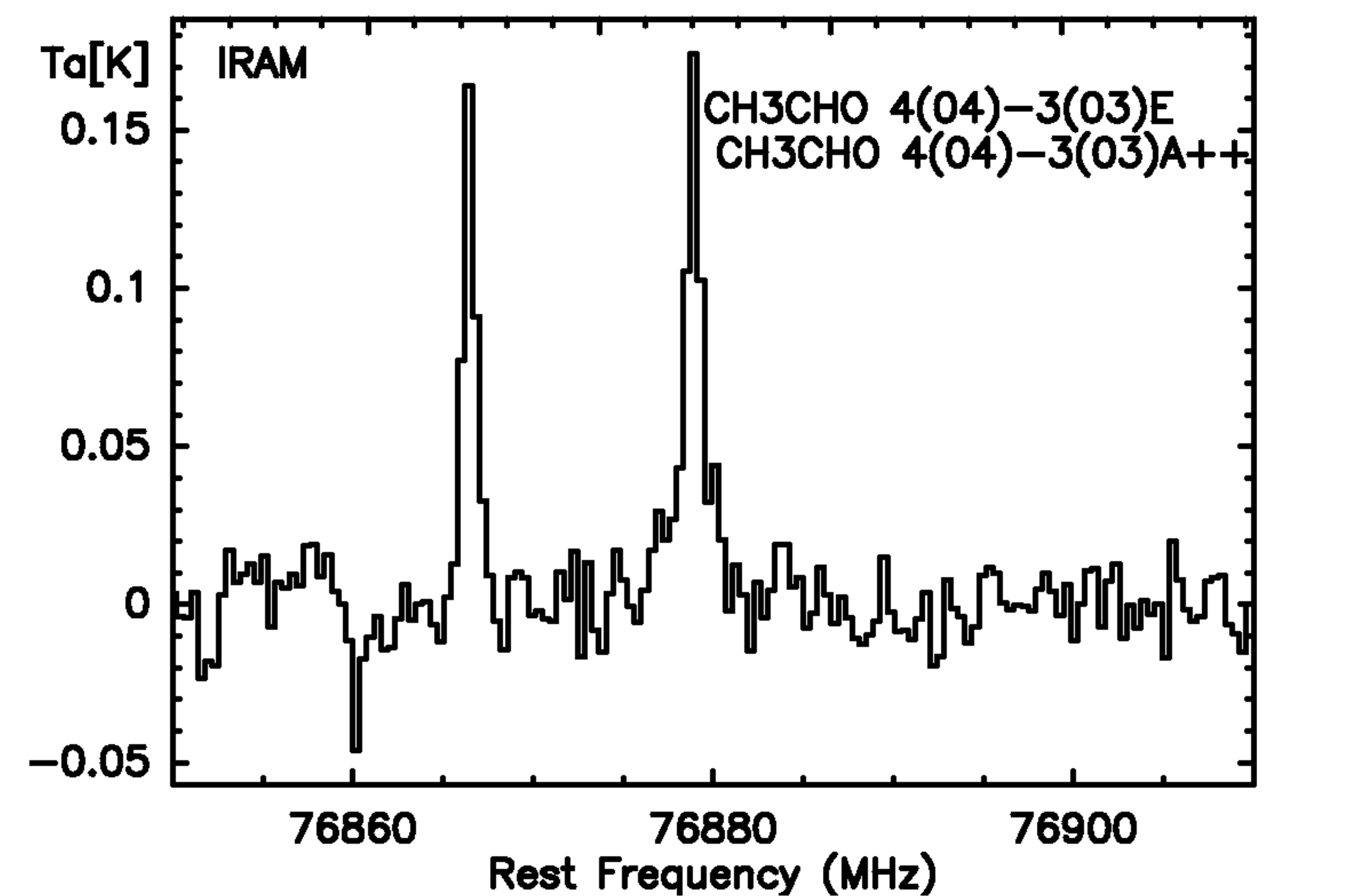}

\includegraphics[width=5.3cm]{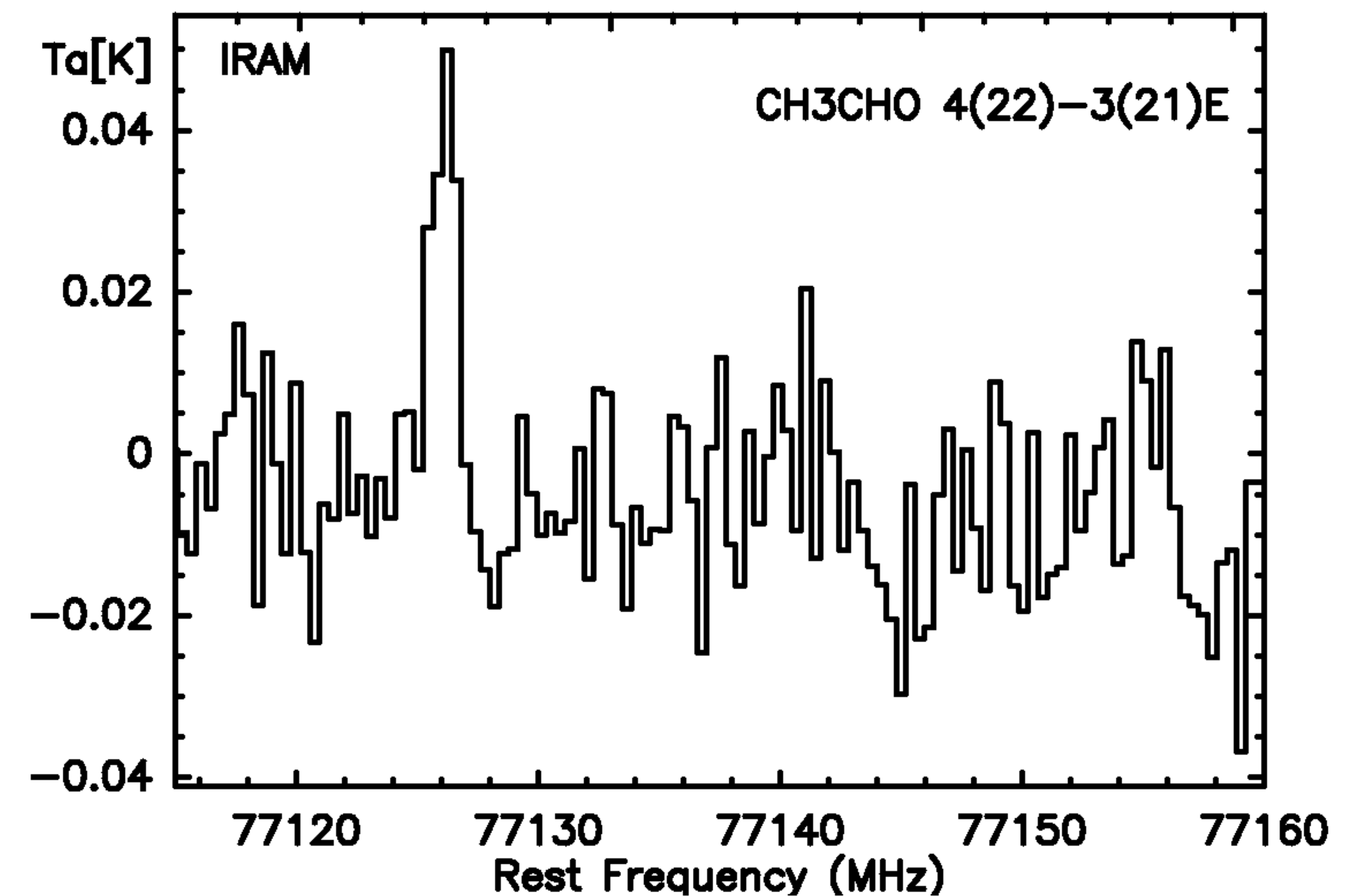}
\includegraphics[width=5.3cm]{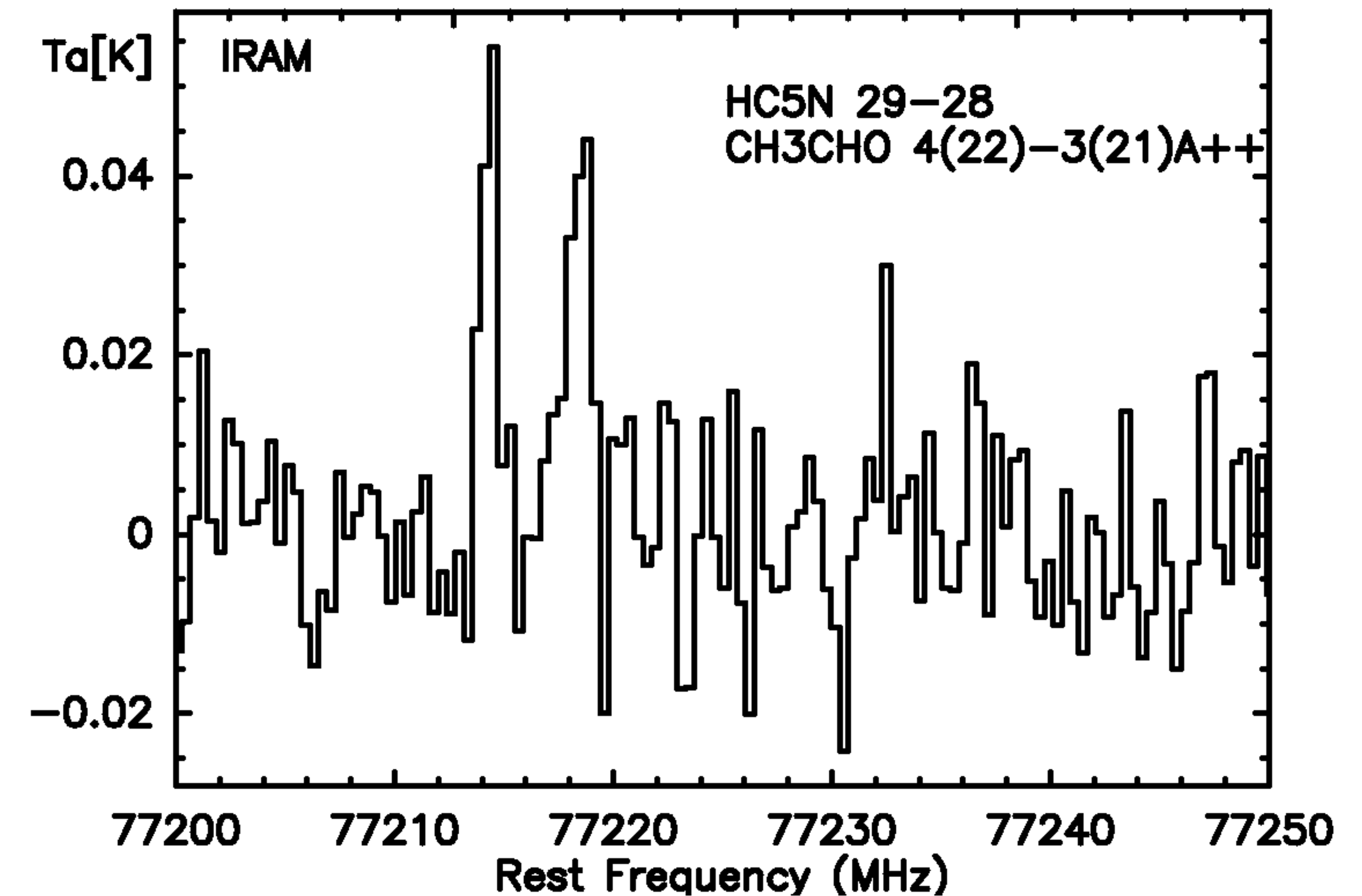}
\includegraphics[width=5.3cm]{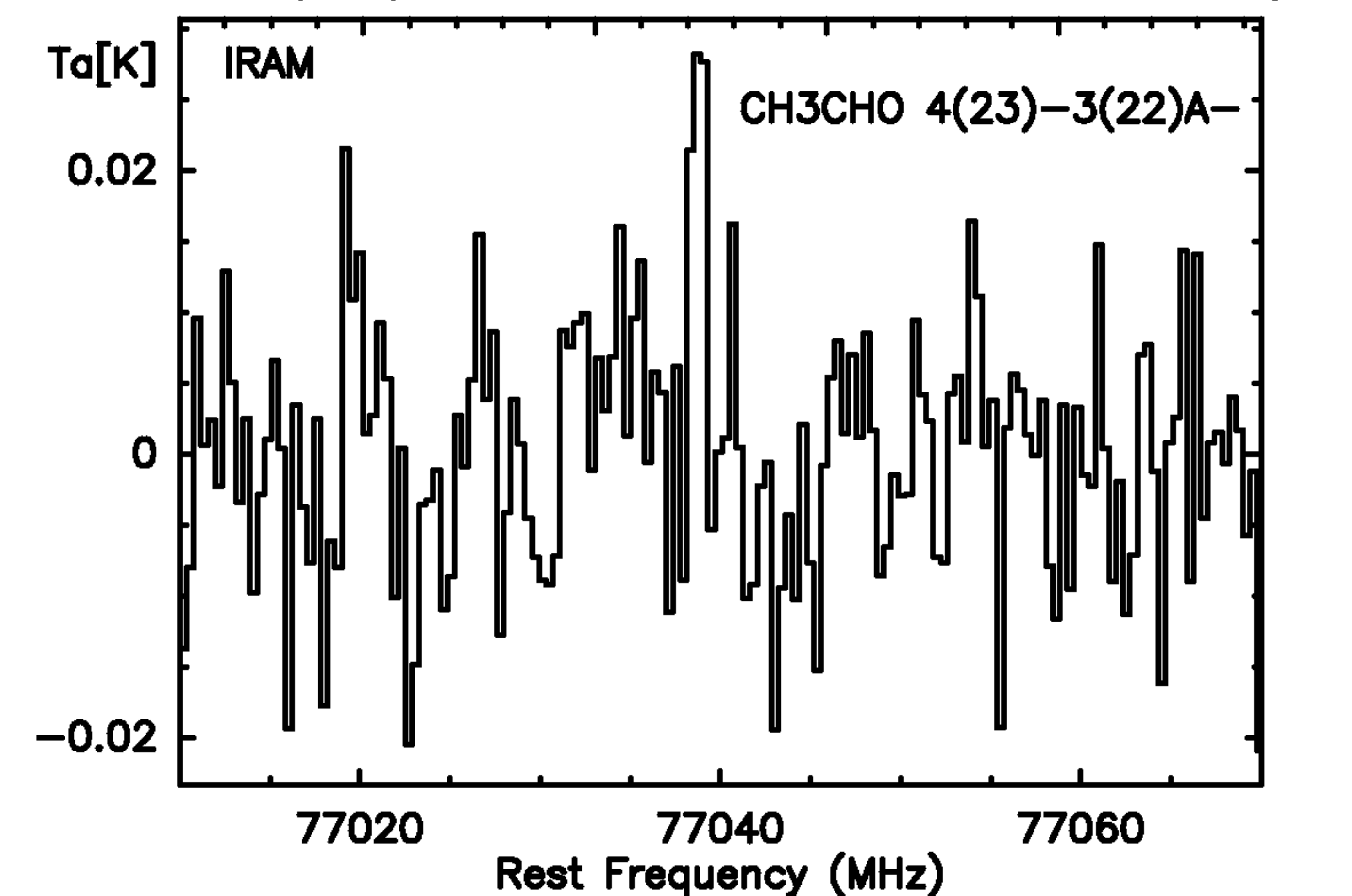}

\includegraphics[width=5.3cm]{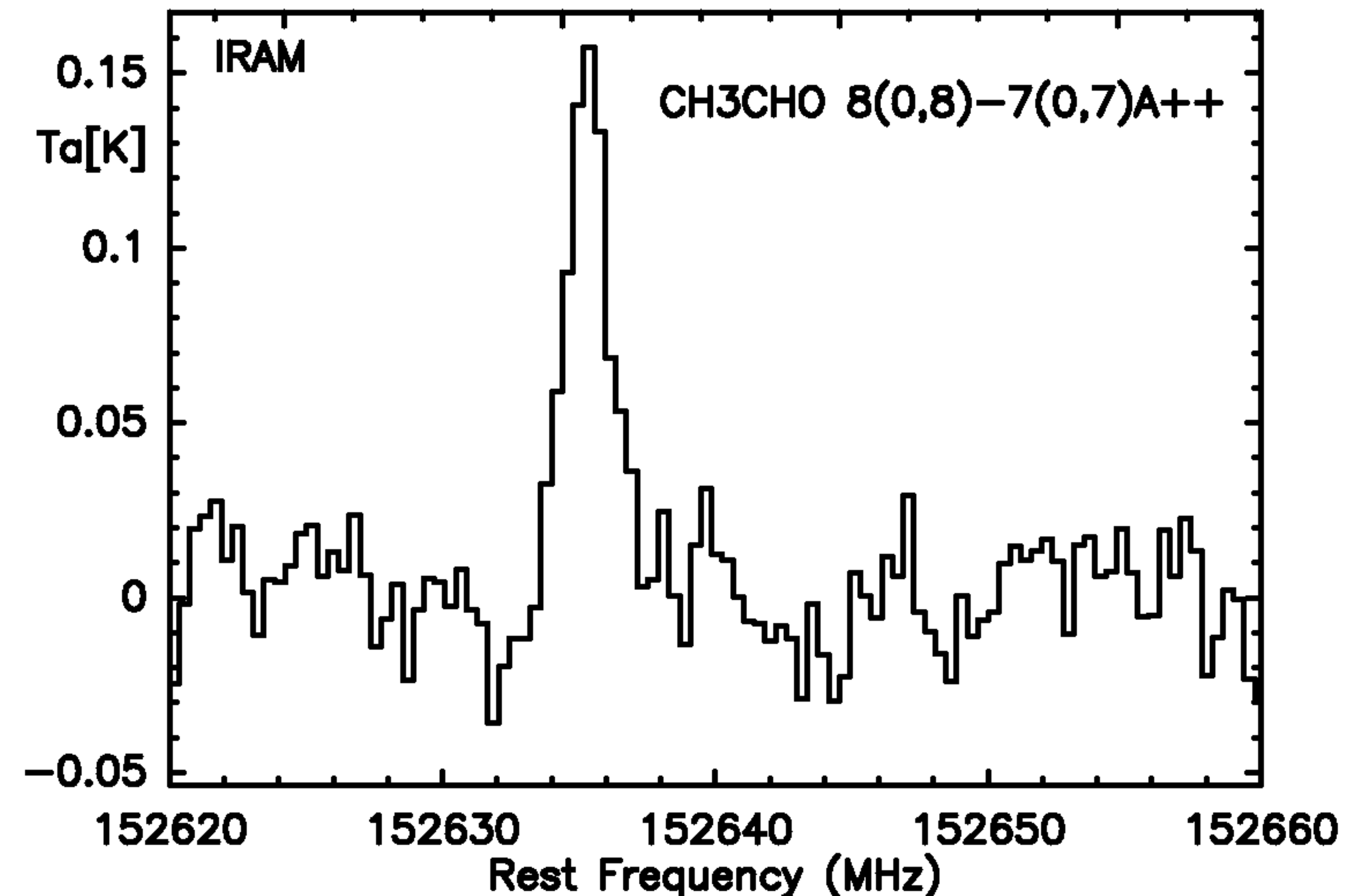}
\includegraphics[width=5.3cm]{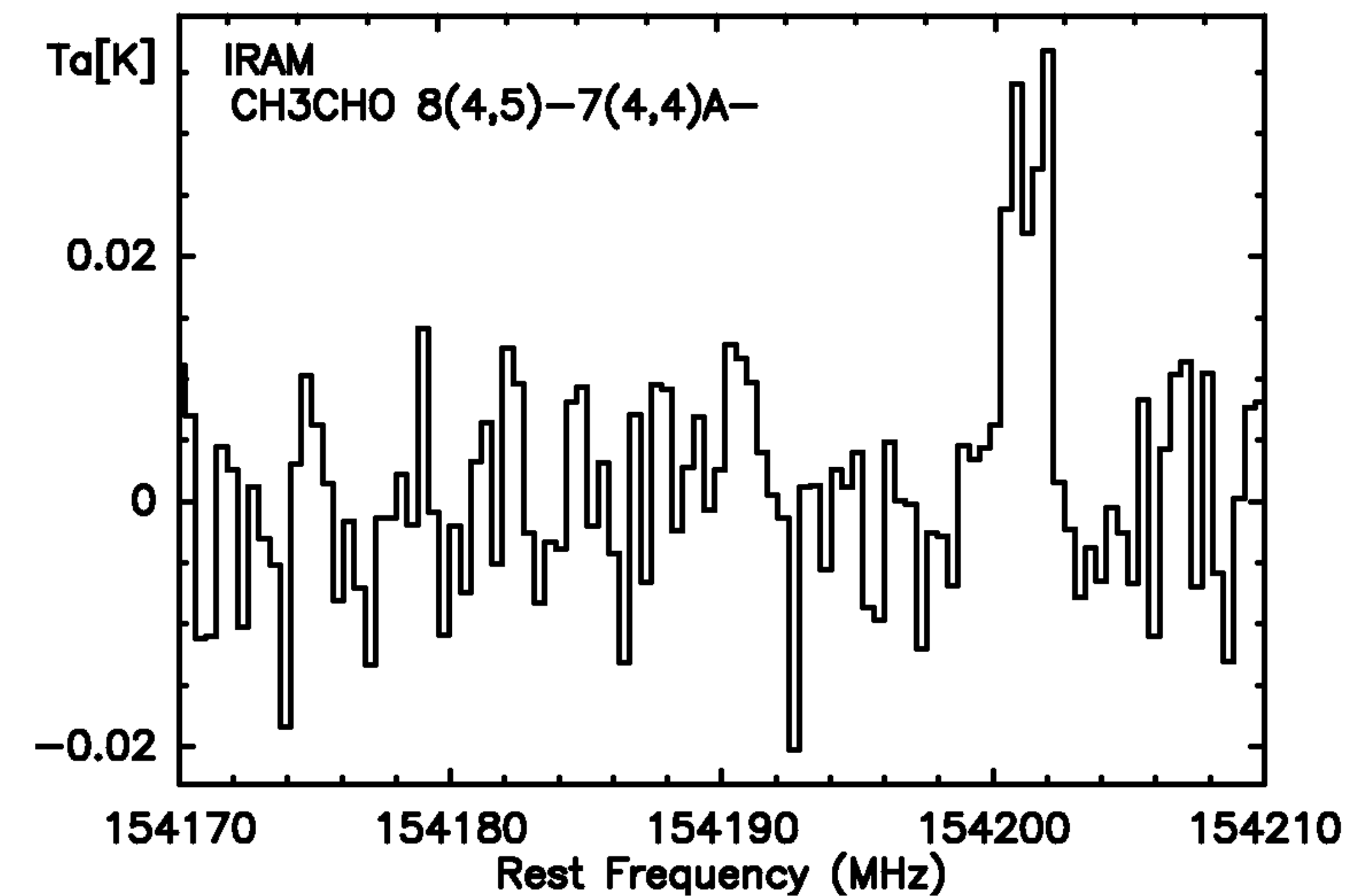}
\includegraphics[width=5.3cm]{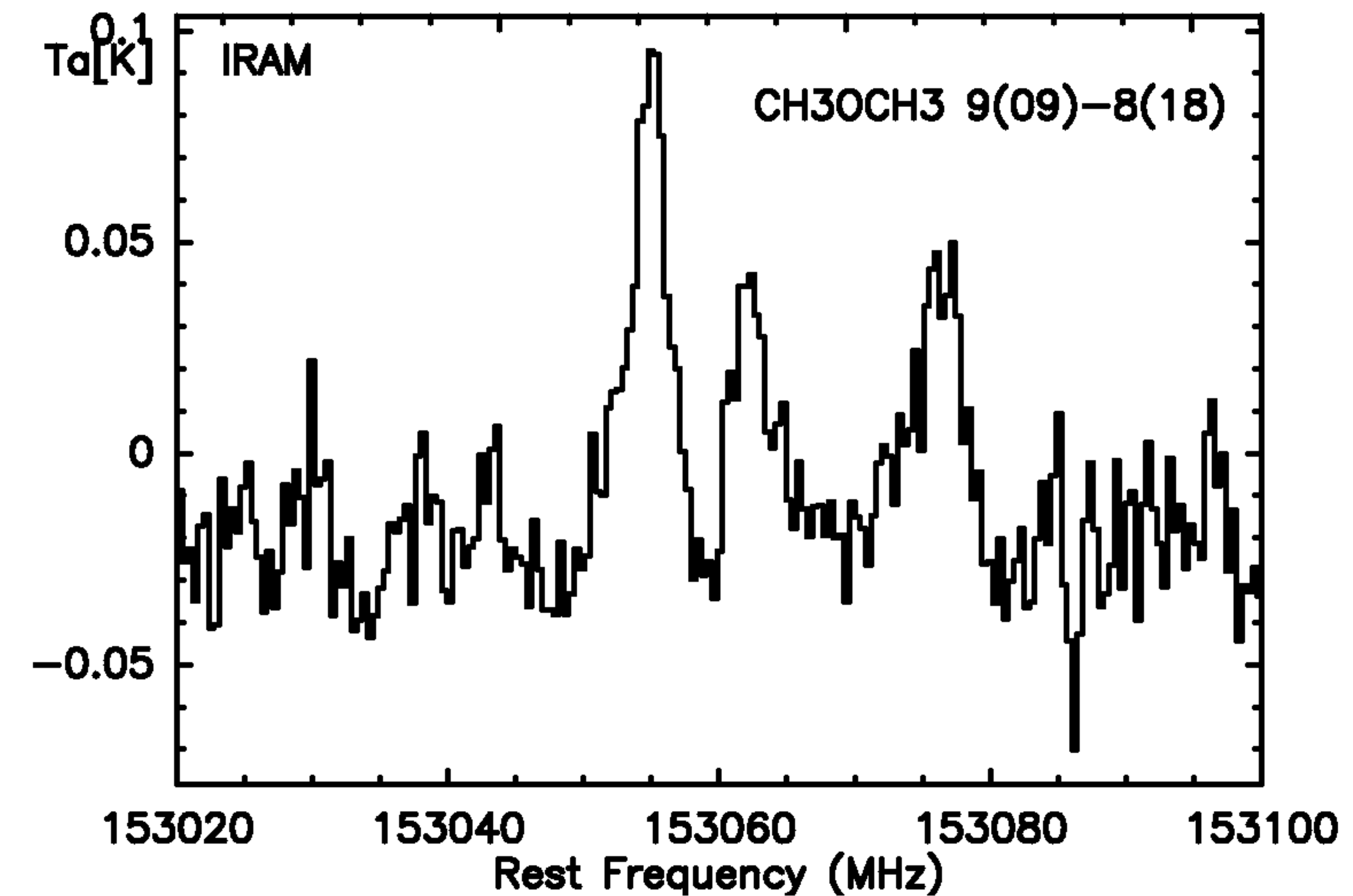}

\includegraphics[width=5.3cm]{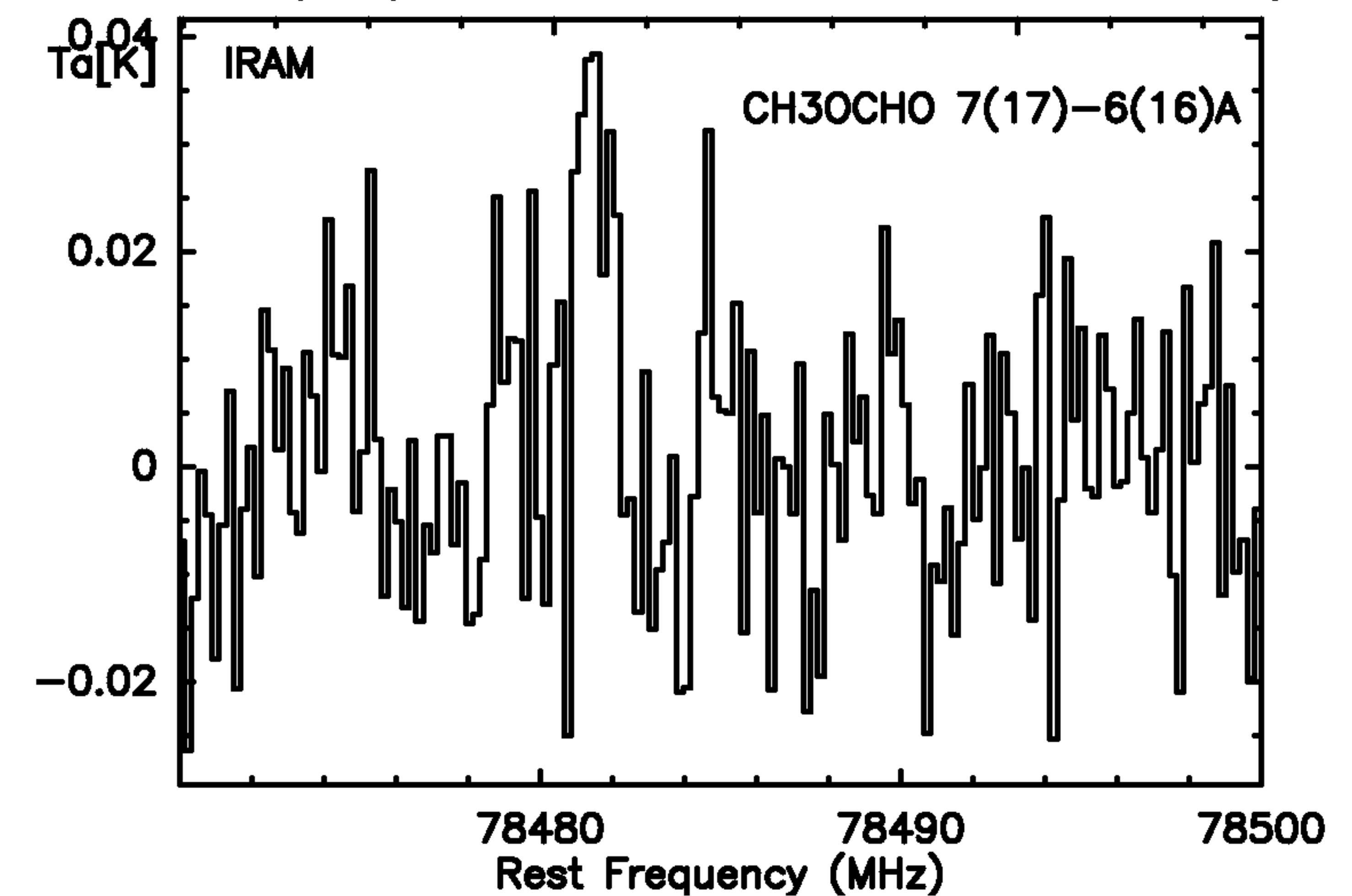}
\includegraphics[width=5.3cm]{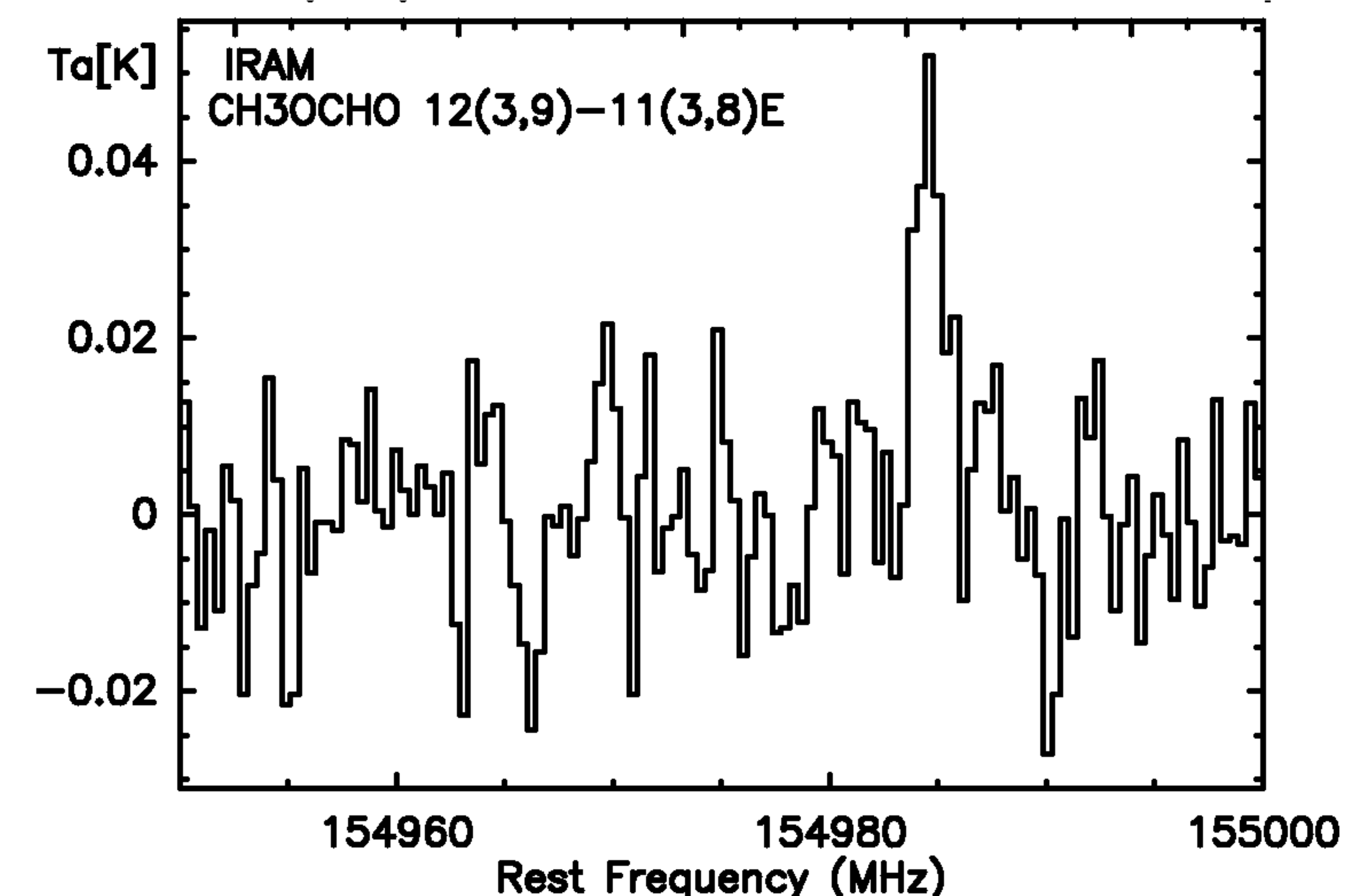}
\includegraphics[width=5.3cm]{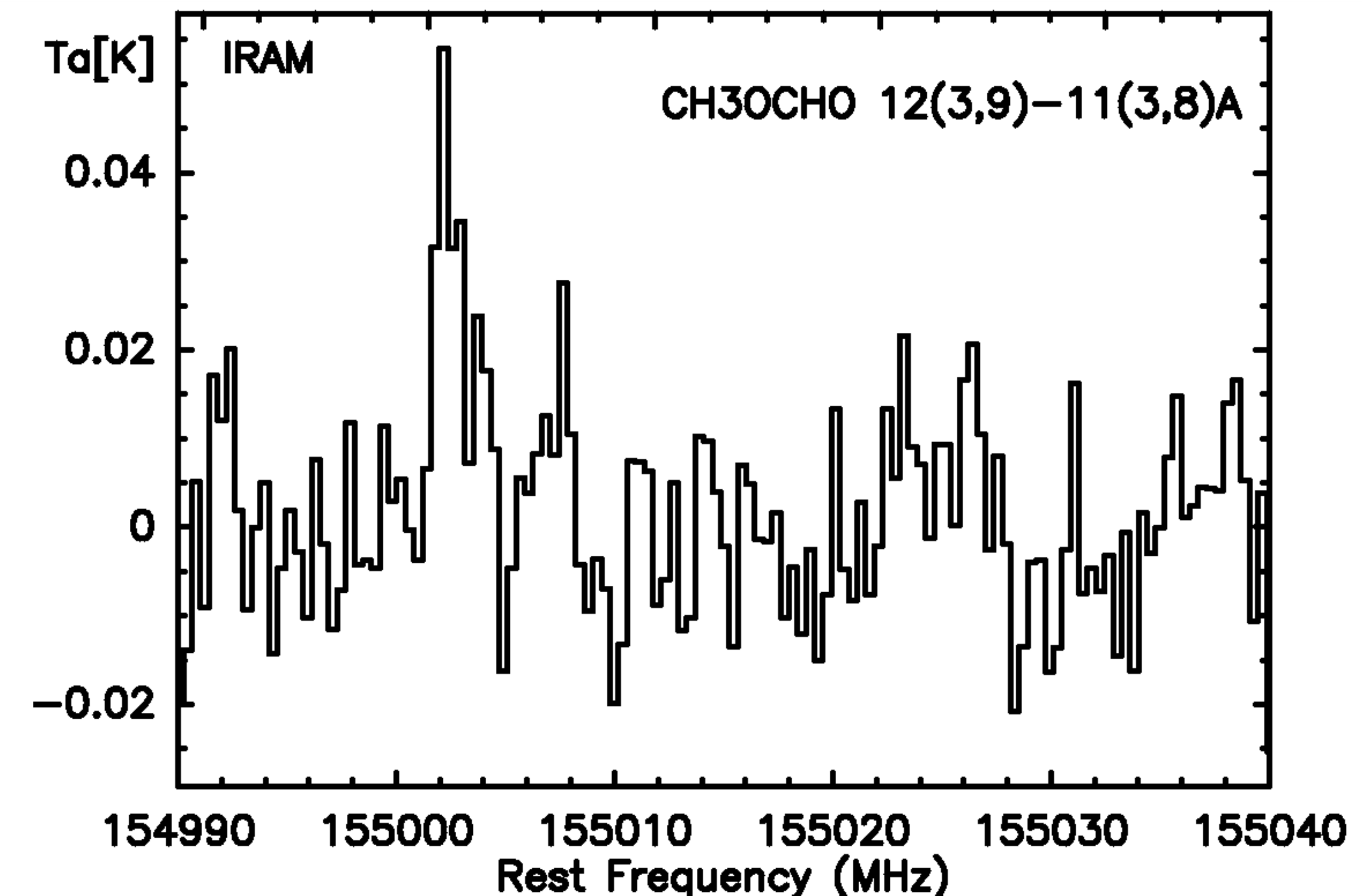}

\includegraphics[width=5.3cm]{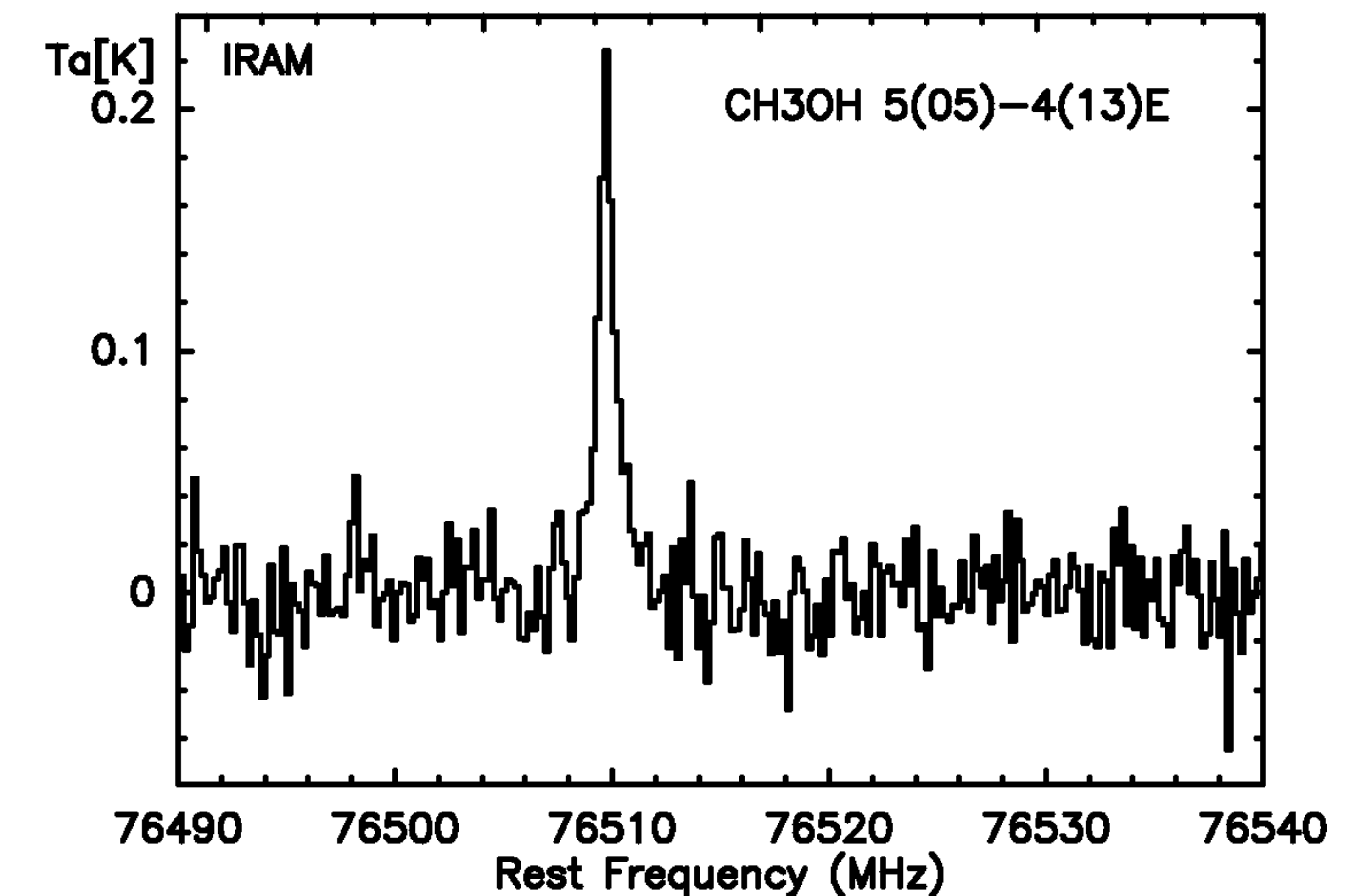}
\includegraphics[width=5.3cm]{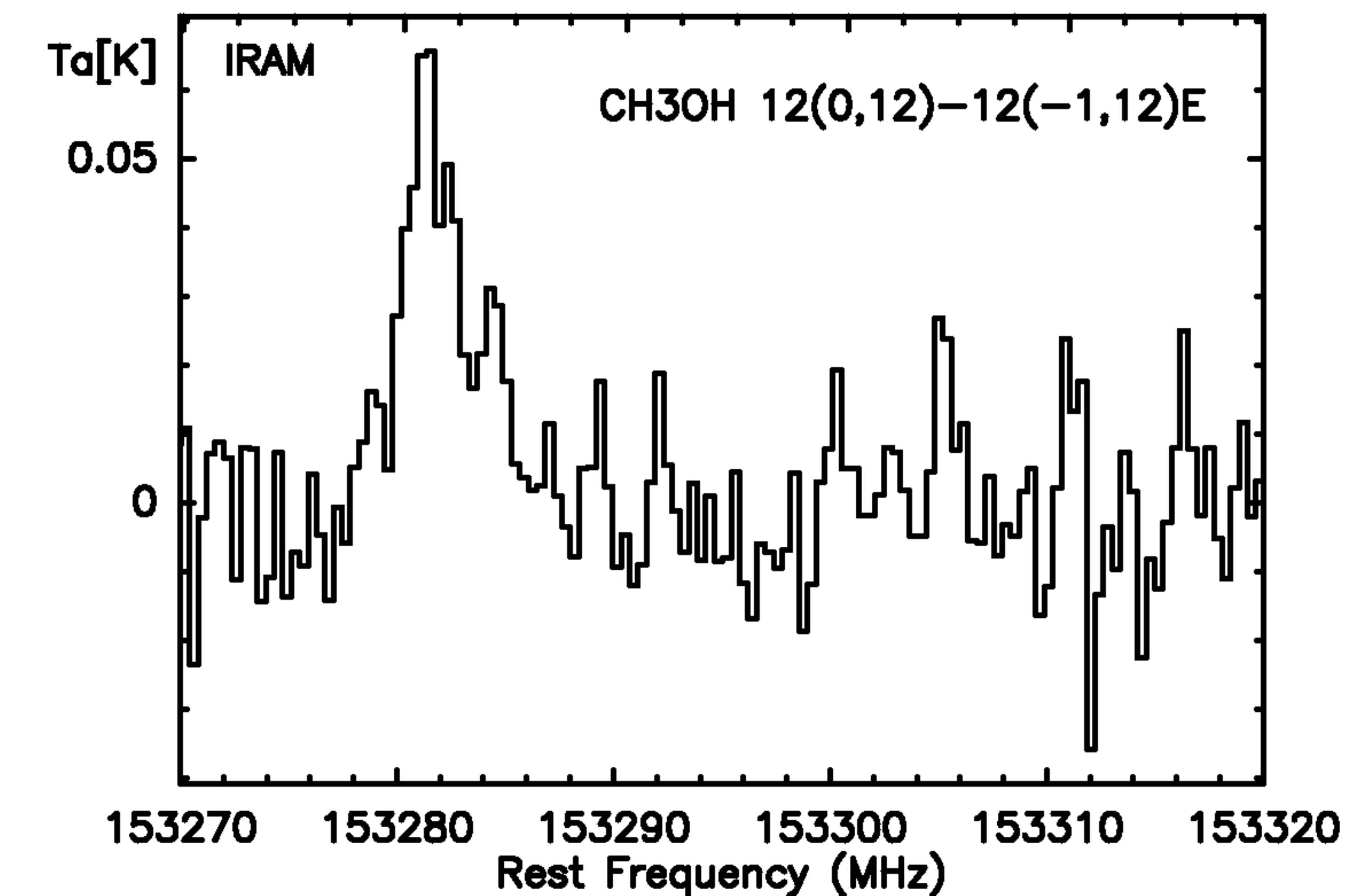}
\includegraphics[width=5.3cm]{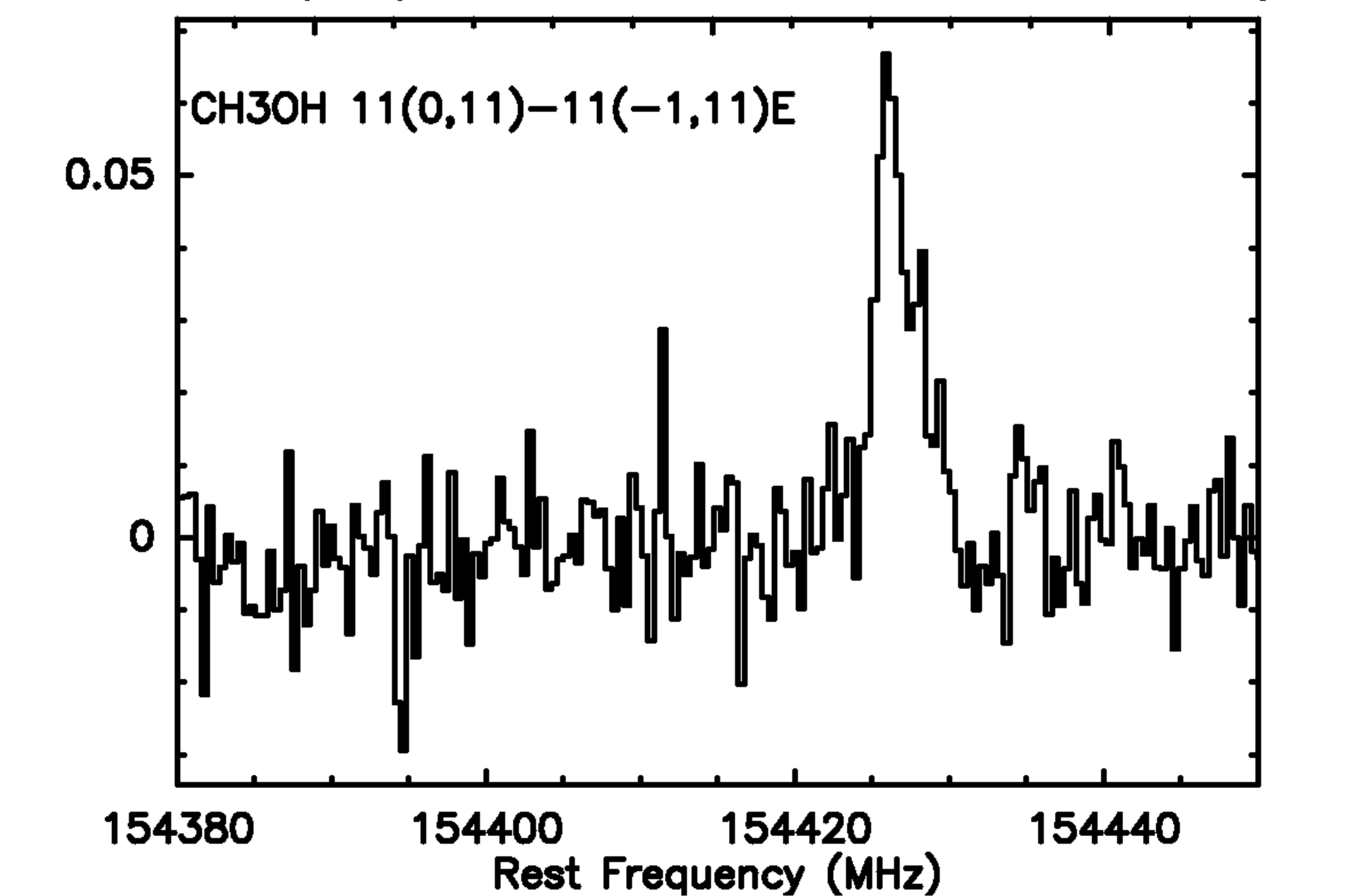}

\includegraphics[width=5.3cm]{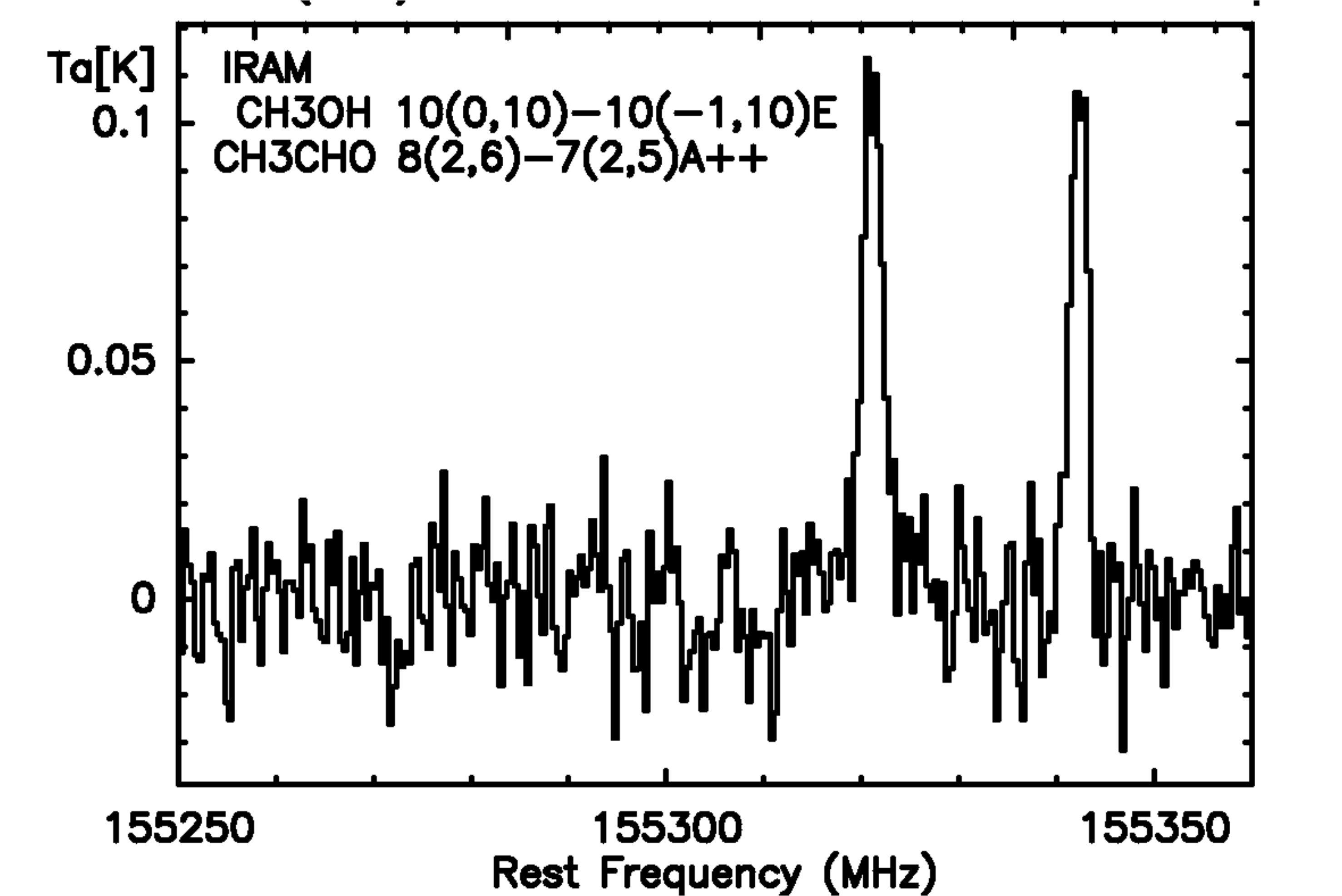}
\includegraphics[width=5.3cm]{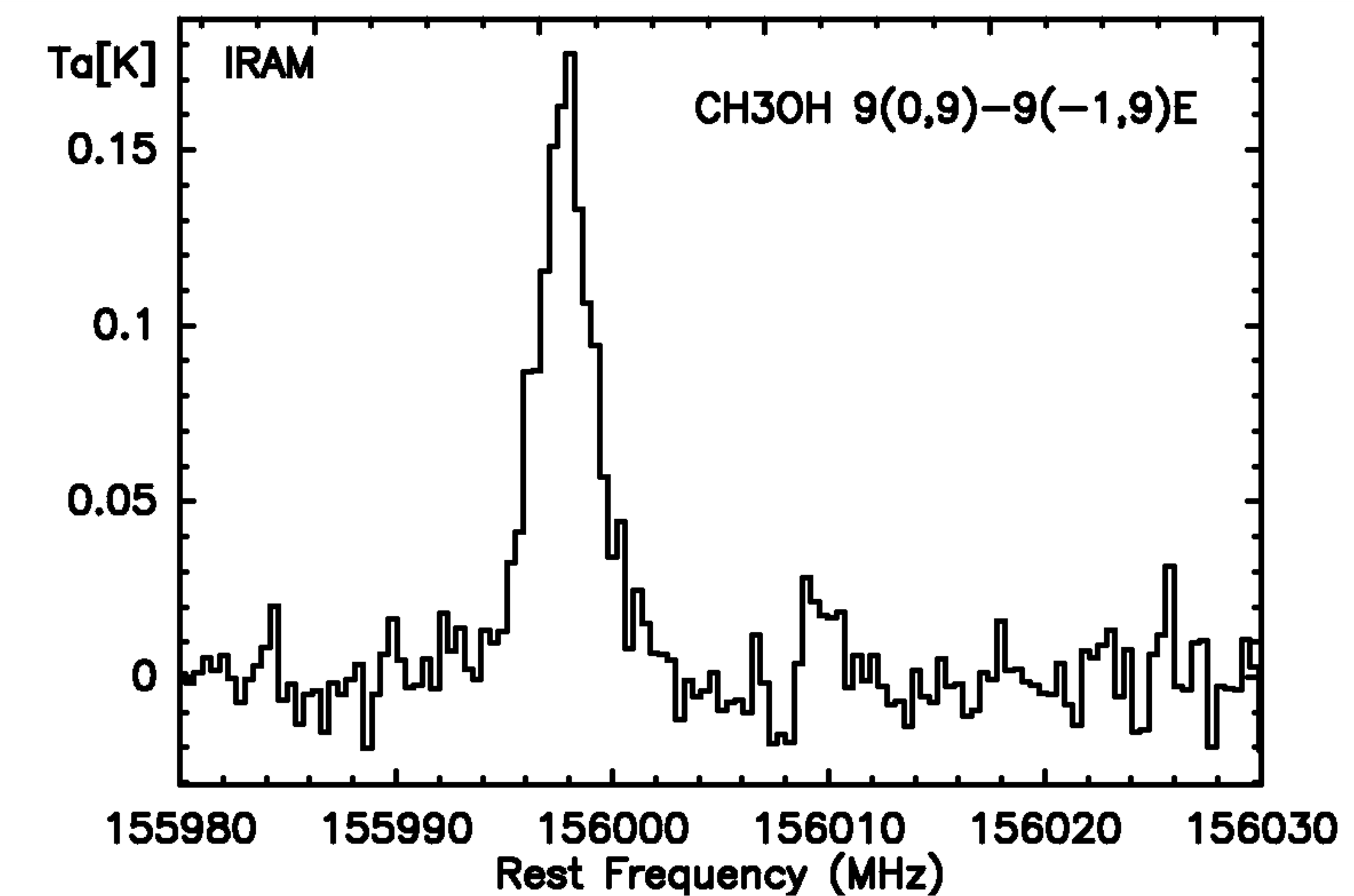}
\includegraphics[width=5.3cm]{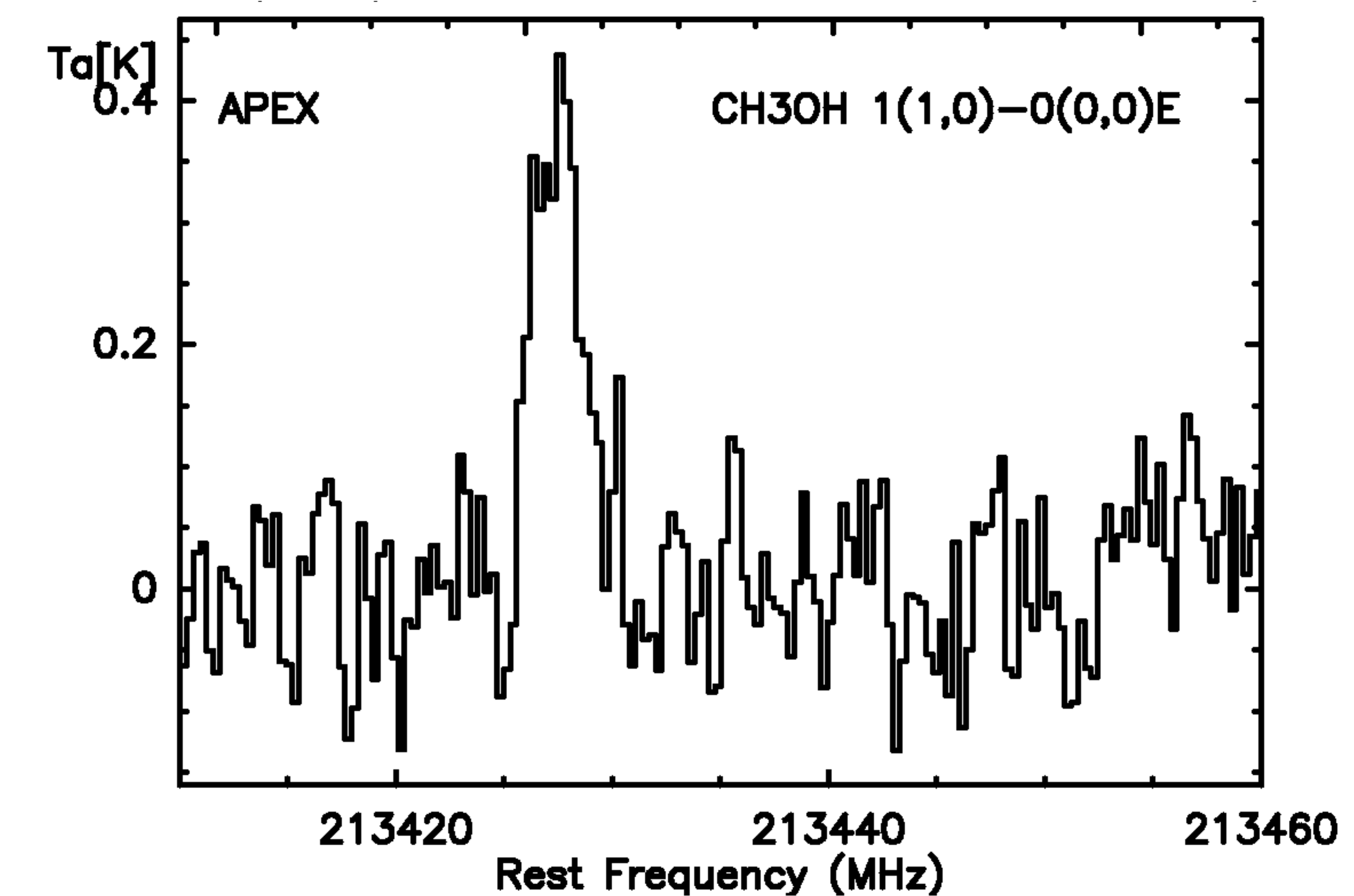}

\caption{Line spectra for IRDC028.34-6.}
\label{spectra6.a}
\end{figure}

\begin{figure}
\includegraphics[width=5.3cm]{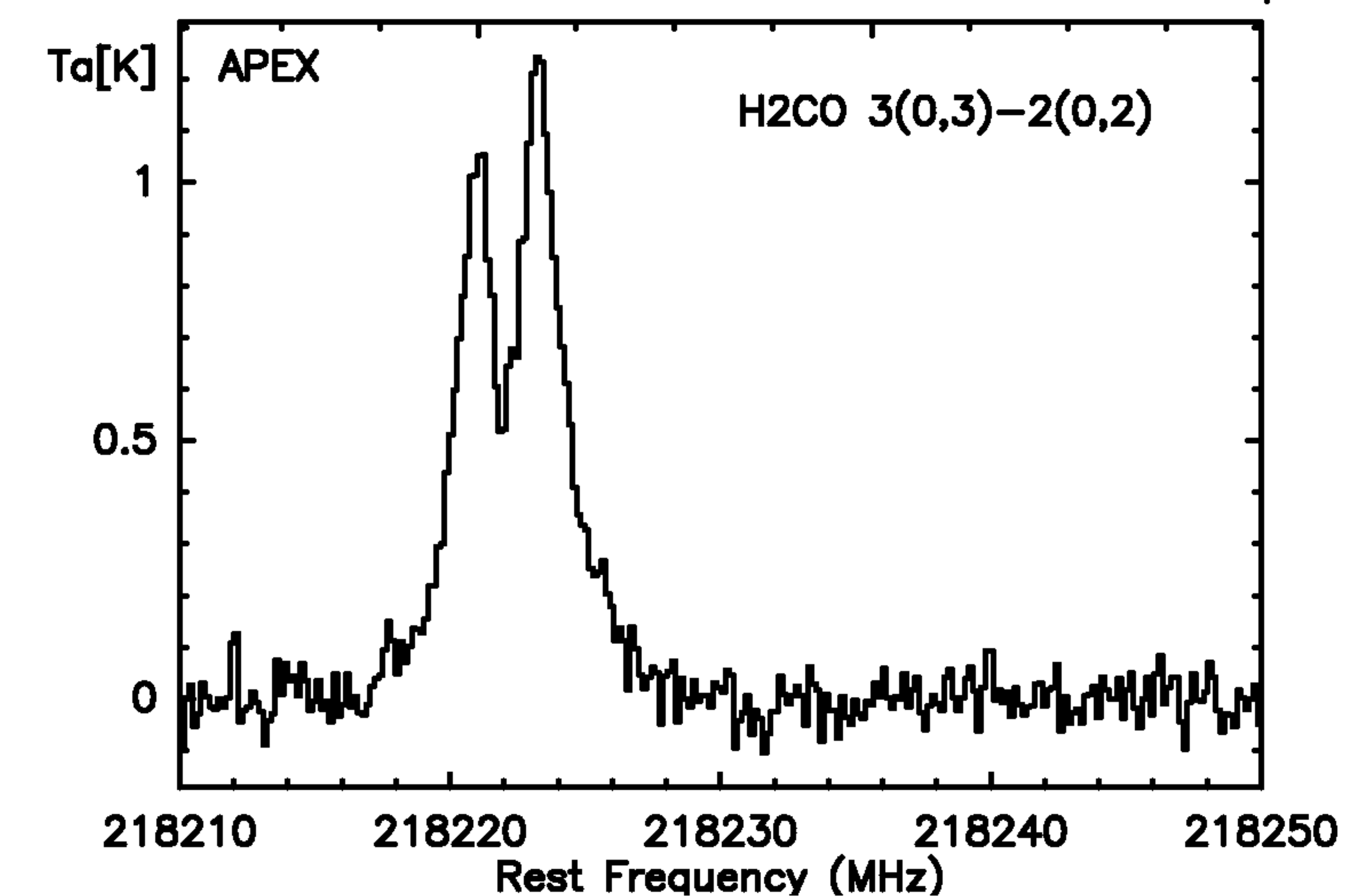}
\includegraphics[width=5.3cm]{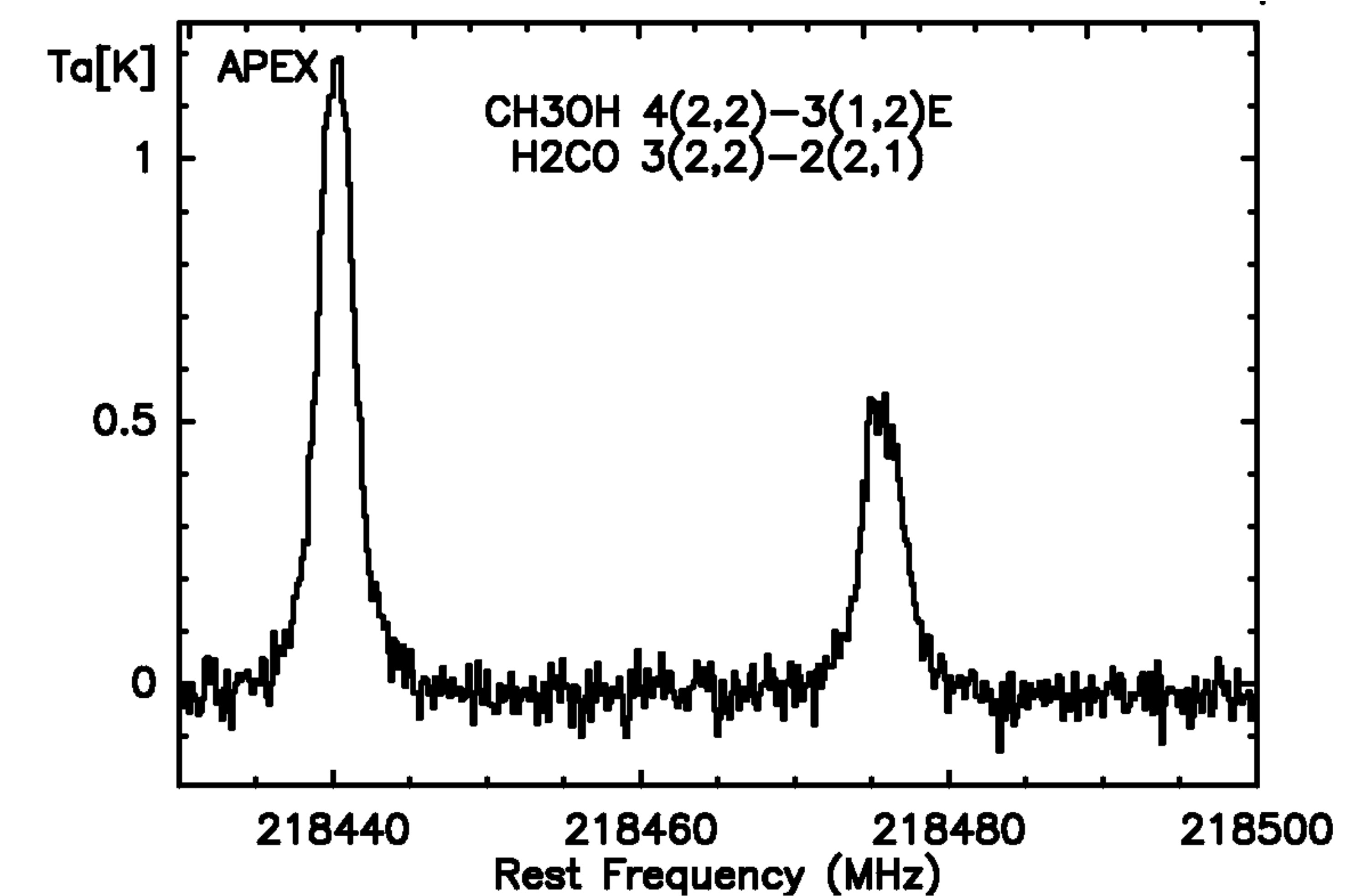}
\includegraphics[width=5.3cm]{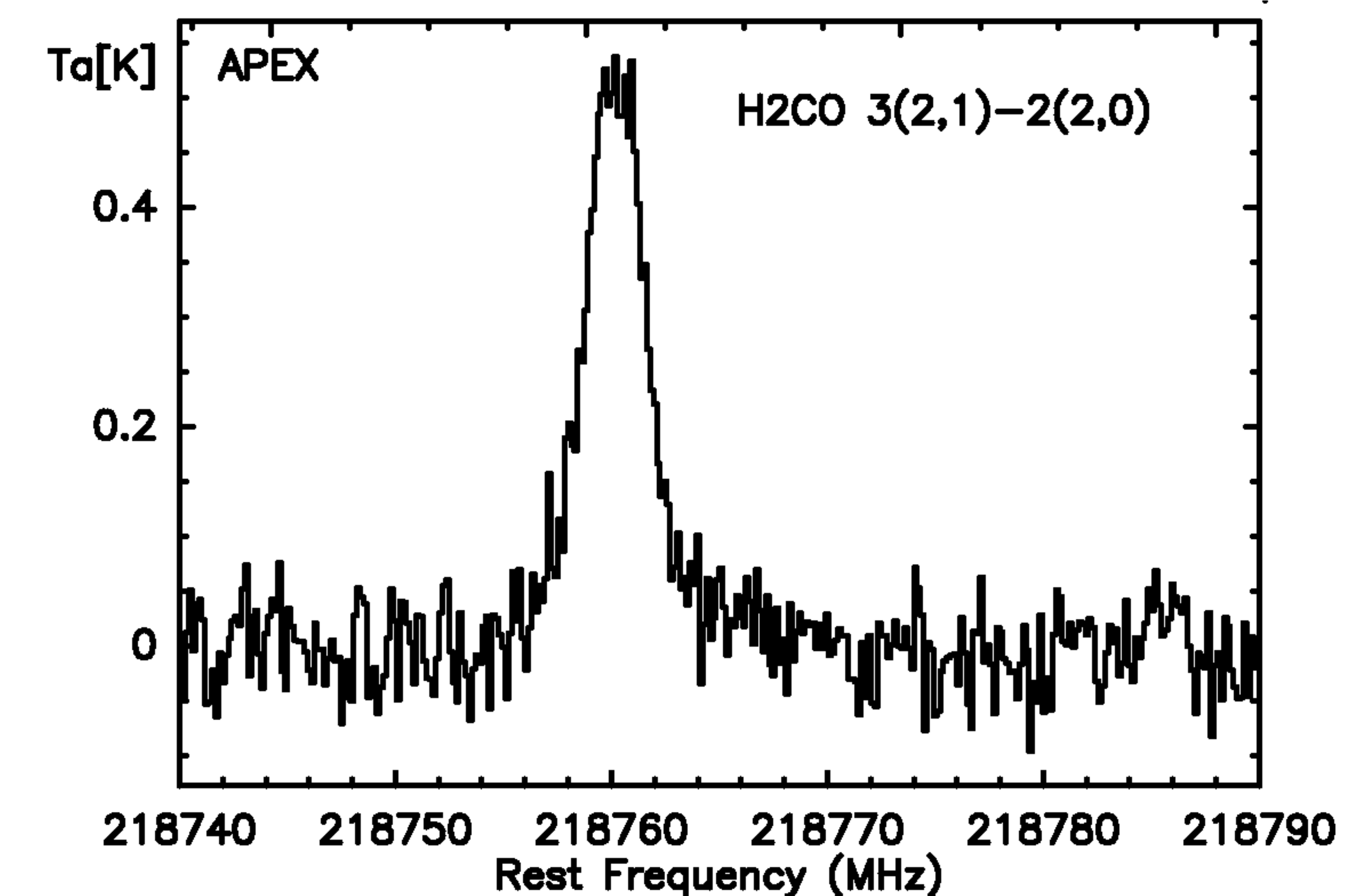}

\includegraphics[width=5.3cm]{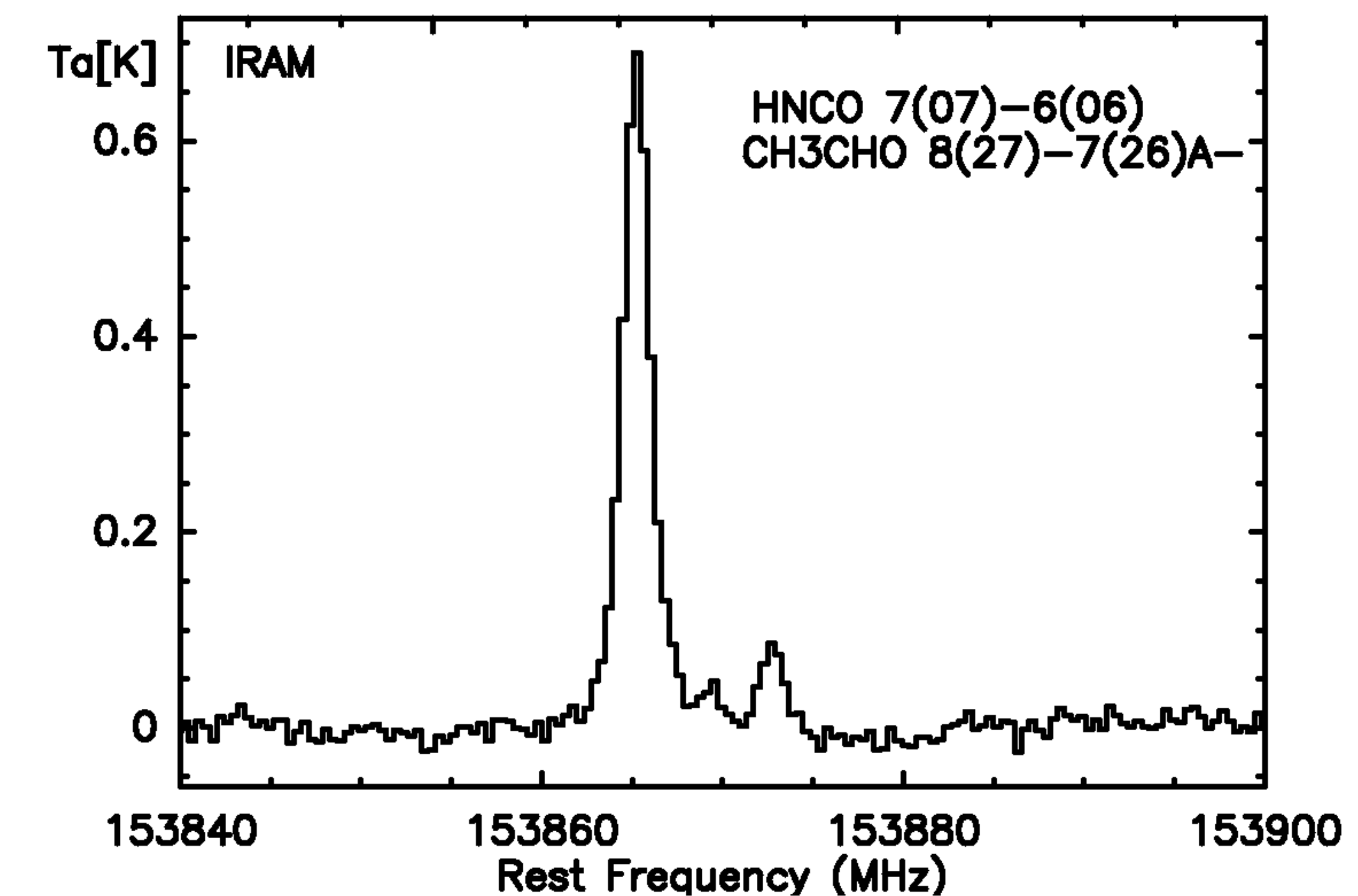}
\caption{Line spectra for IRDC028.34-6 (Continued).}
\label{spectra6}
\end{figure}

As an example, observed spectra for six objects: ``quiescent", ``middle" and ``active" IRDCs for both the southern and northern groups of clouds 
are presented in Figures~\ref{spectra1}-\ref{spectra6}.

\subsection{Mopra observations}

With Mopra, several observational setups were employed in position switching mode.
In all cases, we used the MOPS spectrometer, which allowed us to place 16 ``zoom"
windows along the 8.3 GHz bandpass and reach a resolution of 0.1 km/s, or 30 kHz. 
Since for the present study we are interested mainly in 
organic species, we will focus only on the 
CO, HNCO, CH$_3$CCH, and CH$_3$OH molecules observed with Mopra.
 Four CO isotopologues:
$^{12}$CO, $^{13}$CO, C$^{18}$O, C$^{17}$O, and CH$_3$OH lines  were observed on 27 April-2 May 2011 with the central frequencies 
of 111.6 GHz and 81 GHz, respectively.
The HNCO  and CH$_3$CCH data were obtained by adopting a central frequency of 89.27 GHz, and
have been already presented in \citet{2011A&A...527A..88V}.
Parameters for observed lines can be found in Table~\ref{table:lines_APEX}.  

At all frequency setups, we spent in total  15 min ``on"  source and 15 min on the ``off" position. 
Since for the southern sources no molecular line data were available at the time of the observations, 
for every source, the ``off" position was
chosen individually,  based only on an analysis of the 
 Spitzer mid-IR images and by choosing a position with no signs of mid-IR (MIR) extinction.
The system temperature during the observations was in the range from 200~K to 300~K.
The full-width half-maximum of the beam was measured to be  36$\pm$3$''$ at 86 GHz and
33 $\pm$ 2 $''$ at 115 GHz.
 The measured beam efficiency corresponds to 0.49 at 86 GHz and 0.42 at 115 GHz
 \citep{2005PASA...22...62L}.

Mopra data were originally stored in  a special RPFITS format.
The original RPFITS data were then transferred to FITS format using the ATNF
spectral line analysis package (ASAP). They were then fed into GILDAS/CLASS
for further data reduction and analysis.

\subsection{APEX observations}

Observations with the APEX telescope in position switching mode were conducted on 25-26 September 2011,
15-18 November 2011 and 12-14 September 2012.
Transitions of organic species that were observed during these setups for 
southern and northern sources are presented in 
Tables~\ref{table:lines_APEX}-\ref{table:lines_IRAM}.
The total integration time for each point was around 1-2 min. 
The ``off" source positions were chosen individually
for every source, based on the  analysis of the CO Galactic Ring Survey (GRS) and MIR Spitzer data.
We used the APEX-1 receiver of the Swedish Heterodyne Facility
Instrument \citep[SHeFI;][]{2008A&A...490.1157V}
 with  frequency settings at 218.75, 243.25, and 214.77 GHz.
 The backend for all observations was the
 eXtended bandwidth Fast Fourier Transform Spectrometer (XFFTS)
 with instantaneous bandwidth of 2.5 GHz and 32768 spectral channels
 \citep{2012A&A...542L...3K}.
This allowed us to cover the intervals
213 - 220.5 GHz, and 241.5 - 243 GHz for the southern objects and 213-220 GHz for the
northern objects with 88.5 kHz spectral resolution.
These intervals have the potential to detect many species, 
including CO, SO, DCO$^+$, CH$_3$OH, etc. 
However, for the current study, only  $^{13}$CO, C$^{18}$O, H$_2$CO and CH$_3$OH 
were selected, as shown in Tables~\ref{table:lines_APEX}-\ref{table:lines_IRAM}. 
At the chosen frequencies,  the main beam of the telescope varies between 30$''$ at 211 GHz 
 and 26$''$ at 242 GHz. The main beam efficiency is equal to 0.75.
The system temperature during the observations was about 140-180 K at 
213-220.5 GHz and 210-230 K at 241.5-243 GHz .

The APEX data were observed in
service mode and were provided to us by the APEX staff in ready-to-use
GILDAS/CLASS format.
Therefore,  all necessary data reduction was accomplished with the CLASS software.

\subsection{IRAM observations}

IRAM observations were performed on 8-12 June 2011 in the position switching mode.
Total times ``on" source and ``off" source were determined individually for every source, based on the 
general strengths of more common lines known from previous measurements, and were between 30 and 60 min in total. 
The ``off" source position was selected to be 600$''$ away from the ``on" position  in the right ascension direction.
Pointing was updated every hour, giving a pointing accuracy of around 3$''$.
The typical system temperature during observations was about 200~K.

For these observations, the EMIR receiver with the Fast Fourier Transform Spectrometer (FTS) as
a backend was used. During the observing run, the frequency intervals 
76 - 79 GHz and 152.5 - 156.5 GHz were covered with a spectral resolution of 195~kHz.
These ranges include lines of several complex organic species, as listed in Tables~\ref{table:lines_IRAM} and \ref{table:lines_IRDC028.34-6}. 
According to the IRAM official website, at 86 GHz the beam size is 29$''$, and the beam efficiency is 0.81,
while at 153 GHz, the telescope beam size is 16$''$, and the beam efficiency is 0.74.

The data were analyzed using the GILDAS/CLASS software.

\section{Data reduction and analysis}
\label{results}

\subsection{Observed transitions and detection rates}


Among the detected molecules, CO
is most abundant gas-phase species, which allows us to detect
not only the main CO isotopologue, but also $^{13}$CO and C$^{18}$O.
In the southern sources,  
$^{12}$CO (1-0), $^{13}$CO (1-0),  C$^{18}$O  (1-0) , and C$^{17}$O  (1-0)
were observed with the Mopra telescope, 
and $^{13}$CO (2-1), and C$^{18}$O  (2-1) were observed with APEX.
For northern sources, only APEX observations were performed.
Since  $^{12}$CO and $^{13}$CO are most likely optically thick and 
show complicated line shapes, and C$^{17}$O was detected only in three 
regions, C$^{18}$O was used in all cases 
to estimate the CO column density and relative abundance  with respect to H$_2$. 
It is worth noting that C$^{18}$O has indeed been found to be optically thin in IRDCs \citep[see, e.g.,][]{2011ApJ...738...11H}.
For southern sources, we used  C$^{18}$O (1-0), as obtained with Mopra, while
for northern sources we used C$^{18}$O  (2-1), as obtained with APEX. 
To reconstruct the CO column density from its isotopologue, we 
adopted  [$^{16}$O]/[$^{18}$O]=500 from  \citet{1997A&AS..124..205H}.
To derive the CO molecular abundances, we used the H$_2$ column densities shown
 in Tables~\ref{table:main_Mopra} and~\ref{table:main_IRAM}.
 
 HNCO
data for our southern clouds were taken from the previous study 
of \citet{2011A&A...527A..88V},
where the detection of HNCO~(4$_{0,4}$-3$_{0,3}$) was reported at 87.925 GHz.
The detection rate was 40\% among all of the observed clouds.
For the northern objects, HNCO~(7$_{0,7}$-6$_{0,6}$) was observed with IRAM at 153.865 GHz.
Despite the higher upper state energy 
related to this transition, HNCO shows a higher detection rate than in the southern sample.
In 16 objects out of 18, the detection was stronger than 3$\sigma$.
Such a high detection rate can be explained by 
 much higher sensitivity of the EMIR receivers at the IRAM 30m telescope.

As shown in Tables~\ref{table:lines_APEX}-\ref{table:lines_IRAM},  three transitions of H$_2$CO
were observed in southern and northern clouds around 218 GHz with APEX.
All three transitions were detected only in five southern and four northern objects.
For the rest of the targets, only the (3$_{0,3}$ - 2$_{0,2}$) transition  was detected.
 However,  even in the sources with all three detections, it was not
possible to use the results of the excitation analysis 
 because the lower energy levels were too close for
(3$_{2,2}$ - 2$_{2,1}$) and (3$_{2,1}$ - 2$_{2,0}$) transitions.
Therefore, for all clouds we used H$_2$CO~(3$_{0,3}$ - 2$_{0,2}$) to determine molecular
column densities and abundances.

\begin{figure}
\centering
\includegraphics[width=3.5cm,angle=90]{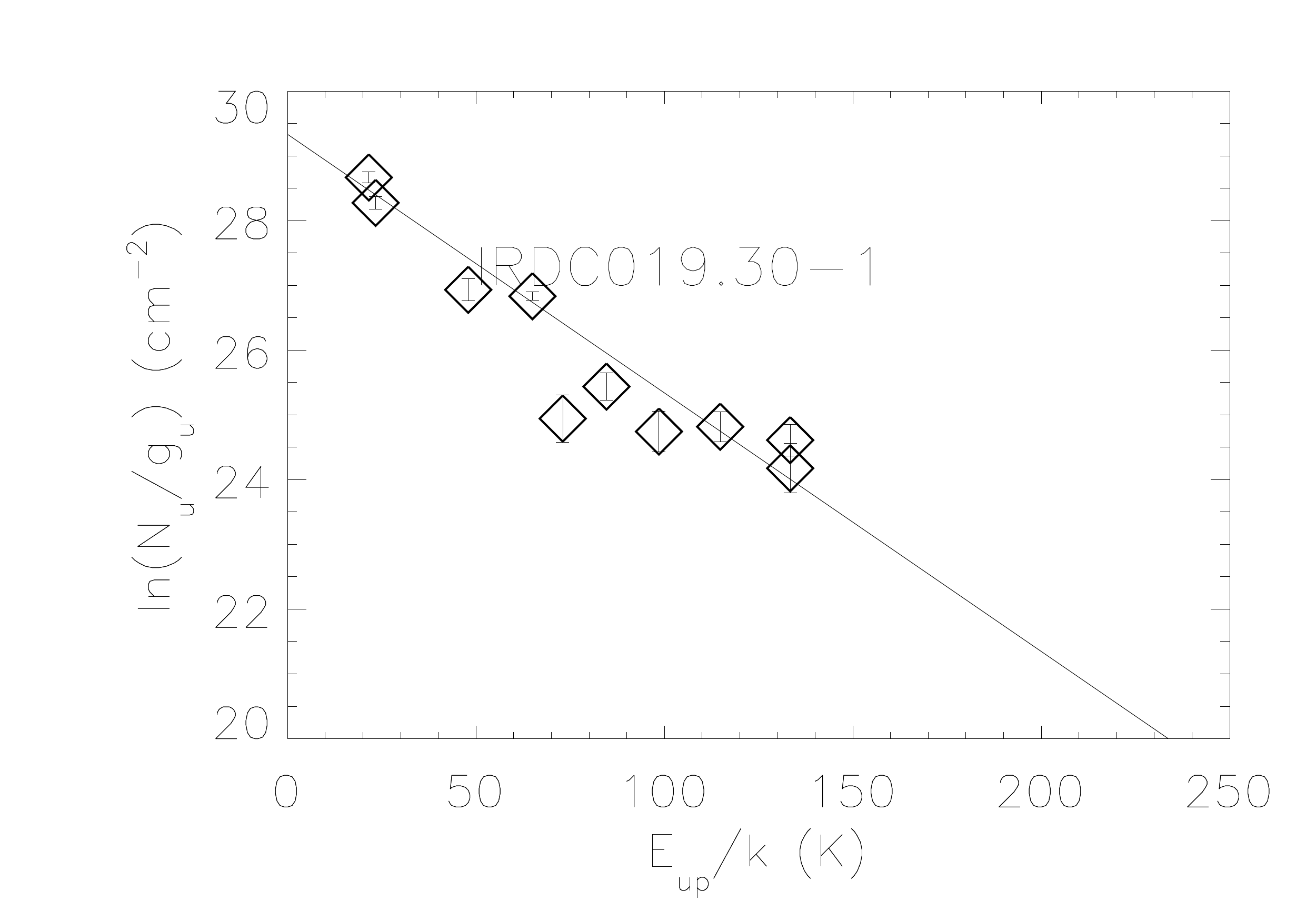}
\includegraphics[width=3.5cm,angle=90]{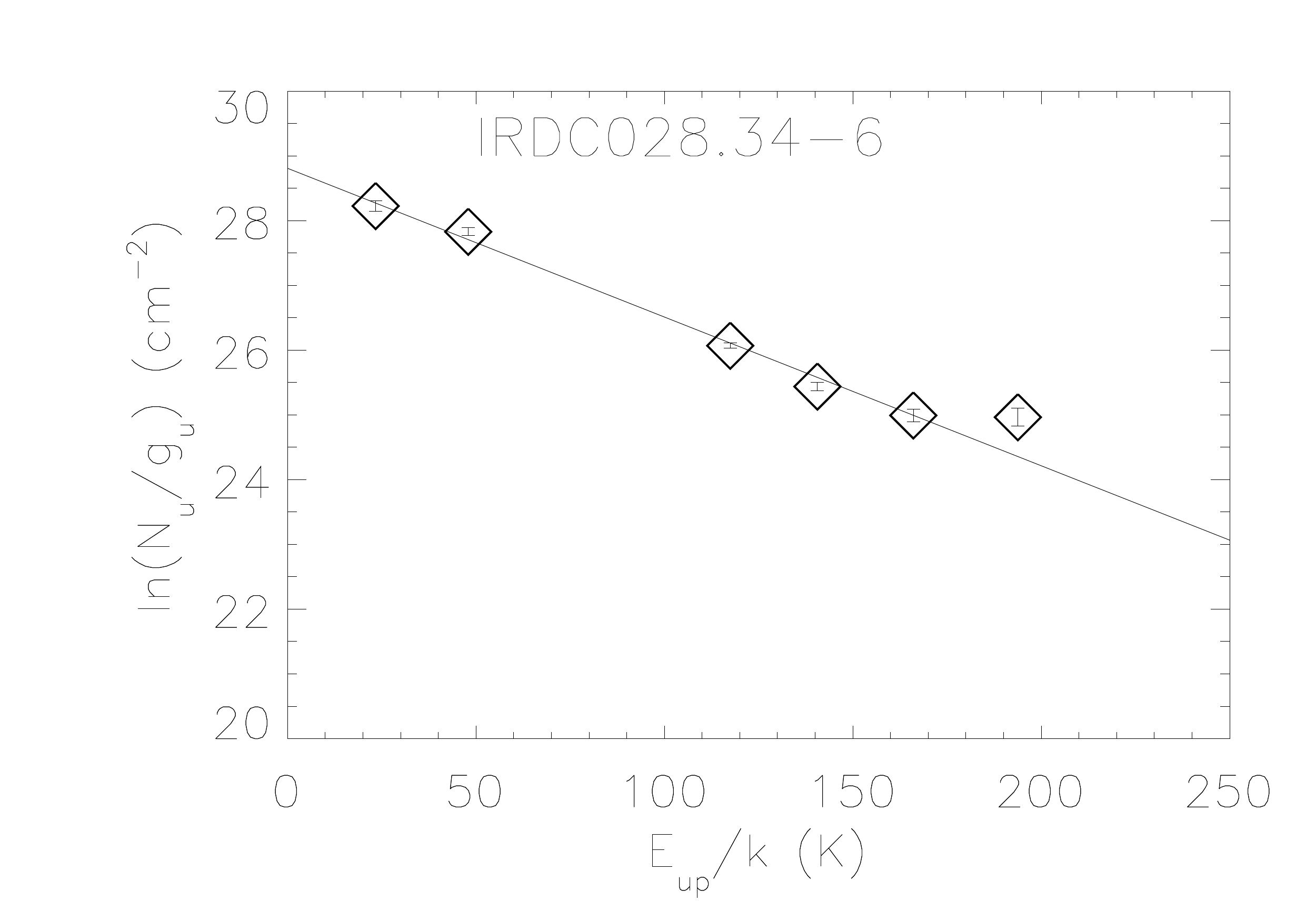}
\includegraphics[width=3.5cm,angle=90]{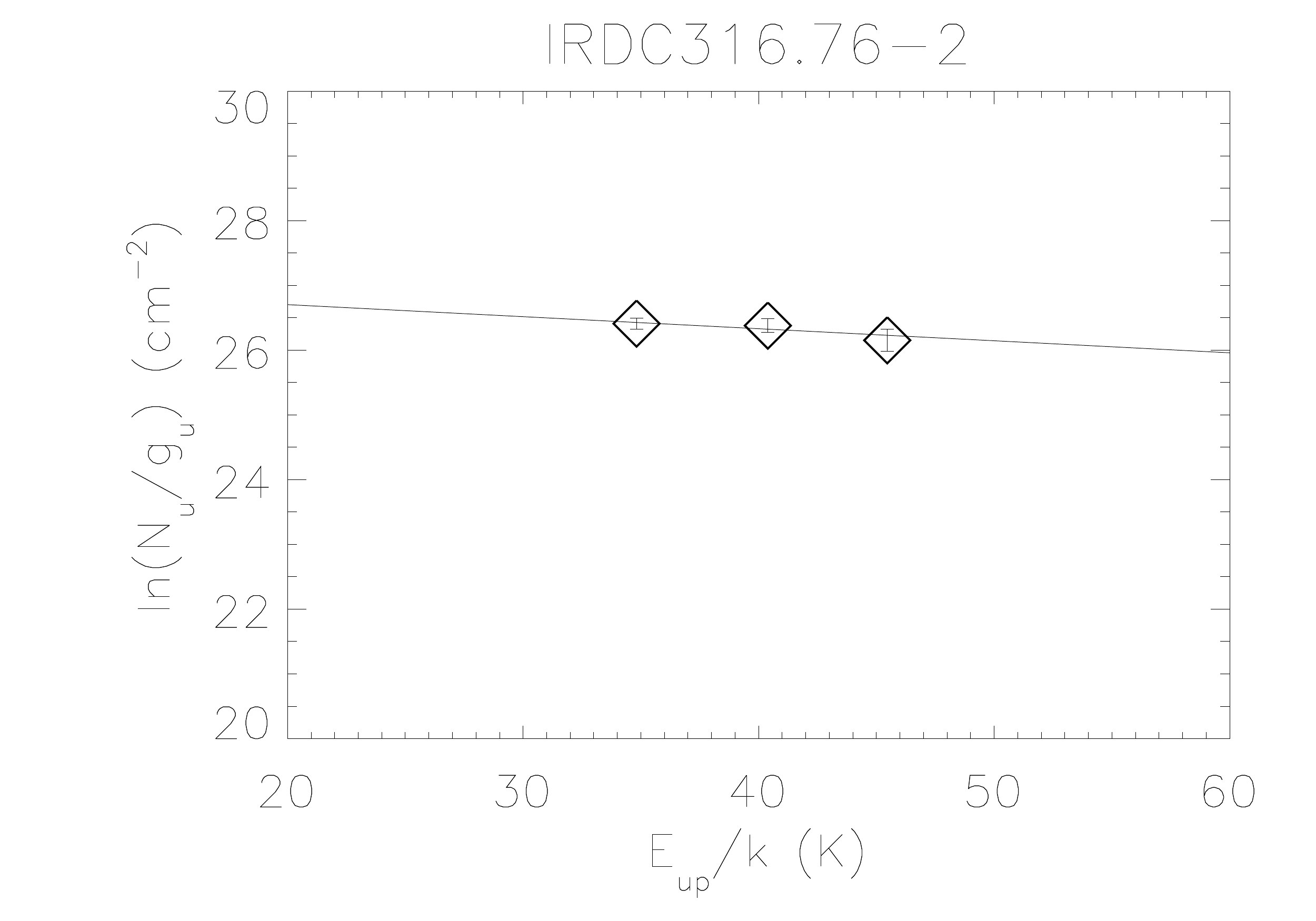}

\includegraphics[width=3.5cm,angle=90]{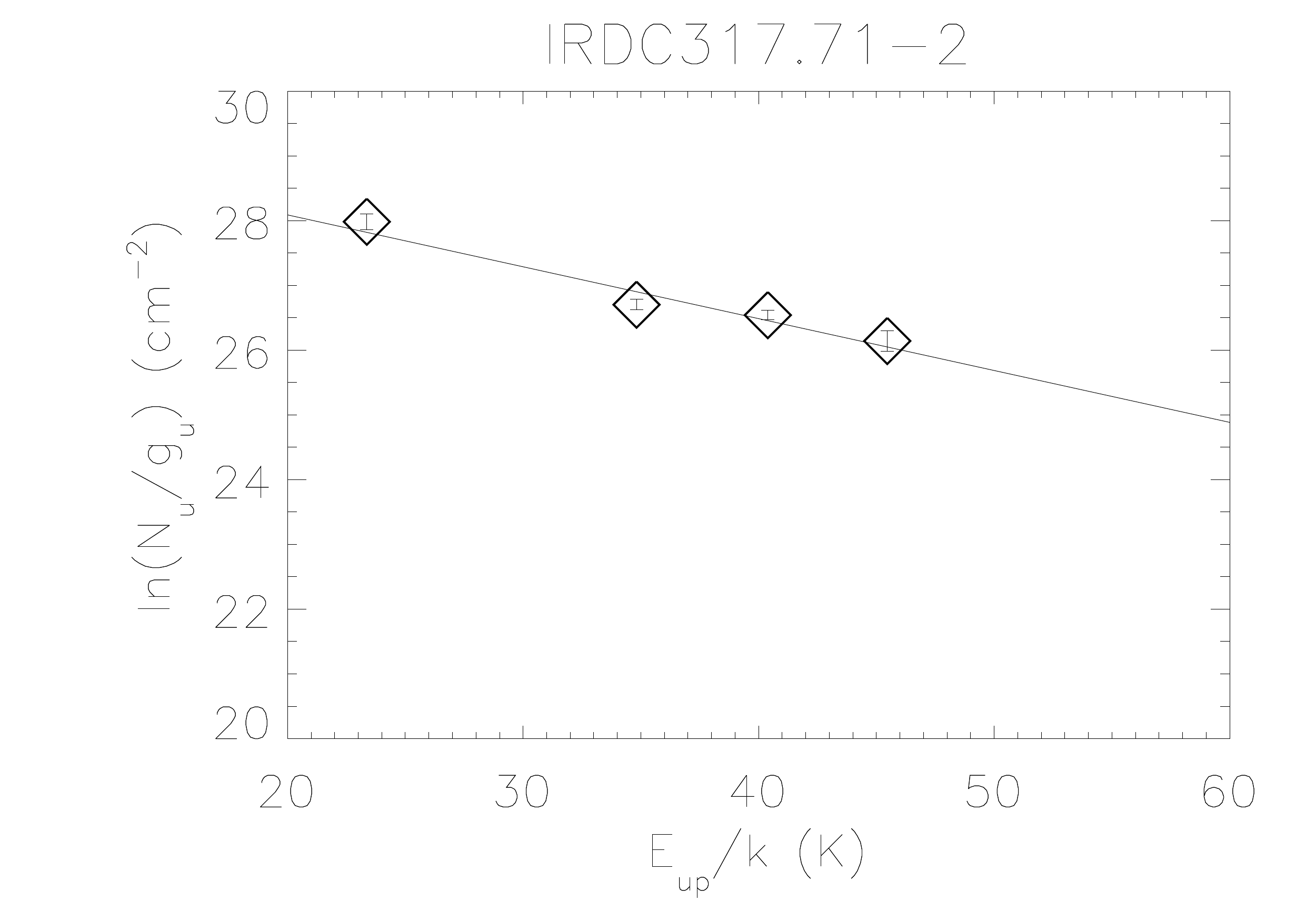}
\includegraphics[width=3.5cm,angle=90]{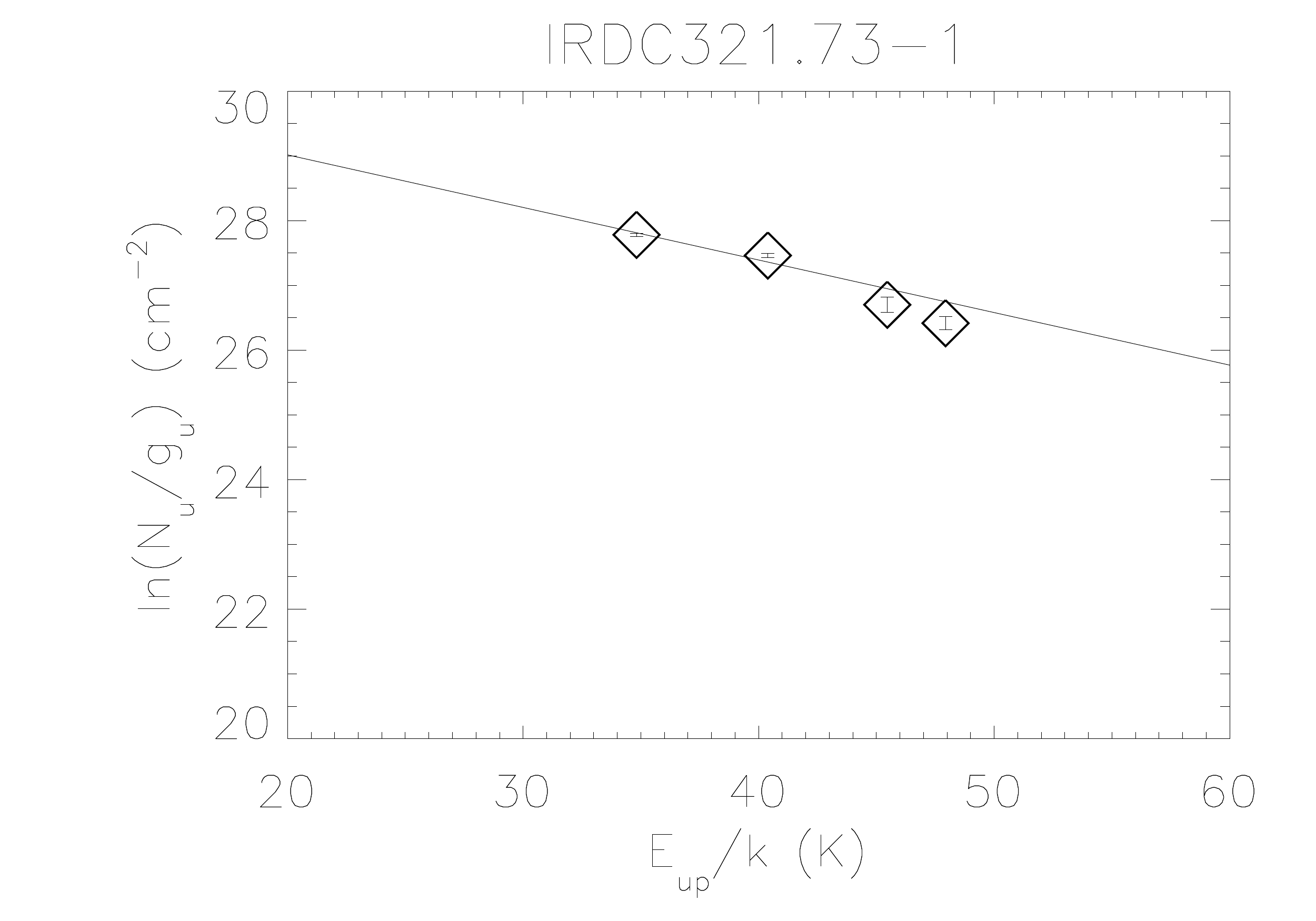}
\caption{Excitation diagrams for CH$_3$OH.}
\label{excitation_CH3OH}
\end{figure}

CH$_3$OH
 showed a high detection rate in the southern clouds; however
more than three transitions were detected only in six objects, for which 
we performed excitation analyses.
Among these six sources the shape of  the excitation diagram allowed  us to 
estimate the methanol column densities
and abundances only in three cases:
IRDC316.76-2, IRDC317.72-2, and IRDC321.73-1.
The other three sources showed a large spread of the points on the population diagram, which might indicate non-LTE behavior. 
For these sources additional observations are required to build an excitation diagram 
and extract more realistic excitation temperatures and molecular column densities.  
In the northern objects,  CH$_3$OH was detected in six clouds
out of 18; for two of which, IRDC019.30-1 and IRDC028.34-6,
we have enough transitions detected to build a population diagram and estimate an excitation temperature.
For the sources with fewer than three detections, molecular abundances were calculated 
based on CH$_3$OH 5$_{-1,5}$ - 4$_{-1,4}$ E for the southern sources and CH$_3$OH 4$_{2,2}$-3$_{1,2}$ E for the northern sources
using Equation~\ref{Equ:column} from Section~\ref{abundance}. 
Figure~\ref{excitation_CH3OH} shows  CH$_3$OH excitation diagrams for northern and southern clouds.
Estimated column densities are presented in Tables~\ref{table:ab_Mopra} -~\ref{table:ab_IRAM}.



 
For the southern sources, CH$_3$CCH J=5-4 (methyl acetylene) was detected during previous studies in five  
out of 37 sources \citep{2011A&A...527A..88V}. 
For two sources, with K=0, 1, and 2 lines, a transition excitation analysis was performed.
For the rest, only K=0 and 1 transitions had intensities higher than 3 $\sigma$. 
Therefore, column densities were estimated using equation~(1) from \citet{2011A&A...527A..88V}.

\begin{figure}
\centering
\includegraphics[width=3.5cm, angle=90]{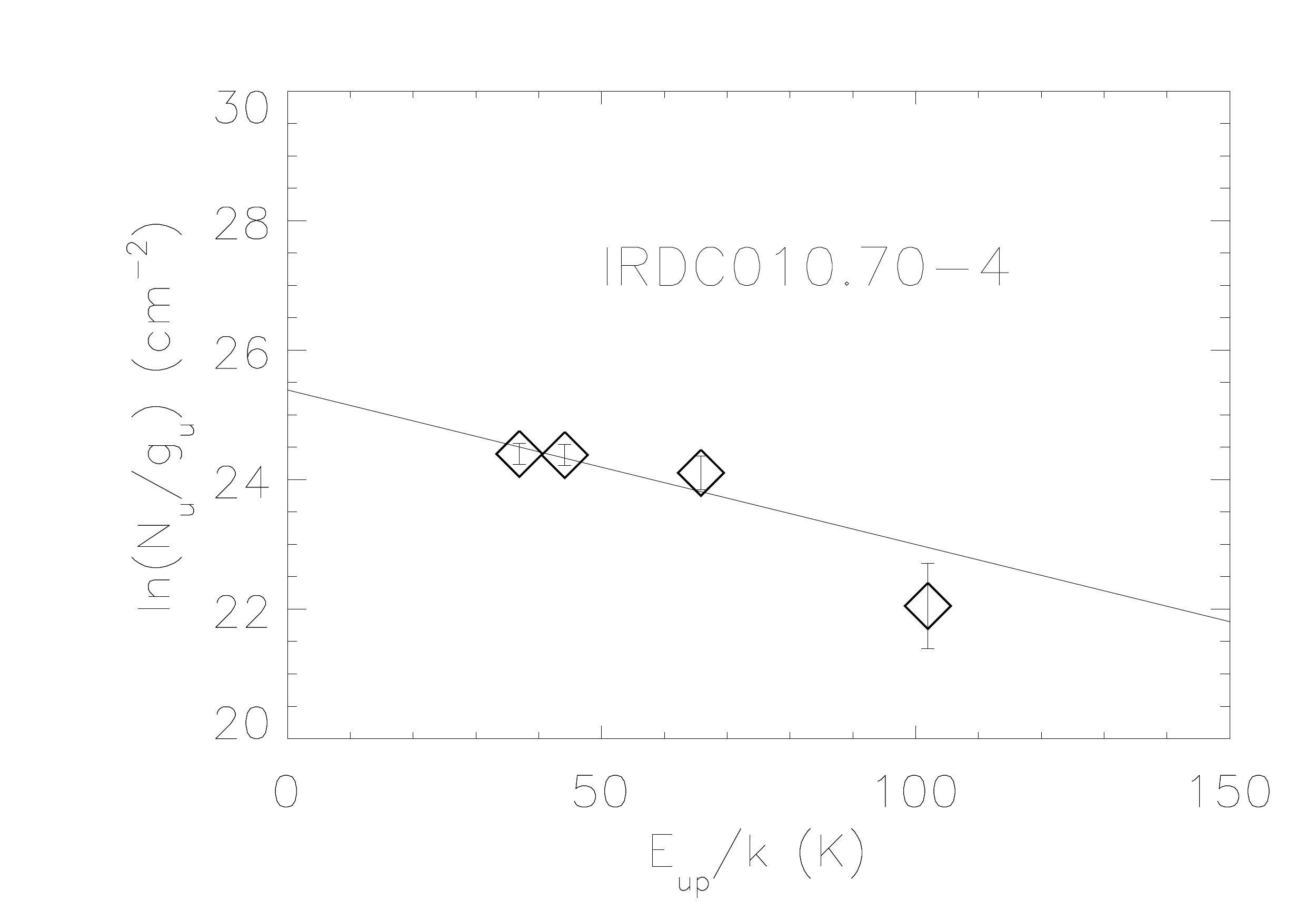}
\includegraphics[width=3.5cm, angle=90]{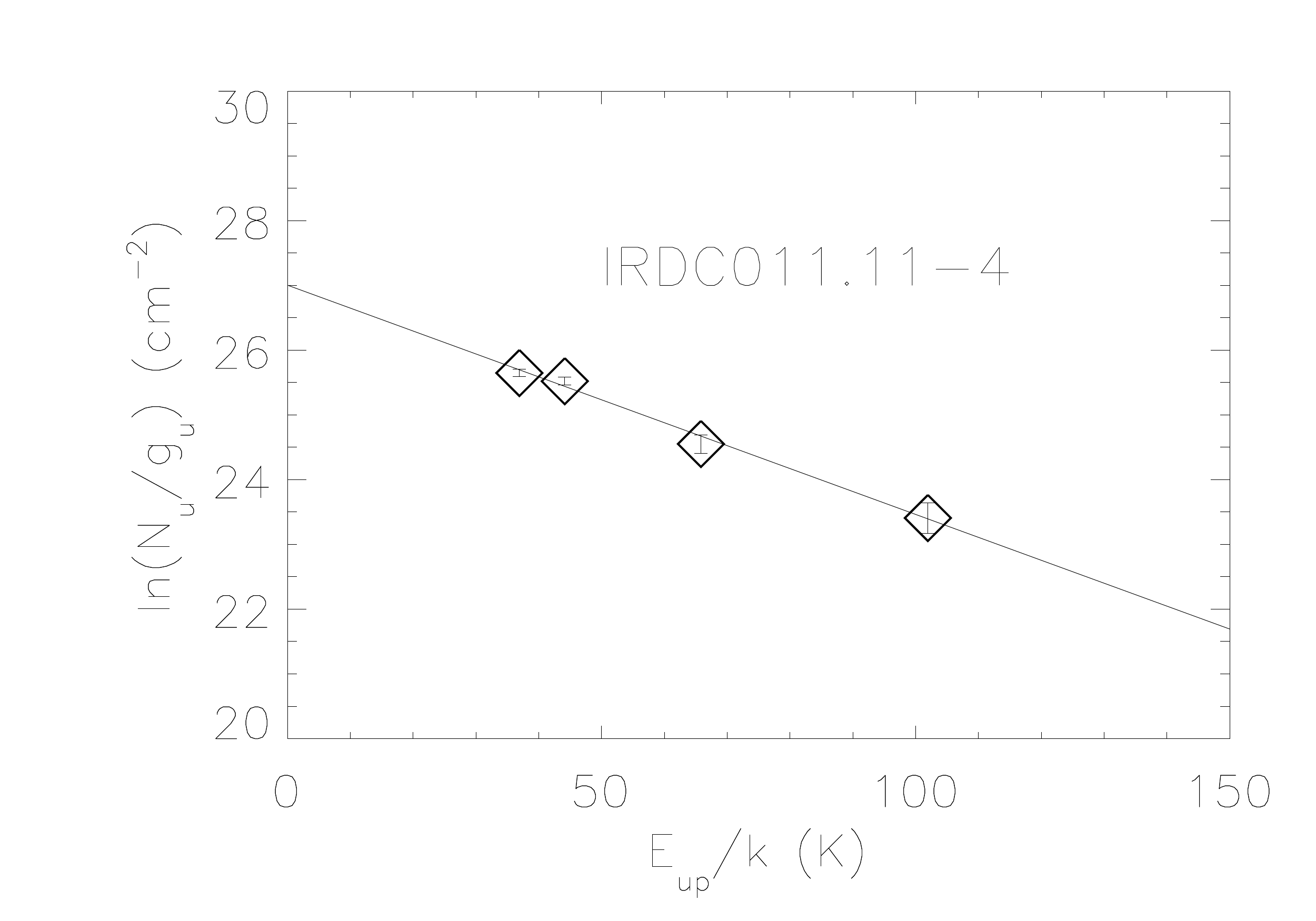}
\includegraphics[width=3.5cm, angle=90]{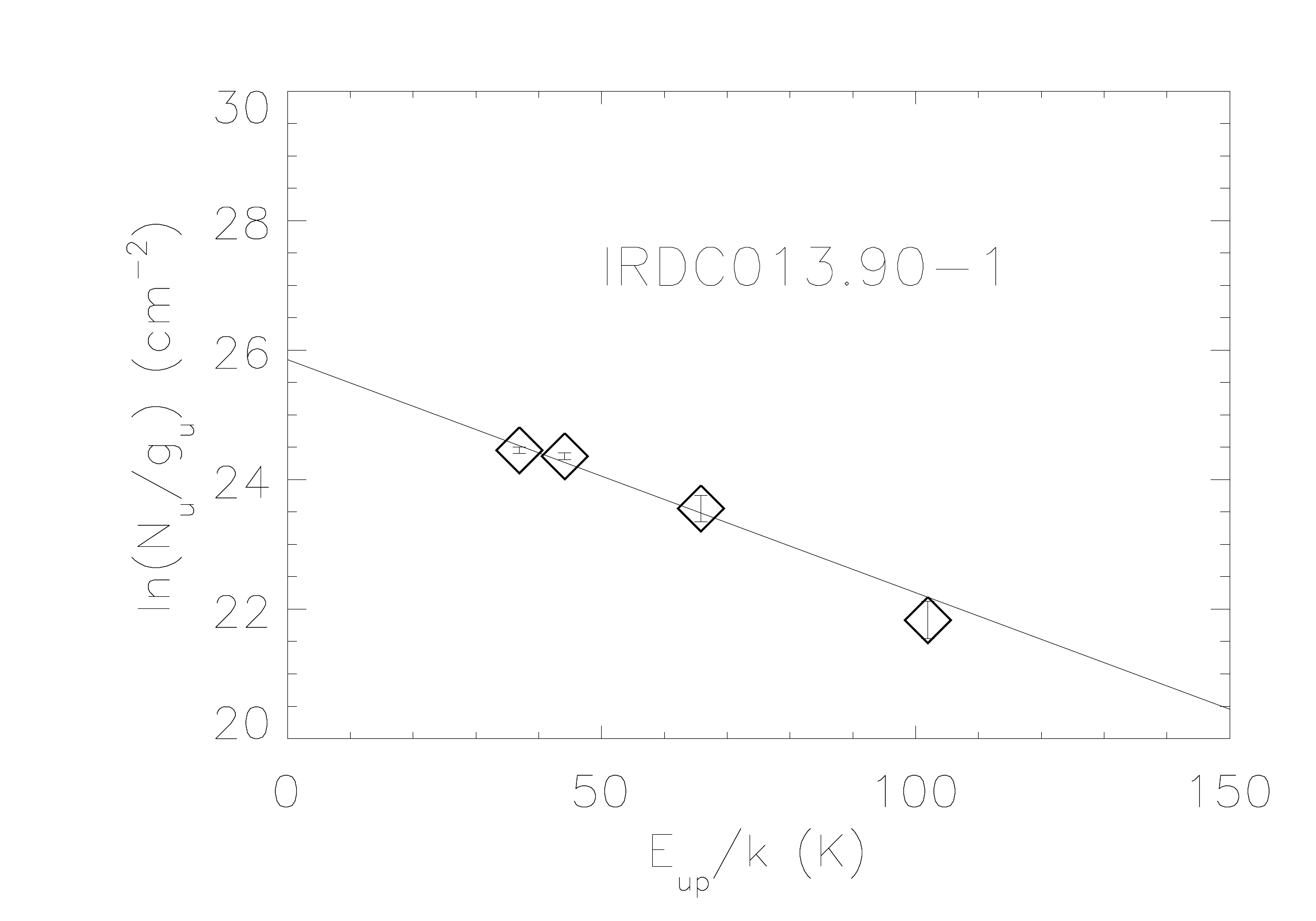}

\includegraphics[width=3.5cm, angle=90]{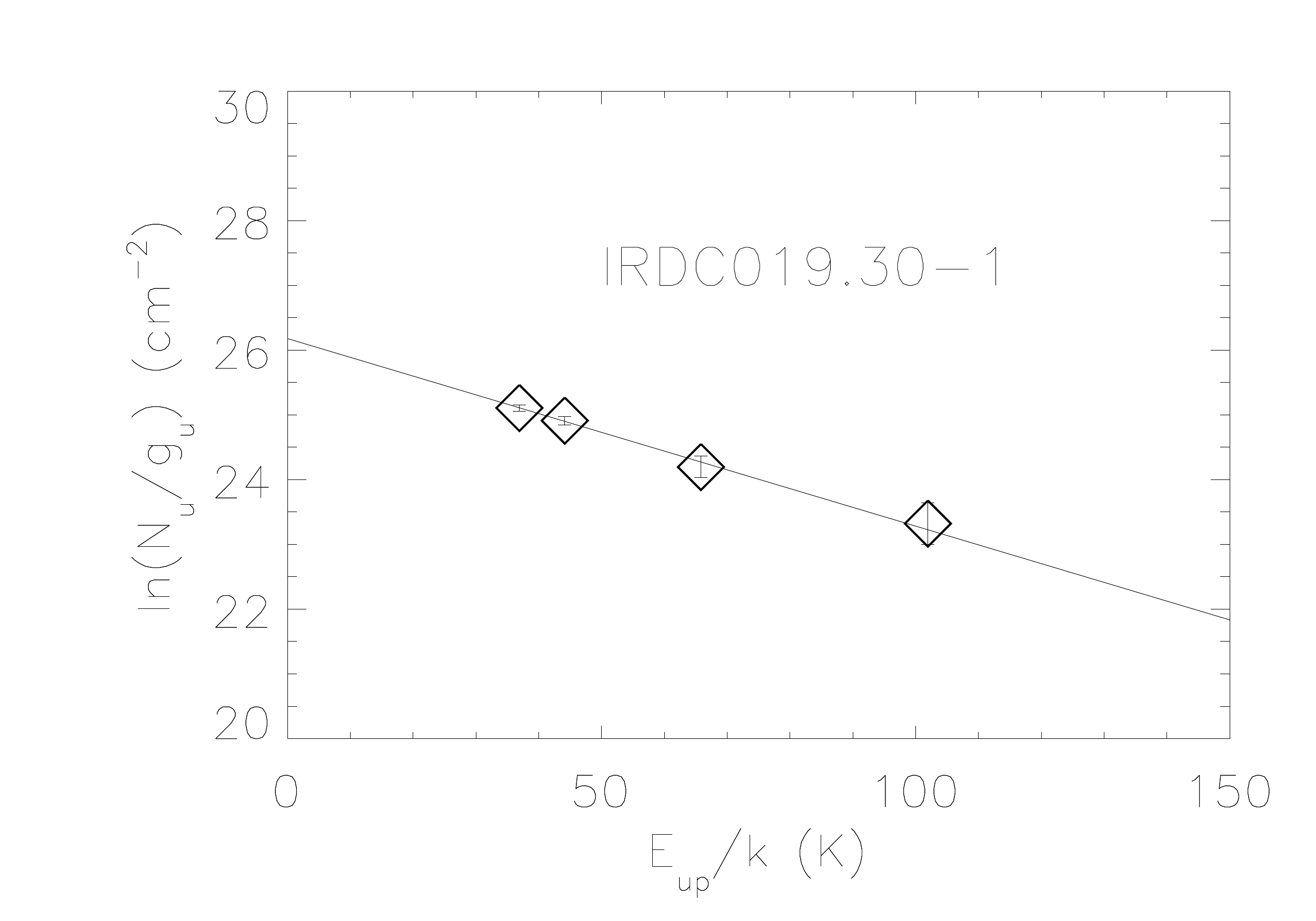}
\includegraphics[width=3.5cm, angle=90]{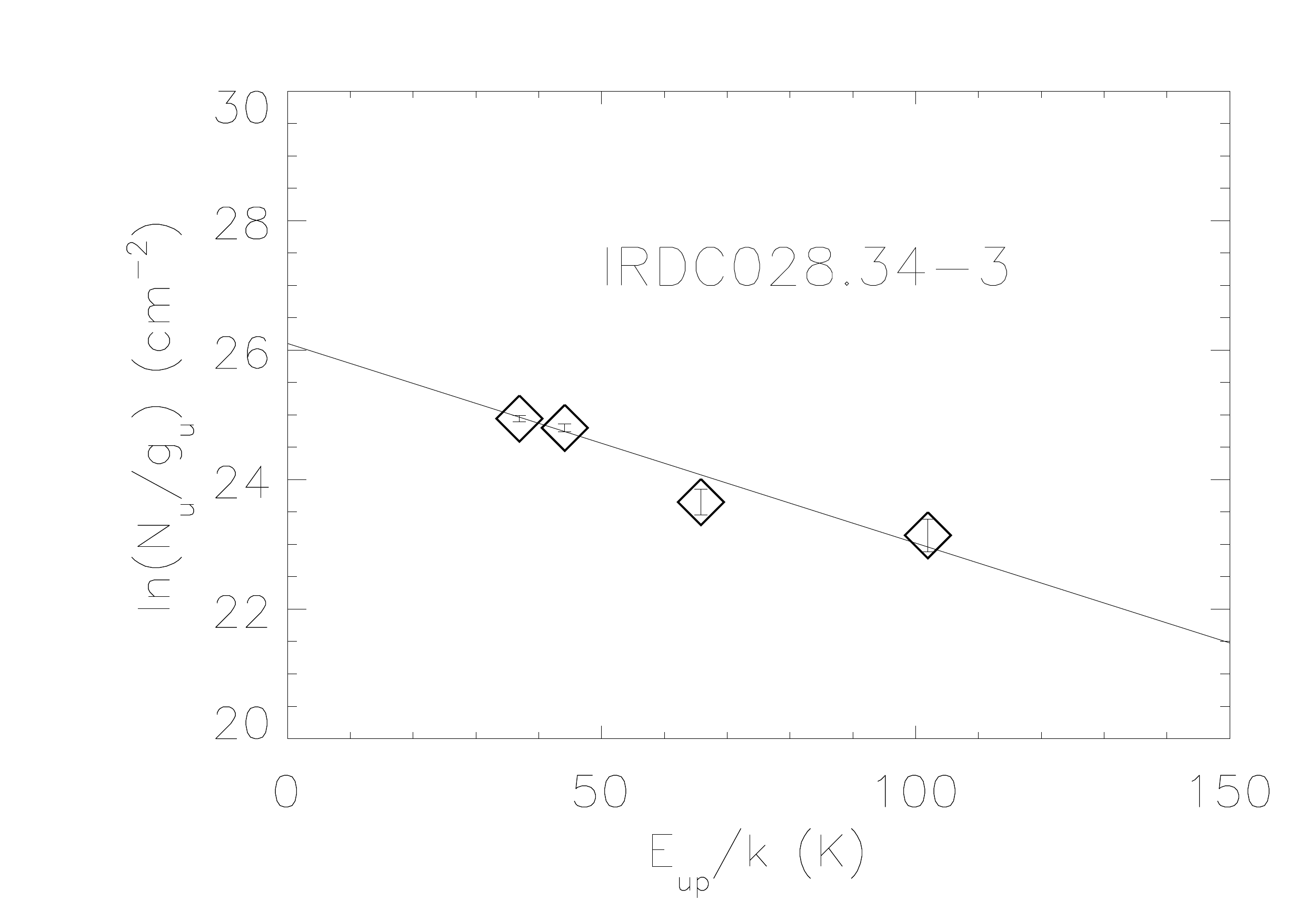}
\includegraphics[width=3.5cm, angle=90]{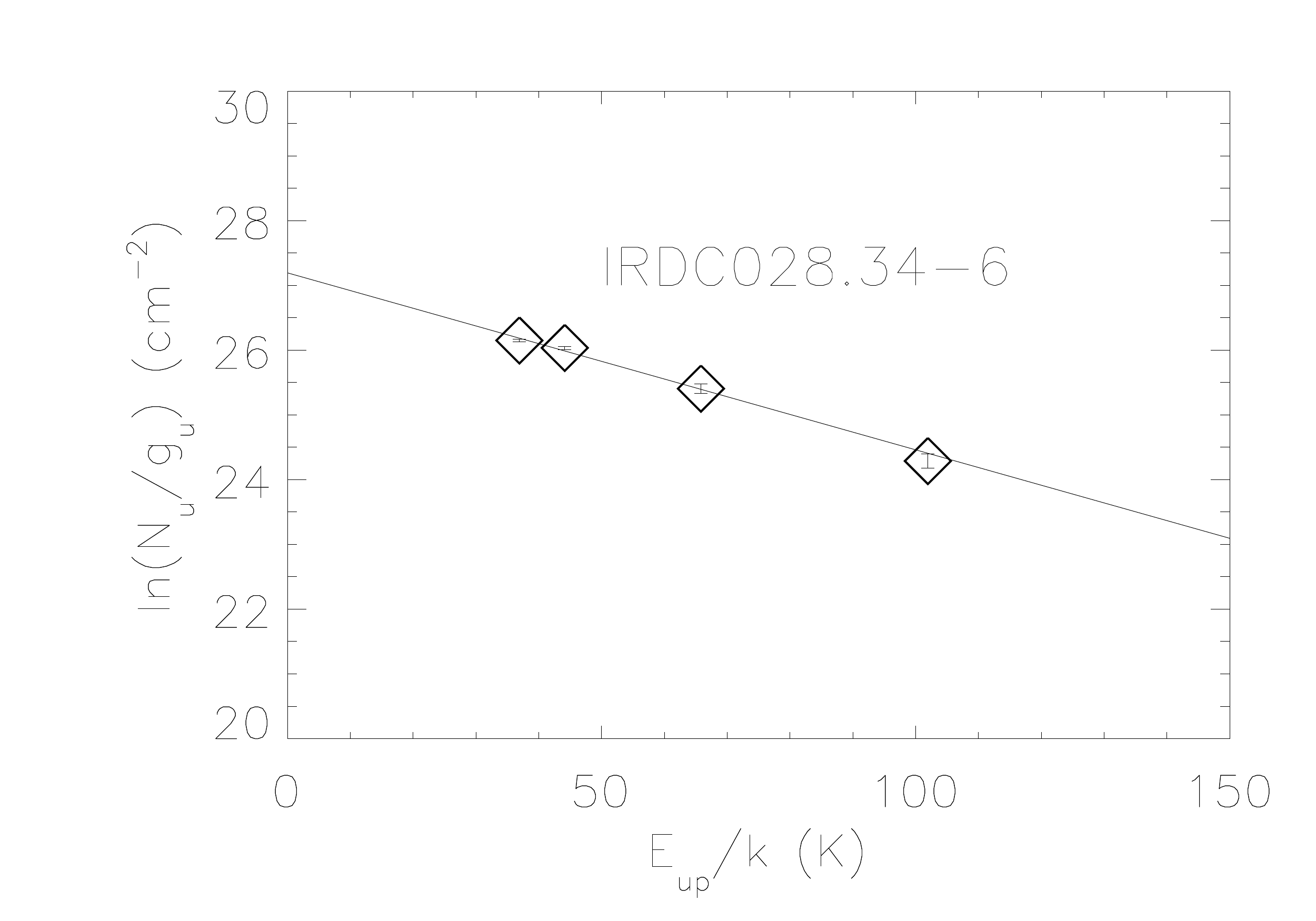}

\includegraphics[width=3.5cm, angle=90]{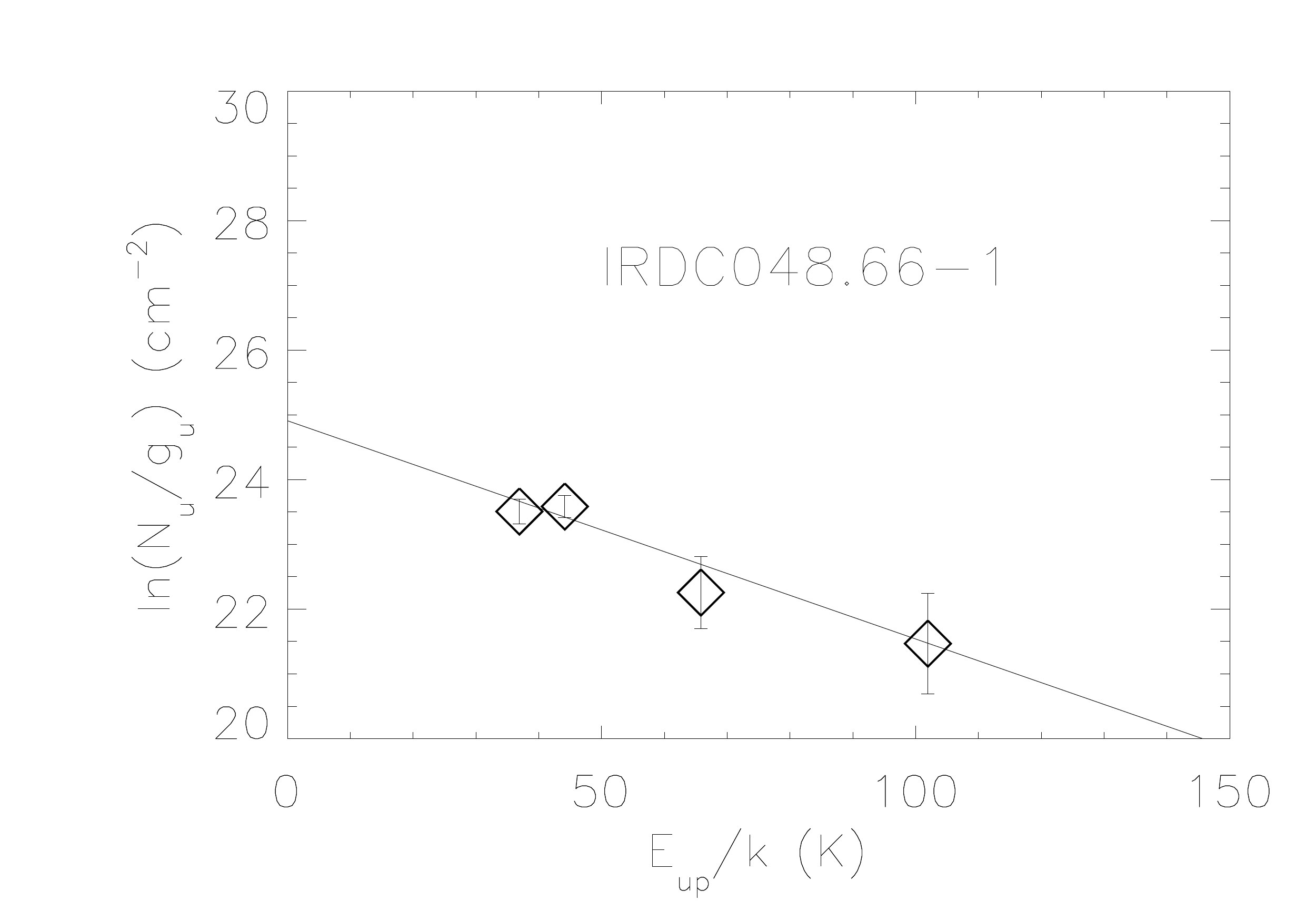}
\includegraphics[width=3.5cm, angle=90]{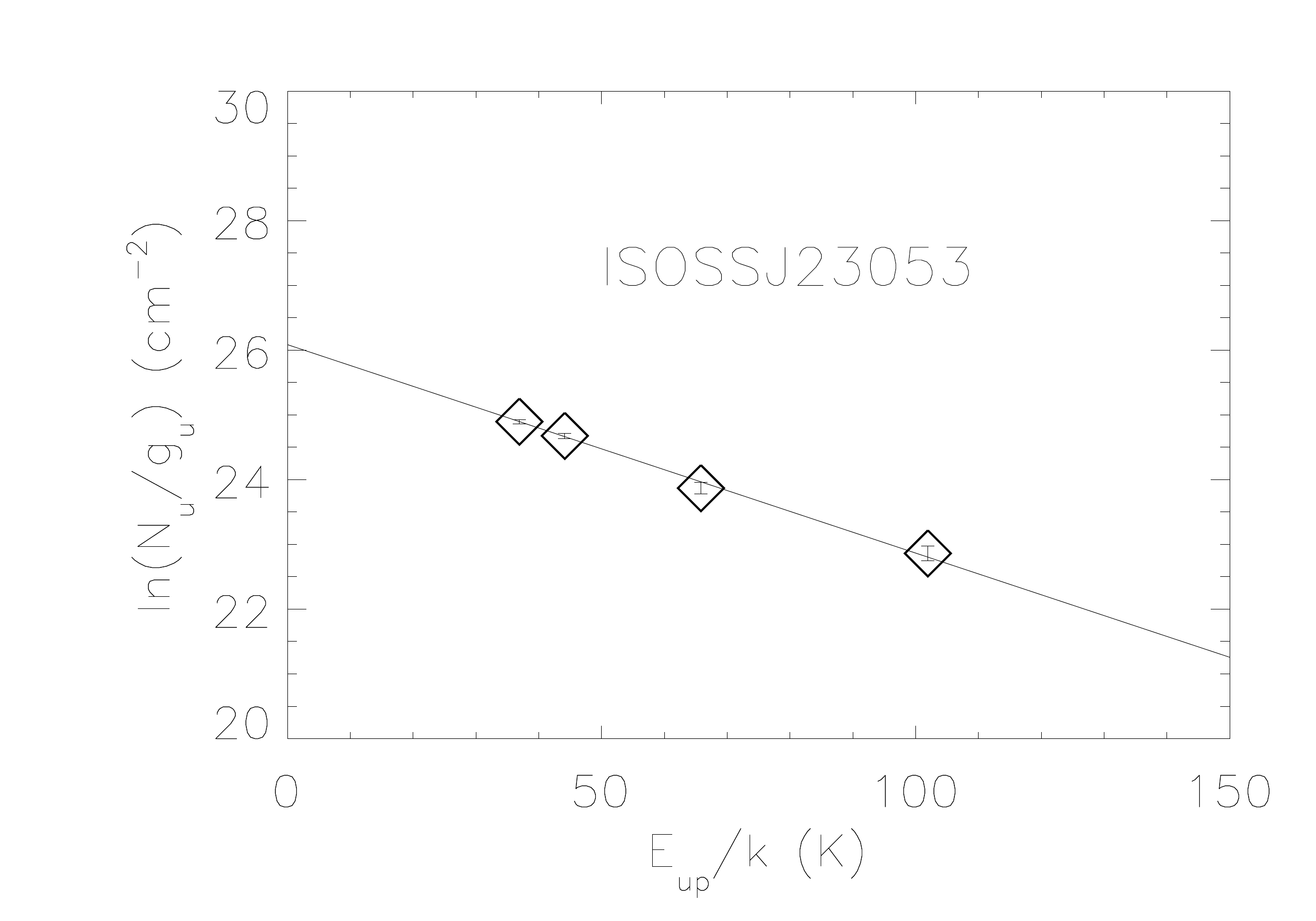}
\caption{Excitation diagrams for CH$_3$CCH.}
\label{excitation_CH3CCH}
\end{figure}

In the northern sources, we observed CH$_3$CCH J=9-8. 
Eight sources show the presence of K=0, 1, 2, 3 components.
For these sources, we performed the excitation analysis and estimated CH$_3$CCH column densities and
excitation temperatures. 
Excitation diagrams for these eight sources are presented in Figure~\ref{excitation_CH3CCH}.
Eight  more sources show the presence of only K=0 and K=1 transitions.
For these cases, 
only the K=0 transition and the simple formula  from \citet{2011A&A...527A..88V} were used to estimate column densities and abundances.

\begin{figure}
\centering
\includegraphics[width=4cm,angle=90]{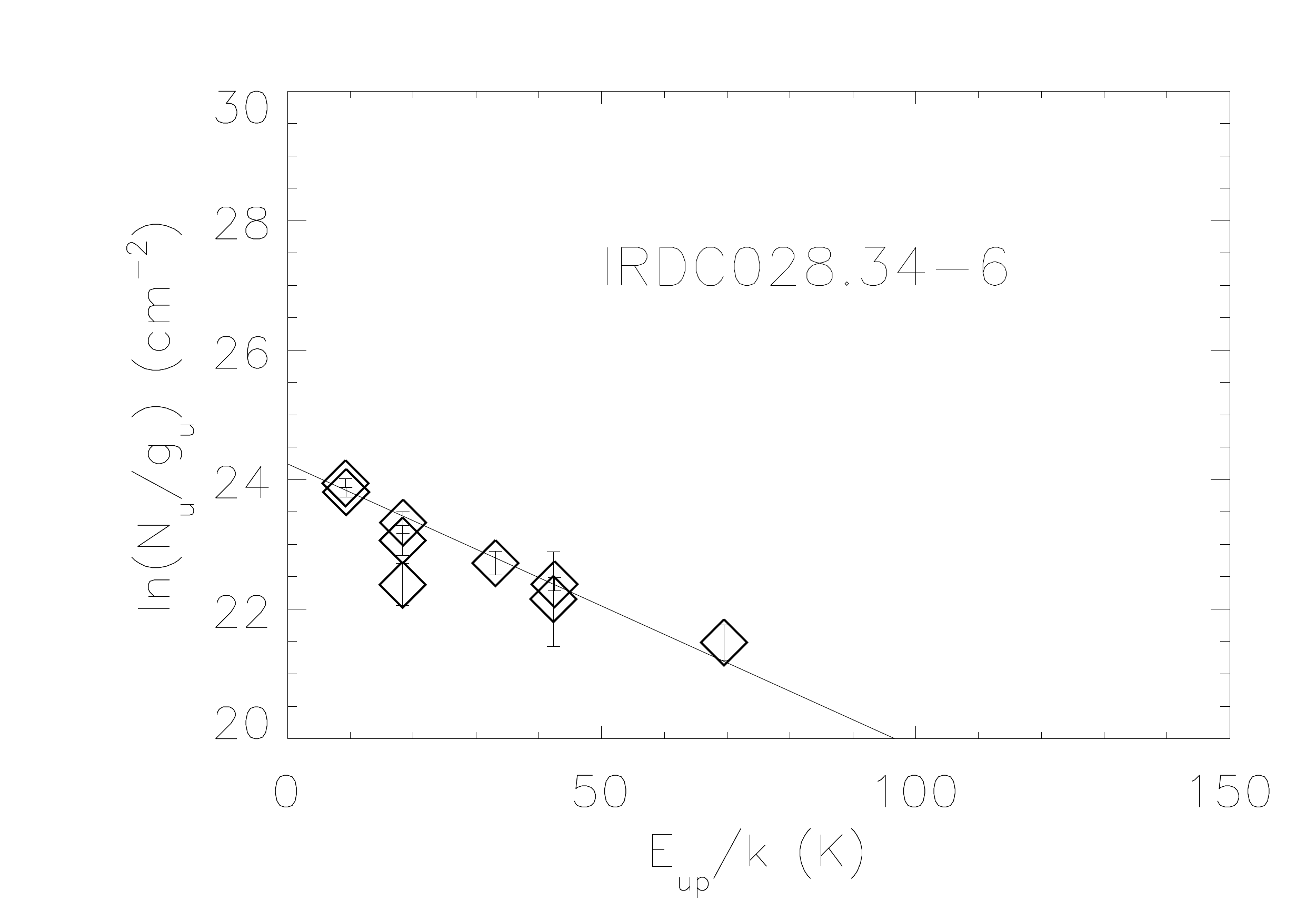}
\includegraphics[width=4cm,angle=90]{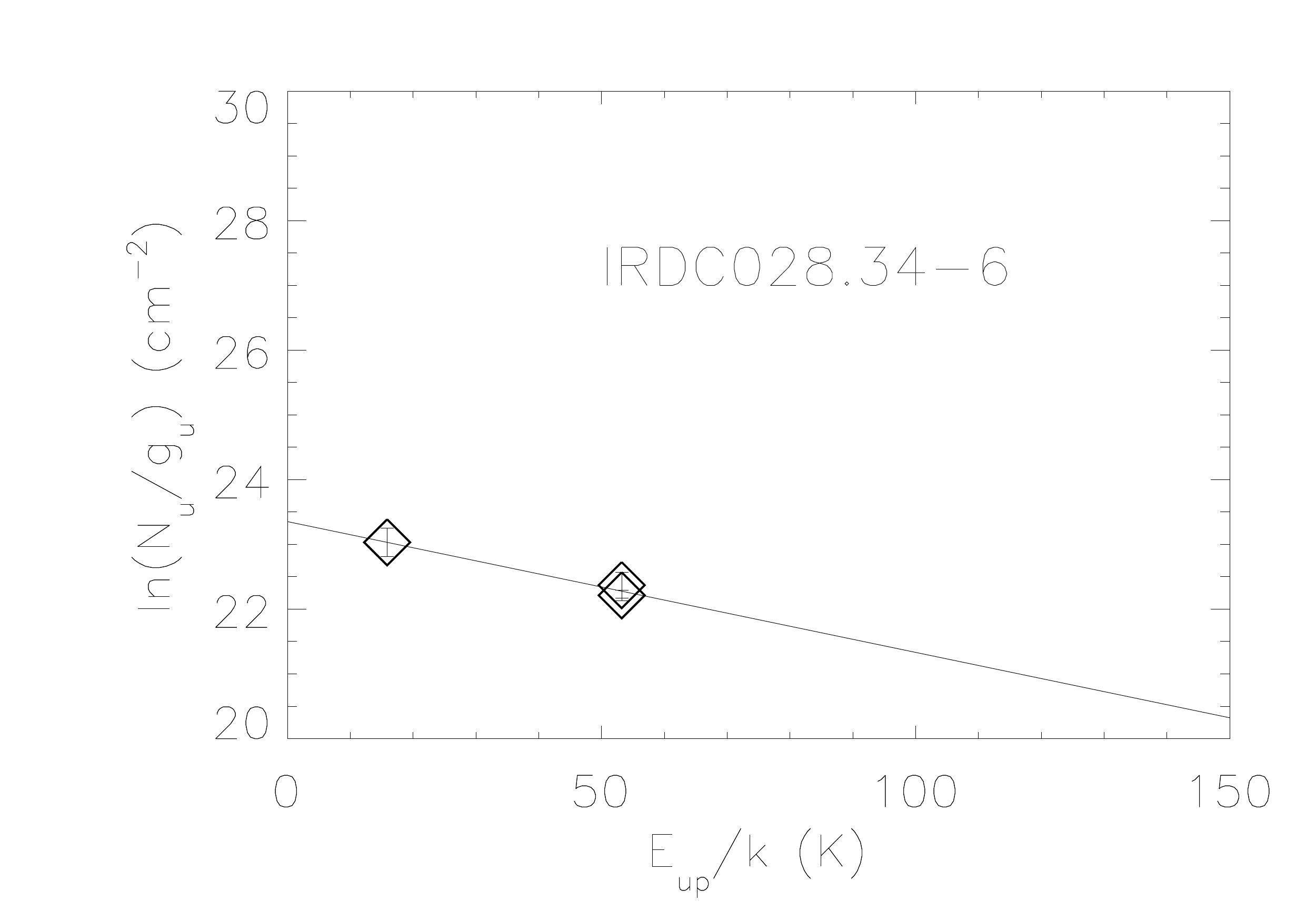}
\caption{Excitation diagrams for CH$_3$CHO (left) and CH$_3$OCHO (right).}
\label{excitation_CH3CHO}
\end{figure}

In the present study,  CH$_3$CHO (acetaldehyde) was covered only in northern  targets.
The 4$_{04}$-3$_{03}$  A++ transition  was detected in more than 50\% of the clouds, including 
those that can be qualified as quiescent.   
The 4$_{04}$-3$_{03}$  E transition,  despite a similar upper energy level, was
detected only in the six most active regions.
This can indicate non-LTE behavior since the two upper states are not degenerate and are of different symmetries.
A unique behavior was noticed for IRDC028.34-6, where we
detected  seven more transitions of CH$_3$CHO around 77 GHz and 153 GHz
and were able to perform an excitation analysis, as shown in Figure~\ref{excitation_CH3CHO}.
See Table~\ref{table:lines_IRDC028.34-6} and Section~\ref{sec:IRDC028.34-6} for more detail.

Our IRAM setup included several transitions of CH$_3$OCH$_3$ (dimethyl ether) around 153055 MHz. 
In five northern objects ,we observed quite strong detections.
Since these transitions are blended, we used a method described in \citet{2006A&A...455..971R} to get a rough estimate
of the CH$_3$OCH$_3$ abundance. 
The obtained values can be found in Table~\ref{table:ab_IRAM}.
The derived abundances of CH$_3$OCH$_3$ are listed as upper limits because of the possibility of line-blending with other species.

In the present study, CH$_3$OCHO  (methyl formate)  was detected only in the active region IRDC028.34-6, 
where three transitions were observed, as shown in Table~\ref{table:lines_IRDC028.34-6}.
These transitions allowed us to perform an excitation analysis
and to estimate the
molecular column density and excitation temperature  (see Figure~\ref{excitation_CH3CHO}).

\subsection{Molecular Abundance Calculations}
\label{abundance}

Molecular column densities and molecular abundances 
have been estimated for all observed species. 
We are assuming that the emission from the COMs is extended and uniformly distributed within the beam of the single-dish observations. 
Therefore, the beam filling factor is $\sim$~1.
To estimate column densities we used two methods.
For both cases, we assumed LTE conditions and
optically thin emission. 
In the case where only one transition was detected for a particular species
(e.g. HNCO, H$_2$CO),
we used the following formula to estimate molecular column densities:

\begin{equation}
\begin{array}{lll}
	N_{\rm tot} &=&\frac {8 \pi }{\lambda^3 A}  \frac {1}{J_\nu (T_{ex})-J_\nu(T_{bg})}  \frac {1}{1-exp(-h \nu /kT_{ex})} \times\\
	&\times &  \frac {Q_{rot}}{g_u exp(-E_l /kT_{ex}) }  \int T_{\rm mb} d \upsilon ,\\
\end{array}
\label{Equ:column}
\end{equation}

\noindent
where $\lambda$ is the rest wavelength of the transition, $A$ is the 
Einstein coefficient,
$g_u$ is the upper state degeneracy, $J_{\nu} (T_{ex})$ and $J_{\nu} (T_{bg})$ are
are values of the Planck function at excitation and background
temperatures, respectively,
$Q_{rot}$ is the partition
function, and $E_l$ is the energy of the lower level \citep{2009AJ....138.1101L}. 
For g$_u$, A and $E_l$, we used values from 
The Cologne Database for Molecular Spectroscopy (CDMS)
\citep{2001A&A...370L..49M,2005JMoSt.742..215M} for most of the species.
For CH$_3$CHO and CH$_3$OCHO,  values from the JPL database
\citep{1998JQSRT..60..883P} were employed.
For the excitation temperatures $T_{ex}$, 
we assumed ammonia kinetic temperatures and/or dust temperature 
from Tables~\ref{table:main_Mopra} and~\ref{table:main_IRAM}. 
We calculated  the partition function Q$_{rot}$ for every source by interpolating 
data from the CDMS for the particular source temperature $T_{ex}$. 
Integrated intensities $\int T_{\rm mb} d \upsilon$ for every species
were estimated as areas under the fitted Gaussians (see Tables~\ref{table:gauss_Mopra} and~\ref{table:gauss_IRAM}).

If at least three transitions were detected (e.g. for CH$_3$OH, and
CH$_3$CCH),
an excitation analysis was used to estimate the molecular column density and 
excitation temperature.
Since we assumed that all levels are in LTE, 
the total column density is related to the single excitation temperature via the
equation

\begin{equation}
 {\rm ln} \left(  \frac {N_{\rm u}}{g_{\rm u}} \right) = {\rm ln} \left( \frac {N_{\rm tot}}{Q (T_{ex})} \right) - \frac {E_{\rm u}}{kT_{ex} },\\
\label{Equ:column2}
\end{equation}

\noindent
where $N_{\rm u}$, $g_{\rm u}$, and  $E_{\rm u}$ are the column density, degeneracy,
and upper state energy for the transition. 
$N_{\rm tot}$ is the total column density,
Q (T$_{ex}$) is the partition function at a given temperature,  and T$_{ex}$ is the
excitation temperature.
As in the first case, all necessary parameters were taken from the
CDMS and JPL databases. 
The population diagram for each molecule with at least three detected transitions was formed by plotting 
${\rm ln} \left(  N_{\rm u}/g_{\rm u} \right)$ against the upper state energy,  $E_{\rm u}/k$, in
Kelvin.
 When the populations of all levels are in LTE,
it should be possible to fit a single straight line to the plotted points.
In this case, the slope of the line will be proportional to $1/T_{\rm ex}$.
Then the determined excitation temperature can be used 
for calculating a total column density.
The method is described in more detail in, e.g., \citet{1999ApJ...517..209G}.


The  H$_2$ column densities required to convert molecular column densities into fractional abundances
were calculated based on 1.2 millimeter SIMBA/SEST (Swedish-ESO Submillimetre Telescope)  continuum
data for  the southern sample, and 870 $\mu$m ATLASGAL (The APEX Telescope Large Area Survey of the Galaxy) 
continuum data for  the northern sample. The continuum maps were convolved to the coarser spatial
resolution of the respective line observations before extracting the peak fluxes. Then, the following
general equation was used:

\begin{equation}\label{Equ:columndens}
N_{\rm H_2} =   \frac {F_{\rm peak} \ R}{\Omega \ B_v(T) \ \varkappa_v \ {\rm m}_{\rm H_2}} .
\end{equation}

\noindent
Here  the measured source peak flux density is given by $F_{\rm peak}$, 
$\Omega$ is the beam solid angle in steradians, $m_{\rm H_2}$ is the mass of one hydrogen
molecule, $R$ is the gas-to-dust ratio, $\varkappa_v$ is the dust opacity per 
gram of dust, and $B_v(T)$ is the Planck function at the dust temperature $T$. 
We adopt a gas-to-dust mass ratio of 100,
an  $\varkappa_v$ value of 1.0 cm$^2$ g$^{-1}$ and 1.85 cm$^2$ g$^{-1}$
for the 1.2 mm and  870 $\mu$m data respectively,
which are values appropriate for cold dense cores 
\citep{1994A&A...291..943O}. 
Obtained  H$_2$ column density values are presented in 
Tables~\ref{table:main_Mopra}-\ref{table:main_IRAM}.
Tables~\ref{table:gauss_Mopra} and \ref{table:gauss_IRAM} show all necessary line parameters.
The resulting molecular abundances for the southern and northern IRDCs 
are listed in Tables~\ref{table:ab_Mopra} and \ref{table:ab_IRAM}, respectively.

\subsection{Temperatures and their influence on molecular abundances}

Temperature determination is important  for the calculation of observed 
abundances as well as for the results of models.
For example, temperatures that differ by a factor of 2 will give different results for H$_2$ and molecular 
column densities and, as
a result, abundances will be uncertain by at least  
a factor of 2-4. 
As shown in \citet{2012ApJ...751..105V},
temperature variations in the chemical model
can also significantly change the abundances of most species.  

As was mentioned in Section~\ref{source_description}, 
for northern sources we use  temperature values from 
\citet{2006A&A...450..569P, 2008ApJ...678.1049S, 2012A&A...544A.146W, 2012ApJ...756...60S} and \citet{2013ApJ...766...68P}. 
Two ways to estimate temperatures have been used for these sources.
The first provides us with the gas temperature and based on the  ratio of NH$_3$ (1,1) and (2,2) inversion transitions,
the second enables IR and sub-millimeter data to estimate the dust temperature.
For our sample, ammonia kinetic temperatures are typically lower than
dust temperatures, as shown in  Table~\ref{table:main_IRAM}. 
Excitation temperatures that we obtained by performing an excitation analysis for CH$_3$CCH 
in eight sources are around 30 K; this value is closer to the dust temperatures  than to ammonia gas temperatures. 
Therefore, for the northern clouds, we estimated an upper and a lower limit for the molecular column densities and abundances,
 using both higher and lower temperatures. 
 In the case where there is no information about temperature values in a 
 particular source, we assume 10~K as a temperature lower limit  and 30~K as a temperature upper limit.
 Both abundance values are presented as upper and lower limits in Table ~\ref{table:ab_IRAM}.
For the southern sample, 
for most of the clouds we only had temperature estimates based on the ammonia data,
Alternative estimates based on the CH$_3$CCH and CH$_3$OH
data are available for a few sources.
Similar to the northern IRDCs case, these few objects show elevated temperatures,  as presented in Table~\ref{table:ab_Mopra}.
For those sources, where alternative temperature estimate is available, 
we estimated upper and lower limits for the molecular abundance values.

The differences between calculated gas and dust temperatures might be  
partially explained by the different telescope beam sizes that were used to
obtain these data.
Our ammonia Parkes data and data from \citet{2008ApJ...678.1049S} were obtained with an $\sim$~70$''$ beam,
ammonia observations by \citet{2006A&A...450..569P} and \citet{2012A&A...544A.146W} were performed with an $\sim$~40$''$ beam,
while the 16$''$ IRAM beam size for the data reported in the current paper is much smaller.
In \citet{2013ApJ...766...68P}, dust temperatures based on Herschel data were obtained with the data smoothed to a beam size of 36$''$ and temperatures
reported in \citet{2010ApJ...715..310R} and \citet{2012ApJ...756...60S} were estimated based on IR and sub-millimeter data with 
beam sizes from 8$''$ to 24$''$.
However, different temperature measurements may also indicate the presence of both cold and warm regions within IRDCs, which are traced by different species.

\subsection{Molecular abundance comparisons  among IRDCs, HMPOs, hot cores, and low-mass pre-stellar cores}



For all  molecular species discussed in the current study, 
there is an abundance variation within one order of magnitude for  the
northern and southern samples.
Such differences are in agreement with our previous work 
\citep{2011A&A...527A..88V}
and can be explained by 
the slight variations in the physical conditions and/or age of the objects. 

The detected CO abundance  is $\sim$ 10$^{-4}$ for  our southern targets, 
and, depending on the excitation temperature, $\sim$ 10$^{-5}$--10$^{-4}$ for the northern targets.
The higher CO values of 10$^{-4}$ are in agreement with the \citet{2006ApJS..166..567R} results for IRDCs 
and the \citet{1997A&AS..124..205H} results for HMPOs, whereas lower CO values of 10$^{-5}$ lie closer to
the typical  estimate for low-mass pre-stellar cores from \citet{1992IAUS..150..171O} and \citet{2006A&A...455..577T}. 

The abundances for HNCO and CH$_3$CCH are slightly higher for  our southern than 
for  our northern clouds. 
Taking into account the lower Mopra sensitivity, such  a difference can be explained by
a selection effect.
With the Mopra telescope, we were able to detect these molecules only 
in the regions with their highest abundance, whereas the high sensitivity of the IRAM telescope
allows us also to detect HNCO and CH$_3$CCH also in  regions with lower concentration.
In general, the calculated HNCO and CH$_3$CCH abundances are  still close to 
low-mass pre-stellar core,  HMPO, and hot core values   
\citep{1995ApJS...97..455S, 2000A&A...361.1079Z, 2009ApJ...690L..27M}.   


 The formaldehyde abundances lie in a range between 10$^{-10}$ and 10$^{-9}$ for both
southern and northern IRDCs.
These values are in good agreement with previous estimates for IRDCs and 
HMPOs  \citep{1997A&AS..124..205H,1998ApJ...508..721C}.
Previous values for  low-mass pre-stellar cores have a larger spread 
from 10$^{-10}$ in L1517B \citep{2006A&A...455..577T}
to 10$^{-8}$ in L134N and TMC-1CP  \citep{1992IAUS..150..171O}.
As  was  first shown by \citet{1987ApJ...318..392M} and later adopted to the case of IRDCs by 
\citet{1998ApJ...508..721C}, 
high abundances of H$_2$CO are typical for objects with  lower densities of $\sim$10$^4$ cm$^{-3}$,
whereas regions with higher densities of $>$ 10$^5$ cm$^{-3}$ show lower formaldehyde abundances. 
This result  will be discussed in more detail in Section~\ref{discussion}.
Such behavior might indicate the presence of regions with greater density in high-mass star forming regions,
in comparison with some low-mass pre-stellar cores. 




For CH$_3$OH, 
we observe a large spread from 10$^{-10}$ to 10$^{-8}$.
 Such  a broad interval is in agreement with previously estimated values for IRDCs by
 \citet{2007A&A...466..215L, 2010ApJ...714.1658S, 2012A&A...540A.104M}, and Sanhueza et al. (submitted), and
 with typical 10$^{-10}$-10$^{-9}$ values for low-mass starless cores \citep{1992IAUS..150..171O,2006A&A...455..577T}.  
 Active regions such as Orion and W3 IRS4 and IRS5 also show a large spread in methanol 
 abundances,
 but with a tendency to higher values: 10$^{-9}$ - 10$^{-7}$
 \citep{1995ApJS...97..455S, 1997A&AS..124..205H}. 
Abundance values of 10$^{-9}$  and lower can be easily explained by non-thermal desorption and
therefore, might correspond to cold ``quiescent'' regions \citep{2007A&A...467.1103G}.  
The higher CH$_3$OH values of  10$^{-8}$ - 10$^{-7}$ require  an additional
warm-up phase
induced by stellar heating or the interaction of outflows, as will be discussed in Section~\ref{section:warmup2}.

While the detection of CH$_3$CHO  and CH$_3$OCHO in EGOs associated with IRDCs
has already been reported  in a recent paper by \citet{2012ApJS..202....1H}, we estimate
abundances of these species  for IRDCs for the first time.
The abundance of CH$_3$CHO lies 
 in the range from 10$^{-9}$ to 10$^{-8}$, depending on the assumed excitation temperature,
for all IRDCs where it was detected (see Table~\ref{table:ab_IRAM}), while
the abundance of CH$_3$OCHO in IRDC028.34-6 is 3.7$\times$10$^{-10}$ for
$T_{ex}$=30~K.
In more evolved high-mass star forming regions such as HMPOs and hot cores, 
typical values for CH$_3$CHO  are in agreement with our result \citep[][and references therein]{2008ApJ...682..283G}, and 
CH$_3$OCHO abundances  
are similar to our single result or somewhat higher, in the range  10$^{-10}$ - 10$^{-8}$ \citep{2004ApJ...615..354B, 2009A&A...499..215B}.
However, 
abundances of both species for  
low-mass  pre-stellar objects lie in the range from   10$^{-11}$ to 10$^{-10}$
\citep{2012A&A...541L..12B,2012ApJ...759L..43C}, 
which is  lower than in IRDCs and hot cores
by a factor of 10$-$10$^3$.


 In summary, these comparisons show  that  there is an overlap in abundances  of carbon-bearing species among
IRDCs, low-mass  pre-stellar cores, HMPOs, and hot cores.
However, the general trend,
 especially for CH$_3$OH, CH$_3$CHO and CH$_3$OCHO,
 is to have a higher  value in those regions
where the temperature is higher, as in HMPOs and especially hot cores.

\subsection{The case of IRDC028.34-6}
\label{sec:IRDC028.34-6}

\begin{figure}
 \centering
   \includegraphics[width=12cm]{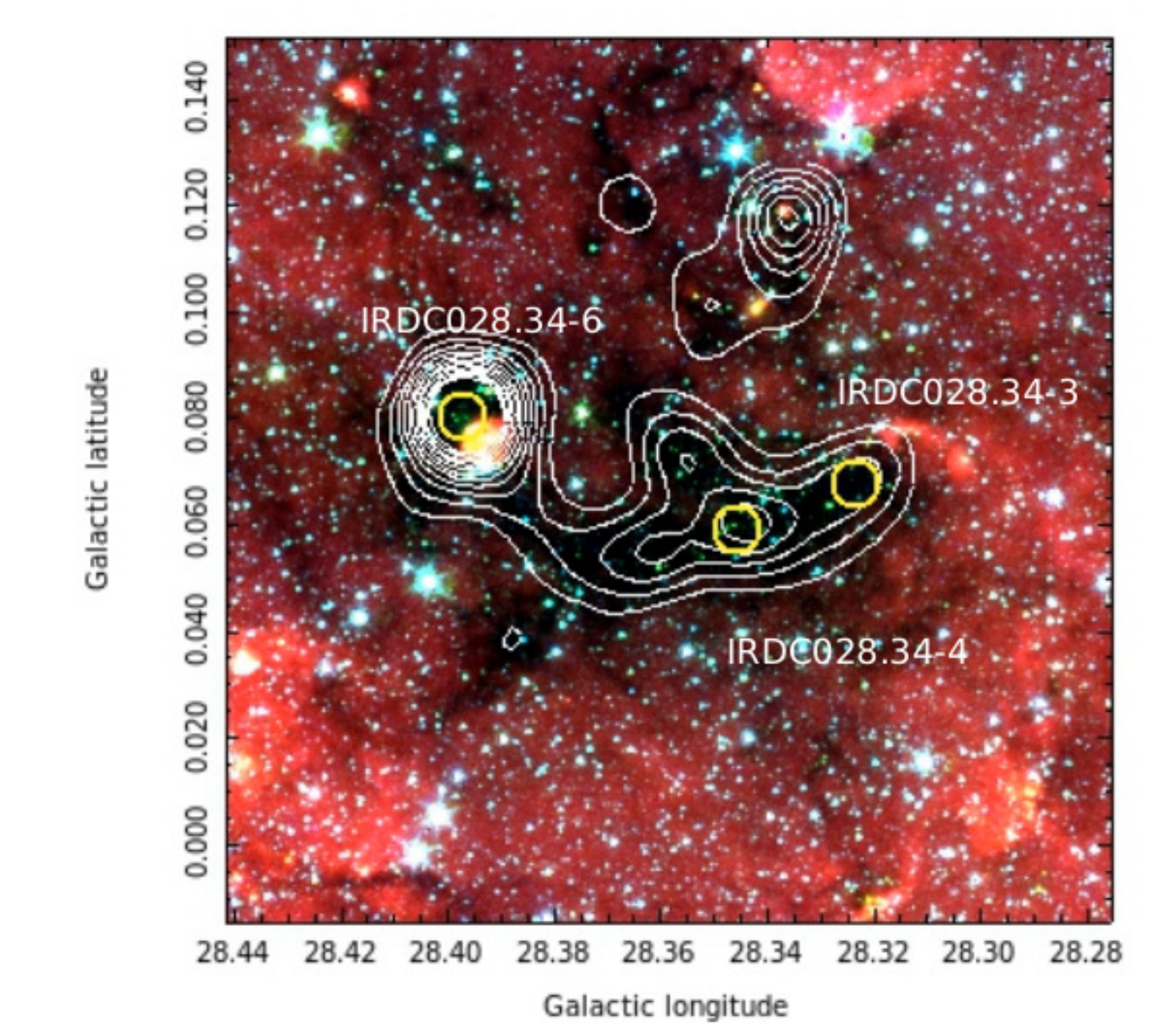}	   
   \caption{Spitzer/GLIMPSE 3-color image or IRDC028.34+0.06, where 3.6$\mu$m is shown in blue, 4.5$\mu$m in green, and 8.0$\mu$m
		in red. White contours show APEX 870$\mu$m emission. Circles mark observed positions and show the IRAM beam size at 154 GHz. }
  \label{3color_IRDC028.34}
\end{figure}

IRDC028.34+0.06 is a relatively well studied and highly structured cloud, with
much continuum and molecular line data collected over the past 15 years. It
was first mentioned in the pioneering H2CO study by \citet{1998ApJ...508..721C},
where basic parameters as distance and molecular abundance were estimated for
IRDCs for the first time. 
In millimeter and sub-millimeter surveys by \citet{2000ApJ...543L.157C} and \citet{2006ApJ...641..389R}, 
the entire cloud shows the
strongest emission in comparison with other objects of their IRDC samples, and
has the highest mass ($>4$000 M$_{sun}$) and H$_2$ column density ($>$ 10$^{23}$ cm$^{-2}$).
First follow-up studies revealed  that the sub-mm continuum emission
associated with the northern tip of G28.34+0.06 \citep[component P2 in][]{1998ApJ...508..721C}
does not spatially coincide with the bright slightly extended MIR
emission of this part of the cloud (see Figure~\ref{3color_IRDC028.34}). Indeed, the peak of the sub-millimeter
emission is about 8" away from the border of the aforementioned MIR emission
blob. The sub-millimeter peak does not show emission at 8 micron in the
Spitzer/GLIMPSE data but, however, is coincident with a faint 24 micron point
source in the Spitzer/MIPS data \citep[cf.][]{2012A&A...547A..49R}. Such a behavior
indicates the presence of different evolutionary stages in relatively close
vicinity. High resolution Submillimeter Array (SMA) observations performed by
\citet{2009ApJ...696..268Z} revealed the presence of CH$_3$CN emission associated with
IRDC028.34-6 and provided an excitation temperature estimate of 120 K. This
detection might indicate the presence of a small hot core associated with this
sub-millimeter source when observed with arcsecond resolution.

In IRDC028.34-6 we detect more molecular lines than in any other target.  
All obtained spectra are presented in Figure~\ref{spectra6}.
Moreover, it is the only source where  methyl formate was detected.
Additional transitions for COMs and smaller carbon-containing species,
detected only in IRDC028.34-6 during the current study
 within the passband of the IRAM 30m observations at 76 GHz and 153 GHz, 
are presented in Table~\ref{table:lines_IRDC028.34-6}.

Since two temperature estimates are available in the literature for IRDC028.34-6,
a value of $\approx$~16~K based on the \citet{2006A&A...450..569P} ammonia data and 
a higher value of 33~K based on  the  \citet{2012ApJ...756...60S} dust analysis,
both values were used to estimate abundances of CO, HNCO, and H$_2$CO,  as shown in Table~\ref{table:ab_IRAM}.
Detection of several transitions for CH$_3$OH, CH$_3$CCH, CH$_3$CHO, and CH$_3$OCHO 
allowed us to perform excitation analyses and obtain a more reliable estimate of the excitation temperature and relative abundances, which are
 presented in Table~\ref{table:ab_IRAM}.


In the next section, we will 
report a comparison between model results and molecular observations for IRDC028.34-6.
In order to make a model that is more reliable and to analyze the behavior of organic species and 
cold and dense gas tracers simultaneously, 
the molecular abundances of HNC, HCO$^+$, N$_2$H$^+$, NH$_3$, C$_2$H, and HC$_3$N 
 were used in addition to the organic molecules presented in the current paper.  The additional abundances 
were taken from previous studies by
\citet{2006A&A...450..569P} and \citet{2012ApJ...756...60S}.
All observational abundances used for the comparison with model results 
are presented in Table~\ref{table:ab_IRDC028.34-6}.

\section{Modeling of IRDC028.34-6}
\label{modeling}

\subsection{1-Phase 0-D Model with constant temperature}

In \citet{2012ApJ...751..105V}, we reproduced  abundances of eight cold and dense gas tracers,
using a 0-D gas-grain pseudo time-dependent model. 
As a test of this simple approach, we have attempted to reproduce abundances of additional organic species with four different sets of conditions. 
In all cases, the H$_2$ density is 10$^5$ cm$^{-3}$, while the temperatures are 15, 25, 35 and 120~K. 
These correspond to the ammonia kinetic temperature,  the excitation temperatures of acetaldehyde  (CH$_3$CHO), and methyl acetylene  (CH$_3$CCH),
the dust temperature, and the CH$_3$CN excitation temperature obtained by  \citet{2009ApJ...696..268Z}
 respectively. 
In this study, we adopted the same gas-grain network as in \citet{2013ApJ...762...86V}, 
which includes gas-phase, gas-grain interactions and reactions on the grain surface. 
This is a version of the KIDA network with an addendum of grain-surface chemical reactions originally published in \citet{2010A&A...522A..42S}.\footnote{kida.obs.u-bordeaux1.fr/uploads/models/benchmark\_2010.dat} 
The granular chemistry was treated by rate equations, rather than the more advanced and computer-intensive stochastic approaches \citep{2013ApJ...762...86V}.
Elemental abundances and initial abundances were taken from 
\citet{2008ApJ...680..371W} \citep[see also][]{1974ApJ...193L..35M, 1982ApJS...48..321G} 
for the low metal case.  
Other than hydrogen, which starts out totally in molecular form, the initial species are atomic in nature.  
The grains are initially bare. 
The cosmic ray ionization rate is assumed to be the canonical value for atomic hydrogen of 1.3$\times$10$^{-17}$ s$^{-1}$.
The observed H$_2$ column density of  8.3$\times$10$^{22}$ cm$^{-2}$  corresponds 
to a visual extinction A$_{\rm V}$ $\approx$ 90 \citep[according to][]{1996Ap&SS.236..285R}.
Unlike the previous study, together with thermal and cosmic ray desorption, we 
include photo-desorption processes, with a yield per photon of 10$^{-3}$ for all species, 
according to suggestions by \citet{2007ApJ...662L..23O}.   
Reactive desorption, as described in \citet{2007A&A...467.1103G}, was not taken into account in \citet{2012ApJ...751..105V}
and was not included in the initial calculations for the current study.

\begin{figure}
 \centering
\includegraphics[width=10cm,angle=90]{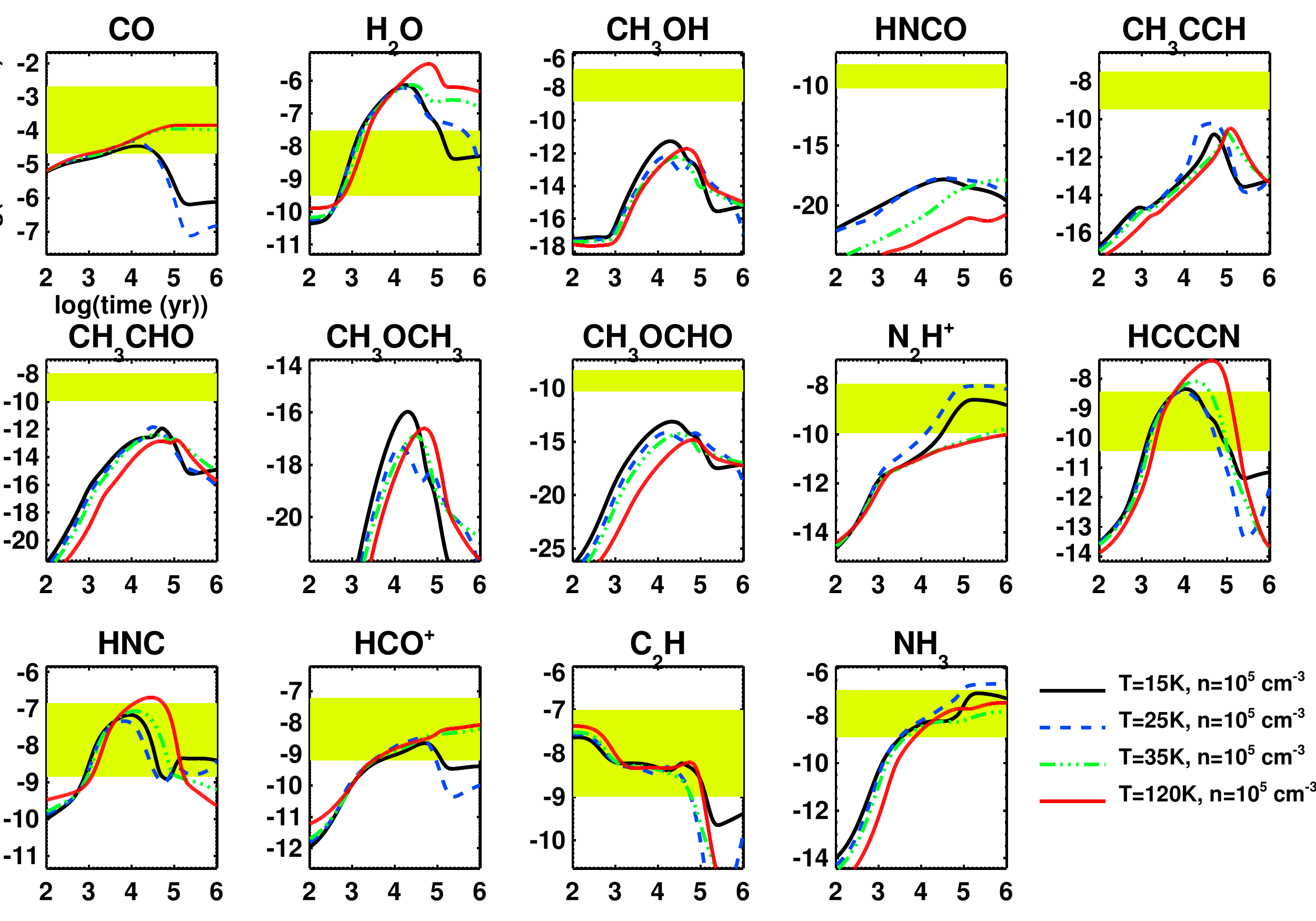}  
   \caption{Calculated fractional abundances relative to H$_2$ as a function of time for 14 gas-phase species. Different lines correspond to
   		different temperatures and boxes correspond to the observed values  in IRDC028.34-6 $\pm$ one order of magnitude.
		For these calculations, we used a 0-D pseudo time dependent model. }
  \label{0D_IRDC028.34}
\end{figure}

Figure~\ref{0D_IRDC028.34} shows modeled abundance profiles for times up to
10$^6$ yr for 14 species.
The boxes in the panels represent observational values for IRDC028.34-6
with respect to H$_2$ $\pm$ one order of magnitude.
We reach agreement within an order of magnitude between model and 
observations for
HNC, HCO$^+$, CCH, HC$_3$N, H$_2$CO,  NH$_3$, and CO
at all four temperatures, and for N$_2$H$^+$ at 15 and 25 K,  in a time range between 10$^4$ and 10$^5$ yr.
However, we cannot reproduce the abundances of the five organic species CH$_3$OH, HNCO, CH$_3$CCH, CH$_3$CHO, and CH$_3$OCHO in the gas phase with either of the four chosen temperatures.
Indeed, differences between model and observational abundances can reach more than two orders of magnitude
at all times.   The chemistry leading to these differences is discussed below.

The primary formation route of methanol is the hydrogenation sequence on surfaces of interstellar grains:
\begin{equation}\label{ch3oh_form}
{\rm grCO}\rightarrow {\rm grHCO}\rightarrow {\rm grH_{2}CO}\rightarrow {\rm grCH_{3}O}\rightarrow {\rm grCH_{3}OH}
\label{meth}
\end{equation}
which is proved to be efficient at a temperature of 10~K \citep[][]{Charnley_ea97, WatanabeKouchi02} 
but not at the higher temperatures considered  here (i.e. T$>$15 K) because the residence time of H atoms on grains becomes too short and reaches just a few minutes. 
By 10$^{6}$ yr at 10 K, the fractional abundance of methanol on grain surface reaches $\sim$10$^{-5}$. 
However, despite being efficiently formed on grain surface at 10~K, methanol has no efficient way to be transferred to the gas phase at this temperature, because the non-thermal desorption processes in the considered model (see above) are all inefficient. 
The most efficient method to produce methanol in the gas phase is via the slow reaction
\begin{equation}\label{hcooch3_form}
{\rm H_{2}CO}+{\rm H_{3}CO^{+}}\rightarrow {\rm H_{2}COHOCH_{2}^{+}+h\nu}
\end{equation}
described in \citet[][]{Horn_ea04}, followed by the subsequent dissociative recombination.   The rate coefficients of the reactions involving the ion H$_2$COHOCH$_2^+$ are highly uncertain because the products involve a major shift in the positions of the hydrogen atoms. Interestingly, the dissociative recombination of ${\rm H_{2}COHOCH_{2}^{+}}$ has two channels in our network: methanol,  with a probability of 0.99, and methyl formate (CH$_3$OCHO) with a probability of 0.01. This fact explains the strong similarity of the abundance profiles between methanol and methyl formate in Figure~\ref{0D_IRDC028.34}. The surface route of formation of methyl formate via the reaction
\begin{equation}
{\rm grHCO}+{\rm grCH_{3}O}\rightarrow {\rm CH_{3}OCHO}
\end{equation}
is inefficient at 10 K due to the negligible mobility of reactant radicals. 
For models with constant temperature greater than 10 K,  the methoxy radical (CH$_{3}$O) cannot be produced efficiently because of the difficulty in hydrogenating CO into methanol via eq.~({\ref{meth}) at these temperatures.   As such, there is no way to produce methyl formate on grain surfaces in models with constant temperature.

Dimethyl ether ${\rm CH_{3}OCH_{3}}$  is also produced only via a gas-phase route in the absence of efficient non-thermal desorption. After production of its protonated ion  ${\rm CH_{3}OCH_{4}^{+}}$, the ether is formed in the subsequent dissociative recombination.   The ion is produced via the slow reaction 
\begin{equation}
{\rm CH_{3}^{+}}+{\rm CH_{3}OH}\rightarrow {\rm CH_{3}OCH_{4}^{+}  +   h\nu.}
\end{equation}
This radiative association is inefficient in the considered isothermal cases, but can be an efficient source of dimethyl ether in hot cores when the gas phase is rich in methanol \citep[e.g.,][]{2006A&A...457..927G, 2013ApJ...762...86V}. The grain surface formation of ${\rm CH_{3}OCH_{3}}$ via the reaction
\begin{equation}
{\rm grCH_{3}}+{\rm grCH_{3}O}\rightarrow {\rm grCH_{3}OCH_{3}}
\end{equation}
can occur at T$\ge$25~K. But again, in models with constant temperature above 10 K, this channel will not work for the reason same as discussed above for  methyl formate; viz., the absence of surface atomic hydrogen to produce methanol 
and other hydrogenated species such as CH$_3$ and CH$_3$O.

 Methyl acetylene (CH$_{3}$CCH) has both gas-phase and grain-surface formation routes. In the gas phase, it is formed via a sequence of ion-molecule reactions involving small  carbon-chain molecules.  On the grain surface, CH$_{3}$CCH is produced via the hydrogenation of C$_{3}$ via reactions with surface hydrogen atoms:
\begin{equation}\label{CH3CCH_form}
{\rm grC_{3}}\rightarrow {\rm grC_{3}H}\rightarrow {\rm grC_{3}H_{2}}\rightarrow {\rm grC_{3}H_{3}}\rightarrow {\rm grC_{3}H_{4}}
\end{equation}
At 10~K, methyl acetylene production occurs via both routes. However, since there is no efficient desorption mechanism at 10~K in the model, CH$_{3}$CCH produced on grains remains locked there. At the same time, gas-phase CH$_{3}$CCH is quickly accreted onto grains due to the relatively high gas density of 10$^{5}$ cm$^{-3}$. 
Thus, the resulting fractional abundance of methyl acetylene in the gas is $\sim$3$\times$10$^{-11}$. Not surprisingly, at higher temperatures of 25--120~K, the peak abundance of CH$_{3}$CCH remains almost unchanged;  the grain-surface formation route becomes inefficient due to the same reason as the route of methanol formation, while the ion-molecule gas-phase reactions exhibit only a weak dependence on temperature. 

Acetaldehyde (CH$_{3}$CHO) in the gas phase is only  produced directly via the dissociative recombination of its protonated precursor, which  is synthesized in the gas-phase reaction. 
\begin{equation}
{\rm H_{3}O^{+}}+{\rm C_{2}H_{2}}\rightarrow {\rm C_{2}H_{5}O^{+} + h\nu.}
\end{equation}
The reactants are  produced via simple ion-molecular chemistry.  The efficiency of this channel is low, and the peak fractional abundance of acetaldehyde only reaches $\sim$10$^{-12}$ with respect to hydrogen at all temperatures studied. 
On the grain surface, CH$_{3}$CHO can be formed via the reaction
\begin{equation}\label{ch3cho_form}
{\rm grCH_{3}}+{\rm grHCO}\rightarrow {\rm grCH_{3}CHO};
\end{equation}
however, at 10~K this reaction is inefficient because the grCH$_{3}$ and grHCO radicals have low mobilities. At  temperatures of 25--120~K, the mobility of the reactants  is appreciable, but the abundances of the radicals is low, because they are produced via the photodissociation of methanol and formaldehyde.  These species are in turn produced by the hydrogenation of CO,  which is itself efficient only at low temperatures near 10~K. Consequently,  there is no way to produce acetaldehyde in appreciable amounts in our models with constant temperature.

For gaseous HNCO, the formation route is a grain-surface reaction
\begin{equation}\label{hnco_form}
{\rm grH}+{\rm grOCN}\rightarrow{\rm grHNCO}
\end{equation}
This reaction is efficient even at 10~K, and the surface abundance of HNCO in our model reaches a relatively high value of 10$^{-6}$. 
Due to the lack of an efficient desorption mechanism  for HNCO at T$<$45~K in our model, the gas-phase abundance of HNCO is low. 

These examples illustrate the fact that the constant temperature and density approach works
only when we analyze simple molecules, which all trace cold and dense material. 
In the case where we seek to explain the behavior of more complex species, we need to  employ a more complex model with time-dependent physical conditions and additional processes such as non-thermal reactive desorption \citep{2007A&A...467.1103G}.}

\subsection{1-Phase 0-D Model with reactive desorption and constant temperature}
\label{section:warmup1}


To improve our results for complex species in IRDC0028.34-6, we included an additional process,  reactive desorption, and, subsequently,  a warm-up phase in the model.
The reactive desorption is 
non-thermal desorption from interstellar dust grains via exothermic surface reactions, which
was described and applied to the case of low-mass pre-stellar  cores by \citet{2007A&A...467.1103G}.
There it was shown that reactive desorption plays an especially important role in
reproducing the abundances of gaseous organic species such as CH$_3$OH.
However, in many studies this process is not taken into account, because extensive laboratory evidence for it is lacking.
In the current study, we assume that the efficiency of the reactive desorption is 1\%; 
 that is, 1\% of the products of a chemical reaction on a grain desorb into the gas while 99\% remain.

\begin{figure}
 \centering
\includegraphics[width=10cm,angle=90]{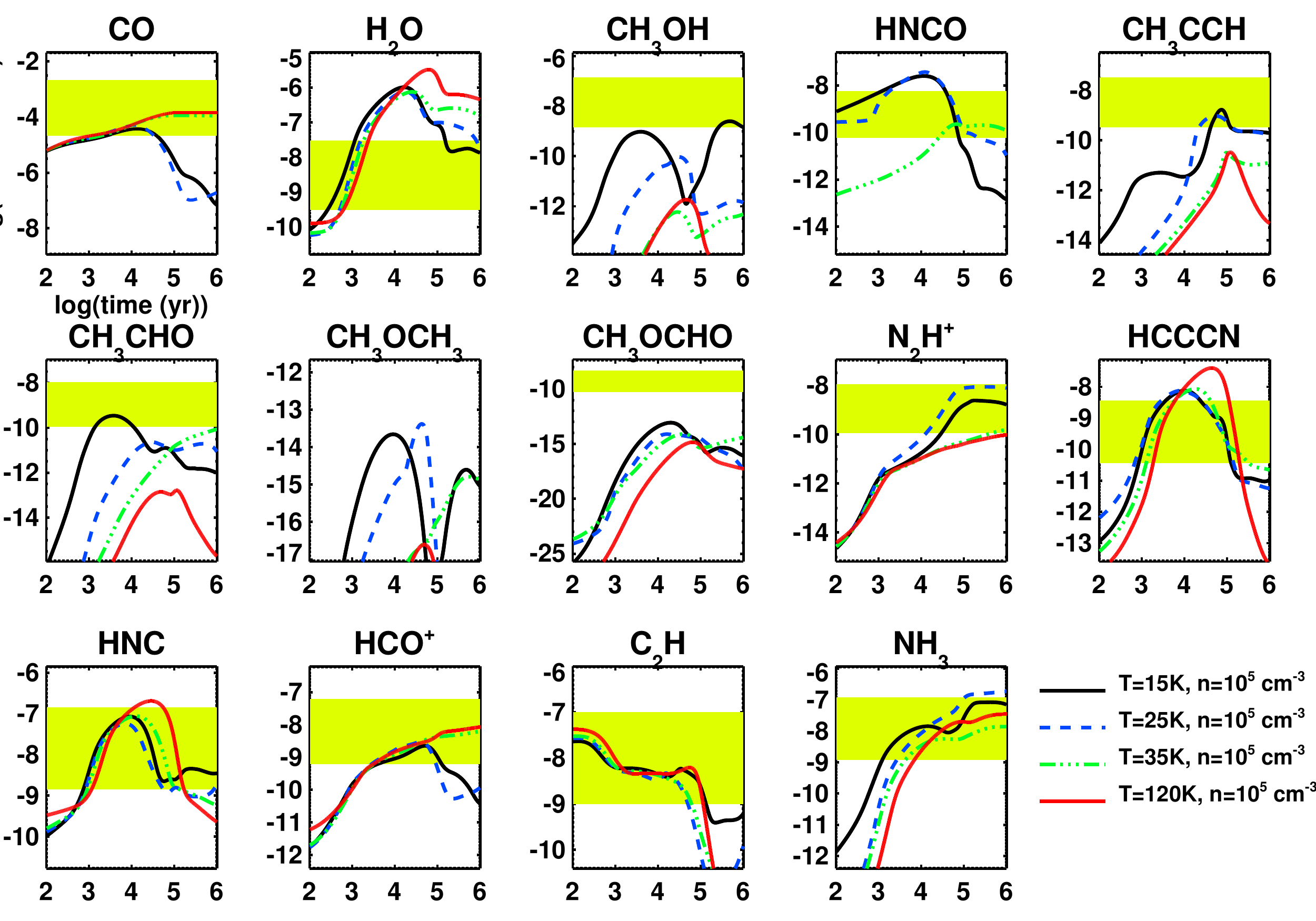}    
   \caption{Calculated fractional abundances relative to H$_2$ as a function of time for 14 gas-phase species. Different lines correspond to
   		different temperatures and boxes correspond to the observed values  in IRDC028.34-6  $\pm$ one order of magnitude.
		For these calculations, we used a 0-D pseudo time dependent model including reactive desorption.}
  \label{0D_IRDC028.34_des1}
\end{figure}

Figure~\ref{0D_IRDC028.34_des1} shows the improvement that we obtain
after adding reactive desorption to the simple model discussed previously. 
The cold dense gas tracers and CO are still reproduced well in the time range 10$^4$-10$^5$ yr at all four temperatures,
except N$_2$H$^+$, which exhibits agreement between model and observation only at 15 and 25 K.
All COMs and smaller carbon-containing species, except CH$_3$OCHO, show at least a 1-2 order-of-magnitude  increase in abundance 
at 15 and 25~K,  the two lower temperatures studied.
The gas-phase abundance of methyl formate remains almost unchanged in comparison with Figure~\ref{0D_IRDC028.34} because it is still mostly produced via the dissociative recombination of its protonated form made in  reaction (\ref{hcooch3_form}). The abundances of the reactants of this reaction, formaldehyde and protonated formaldehyde, are governed by gas-phase chemistry, and only slightly affected by the introduction of reactive desorption in the model. 
The abundances of HNCO, CH$_3$CHO and CH$_3$OCH$_3$ increase also at 35~K.  


 After we add reactive desorption, the dominant chemical processes of the organic species on the grain surface and in the
gas phase become more dependent on the temperature.
At 15 and 25~K, CH$_3$CHO, CH$_3$OCHO and CH$_3$OCH$_3$ still form mainly in the gas phase via reactions discussed in the previous section.
Moreover, including reactive desorption increases the abundances of their parent ions, which
 leads to at least some increase in the abundance of the resulting neutral species \citep{2013ApJ...769...34V}. 
At the same temperatures, reactive desorption becomes the main formation mechanism  for CH$_3$OH, CH$_3$CCH and HNCO
that are formed in reactions (\ref{ch3oh_form}), (\ref{CH3CCH_form})  and (\ref{hnco_form}), correspondingly, and then ejected to the gas.

As the temperature increases to 35~K, the gas-phase abundances of all considered COMs, except  CH$_3$CHO and CH$_3$OCHO, are significantly enhanced by reactive desorption, which is  their dominant formation mechanism. 
However, at this temperature the abundances of  parent radicals such as grCH$_3$ and grCH$_3$O  are
lower than at  10 and 25~K. 
Hence, although reactive desorption becomes a dominant formation route, it is not efficient enough to produce the
required amount of the observed COMs. 
A further increase in temperature to 120 K entirely kills grain-surface chemistry, which eliminates the effects of reactive desorption.  Correspondingly, the abundances of species at 120 K become very similar to those in the model without reactive desorption.
Only for 
HNCO  at 10, 25 and 35~K,  and CH$_3$CCH at 10 and 25~K can the observational abundances be explained by  including reactive desorption.
Modeled abundances for CH$_3$CHO and CH$_3$OCHO
are still too low  at all four temperatures, and CH$_3$OH is reproduced only at 10~K at the very early time range 10$^3$-10$^4$ yr
 with predicted abundances reaching values of $\sim$10$^{-10}$.
Clearly, further improvement is required to explain the abundances of  the COMs and smaller organic species.

\subsection{2-Phase 0-D Model: Cold and Warm-up Phase}
\label{section:warmup2}

 We next include a warm-up phase in the model.  As stated previously, the phase can be due to stellar heating or the interaction of shock waves in outflows \citep{2008ApJ...681L..21A,2004MNRAS.350.1029V}.
For IRDCs, even ammonia temperatures are elevated to 15-25~K, in comparison with the usual 10~K  
for low-mass pre-stellar cores. 
Dust temperatures  from \citet{2012ApJ...756...60S}
and excitation temperatures obtained from the CH$_3$CCH analysis 
are even higher  at $\sim$~30~K. 
In the first phase of our warm-up model, which lasts for 10$^6$ yr,  the temperature and density
remain at 10~K and 10$^5$ cm$^{-3}$, respectively.
In the second phase, the density stays the same, while
 the temperature increases from 10 to 30 K over 6.5$\times$10$^4$ yr with a T$\propto t^2$  time dependence
\citep[see][]{2006A&A...457..927G},
 and remains constant after it reaches a maximum value of 30~K.  
 The cold phase is notable for the production of methanol and formaldehyde on the ice.

\begin{figure}
 \centering
  \includegraphics[width=8cm,angle=90]{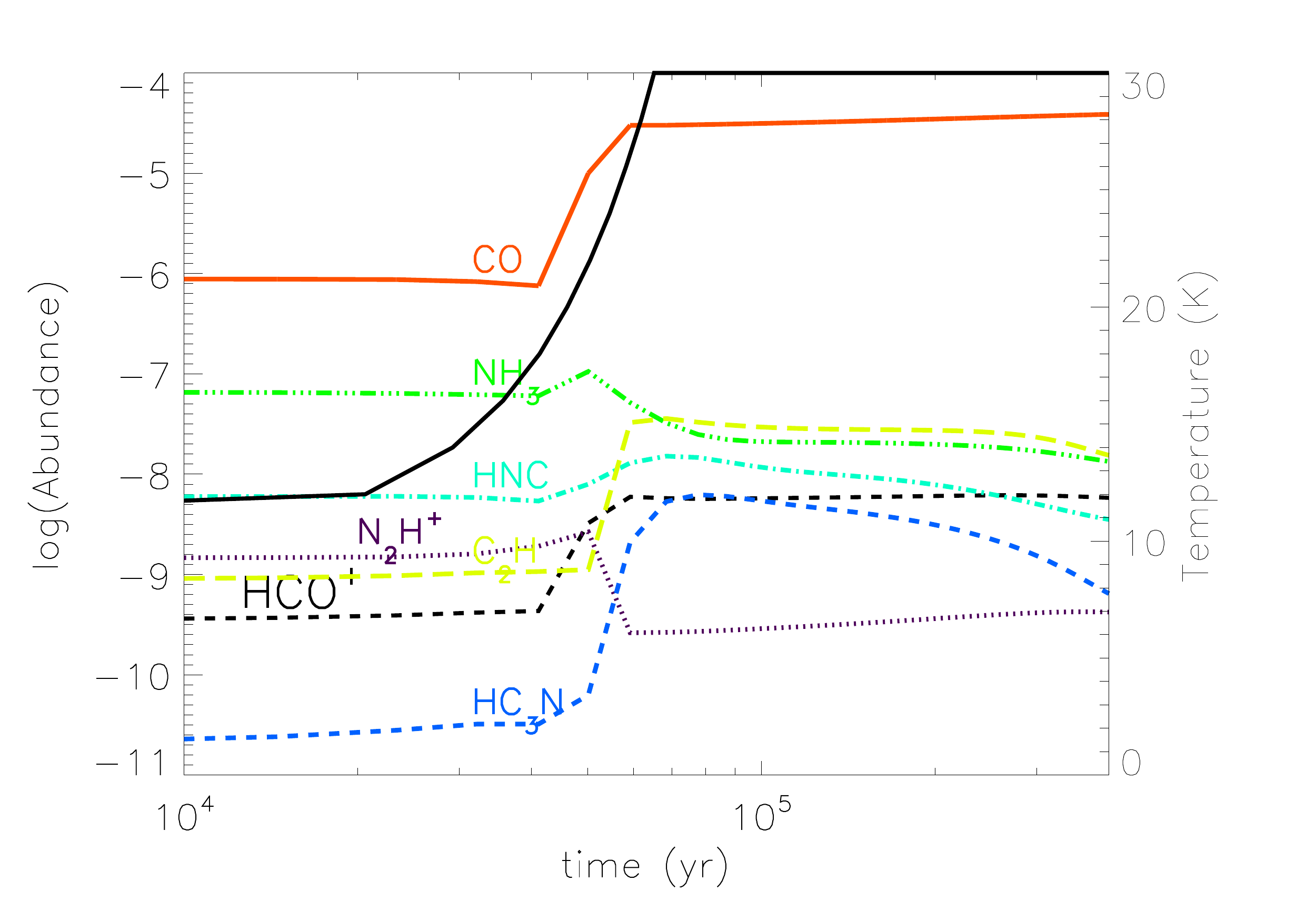}    
  \includegraphics[width=8cm,angle=90]{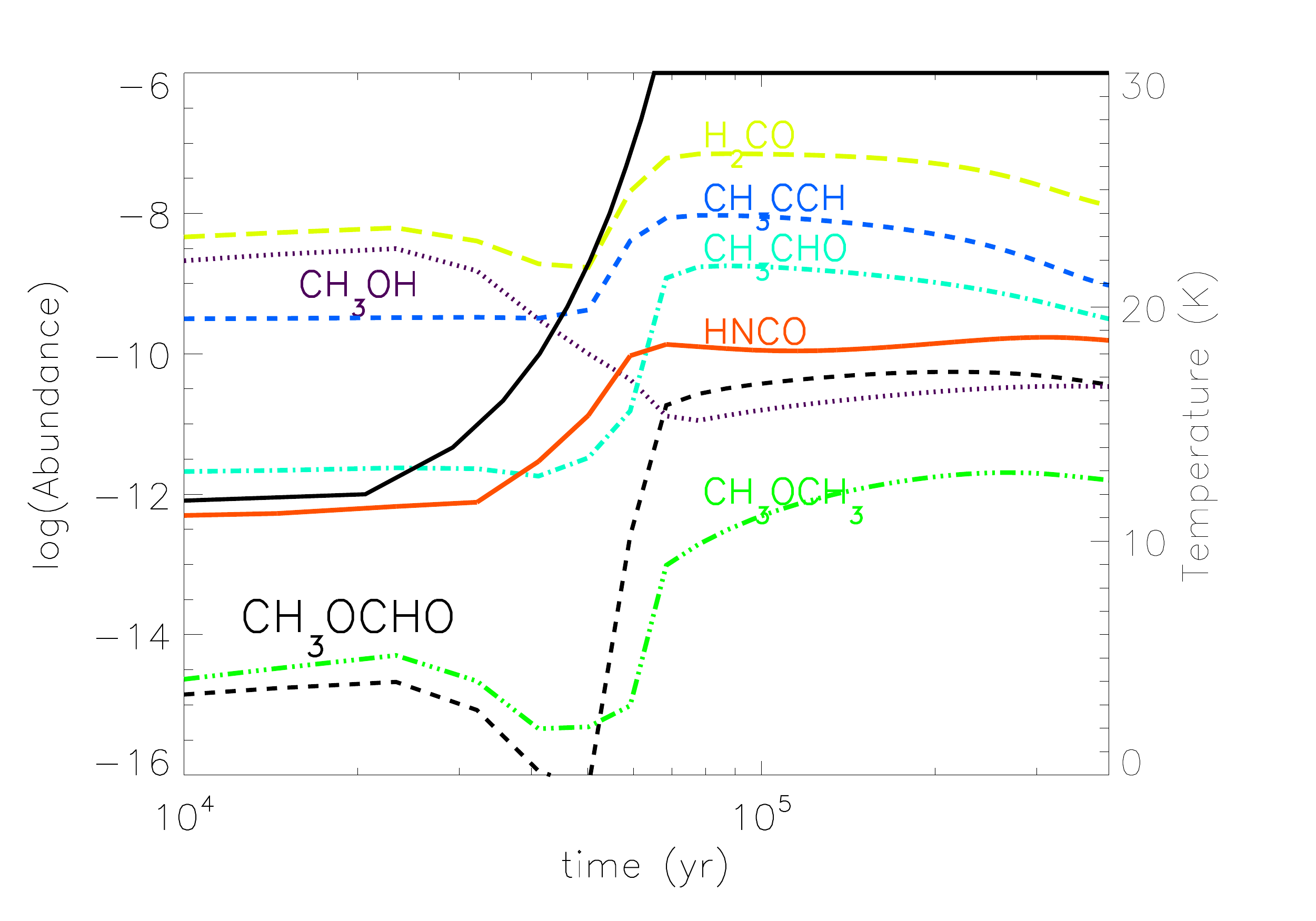}      
   \caption{Upper panels: Calculated fractional abundances relative to H$_2$ for our standard warm-up model
   		 as a function of time for 14 gas-phase species  starting 10$^{4}$ yr into the warm-up phase and ending at the 4$\times$10$^5$ yr.  
		 The duration of the 10 K cold phase is 10$^6$ yr and that of the warm-up phase is 6.5$\times$10$^4$~yr, with $T \propto$ time$^2$, 
		a maximum temperature of 30~K, and a constant density of 10$^5$ cm$^{-3}$. 		
		On all panels, the black solid line indicates
		the temperature evolution with time.}
  \label{IRDC028.34_warmup}
\end{figure}

Fractional abundances relative to H$_2$ as a function of time for 14 gas-phase species  starting at 10$^4$ yr into  the warm-up phase
are presented in Figure~\ref{IRDC028.34_warmup}. 
The gradual temperature increase after the 10$^6$-yr  cold phase leads to an enhancement of all
molecules discussed in the paper, including
CH$_3$OH, CH$_3$CHO, and CH$_3$OCHO. 
 During the cold phase,  precursor species such as formaldehyde and methanol collect on the grain surface. As the temperature begins to increase, surface radicals are formed mainly by photodissociation processes and then begin to diffuse and react to form larger amounts of more complex species.   
 \citep[e.g.][]{2008ApJ...682..283G}.
Reactive desorption then results in greater abundances of gas-phase COMs
compared with the same process at all four constant temperatures discussed above. 

The abundances of 12 species covered by the observations can be reproduced simultaneously by the model described above
at 3$\times$10$^5$ yr,  which is somewhat after the final temperature of 30 K is reached.
Methanol is an exception; in this case,  the agreement  occurs before
6$\times$10$^4$ yr;  i.e., when the temperature is still below 20~K.     
When the temperature exceeds 20~K, CO evaporates from the grain surface.  This process leads to an increase in the abundance of HCO$^{+}$, which destroys methanol in the gas. Additionally, at 30~K, the accretion rates of gas-phase molecules including methanol are faster than at 10~K, which also suppress their abundance.
In Section~\ref{discussion}, we will discuss physical conditions that might
better  explain the observed CH$_3$OH values.

\section{Discussion}
\label{discussion}

\begin{figure}
 \centering
   \includegraphics[width=5.7cm,angle=90]{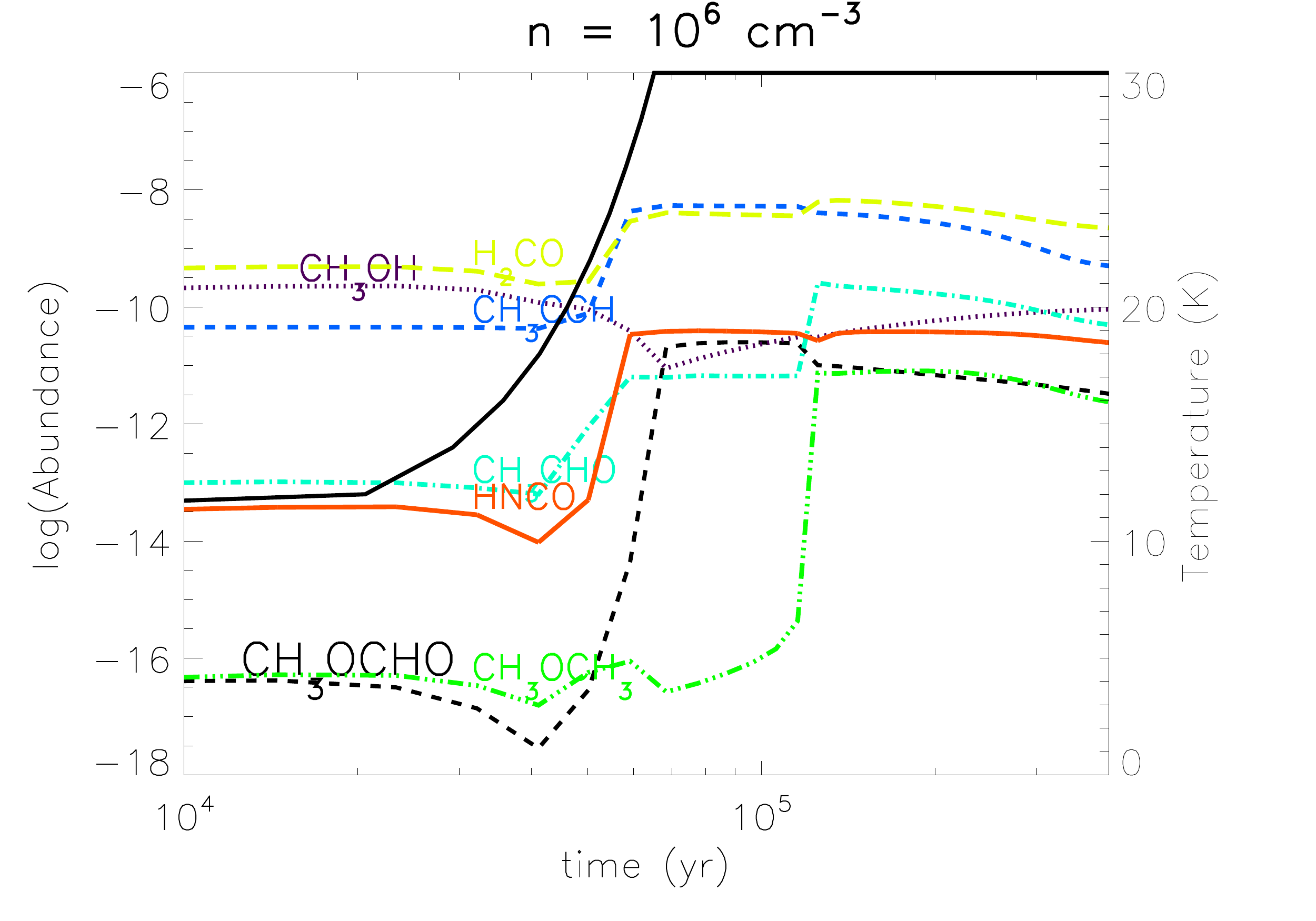} 
  \includegraphics[width=5.7cm,angle=90]{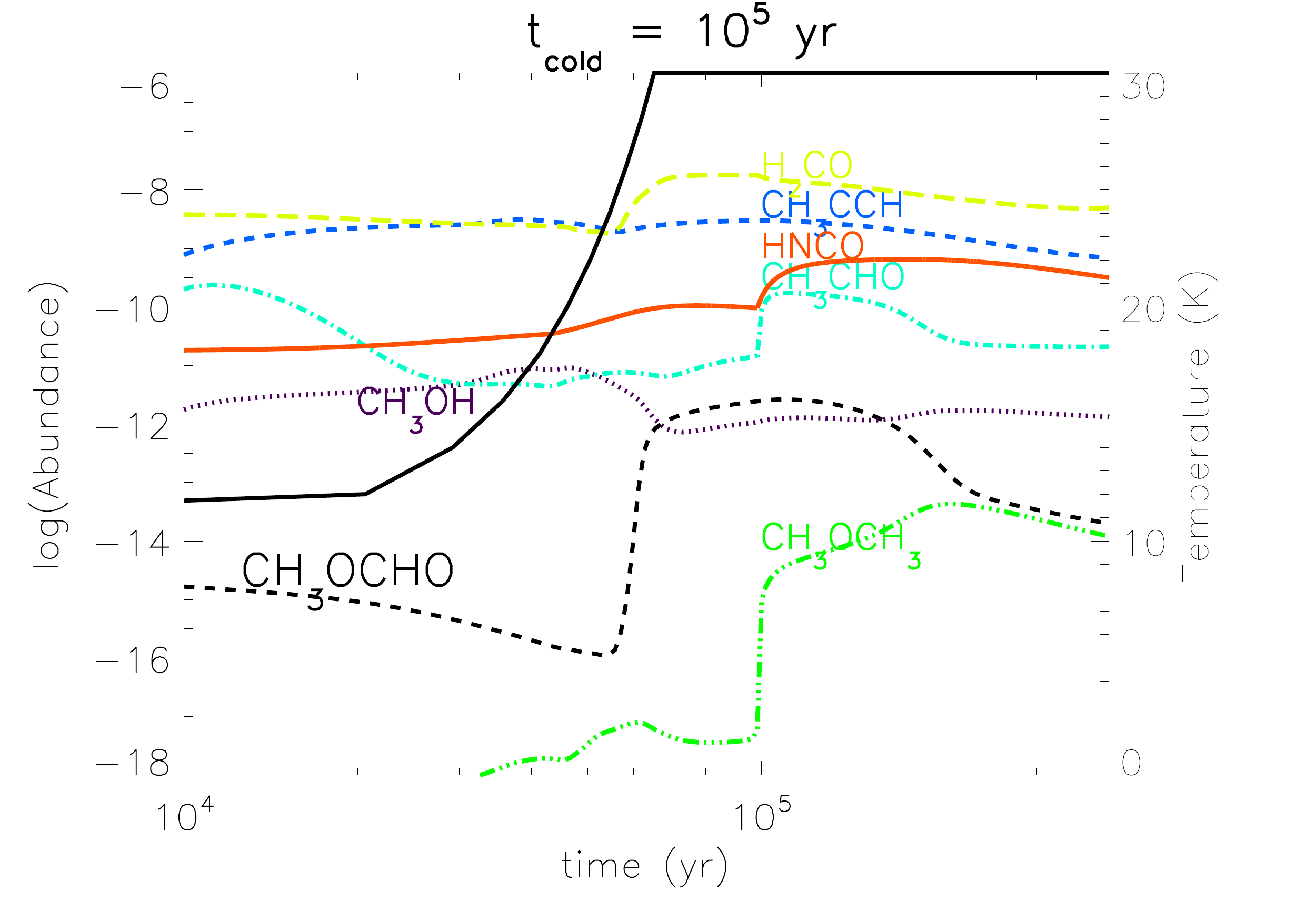}    
  \includegraphics[width=5.7cm,angle=90]{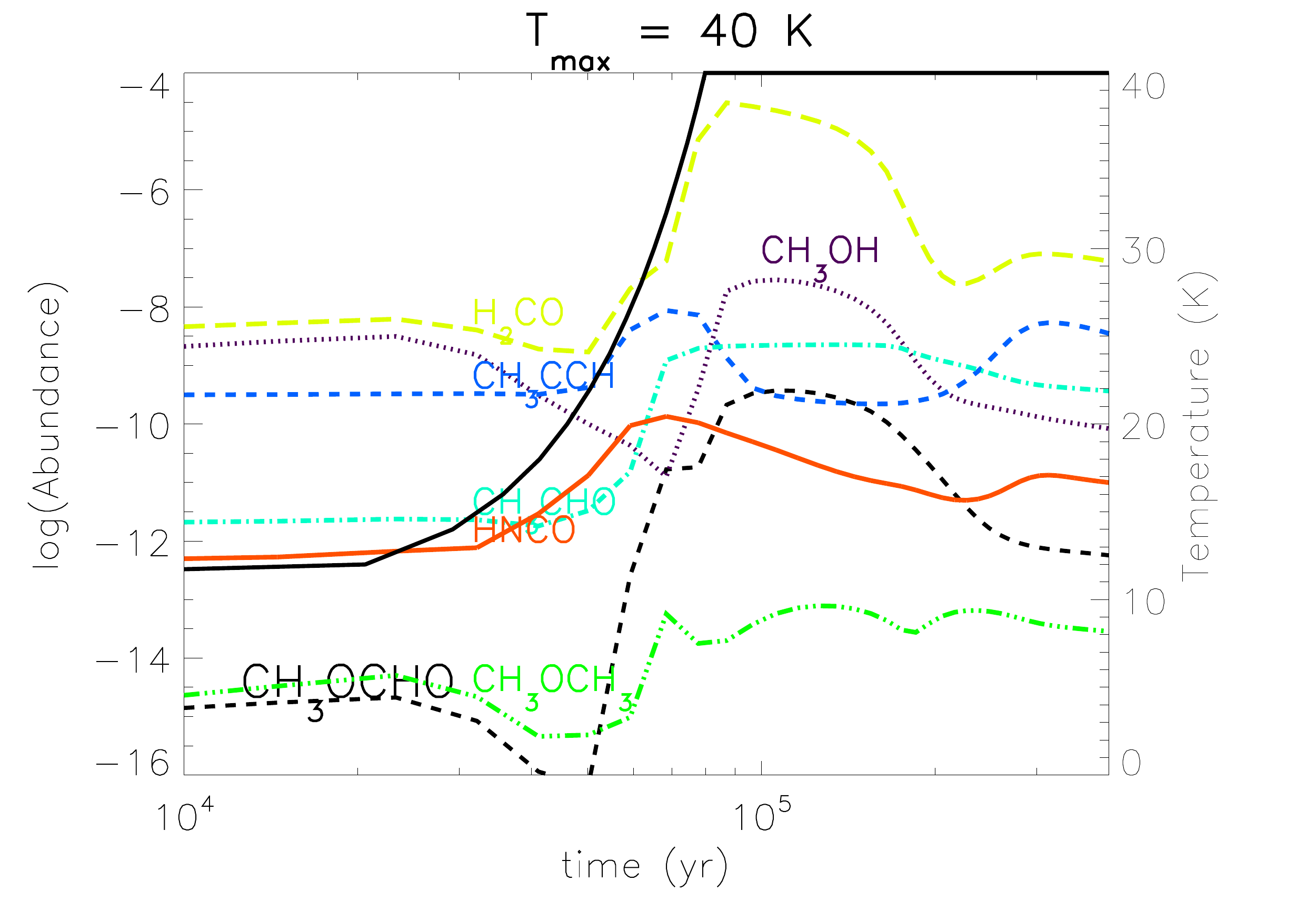}   
   \includegraphics[width=5.7cm,angle=90]{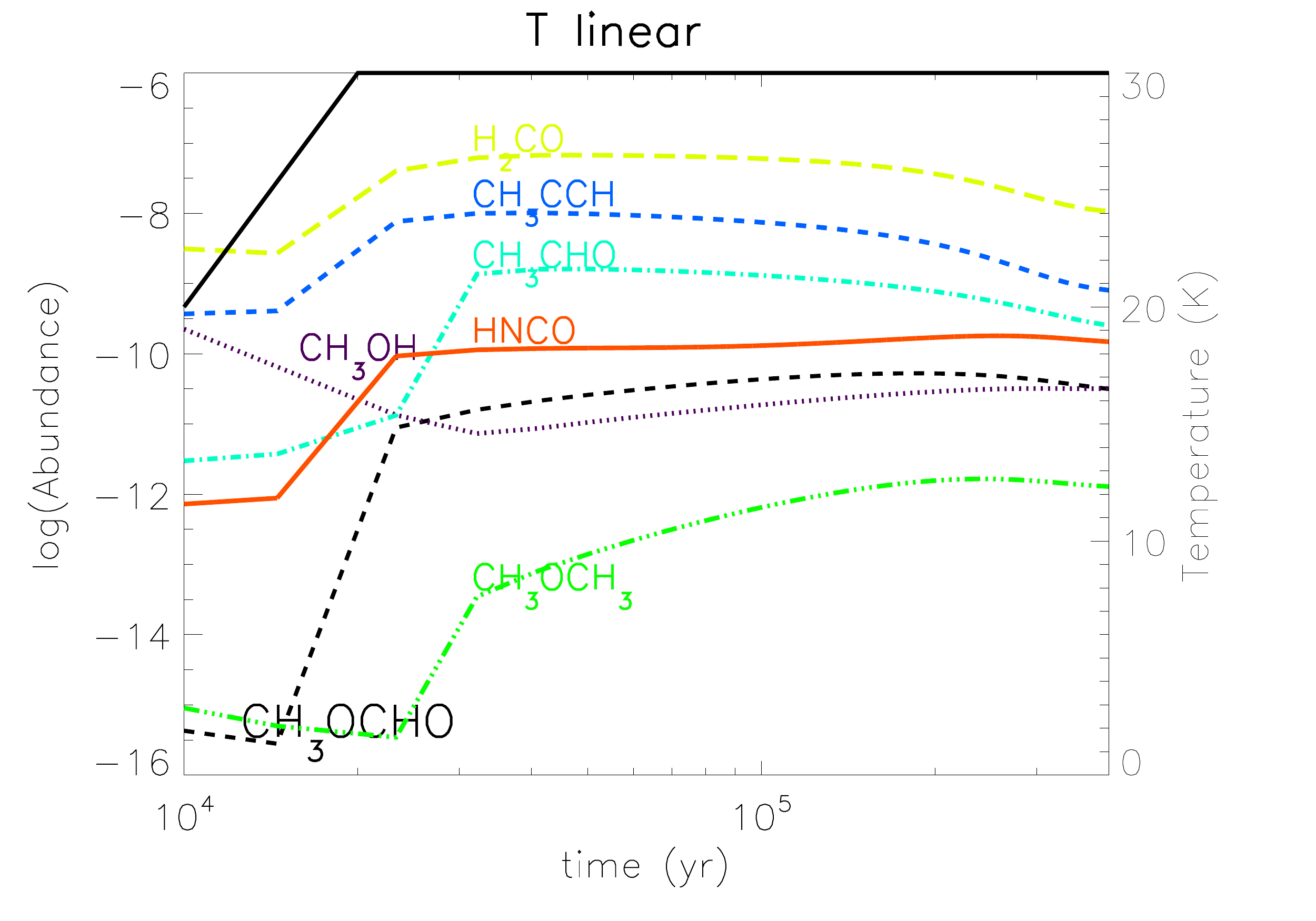}  
     \includegraphics[width=5.7cm,angle=90]{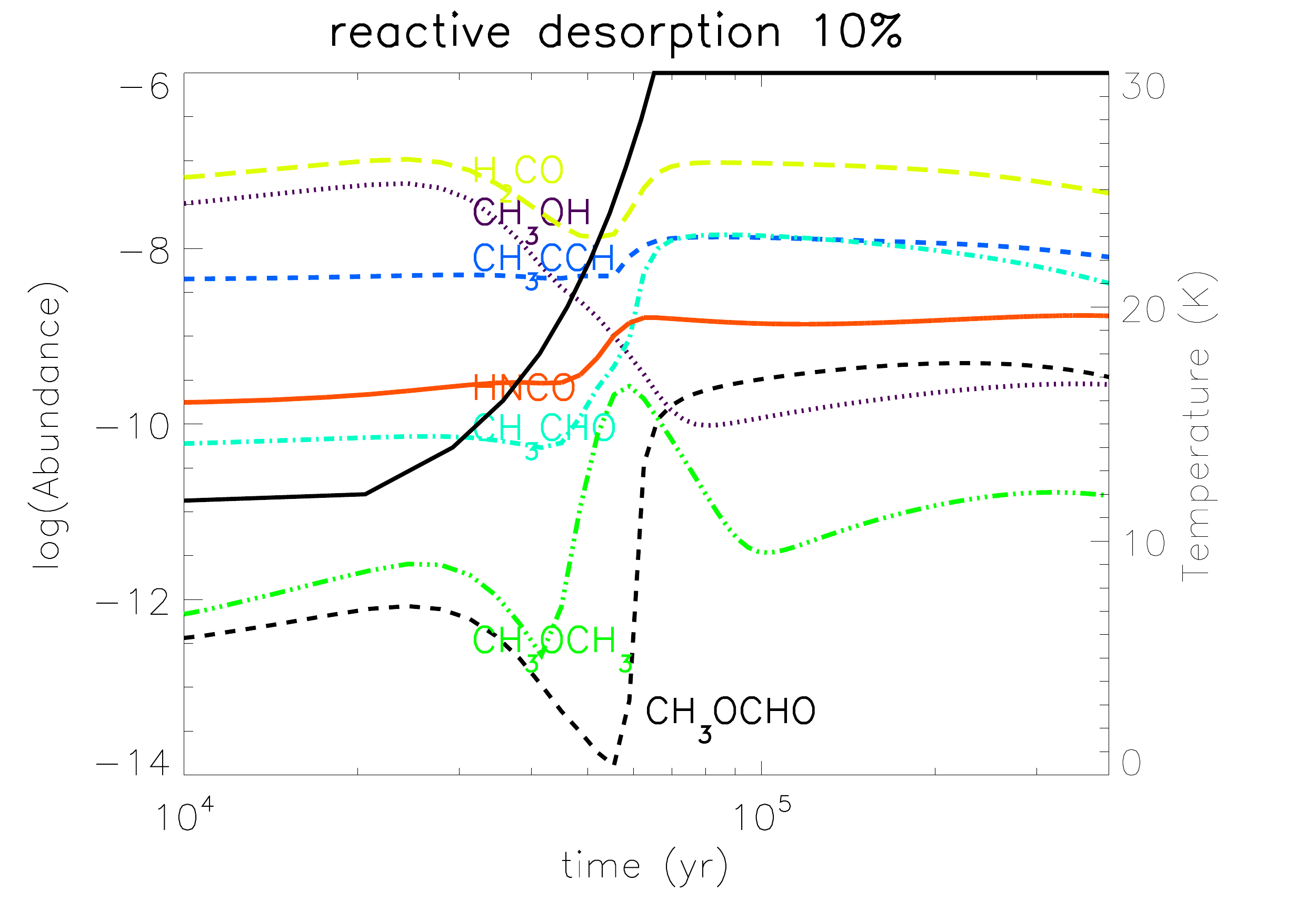}    
       \includegraphics[width=5.7cm,angle=90]{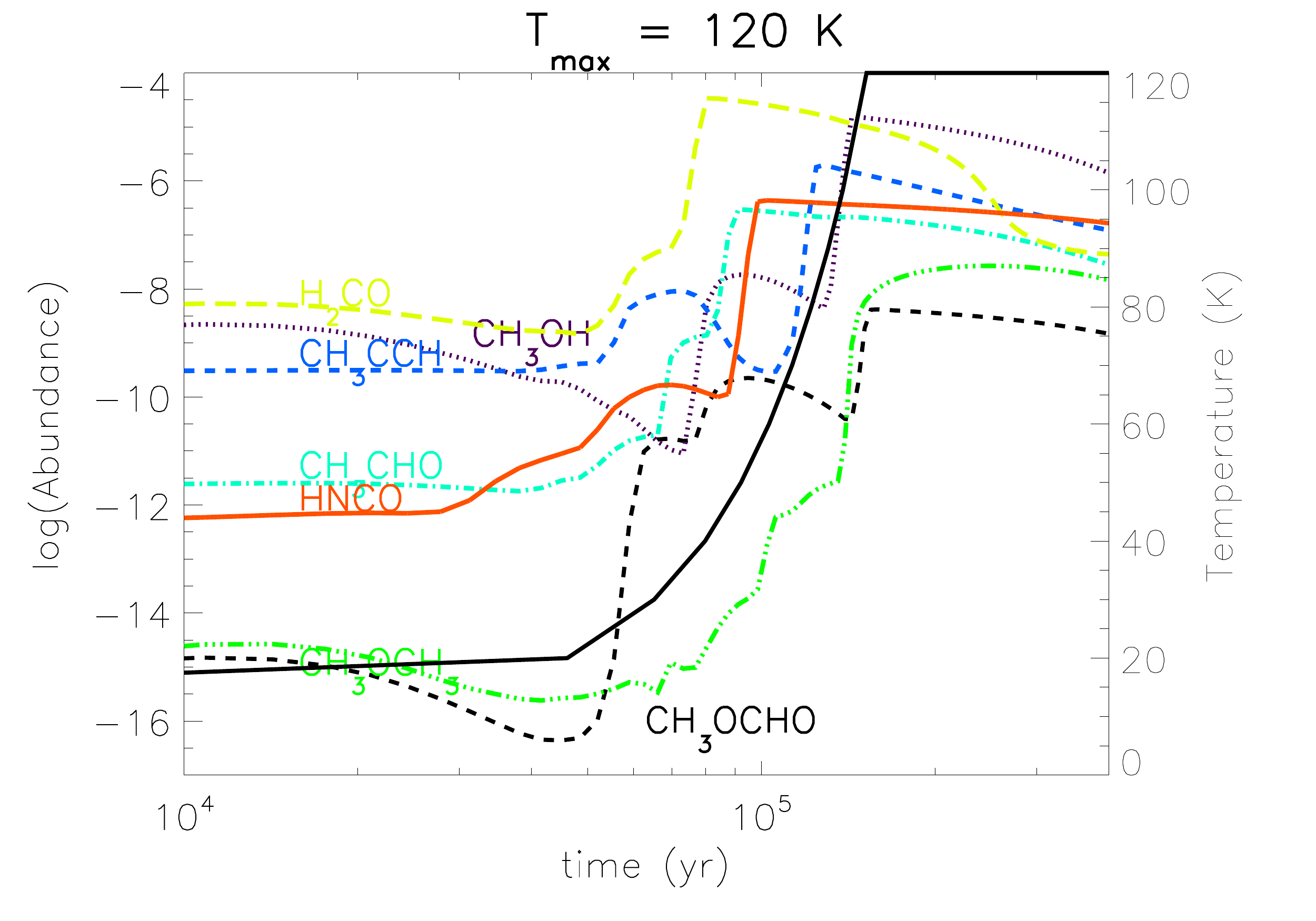}
   \caption{Calculated fractional abundances relative to H$_2$ as a function of time for seven species discussed in the paper. 
                  Different panels correspond to  models with specific parameters different from those of the standard model.
		Upper left panel: density n = 10$^6$ cm$^{-3}$; upper right panel:
		cold phase of 10$^5$ yr; middle left panel: maximum temperature of 40 K;
		middle right panel: linear increase of the temperature from 10 to 30 K. On all panels, the black solid line indicates
		the temperature evolution with time;
		lower left panel: reactive desorption efficiency is 10\% instead of 1\%;
		lower right panel: maximum temperature of 120 K.}
  \label{cold_phase_CH4O}
\end{figure}

 The results of our warm-up model with
a prior cold phase show that both phases are required to reproduce COMs in IRDC028.34-6 , 
 although there is no single period of time where all of the species are reproduced.  
 In an attempt to improve our model, we have varied some of the parameters.  
 Figure~\ref{cold_phase_CH4O} shows the results of six warm-up models with a variety of different parameters, 
 while Table~\ref{table:ab_IRDC028.34-6} lists a larger number of abundances for these 
 different models and our standard model at a time of $3 \times 10^{5}$ yr into the warm-up phase, when the best fit
 between model and observations is reached.

To investigate the influence of  reactive desorption on our results,
we ran the standard chemical model with a higher efficiency of 10\%.
The results of this model are presented in Figure~\ref{cold_phase_CH4O}.
The more effective reactive desorption significantly increases the abundances of
most of the COMs. 
The increase leads to much better agreement between model and observations
for CH$_3$OCHO at 3$\times$10$^5$ yr and CH$_3$OCH$_3$ at 6$\times$10$^4$ yr.
However, the methanol abundance still decreases after the temperature becomes higher than 20~K
and the formaldehyde abundance becomes one order of magnitude too high in the model.
Taking into account the overall agreement between model and observations, 
we can conclude that the higher efficiency for reactive desorption 
does not provide us with a better fit to the abundances
if we consider abundances for all available species and not only COMs such as CH$_3$OCHO and CH$_3$OCH$_3$.

Since some of the species might also trace denser regions,
 we ran an alternative chemical model with the higher density of 10$^6$ cm$^{-3}$.
The upper left panel of Figure~\ref{cold_phase_CH4O} shows that the increase in density
leads to a reduction of the abundances for most of the species.
As a result, the modeled abundances of HNCO, CH$_3$CHO, CH$_3$OCHO, N$_2$H$^+$ and HNC
become more than one order of magnitude different from the observational values. 
However, in the case of formaldehyde, the abundance drop leads to
a better agreement between model and observations.
Such  behavior might  suggest that we detect species that 
actually  are located in different parts of the cloud.
For example, H$_2$CO might trace more dense regions,
while  HNCO, CH$_3$OCHO, N$_2$H$^+$ and HNC might
dominate in regions with lower density. 

In a similar way, 
the standard model with a shorter cold phase lasting only 10$^{5}$ yr leads to a reduction of the gas-phase abundances for
the most of the discussed species, as shown in the upper right panel of Figure~\ref{cold_phase_CH4O}.
An extended cold phase of 10$^7$ yr also results in lower COM abundances.  
That indicates that the duration of the cold phase 
plays an 
important role in the formation of the 
COMs considered in the present paper.


 In addition to the cold phase, we need to look at the results of variations in the parameters for the warm-up phase, such as
 the shape of the temperature profile, the warm-up time,  
and the final temperature value.
For our standard warm-up model,  we used a quadratic temperature profile from \citet{2006A&A...457..927G}.
However, unlike their study, we increased the temperature to only 30 K, because this
value is consistent with dust  temperatures reported in 
\citet{2012ApJ...756...60S} and excitation temperatures obtained from 
the excitation analysis of CH$_3$OH and CH$_3$CHO.
An increase in the final temperature to 40 K results in  a drop of 
2 orders of magnitude for the gas-phase abundances of HNCO and CH$_3$OCH$_3$,
and a behavior very similar to that at 30 K for CH$_3$CHO and CH$_3$OCHO, as can be seen in the  lower left panel of
Figure~\ref{cold_phase_CH4O}.
The only species where we achieve significant improvement at 40~K
is CH$_3$OH. 
 At this higher final temperature,  the methanol fractional abundance 
reaches 3$\times$10$^{-8}$ at 10$^5$ yr after the warm-up begins.  
This value is in  a good agreement  with  the value of 4$\times$10$^{-8}$ obtained from observations
and shows that methanol can at least partly come from regions with a higher temperature.  
As can be seen in  Figure~\ref{cold_phase_CH4O}, the further increase of the final temperature to 120~K will lead to 
more than two orders of overprediction for most of the species including
CH$_3$OH, HNCO, CH$_3$CCH, CH$_3$CHO, HCCCN, HNC, and NH$_3$.
However, beam dilution in a single-dish observation is also possible.
In this case, when a compact hot core is deeply embedded within an IRDC, the estimated observational abundances
would be at least two orders of magnitude higher, and hence, show
good agreement with a higher final temperature model.
The tracers of cold and dense gas, in such a model, might exist in the outer shell with lower temperature.
Nevertheless, with a hot core temperature in IRDC028.34-6, we will not be able to reproduce the abundances of most of the observed molecules
with the homogeneous warm-up model. 

An extensive comparison between observations and 
 modeling results with the shorter duration of the cold phase, higher final temperature 
and higher density shows  that the abundances of some species at 3$\times$10$^5$ yr,
such as CO, CH$_3$CCH, HCO$^+$, C$_2$H and NH$_3$, 
do not depend on these variations in conditions.
 Species as H$_2$CO, CH$_3$OH, CH$_3$OCH$_3$ and HNC, on the contrary, are very sensitive to these changes.
This conclusion has to be taken into account in future studies on ``IRDC-like''  environments; i.e.,
temperatures between 15 and 30~K,  and densities in the range of 10$^5$-10$^6$ cm$^{-3}$.

Apart from the T$\propto$t$^2$ temperature warm-up, we also ran the code with a more
rapid  and less rapid (linear) warm-up.
For the rapid warm-up,  we increased the temperature 
as a step function from 10 to 30 K immediately after the 10$^6$ yr of the cold phase.
This  warm-up gives reasonable abundances for most of
the species except formaldehyde and methanol, which become significantly under-abundant in the model,  even though
the methanol abundance on the grain surface after  the very rapid warm-up and after the original T$\propto$t$^2$ temperature rise
are similar. 
For the linear warm-up,
the maximum temperature was reached at  6$\times$10$^4$ yr after the start of the warm up.  
As shown in the lower right panel of Figure~\ref{cold_phase_CH4O},   the final results of this model lie very close to those of our standard model. 
 Consequently, we are once again able to
reproduce the abundances of most of the species at 3$\times$10$^5$ yr, except for  methanol.
Hence, at the considered temperature, density and time ranges, the shape of the temperature profile 
does not play as significant  a role for the selected molecular species as the density value or the duration of the cold phase.
Nevertheless, the shape of the temperature profile might help to 
explain the composition of warmer and more evolved regions, which we do not 
consider here. 
To better test the shape of the warm-up profile at the current conditions,  we need to  include  observational abundances for additional  
species that might be more sensitive to such changes. 


\section{Summary and Conclusions}
\label{summary}

In the current paper,  we have presented new molecular line observations for  a variety of IRDCs.
In total, eight COMs and smaller carbon-containing species were detected:
CO, HNCO, H$_2$CO, CH$_3$CCH, CH$_3$OH,
CH$_3$CHO, CH$_3$OCHO and CH$_3$OCH$_3$.
For all species, 
molecular column densities and abundances were estimated.
The calculated molecular abundances are typically in agreement with
 previously obtained IRDC values if available.
A direct comparison with abundances for low-mass pre-stellar cores,
HMPOs, and hot cores  reveals that 
COM abundances in IRDCs tend to be higher 
than in low-mass pre-stellar cores,
although they do not reach the maximum HMPO and hot core values.  


To better understand the chemical processes in IRDCs,
we built a chemical model with warm-up and including reactive desorption, and compared  model and observational results  for IRDC028.34-6.
Our study shows that one promising way to explain the abundances of organic species in IRDCs is to 
quadratically increase the temperature
from 10 to 30 K  after  a 10$^6$ yr duration of the cold phase.
Although the results depend more on the duration of the cold phase  than on the 
temperature profile, 
too fast a warm-up can  cause a drop in the abundance of some organic species  when the final temperature is reached.
Our model represents the general trend in observed molecular abundances very well.
However, it has to be noticed that since some species are very sensitive to changes in the temperature and
density conditions, to get better agreement between the model and observations, it is better to run a separate
chemical model for each object with  physical parameters different from what was discussed in Sections~\ref{section:warmup1} and ~\ref{section:warmup2}. 


The temperature increase that occurs in the IRDC environment can be caused by 
the beginning of the star formation processes in the cloud, or by the influence of  
outflows and shocks from  nearby regions.
The results of our study show that the analysis of organic species allows us not only to understand 
 the chemical processes, but also to reconstruct any change in physical conditions, such as warm-up, that may  take place in IRDCs.

\acknowledgements

EH wishes to  acknowledge the support
of the National Science Foundation for his astrochemistry program, and his program in chemical kinetics through the Center for the
Chemistry of the Universe. He also acknowledges support from the NASA Exobiology and Evolutionary Biology program through a subcontract from Rensselaer Polytechnic Institute.
 IZ was partially supported by the Russian Academy of Sciences, the Ministry of education and science of Russian Federation (project 8421) and Russian Foundation for Basic Research (grant RFBR 12-02-00861).

We would like to thank an anonymous referee for valuable comments and suggestions, which helped to improve the paper significantly.
This research has made use of the NASA/ IPAC Infrared Science Archive, which is operated by
the Jet Propulsion Laboratory, California Institute of Technology, under contract with the
National Aeronautics and Space Administration.
The 22-m Mopra antenna is part of the Australia Telescope,
which is funded by the Commonwealth of Australia for operations as a National
Facility managed by CSIRO. The University of New South Wales Digital Filter
Bank used for the observations with the Mopra Telescope was provided with
support from the Australian Research Council.

{\it Facilities:} \facility{Nickel}, \facility{HST (STIS)}, \facility{CXO (ASIS)}.

\begin{deluxetable}{r c c c c c c c }    
\tabletypesize{\tiny}
\tablecolumns{8}
\tablewidth{0pc}
\tablecaption{List of observed southern IRDCs. \label{table:main_Mopra} } 
\tiny 
\tablehead{
\colhead{Name} & \colhead{R.A.} & \colhead{Decl.} & \colhead{Distance\tablenotemark{a}} &
\colhead{T \tablenotemark{a}} & \colhead{N(H$_2$)\tablenotemark{a}}   &
\colhead{category\tablenotemark{c}}\\
\colhead{ }&\colhead{(J2000.0)} & \colhead{(J2000.0)} &\colhead{(kpc)} & \colhead{(K)} & \colhead{$\times$10$^{22}$ cm$^{-2}$}   &\colhead{ }}

\startdata
IRDC309.13-1 &    13 45 17.521 &  -62 22 02.84 &3.9    & 16.3  &   0.6 & M    \\
IRDC309.13-2 &    13 45 22.610 &  -62 23 27.48 &3.9    & 14.7  &   0.5 & M    \\ 
IRDC309.13-3 &    13 45 16.775 &  -62 25 37.25 &3.9    & 35.4  &   0.4 & A    \\ 
IRDC309.37-1 &    13 48 38.532 &  -62 46 17.55 &3.4    & 31.4  &   0.9 & A    \\ 
IRDC309.37-2 &    13 47 56.116 &  -62 48 33.46 &3.4    & 15.7  &   0.8 & A    \\ 
IRDC309.37-3 &    13 48 39.383 &  -62 47 22.39 &3.4    & 15.7  &   0.7 & Q    \\ 
IRDC309.94-1 &    13 50 54.970 &  -61 44 21.00 &5.3\tablenotemark{b}& 48.8  &   5.2 & A    \\ 
IRDC310.39-1 &    13 56 01.359 &  -62 14 18.29 &4.2    & 27.4  &   1.2 & A, SiO  \\ 
IRDC310.39-2 &    13 56 00.759 &  -62 13 59.80 &4.2    & 27.4  &   1.2 & A    \\ 
IRDC313.72-1 &    14 22 53.158 &  -61 14 41.00 &3.3    & 19.9  &   0.4 & A, SiO  \\ 
IRDC313.72-2 &    14 22 57.151 &  -61 14 10.84 &3.3    & 19.9  &   0.4 & A, SiO  \\ 
IRDC313.72-3 &    14 23 02.720 &  -61 13 39.64 &3.3    & 19.9  &   0.3 & Q, SiO  \\ 
IRDC313.72-4 &    14 23 04.533 &  -61 14 46.00 &3.3    & 19.9  &   0.4 & Q, SiO  \\ 
IRDC316.72-1 &    14 44 19.000 &  -59 44 29.00 &2.7    & 26.1  &   1.2 & M    \\ 
IRDC316.72-2 &    14 44 15.400 &  -59 43 20.00 &2.7    & 24.3  &   1.3 & Q    \\ 
IRDC316.76-1 &    14 44 56.000 &  -59 48 08.00 &2.7    & 22.6  &   4.1 & A, SiO  \\ 
IRDC316.76-2 &    14 45 00.500 &  -59 48 44.00 &2.8    & 23.2  &   4.8 & A    \\ 
IRDC317.71-1 &    14 51 06.905 &  -59 16 11.03 &3.0    & 15.6  &   1.2 & Q    \\ 
IRDC317.71-2 &    14 51 10.975 &  -59 17 01.73 &3.0    & 16.6  &   3.5 & A    \\ 
IRDC317.71-3 &    14 51 19.667 &  -59 17 43.77 &3.2    & 15.6  &   0.6 & Q, SiO  \\ 
IRDC321.73-1 &    15 18 26.387 &  -57 22 00.14 &2.2    & 22.0  &   1.0 & M, SiO  \\ 
IRDC321.73-2 &    15 18 01.693 &  -57 22 02.00 &2.2    & 11.7  &   1.7 & M    \\ 
IRDC321.73-3 &    15 18 01.065 &  -57 21 24.48 &2.1    & 11.7  &   1.7 & A    \\ 
IRDC013.90-1 &    18 17 33.378 &  -17 06 36.70 &2.5    & 12.9  &   2.6 & M    \\
IRDC013.90-2 &    18 17 19.350 &  -17 09 23.69 &2.4    & 13.4  &   1.1 & Q    \\ 
\enddata
	
\tablenotetext{a}{ distance, kinetic temperature, and H$_2$ column density are taken from \citet{2011A&A...527A..88V}}
\tablenotetext{b}{ from \citet{2001PASJ...53.1037S}}
\tablenotetext{c}{  MIR classification:  "A" indicates "active" cores, "M" - "middle", "Q" - "quiescent", "SiO" - detected SiO emission
 as described in  \citet{2011A&A...527A..88V} and Linz et al. (in prep.) (see text in Section 2.1).}

\end{deluxetable}

\begin{table}
\tabletypesize{\tiny}
\tiny
\tablewidth{0pc}
\caption{List of observed northern IRDCs.}
\label{table:main_IRAM}      
\begin{tabular}{l c c c c c c  l}
\hline            
\noalign{\smallskip}
Name & R.A.&  Dec. & Distance & T$_{kin}$ [T$_{dust}$] & N(H$_2$) &  Category & Ref.\\
& (J2000.0) & (J2000.0) &(kpc) & (K)  [(K)]  & $\times$10$^{22}$ cm$^{-2}$ &  &  \\  
\noalign{\smallskip}
\hline                        
\noalign{\smallskip}

IRDC010.70-2  &   18:09:45.6  & -19:42:07.9  & 3.46 & 15	           & 2.5  & A,SiO  & 1   \\
IRDC010.70-4  &   18:09:58.1  & -19:48:43.4  & 3.46 & 16	           & 1.8  & A,SiO  & 1   \\

IRDC011.11-2  &   18:10:33.5  & -19:21:47.0  & 3.6  &14             & 2.3 & Q,SiO  & 2,3 \\
IRDC011.11-4  &   18:10:28.3  & -19:22:31.5  & 3.6  &13             & 4.2 & A,SiO  & 2,3 \\
IRDC011.11-5  &   18:10:18.1  & -19:24:37.8  & 3.6  &	           & 3.1 & Q      & 2,3 \\

IRDC013.90-1  &   18:17:34.9  & -17:06:48.1  & 2.5  &13             & 3.5 & Q/M    & 4   \\

IRDC015.05-2  &   18:17:38.8  & -15:48:47.3  & 3.2  & [30]          & 1.8 & Q,SiO  & 5,6   \\
IRDC015.05-3  &   18:17:50.4  & -15:53:37.4  & 3.2  &15 [23]      & 1.5 & Q      & 5,6   \\

IRDC018.48-7  &   18:25:22.89 & -12:54:50.7  & 3.5  &	            & 2.3 & M      & 7,8\\
IRDC018.48-8  &   18:25:14.46 & -12:54:08.8  & 3.5  &	            & 1.1 & Q,SiO  &7,8\\

IRDC019.30-1  &   18:25:58.5  & -12:03:59    & 2.2     & 15            & 2.0  & A,SiO  & 2,3,6,9 \\

IRDC028.34-3  &   18:42:50.9  & -04:03:14     & 4.8     & 16   [32]   & 2.2 & M,SiO  & 2,3,6 \\
IRDC028.34-4  &   18:42:46.6  & -04:04:11.5  & 4.8     &  13  [24]    & 2.5 & M,SiO? & 2,3,6 \\
IRDC028.34-6  &   18:42:52    & -03:59:54       & 4.8     &  16  [33]    & 8.3 & A,SiO  & 2,3,6 \\

IRDC048.66-1  &   19:21:49.9  &  13:49:34.7  & 2.5      &   [19]           & 1.6 & A,SiO  & 5,10   \\
IRDC048.66-3  &   19:21:44.7  &  13:49:25.7  & 2.5      &	[17]           & 1.5 & Q,SiO  & 5,10   \\

IRDC079.31-2  &   20:31:57.5  & +40:18:30    &1/2.36 &13                 & 14.2$^a$ &	Q    & 2,3   \\

ISOSSJ23053   &   23:05:21.7  & +59:53:43    & 4.31   & 14[17]                     & 4.5$^b$ &	M   & 9,11,12 	  \\

\noalign{\smallskip}
\hline                                   
\end{tabular}

Notes:  Kinetic temperature estimated based on ammonia measurement and dust temperature;
	H$_2$ column density calculated adopting 29 $''$ beam size and ammonia kinetic temperature; 
	category is determined according to MIR classification (see Table~\ref{table:main_Mopra}).
	
	(a) H$_2$ column density adopted from  \citet{2006A&A...450..569P},
	(b) H$_2$ column density calculated based on SCUBA 850 $\mu$m flux estimate from \citet{2007A&A...474..883B}.

(1) \citet{2012A&A...544A.146W},
(2) \citet{2006A&A...450..569P},
(3) \citet{1998ApJ...508..721C},
(4) \citet{2011A&A...527A..88V},
(5) \citet{2006ApJ...641..389R},
(6)\citet{2012ApJ...756...60S},
(7) \citet{2008ApJ...678.1049S},
(8)\citet{2002ApJ...566..931S},
(9)\citet{2012A&A...547A..49R},
(10)\citet{2013ApJ...766...68P},
(11)\citet{2007A&A...474..883B},
(12)\citet{1988A&A...203..367W}.

\end{table}

\begin{table}
\tabletypesize{\tiny}
\tiny
\caption{Transitions of organic species observed for southern IRDCs.} 
\label{table:lines_APEX} 
\begin{tabular}{l c l l l l l c} 
\hline\hline
\noalign{\smallskip} 
Molecule  & Transition & Rest frequency &  A   & g$_u$    & E$_{up}$/k & Telescope & Beam \\
          &            & (MHz)          & ($\times$10$^{-5}$ s$^{-1}$) &    & (K) & & ($''$)  \\
\noalign{\smallskip}
\hline\hline
\noalign{\smallskip} 

CH$_3$OH   & 5$_{-1,5}$-4$_{0,4}$ E     &   84521.206 & 0.196  &  11     &40.39  & Mopra & 36 \\
\noalign{\smallskip}

CH$_3$CCH      & 5$_2$-4$_2$		  &   85450.7 &  0.170    &  22     & 41.21 & Mopra & 36  \\
CH$_3$CCH      & 5$_1$-4$_1$		  &   85455.6 &  0.194  &  22     & 19.53 & Mopra & 36  \\
CH$_3$CCH      & 5$_0$-4$_0$                  &   85457.2 &  0.202  &  22     &   12.30 & Mopra & 36\\
\noalign{\smallskip}

HNCO       & 4$_{04}$-3$_{03}$                  &   87925.238 &  0.878 & 9       & 10.54  & Mopra &  36  \\
\noalign{\smallskip}

C$^{18}$O   & 1-0                                            &   109782.176 & 0.0062  &  3     &5.26        & Mopra & 33  \\
\end{tabular}

Notes: 
Einstein A coefficients, degeneracy, and 
lower state energy levels for 
all  molecules talen from CDMS Catalog (http://www.astro.uni-koeln.de/cdms/catalog). 

\end{table}

\begin{table}
\caption{Transitions of organic species observed for northern IRDCs.} 
\label{table:lines_IRAM} 
\begin{tabular}{l c l l l l l c} 
\hline\hline
\noalign{\smallskip} 
 
Molecule  & Transition & Rest frequency &  A   & g$_u$    & E$_{up}$/k & Telescope  & Beam \\
          &            & (MHz)          & ($\times$10$^{-5}$ s$^{-1}$) &    &(K) &  & ($''$) \\
\noalign{\smallskip}
\hline\hline
\noalign{\smallskip}

CH$_3$OH       & 5$_{05}$-4$_{13}$  E     &   76509.6 & 0.089   &  11     &47.93 & IRAM & 33  \\
\noalign{\smallskip}

CH$_3$CHO      & 4$_{04}$-3$_{03}$  E     &   76866.4 & 1.485  &  18     &9.32 & IRAM & 33 \\
CH$_3$CHO      & 4$_{04}$-3$_{03}$  A++	  &   76878.9 & 1.485  &  18     &9.23 & IRAM & 33 \\
\noalign{\smallskip}

CH$_3$OCH$_3$  & 9$_{09}$-8$_{18}$ AA     &   153054.49& 2.191   & 190      & 40.39 & IRAM & 16  \\
CH$_3$OCH$_3$  & 9$_{09}$-8$_{18}$ EE     &   153054.84& 2.191   & 304      & 40.39 & IRAM & 16  \\
CH$_3$OCH$_3$  & 9$_{09}$-8$_{18}$ AE     &   153055.18& 2.191   & 114      & 40.39 & IRAM &  16 \\
\end{tabular}

Notes: 
Einstein A coefficients, degeneracy, and 
lower state energy levels for CH$_3$CHO taken from JPL data base 
(http://spec.jpl.nasa.gov), for 
all other molecules from CDMS Catalog (http://www.astro.uni-koeln.de/cdms/catalog). 

\end{table}

\begin{table}
\caption{Transitions of organic species detected with IRAM telescope only in  IRDC028.34-6.} 
\label{table:lines_IRDC028.34-6} 
\begin{tabular}{l c l l l l } 
\hline\hline
\noalign{\smallskip} 
 
Molecule  & Transition & Rest frequency &  A   & g$_u$    & E$_{up}/k$  \\
          &            & (MHz)          & ($\times$ 10$^{-5}$ s$^{-1}$) &    &(K)   \\
\noalign{\smallskip}
\hline\hline
\noalign{\smallskip}

CH$_3$OH       & 12$_{0,12}$-12$_{-1,12}$  E     &   153281.207 & 1.288   &  25     &193.78  \\
\noalign{\smallskip}

CH$_3$OH       & 11$_{0,11}$-11$_{-1,11}$  E     &   154425.765 & 1.422   &  23     &166.04  \\
\noalign{\smallskip}

CH$_3$OH       & 10$_{0,10}$-10$_{-1,10}$  E     &   155320.834 & 1.549   &  21     &140.60   \\
\noalign{\smallskip}

CH$_3$OH       &  9$_{0,9}$-9$_{-1,9}$  E        &   155997.472 & 1.671    &  19     &117.45   \\
\noalign{\smallskip}

CH$_3$CHO      &  4$_{2,2}$-3$_{2,1}$  A++       &   77218.29   & 1.642   &  18     &18.32  \\
\noalign{\smallskip}

CH$_3$CHO      &  4$_{2,3}$-3$_{2,2}$  A--       &   77038.6    & 1.634  &  18     &18.31  \\
\end{tabular}

Notes: 
Einstein A coefficients, degeneracy, and 
lower state energy levels for CH$_3$CHO and CH$_3$OCHO taken from JPL data base 
(http://spec.jpl.nasa.gov), for 
all other molecules from CDMS Catalog (http://www.astro.uni-koeln.de/cdms/catalog). 

\end{table}

\begin{deluxetable}{r c c c c c c c}  
\tabletypesize{\tiny}
\tablecolumns{7} 
\tablewidth{0pc}
\tablecaption{Abundances estimated for southern IRDCs. \label{table:ab_Mopra} } 
\tiny 
\tablehead{
\colhead{Name}  & \colhead{CO\tablenotemark{a}}  &  \colhead{CO\tablenotemark{b}} & \colhead{HNCO\tablenotemark{c}} & \colhead{H$_2$CO\tablenotemark{d}}  
& \colhead{CH$_3$OH\tablenotemark{e}} & \colhead{CH$_3$CCH}  &\colhead{category\tablenotemark{g}}\\
\colhead{ } &\colhead{ } &\colhead{ } &\colhead{ } & \colhead{ }&\colhead{[T$_{EX}$]} & \colhead{[T$_{EX}$]}  &\colhead{ }}

\startdata
IRDC309.13-1 & 1.4(-04)  &      3.6(-04)  & 	  &   7.6(-10) & 7.2(-09)\tablenotemark{f} &	  & M	 \\
IRDC309.13-2 &  2.0(-04) &        n/a    & 	  &	       &  &	  & M	 \\ 
IRDC309.13-3 &  3.7(-04) &         n/a    & 	  &	       &	  &	  & A	 \\ 
IRDC309.37-1 & 2.7(-04)  &       6.8(-04)  & 3.7(-09)  &   1.7(-09) &   &	  & A	 \\ 
IRDC309.37-2 & 1.3(-04)  &          n/a    & 	  &	       & 6.1(-09)\tablenotemark{f} &	  & A	 \\ 
IRDC309.37-3 &  2.5(-04) &      6.6(-04)  & 	  &	       &	  &	  & Q	 \\ 
IRDC309.94-1 &  9.3(-05) &      1.5(-04)  & 7.3(-10)  &   1.3(-09) & 1.3(-08)\tablenotemark{f} &   & A	 \\ 
IRDC310.39-1 &  2.8(-04) &         n/a    & 3.5(-10)  &   1.2(-09) & 1.7(-08)\tablenotemark{f} &    & A, SiO  \\
\enddata


\tablecomments{a(b) = a$\times$10$^b$. Blank spaces indicate non-detections and n/a indicates absence of data}

\tablenotetext{a}{CO abundance estimated based on C$^{18}$O(2-1) APEX data. }	
\tablenotetext{b}{CO abundance estimated based on C$^{18}$O(1-0) Mopra data. }
\tablenotetext{c}{HNCO data taken from  \citet{2011A&A...527A..88V}.}
\tablenotetext{d}{H$_2$CO abundance estimated based on H$_2$CO(3$_{0,3}$-2$_{0,2}$) APEX data.} 
\tablenotetext{e}{CH$_3$OH abundance estimated based on APEX data.}
\tablenotetext{f}{Abundances calculated based on CH$_3$OH 5$_{-1,5}$ - 4$_{-1,4}$ E  transition.}
\tablenotetext{g}{ MIR categories; see Table 1.}

\end{deluxetable}

\begin{deluxetable}{r c c c c c c c c}  
\tabletypesize{\tiny}
\tablecolumns{9} 
\tablecaption{Abundances estimated for northern  IRDCs.\label{table:ab_IRAM}} 

\tiny     
\tablehead{
\colhead{Name}  & \colhead{CO\tablenotemark{a}} & \colhead{HNCO} & \colhead{H$_2$CO\tablenotemark{b}}  
& \colhead{CH$_3$OH} &  \colhead{CH$_3$CCH} & \colhead{CH$_3$CHO} & \colhead{CH$_3$OCH$_3$} &
\colhead{category\tablenotemark{f}}\\
\colhead{ } &\colhead{ } &\colhead{ } & \colhead{ }&\colhead{ [T$_{ex}$] }  &\colhead{ [T$_{ex}$]}&\colhead{  [T$_{ex}$]}  &\colhead{ }}

\startdata
IRDC010.70-2  &   0.9-2.7(-04)  & 1.2-3.2(-10)  & 6.3-25.1(-10)   & 4.8-9.8(-09)\tablenotemark{e} &    0.6-1.3(-09)      & 1.9-10.3(-09)   &                       & A,SiO   \\
IRDC010.70-4  &   1.4-3.9(-04)  & 0.9-2.4(-10)  & 4.5-16.0(-10)  &   & 4.2$\pm$1.5(-09) &                            &                       & A,SiO   \\
IRDC010.70-4  &   	                    &                        &                           &     & [42$\pm$13 K]      &                            &                       &         \\

IRDC011.11-2  &   0.7-2.6(-04)  & 0.5-1.4(-10)  &  1.4-6.3(-10) &    &      0.5-1.1(-09)      &0.4-2.5(-09)     & $<$2.4(-09) & Q,SiO  \\
IRDC011.11-4  &   0.7-2.5(-04)  & 1.3-4.0(-10)  &  4.1-20.8(-10) & 3.6-7.4(-09)\tablenotemark{e} & 4.5$\pm$0.8(-09) & 0.4-2.9(-09)      &	           & A,SiO   \\
IRDC011.11-4  &   	                    &                        &                            &     &  [28$\pm$2 K]      &                           &	       &        \\
IRDC011.11-5  &   0.5-3.0(-04)  & 0.3-1.0(-10)  &  0.3-3.0(-10) &     &		                     &	                        &	       & Q	 \\

IRDC013.90-1  &   1.0-4.0(-04)  & 0.4-1.2(-10)  & 2.8-14.4(-10)  &                                  & 1.7$\pm$0.3(-09) &	                        & $<$1.4(-09) & Q/M	    \\
\enddata


\tablecomments{a(b) = a$\times$10$^b$.  Blank spaces indicate non-detections and n/a indicates absence of data}
	
\tablenotetext{a}{CO abundance estimated based on C$^{18}$O(2-1) APEX data. }
\tablenotetext{b}{H$_2$CO abundance estimated based on H$_2$CO(3$_{0,3}$-2$_{0,2}$) APEX data.} 
\tablenotetext{c}{CH$_3$OH abundance estimated based on \citet{2012ApJS..202....1H} data and APEX data from the current survey.}
\tablenotetext{d}{CH$_3$OH abundance estimated based on APEX and IRAM data.}
\tablenotetext{e}{Abundances calculated based on CH$_3$OH 4$_{2,2}$ - 3$_{1,2}$ E  transition.}
\tablenotetext{f}{ MIR categories;  See Table 1.}

\end{deluxetable}


\begin{deluxetable}{r c c c c c}  
\tabletypesize{\tiny}
\tablecolumns{6} 
\tablewidth{0pc}
\tablecaption{Line parameters for southern  IRDCs.
\label{table:gauss_Mopra}} 
\tiny     
\tablehead{
\colhead{Name}  & \colhead{$\int \! T_{A} \, dv $} & \colhead{V$_{lsr}$} & \colhead{$\Delta$ V}  
& \colhead{T$_A$} &  \colhead{1$\sigma$ rms} \\
\colhead{ } &\colhead{(K km/s)} &\colhead{ km/s} & \colhead{ km/s}&\colhead{ K } &\colhead{ K } }

 \startdata
\multicolumn{6}{c}{CH$_3$OH 5$_{-1,5}$-4$_{0,4} E$} \\
IRDC309.13-1  &   n/a	     	   &	   	    &	            &	     &              \\ 
IRDC309.13-2  &   n/a	     	   &	   	    &	            &	     &              \\ 
IRDC309.13-3  &   n/a	     	   &	   	    &	            &	     &              \\ 
IRDC309.37-1  &   n/a	     	   &	   	    &	            &	     &              \\ 
IRDC309.37-2  &   n/a	     	   &	   	    &	            &	     &              \\ 
IRDC309.37-3  &   n/a	     	   &	   	    &	            &	     &              \\ 
IRDC309.94-1  &    1.7      (0.04) &  -60.3 (0.03)  &   3.3 (0.09)&  0.49 & 2.8 (-02) \\
\enddata
\tablecomments{ Blank spaces indicate non-detections and n/a indicates absence of data}

\end{deluxetable}


\begin{deluxetable}{r c c c c c}  
\tabletypesize{\tiny}
\tablecolumns{6} 
\tablewidth{0pc}
\tablecaption{Line parameters for northern  IRDCs.
\label{table:gauss_IRAM}} 
\tiny     
\tablehead{
\colhead{Name}  & \colhead{$\int \! T_{A} \, dv $} & \colhead{V$_{lsr}$} & \colhead{$\Delta$ V}  
& \colhead{T$_A$} &  \colhead{1$\sigma$ rms} \\
\colhead{ } &\colhead{(K km/s)} &\colhead{ km/s} & \colhead{ km/s}&\colhead{ K } &\colhead{ K } }
  
\startdata
\multicolumn{6}{c}{CH$_3$OH 5$_{0,5}$-4$_{1,3}$ E} \\
IRDC010.70-2  & 	       &	           &		     &	          &  2.8 (-02)  \\
IRDC010.70-4  & 	       &	           &		     &	          &  2.9 (-02)  \\ 
IRDC011.11-2  & 	       &	           &		     &	          &  1.4 (-02)  \\ 
IRDC011.11-4  &  0.2  (0.07)   &   29.8  (0.70)    &	5.3 (1.96)   &  4.73(-02) &  1.7 (-02)  \\
IRDC011.11-5  & 	       &	           &		     &	          &  1.3 (-02)  \\
IRDC013.90-1  & 	       &	           &		     &	          &  1.4 (-02)  \\
IRDC015.05-2  & 	       &	           &		     &	          &  1.4 (-02)  \\
\enddata
\tablecomments{Blank spaces indicate non-detections and n/a indicates absence of data}
\end{deluxetable}

\begin{table}
\caption{Observational and modeled abundances for  IRDC028.34-6 with respect to H$_2$. }
\label{table:ab_IRDC028.34-6} 
\tiny 
\begin{tabular}{r  c c c c c c c } 
\noalign{\smallskip} 
\hline\hline
\noalign{\smallskip} 
 
 Species & Observations  &   \multicolumn{6}{c}{Model (time=3$\times$10$^5$ yr$^a$)}  \\
 
 \noalign{\smallskip} 
\hline
\noalign{\smallskip} 
 
 &   & T=30 K$^b$ & T=30 K$^b$ &       T=30 K &  T=30 K & {\bf  T=40 }K &  T=30 K \\
 &   & n=10$^5$ cm$^{-3}$  &n=10$^5$ cm$^{-3}$ &     {\bf  n=10$^6$ cm$^{-3}$ } & n=10$^5$ cm$^{-3}$  & n=10$^5$ cm$^{-3}$  & n=10$^5$ cm$^{-3}$ \\
 &   & t$_{cold}$=10$^6$ yr  &t$_{cold}$=10$^6$ yr  &      t$_{cold}$=10$^6$ yr  & {\bf t$_{cold}$=10$^5$ yr } & t$_{cold}$=10$^6$ yr  & t$_{cold}$=10$^6$ yr  \\
 &   & T$\propto$time$^2$ & T$\propto$time$^2$ &      T$\propto$time$^2$ & T$\propto$time$^2$ & T$\propto$time$^2$ & {\bf T$\propto$time  }\\
 &   & 1\%$^c$  &  {\bf 10\%}  &      1\%  & 1\%  &  1\%  & 1\%   \\ 

\noalign{\smallskip}
\hline\hline
\noalign{\smallskip}
 
CO      & 2.1(-04)     &   4.4(-05)  &  5.0(-05)   & 5.8(-05)   & 7.6(-05) &  7.0(-05)  & 4.8(-05)  \\
\noalign{\smallskip}

H$_2$CO      &  3.1(-09)    &  2.2(-08)   &    {\bf 7.7(-08)}      &  2.8(-09)  & 6.0(-09) &  {\bf 6.6(-08)}  &   1.4(-08)  \\
\hline

\end{tabular}

Note: a(b) = a$\times$10$^b$. Boldface indicate more than one order of magnitude disagreement between model and observations.

$^a$  Model result at moment of the best fit for the standard model .

$^b$   Maximum temperature of warm-up phase.

$^c$  Reactive desorption efficiency.

$^d$   Data taken from \citet{2012ApJ...756...60S}.

$^e$   Data taken from \citet{2012A&A...544A.146W}.

\end{table}


\bibliographystyle{apj}
\bibliography{apj-jour,methanol8eh}

\clearpage





\clearpage


\end{document}